\documentclass{emulateapj}
\usepackage{graphicx,natbib,psfig,textcomp,fontenc,lscape,color}
 \usepackage{enumerate}

\newcommand{\ltsima} {$\; \buildrel < \over \sim \;$}  
\newcommand{\gtsima} {$\; \buildrel > \over \sim \;$}  
\newcommand{\lta} {\lower.5ex\hbox{\ltsima}}  
\newcommand{\gta} {\lower.5ex\hbox{\gtsima}}

\newcommand{\ergs}{\>{\rm erg}\,{\rm s}^{-1}}

\newcommand{\mum}{$\mu$m}
\shorttitle{Study of SEDs of radio galaxies at z$\sim$1-3}
\shortauthors{Baldi R.~D. et al.}
\begin{document}


\title{Spectral Energy Distributions of low-luminosity radio galaxies at \scriptsize{z}\normalsize$\sim1-3$: \\
  a  high-\scriptsize{z}\normalsize view of the host/AGN connection.}

\author{Ranieri D. Baldi\altaffilmark{1,2}, Marco Chiaberge\altaffilmark{2,3,4},
  Alessandro Capetti\altaffilmark{5}, Javier Rodriguez-Zaurin\altaffilmark{6,2},
  Susana Deustua\altaffilmark{2},William B. Sparks\altaffilmark{2}}


\altaffiltext{1}{SISSA, via Bonomea 265, 34136 Trieste, Italy}
\email{rbaldi@sissa.it} 
\altaffiltext{2}{Space Telescope Science Institute,
  3700 San Martin Drive, Baltimore, MD 21218, USA}
\altaffiltext{3}{INAF-Istituto di Radio Astronomia, via P. Gobetti 101,
  I-40129 Bologna, Italy} 
\altaffiltext{4}{Center for Astrophysical Sciences, Johns Hopkins University, 3400 N. Charles Street Baltimore, MD 21218}
\altaffiltext{5}{INAF-Osservatorio Astronomico di
  Torino, Strada Osservatorio 20, 10025 Pino Torinese, Italy}
\altaffiltext{6}{Insituto de Astrofisica de Canarias, Via Lactea s/n, La Laguna 38200, Spain; Departamento de Astrofisica, Universidad de La Laguna, La Laguna 38206, Spain}

\begin{abstract}

  We study the Spectral Energy Distributions, SEDs, (from FUV to MIR bands) of
  the first sizeable sample of 34 low-luminosity radio galaxies at high
  redshifts, selected in the COSMOS field.  To model the SEDs we use two
  different template-fitting techniques: i) the {\it Hyperz} code that only
  considers single stellar templates and ii) our own developed technique {\it
    2SPD} that also includes the contribution from a young stellar population
  and dust emission. The resulting photometric redshifts range from z$\sim$0.7
  to 3 and are in substantial agreement with measurements from earlier work,
  but significantly more accurate. The SED of most objects is consistent with
  a dominant contribution from an old stellar population with an age $\sim 1 -
  3 \times$ 10$^{9}$ years.  The inferred total stellar mass range is
  $\sim$10$^{10} - 10^{12}$ M$_{\odot}$.  Dust emission is needed to account
  for the 24 $\mu m$ emission in 15 objects. Estimates of the dust luminosity
  yield values in the range $L_{\rm dust} \sim10^{43.5} -10^{45.5}$ erg
  s$^{-1}$. The global dust temperature, crudely estimated for the sources
  with a MIR excess, is $\sim$ 300-850 K. A UV excess is often observed with a
  luminosity in the range $\sim 10^{42}-10^{44}$ erg s$^{-1}$ at 2000 \AA\,
  rest frame.

  Our results show that the hosts of these high-z low-luminosity
  radio sources are old massive galaxies, similarly to the local
  FRIs. However, the UV and MIR excesses indicate the possible significant
  contribution from star formation and/or nuclear activity in such bands, not
  seen in low-z FRIs. Our sources display a wide variety of properties: from
  possible quasars at the highest luminosities, to low-luminosity old
  galaxies.

\end{abstract} 


\section{Introduction}

The search for radio-loud Active Galactic Nuclei (AGN) is one of the widely
used tools to study the distant universe (z $>$ 1). Indeed the first objects
found at z $>$ 1 were radio-loud quasars (see \citealt{stern99} and references
therein) and a large number of radio galaxies ($\sim$300) are known to exist
at high redshifts (e.g., \citealt{miley08}).

Flux-limited samples of radio galaxies such as the 3CR and its deeper
successors 6C and 7C catalogs are affected by a tight redshift-luminosity
correlation. This, alongside with the steep luminosity function of radio
sources, gives rise to a selection bias which results in the presence of high
power sources predominantly at high redshifts only and low power sources
exclusively at low redshift. Therefore, our knowledge of radio galaxies and
their hosts at high redshifts is exclusively based on studies of powerful
(edge-brightened, \citealt{fanaroff74}) FR~IIs. Thus, the properties of the
low-luminosity AGN population in the early Universe is essentially
unknown. The missing pieces of the puzzle might be obtained by studying a
sample of low luminosity ('edge-darkened', \citealt{fanaroff74}) FR~Is at high
redshifts. The lower power of these objects with respect to that of high-z
FR~IIs might allow us to analyze with more clarity the properties of the hosts
at high redshifts. Furthermore, in terms of host galaxy properties, low power
sources are both more abundant and most likely more similar to quiescent
galaxies (in terms of host galaxy properties) than FR~IIs.

Only a few FR~Is at high redshift are present in the 7C sample
\citep{heywood07} and two were possibly found in HDF North
\citep{snellen01}. \citet{chiaberge09} obtained the first sizeable sample of
37 FR~I candidates at z $\gtrsim 1$, located in the 2 deg$^{2}$ area of the
sky observed by the COSMOS survey \citep{scoville07}. In this paper we perform
a detailed analysis of the photometric properties of the host galaxies of
these radio sources, by taking advantage of the large multi-wavelength
coverage provided by the COSMOS collaboration. This allows us to derive their
spectral energy distributions (SEDs) from $\sim$ 0.15 \mum\ to 24 \mum. We
then model the obtained SEDs with synthetic stellar population templates with
the aim of both inferring the properties of each galaxy and of deriving an
accurate estimate of the photometric redshifts.

\citet{mobasher07} and \citet{ilbert09} performed a similar study with the aim
of determining the photo-z for the whole COSMOS sample.  However, as already
noted by \citet{chiaberge09}, the faint counterparts of these radiogalaxies
might be easily misidentified or be missed from the COSMOS catalog. Therefore
a careful object-by-object study of the sample is necessary to correctly
identify the genuine counterparts of the radio sources at all
wavelengths. This allows us to obtain reliable SEDs and perform a robust study
of the hosts of these objects.  In this work, using the multi-band data
provided by the COSMOS survey (described in Section~\ref{data}), we obtain the
photometric data from the catalog checking visually the correct identification
of each object of the sample (Section~\ref{identification}). In
Section~\ref{fittingsed} we describe the codes used to model the SEDs, {\it
  Hyperz} and {\it 2SPD}.  We present the results obtained from the SED
modeling in Section~\ref{results}: the photometric redshifts, the radio power
distribution, and the host properties, such as stellar ages, masses and dust
and UV components. In Section~\ref{discussion} we summarize the results and we
discuss our preliminary findings.

We adopt a Hubble constant of H$_{0}$ = 71 km s$^{-1}$ Mpc$^{-1}$,
$\Omega_{m}$ = 0.27 and $\Omega_{vac}$ = 0.73, as given by the {\it WMAP}
cosmology (e.g., \citealt{spergel03,jarosik11}). All the magnitudes are in AB
mag system, if not otherwise specified.

\section{Sample}
\label{sample}

\citet{chiaberge09} selected a sample of 37 high-z low-power radio galaxies at
z $\sim$ 1-2 in the Cosmic Evolution Survey (COSMOS) field
\citep{scoville07}. This is the first sizeable sample of low-power radio
sources at mid- to high-redshifts. This sample has been obtained with a
four-steps multi-wavelength selection process, which is described in detail in
\citet{chiaberge09}.  Here we briefly summarize the selection procedure which
depends on the following assumptions: i) the FR~I/FR~II break luminosity at
1.4 GHz does not change with redshift, and ii) the host properties of distant
low-luminosity sources are similar to those of high-power FR~IIs in the same
redshift range.

The first step consists of selecting radio sources from the FIRST catalog
\citep{FIRST} whose 1.4 GHz total flux corresponds to the luminosities
expected by FR~Is at 1$<z<$2. The second step is based on a radio
morphological classification: the radio sources which show clear
``edge-brightened'' structures are rejected in order to exclude the bona-fide
FR~IIs from the sample. The third step implies the optical identification of
the radio sources in the COSMOS optical images. The objects associated with
host galaxies brighter than $i$ $<$ 22 were rejected to exclude lower
redshifts starburst galaxies.  As a final step, u-band dropouts are also
rejected since these are objects most likely located at redshift higher than
z=2. The resulting sample consists of 37 FR~I candidates. {\it A posteriori},
the photometric redshifts range of most of them is between $\sim$1 and $\sim$2
\citep{mobasher07,ilbert09}, with the exception of 3 objects (namely, 7, 27,
and 66\footnote{The 'naming convention' of the sources used in the paper is
  based on the listing number in the COSMOS catalog.}) that we exclude from
any further analysis because out of the redshift range of interest.  We are
then left with a final sample of 34 objects. The list of the radio
  sources is presented in Table~\ref{1table}.

\begin{table*}
\begin{center}
\caption{Radio positions, redshifts and radio flux of the sample}
\begin{tabular}{c|cc|cc|cc}
\tableline\tableline
ID & RA  & DEC  &  z$_{phot,Ilbert}$ & z$_{spec}$ & F$_{NVSS}$ & F$_{FIRST}$  \\
\tableline
  1  &150.20744 & 2.2818749 &  0.92$^{+0.02}_{-0.06}$ & 0.8827$^{a}$-0.8823$^{b}$   &$<$2.5  &  1.79   \\    
  2  &150.46751 & 2.7598829 &                                  &                                              &   2.6    &  1.08   \\     
  3  &150.00253 & 2.2586310 &  1.96$^{+0.36}_{-0.41}$ &                                             &   5.2    &  4.21    \\     
  4  &149.99153 & 2.3027799 &  1.45$^{+0.08}_{-0.15}$ &                                              &   7.5    &  5.99    \\     
  5  &150.10612 & 2.0144780 &  1.84$^{+0.22}_{-0.12}$ &                                              &   3.4    &  1.30    \\     
 11  &150.07816 & 1.8985500 &  1.31$^{+0.60}_{-0.24}$ &                                             & $<$2.5  &  1.13    \\   
 13  &149.97784 & 2.5042069 &  1.09$^{+0.06}_{-0.06}$ &                                             &   2.4    &  1.51    \\     
 16  &150.53772 & 2.2673550 &  0.95$^{+0.05}_{-0.02}$ & 0.9687$^{b}$                       &   4.4    &  5.70    \\    
 18  &149.69325 & 2.2674670 &  0.92$^{+0.01}_{-0.09}$ &                                            &   5.1    &  4.39    \\    
 20  &149.83209 & 2.5695460 &  1.00$^{+0.02}_{-0.05}$ &                                            &$<$2.5     &  1.33    \\    
 22  &149.89508 & 2.6292144 &                                   &                                            & $<$2.5     &  2.74    \\   
 25  &150.45673 & 2.5597000 &  1.12$^{+0.29}_{-0.03}$ &  0.7917$^{b}$$^{*}$             &   2.7    &  2.18    \\     
 26  &149.62114 & 2.0919881 &  1.20$^{+0.06}_{-0.02}$ &                                            &   3.2    &  1.88    \\     
 28  &149.60064 & 2.0918673 &                                   &                                            &   2.4    &  1.77    \\     
 29  &149.64587 & 1.9529760 &  1.59$^{+0.45}_{-0.30}$ &                                             &   2.3    &  2.12   \\     
 30  &149.61542 & 1.9910541 &  0.90$^{+0.31}_{-0.03}$ &                                             &   2.4    &  1.26   \\     
 31  &149.61916 & 1.9163600 &  0.91$^{+0.02}_{-0.05}$ & 0.9132$^{a}$-0.9123$^{b}$  &   4.1    &  3.71    \\     
 32  &149.66830 & 1.8379777 &                                   &                                             &   3.1    &  1.31    \\     
 34  &150.56023 & 2.5861051 &  1.42$^{+0.65}_{-0.32}$ &                                              &   4.5    &  5.25    \\     
 36  &150.55662 & 1.7913361 &                                  &                                               &   3.3    &  3.19    \\     
 37  &150.74336 & 2.1705379 &  1.27$^{+0.09}_{-0.02}$ &                                              &  2.2     &  1.87    \\     
 38  &150.53645 & 2.6842549 &  1.12$^{+0.10}_{-0.05}$ &                                               &  11.6    & 10.01    \\     
 39  &149.95804 & 2.8288901 &  1.08$^{+0.03}_{-0.03}$ &                                               & $<$2.5     &  1.37    \\   
 52  &149.90590 & 2.3964710 &  0.74$^{+0.02}_{-0.03}$ & 0.7417$^{b}$                           &  $<$2.5    &  1.54    \\  
 70  &150.61987 & 2.2894360 &  2.21$^{+0.57}_{-0.37}$ &                                                &  4.5     &  3.90    \\     
202  &149.99506 & 1.6324950 & 1.19$^{+0.24}_{-0.14}$ &                                                &  3.3     &  1.08    \\     
219  &150.06444 & 2.8754051 & 1.04$^{+0.01}_{-0.09}$ &                                                & $<$2.5     &  1.85    \\   
224  &150.28999 & 1.5408180 &  1.10$^{+0.07}_{-0.03}$ &                                                &  3.2     &  3.31    \\     
226  &150.43864 & 1.5934480 &  1.76$^{+0.61}_{-0.14}$ &                                                & $<$2.5     &  1.19    \\   
228  &149.49455 & 2.5052481 &  1.89$^{+0.74}_{-0.59}$ &                                                &  3.7     &  2.04    \\     
234  &150.78925 & 2.4539680 &  1.10$^{+0.12}_{-0.03}$ &                                                &  5.2     &  4.43    \\     
236  &150.70554 & 2.6296339 &                                   &  2.132$^{c}$                             &  7.0     &  7.10     \\    
258  &149.55934 & 1.6310670 &   0.90$^{+0.02}_{-0.02}$ & 0.9009$^{a}$                          &  3.7     &  2.24   \\     
285  &150.72131 & 1.5823840 &   1.21$^{+0.06}_{-0.08}$ &                                                &   3.5     &  2.95   \\     
\tableline
\end{tabular}
\label{1table}
\tablecomments{Column description: (1) ID number of the object; (2)-(3) right
  ascension and declination of radio source; (4) photometric redshifts found by \citet{ilbert09}; (5) spectroscopic
  redshift: $a$ from zCOSMOS survey \citep{lilly07}, $b$ form Magellan survey
  \citep{trump07}, and $c$ from \citet{prescott06}. The spectroscopic redshift
  of object 25 (marked with $*$) is considered incorrect (see Section~\ref{photored}); (6)-(7) flux at 1.4 GHz from NVSS (from
  http://www.cv.nrao.edu/nvss/NVSSlist.shtml) and FIRST (taken from
  \citealt{chiaberge09}) catalogs in mJy.}
\end{center}
\end{table*}

\section{Data}
\label{data}

\subsection{UV, optical, and IR photometric data} 

The photometric data used to build the SEDs of our sources are taken from the
COSMOS survey \citep{scoville07}. The survey comprises ground based as
well as imaging and spectroscopic observations from radio to
X-rays wavelengths, covering a 2 deg$^{2}$ area. Given the high sensitivity and
resolution of these data, COSMOS provides samples of high redshift objects
with greatly reduced cosmic variance as compared to earlier surveys.

Ground-based UV, optical, and IR observations and data reduction are presented
in \citet{capak07}, \citet{capak08} and \citet{taniguchi07}. A multiwavelength
photometric catalog was generated using SExtractor \citep{bertin96}. The
COSMOS catalog is derived from a combination of the CFHT $i^{*}$ and Subaru
$i^{+}$ images, to which the authors refer as 'I-band images. The catalog
includes objects with total ("mag-auto") I $<$ 25 and searches for
counterparts in a radius of 1\arcsec around the I-band detection. At fainter
magnitudes the catalog begins to be incomplete and have more spurious
detections, and photometric redshifts are poorly constrained.

For this study we use the COSMOS Intermediate and Broad Band Photometry
Catalog 2008
\citep{capak08}\footnote{http://irsa.ipac.caltech.edu/cgi-bin/Gator/nph-dd}
which provides the multiwavelength magnitudes of our sources from FUV to K
bands. Narrow band filters are not considered due to the possible strong
contamination of emission lines from the AGN and their low signal-to-noise
ratio.

HST \citep{koekemoer07} and Spitzer data (both IRAC and MIPS,
\citealt{sanders07}) are also included in the COSMOS survey. The later is
presented in two separate catalogs. S-COSMOS IRAC 4-channel Photometry Catalog
June 2007 is for IRAC data. S-COSMOS MIPS 24 Photometry Catalog October 2008
and S-COSMOS MIPS 24 $\mu$m DEEP Photometry Catalog June 2007 are for MIPS data
at 24 $\mu$m with two different flux limits, 0.15 and 0.08 mJy, respectively.

For the sake of clarity, we summarize here the available datasets from the
COSMOS survey used in this work (Table~\ref{filters}):

\begin{enumerate}
\item {\bf GALEX}: UV data were taken using GALEX \citep{martin05}. The
  observations, performed as part of the GALEX Deep Imaging Survey, are in the
  NUV and FUV bands, respectively with an angular resolution of 5\farcs6 (NUV)
  and 4\farcs2 (FUV) \citep{morrissey07}.

\item {\bf HST}: the COSMOS HST data are single-orbit F814W ACS images. They
  have the highest angular resolution ($\sim0\farcs09$, \citealt{koekemoer07})
  among the COSMOS images.

\item {\bf SUBARU}: The Suprime-Cam instrument on the Subaru telescope
  observed the COSMOS field in six broad bands ($B_{J}$, $g^{+}$, $V_{J}$, $r^{+}$, $i^{+}$,
  $z^{+}$) with an angular resolution of $\sim$0\farcs2 \citep{taniguchi07}.

\item {\bf CFHT}: the Canada-France-Hawaii Telescope (CFHT) provides
$u^{*}$ and $i^{*}$ images using Megacam \citep{boulade03} and
$K$ images with Wircam. 

\item {\bf UKIRT}:  near infrared Wide Field camera (WFCAM) on the  United
    Kingdom Infrared Telescope (UKIRT) provides the $J$-band images
  \citep{casali07}.

\item {\bf NOAO}: $K_{S}$ data are taken at the Kitt Peak National Observatory
  (KPNO) telescope with FLAMINGOS and The Cerro Tololo Inter-American
  Observatory (CTIO) telescope \citep{capak07}. These telescopes belong
  to the National Optical Astronomy Observatory (NOAO).

\item {\bf Spitzer}: Spitzer cycle-2 S-COSMOS is an infrared imaging survey of
  the COSMOS field \citep{sanders07}. They obtained observations with the IRAC
  camera in four channels, at 3.6, 4.5, 5.6, and 8 $\mu$m, and with MIPS in
  the 24, 70, 160 $\mu$m band. We only consider the data at 24 $\mu$m,
    since no object is detected at longer wavelengths (with the exception of
    object 37 which is also detected at 70 $\mu$m).

\end{enumerate}

\begin{table}
\begin{center}
\caption{COSMOS broad bands and their properties.}
\begin{tabular}{ccccc}
\tableline\tableline
Filter & Telescope & $\lambda_{eff}$ & FWHM & sensitivity \\
\tableline
$FUV$    &  GALEX      & 1538.6\AA  &  230.8\AA  & 25.7 \\
$NUV$    &  GALEX      & 2315.7\AA  &  789.1\AA  & 26.0  \\
$u^{*}$  &  CFHT       & 3911.0\AA  &  538.0\AA  & 26.5 \\
$B_{J}$  &  Subaru     & 4439.6\AA  &  806.7\AA  & 27.0 \\
$g^{+}$  &  Subaru     & 4728.3\AA  &  1162.9\AA & 27.0  \\
$V_{J}$  &  Subaru     & 5448.9\AA  &  934.8\AA	 & 26.6 \\
$r^{+}$  &  Subaru     & 6231.8\AA  &  1348.8\AA & 26.8 \\
$i^{*}$  &  CFHT       & 7628.9\AA  &  1460.0\AA & 24.0 \\
$i^{+}$  &  Subaru     & 7629.1\AA  &  1489.4\AA & 26.2 \\
$F814W$  &  HST	       & 8037.2\AA  &  1539.0\AA & 27.2 \\
$z^{+}$  &  Subaru     & 9021.6\AA  &  9021.6\AA & 25.2 \\
$J$      &  UKIRT      & 12444.1\AA &  1558.0\AA & 23.7 \\
$K_{S}$  &  NOAO       & 21434.8\AA &  3115.0\AA & 21.6 \\
$K$      &  CFHT       & 21480.2\AA &  3250.0\AA & 23.7 \\
IRAC1    &  Spitzer    & 35262.5\AA &  7412.0\AA & 23.9 \\
IRAC2    &  Spitzer    & 44606.7\AA &  10113.0\AA& 23.3 \\
IRAC3    &  Spitzer    & 56764.4\AA &  13499.0\AA& 21.3 \\
IRAC4    &  Spitzer    & 77030.1\AA &  28397.0\AA& 21.0 \\
MIPS1    &  Spitzer    & 23.68$\mu$m&  4.7$\mu$m& 29.6 \\
\tableline
\end{tabular}
\label{filters}
\label{log}
\tablecomments{List of the filters used for our analysis, the associated
  telescopes, the effective wavelengths $\lambda_{eff}$, the PSF full-width
  half maximum (FWHM) of the images in each band, and their sensitivities
  at 5$\sigma$ (mag).}
\end{center}
\end{table}

\subsection{Radio data}

\citet{chiaberge09} selected the sample in the COSMOS field using radio fluxes
at 1.4 GHz from the FIRST survey \citep{FIRST}. The data are obtained with the
VLA in B configuration with an angular resolution of $\sim$5\arcsec and reach
a flux limit of $\sim$1 mJy and are listed in \citet{chiaberge09}.

In addition, the NVSS survey \citep{NVSS} provides 1.4 GHz radio data for our
sample, but at lower resolution ($\sim$45 \arcsec) and with a higher flux
density limit ($\sim$2.5 mJy) than the FIRST survey. These differences imply
that seven of our objects are missed in the NVSS catalog.  Nonetheless, the
NVSS data are useful since they are more sensitive to diffuse low surface
brightness radio emission that the FIRST data. The radio fluxes are taken from
the NVSS archive\footnote{http://www.cv.nrao.edu/nvss/NVSSlist.shtml}. A
NVSS/FIRST comparison indicates that the flux ratio between the two catalogs
is usually between $\sim 1$ and 2, and never larger than $\sim$
3. Table~\ref{1table} shows the NVSS and FIRST radio fluxes of the objects.

\subsection{Spectroscopy}

The COSMOS survey provides spectroscopic data from the Very Large
Telescope (VLT) (zCOSMOS, \citealt{lilly07}) and from the Magellan (Baade)
telescope \citep{trump07} 

zCOSMOS is a large-area redshift survey which consists of two parts. The
first, namely zCOSMOS-bright, considers a magnitude-limited (I-band mag
$<$22.5) sample of $\sim$20,000 galaxies located within the central 1 deg$^2$.
Spectra cover the wavelength range 5500 \AA\, $<\lambda<$ 9000 \AA. The
second, namely zCOSMOS-deep, is an ongoing survey (not yet public) of
$\sim$10,000 blue galaxies in the same field filtered with a color selection
to be in the range of 1.4 $<$ z $<$ 3.0. The spectra cover the wavelength
range of 3600 \AA\, $<\lambda<$ 6800 \AA.

The Magellan survey presents spectroscopic redshifts for the first 466 X-ray
and radio-selected AGN candidates. The wavelength coverage of these spectra is
$\sim$ 5500-9200\AA. Their redshift yield is 72\% for i$_{AB}$ $<$ 24 and
$>$90\% for i$_{AB}<$22. 

In this work we use only spectroscopic measurements with a confidence level
greater than 99\%. Six objects of the 34 objects are included in the
spectroscopic surveys described above with the required quality (Table~\ref{1table}).

\section{Multi-band counterparts identification}
\label{identification}

The process of counterparts identification of our radio galaxies in the COSMOS
catalog suffers from the limitations typical of any multiband survey (such as
misidentification of targets with close neighbor or the contamination by
nearby bright sources). We then prefer to perform our multi-band counterpart
identification on each source by visually inspecting its multiband images,
rather than blindly use the data provided by the COSMOS catalog.

We start the process looking for a I-band counterpart to the radio source in
the COSMOS catalog by adopting 0\farcs3 as search radius. 29 identifications
are found in our sample, most at distances smaller than 0\farcs1. Three
objects (22, 28, and 32) are clearly visible in the I-band images, but they
are not found in the COSMOS catalog, since they are below its detection
threshold.  In addition, there are two objects (2 and 36) for which no I-band
counter part is found.

For the I-band detected sources, we search in the COSMOS broadband catalog
that provides photometry (from the FUV to the K band) over an aperture of
3\arcsec\ diameter. The COSMOS catalog associates its counterparts in the
remaining bands with the brightest and closest sources to its I-band position
within a radius of 1\arcsec. We extend this method including also the
Spitzer/IRAC and MIPS catalogs, by using a larger search radius (2\arcsec) due
to their coarser resolutions with respect to the COSMOS broadband images.  For
the 5 sources not present in the COSMOS catalog (because they are too faint in
the I-band) we note that in all cases a clear counterpart is instead present
at longer wavelength providing us with a robust multiband identification.

\begin{figure*}
\centerline{
\includegraphics[angle=0,scale=0.60]{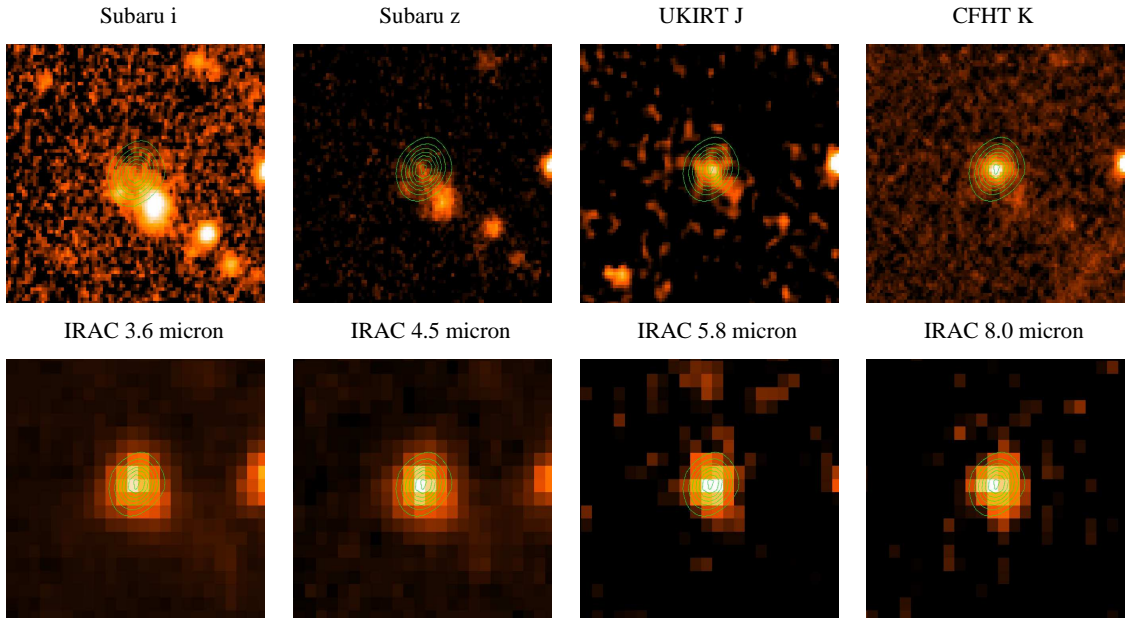}}
\caption{Images of object 28 taken from the COSMOS survey. From upper left to
  lower right panels we show the images from Subaru $i^{+}$, Subaru $z^{+}$,
  UKIRT J, CFHT K and the four channels of Spitzer/IRAC. The size of the image
  is 5\farcs5 $\times$ 5\farcs5. We overlap the radio contours (green lines)
  at 1.4 GHz from VLA COSMOS data on the frames to show the position of the radio
  source.}
\label{28}
\end{figure*}

\begin{figure*}
\centerline{
\includegraphics[angle=0,scale=0.40]{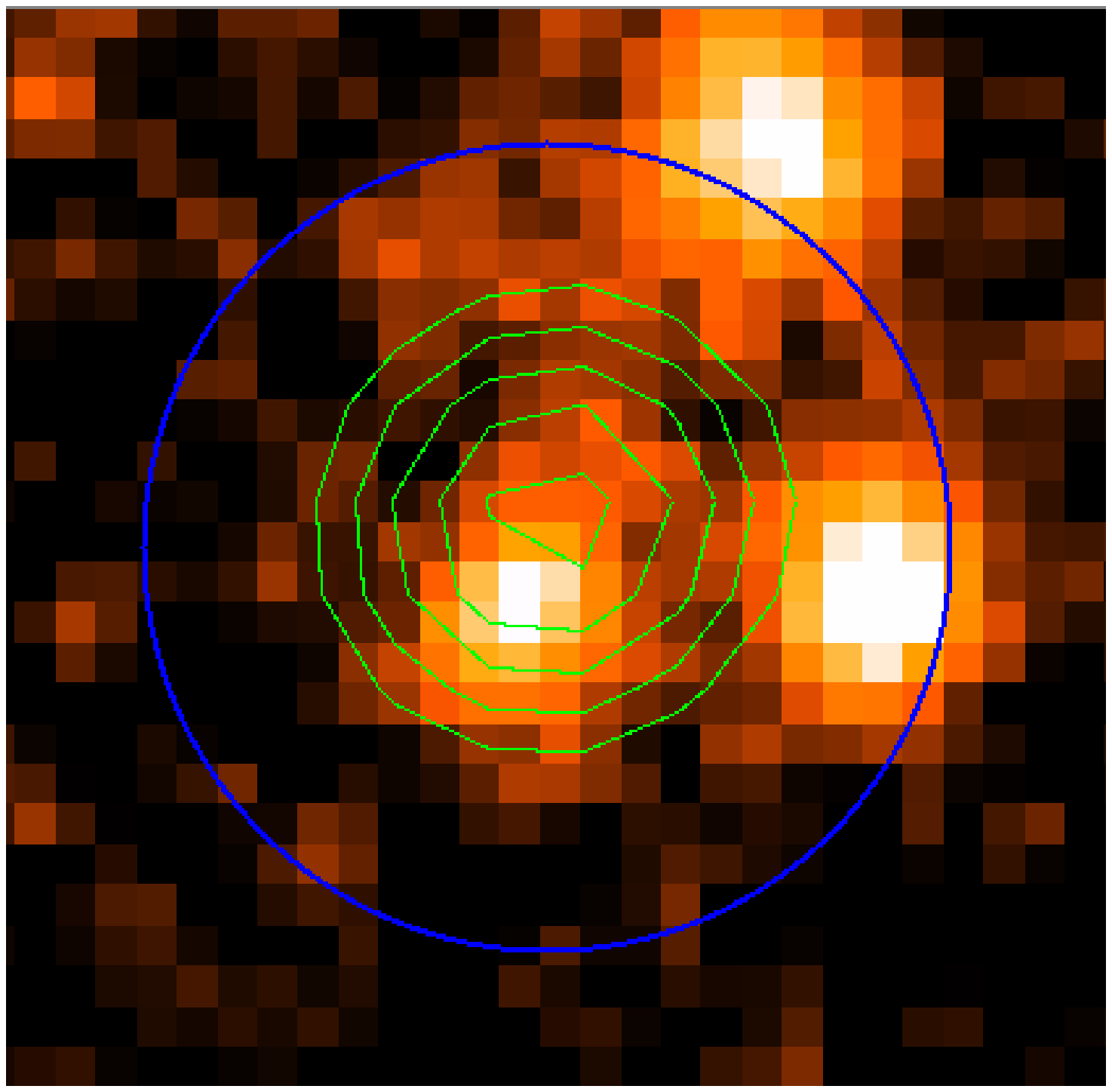}
\includegraphics[angle=0,scale=0.47]{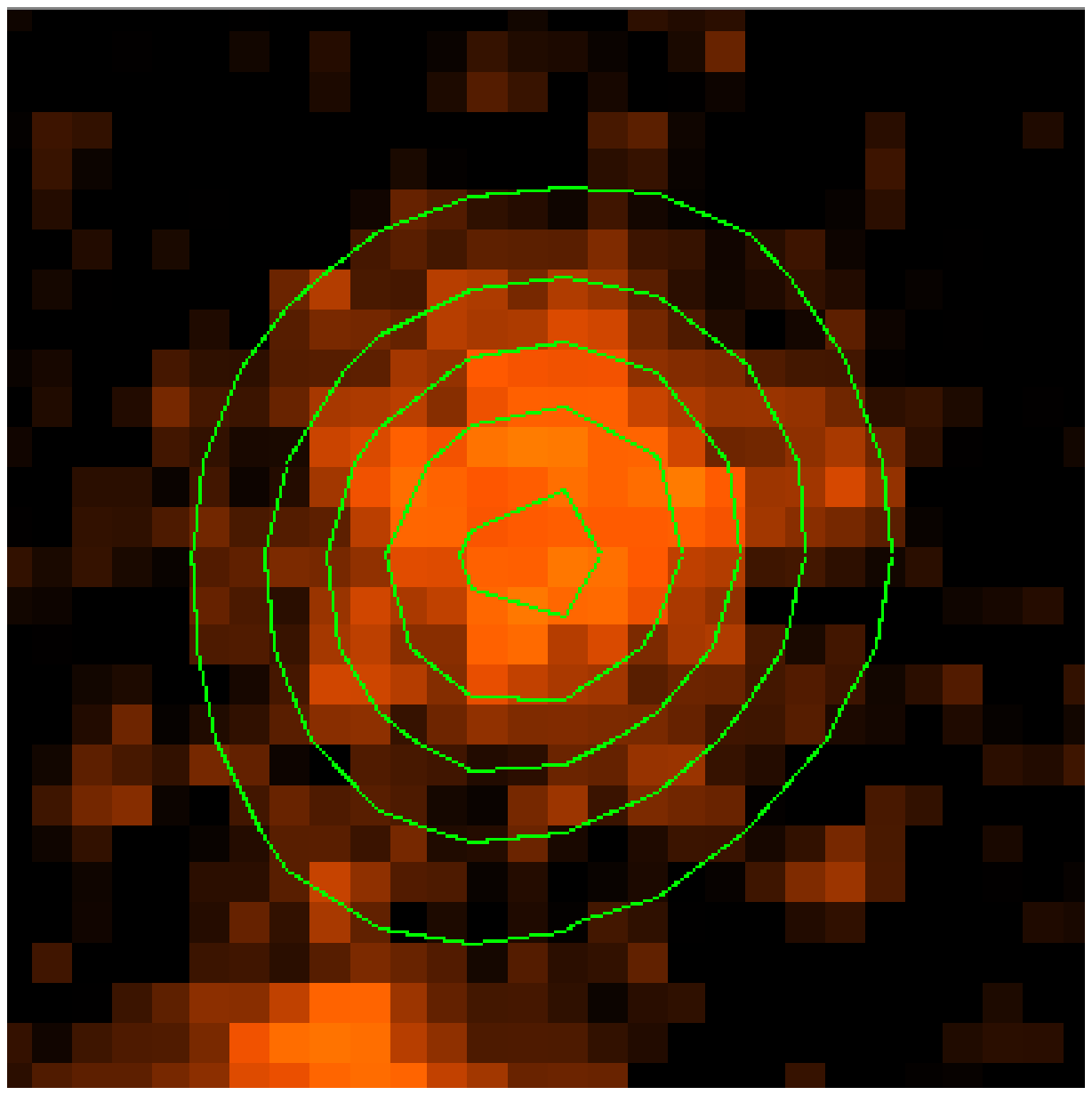}
\includegraphics[angle=0,scale=0.47]{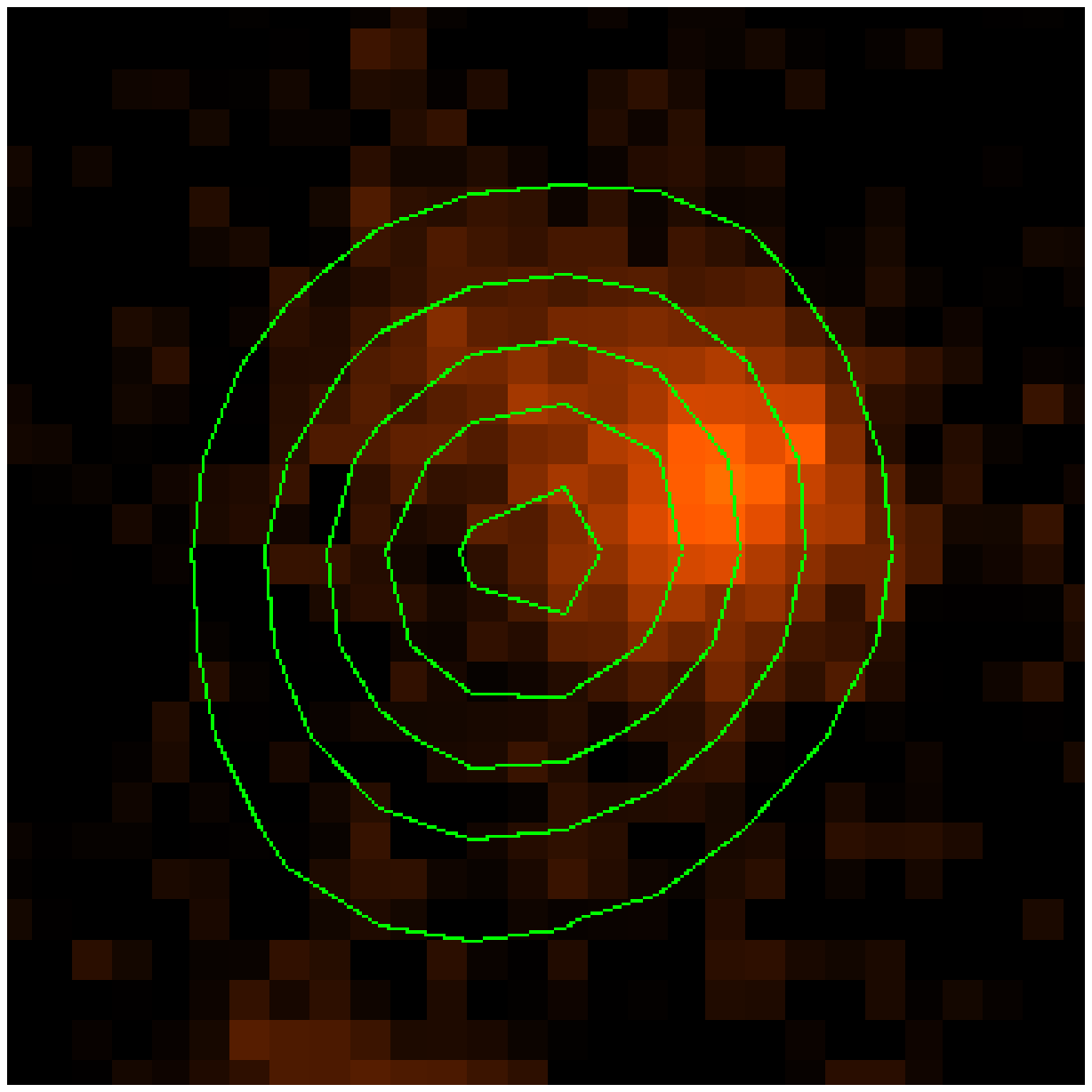}}
\caption{Left panel: Subaru $i^{+}$ image of object 34 with superposed the
  radio contours (green) and the 3\arcsec\ aperture centered on the radio
  source (blue). This is a case in which, in addition to the radio-galaxy,
  other objects lie inside the integration aperture of the target,
  contaminating the COSMOS catalog flux measurement. Middle and right panels:
  Subaru $z^{+}$ and CFHT $u^{*}$ images of object 5 with superposed the radio
  contours. This is a case of misidentification of a source with a
  neighbor included within the 1\arcsec\ from the target) in different
  bands. The size of the images is 2\arcsec $\times$ 2\arcsec. }
\label{misid}
\end{figure*}

During the visual check of the individual counterparts, two main problems
emerge in the COSMOS broadband catalog: i) the presence of nearby sources,
within the 3\arcsec\ radius used for the aperture photometry, contaminating
the broad band measurements of the genuine emission from the radio-galaxies of
our sample; ii) the counterparts to the i-band object does not always
correspond to the same object over the various bands. Object 28 is a clear
example of this situation (see Fig.~\ref{28}): from the J band to 8$\mu$m the
brightest source in the field is coincident with the radio source, while in the
$i^{+}$ and $z^{+}$ bands this is out-shined by a nearby object, causing an
erroneous identification in the COSMOS catalog across the various
bands. Furthermore, the presence of this neighbor also causes a strong (and
dominant short-ward of the J band) contamination to the genuine flux of the
radio-galaxy. Other examples of contamination and mis-identification are
reported in Fig.\ref{misid}.

In order to amend these problems, we first perform a new 3\arcsec\ aperture
photometry properly centered on the position of the radio source, to isolate
the genuine emission of the counterparts. In case of contamination from nearby
source(s), we subtract from the flux resulting from the photometry centered on
the radio source the emission from the neighbor(s), limiting to the fraction
that falls into the 3\arcsec\ aperture. When we cannot separate a source from
a close source or when the counterpart is not visible in a given band, we
measure an 1-$\sigma$ upper limit to the flux. 

At 24 $\mu$m, when the counterpart of a source is not seen at the radio
position, we use the detection limit of MIPS catalogs as upper-limit to the
source flux. However, when a nearby source contaminates the emission of our
target, we prefer to measure the upper-limit directly on the image at the
radio position. In addition, in the NUV/FUV band, if the source is not
detected in GALEX we prefer not to include upper limits in our analysis,
because its corresponding flux is substantially higher than those at larger
wavelengths.

\begin{figure*}
\centerline{
\includegraphics[angle=0,scale=0.60]{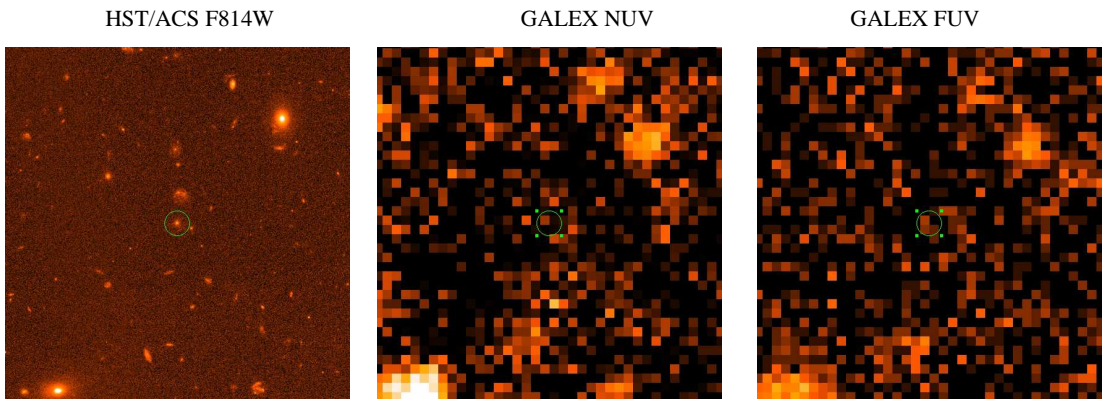}}
\caption{From left to right: HST/ACS F814W, GALEX NUV and FUV bands images of
  object 13. The radio source (whose position is marked by the circles at the
  images center) does not apparently correspond to any UV emission, contrarily
  to the detection found in the catalog. The size of the images is
  7\farcs5$\times$7\farcs5. }
\label{13}
\end{figure*}

\begin{figure}
\centerline{
\includegraphics[angle=0,scale=0.50]{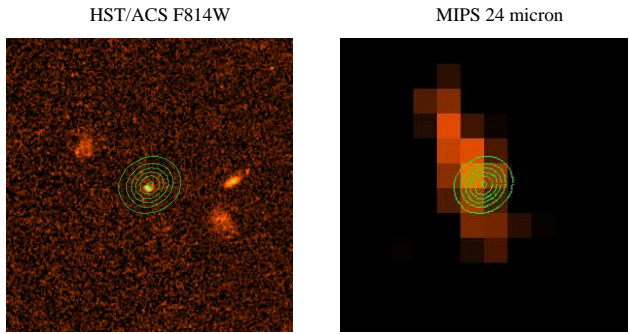}}
\caption{Images of object 70 from ACS (left) and MIPS 24$\mu$m (right). This
  provides an example of a source clearly emitting at 24$\mu$m, but not
  present in the COSMOS Spitzer/MIPS catalog. The position of the radio source
  is represented by the green radio contours from VLA COSMOS. The size of the
  images is 28\arcsec$\times$28\arcsec.}
\label{70}
\end{figure}

In some cases the GALEX photometry returns apparently incorrect results. In
Fig.~\ref{13} we show as an example the HST and GALEX images of object 13;
this source is labeled as detected by the COSMOS catalog in both the FUV and
NUV GALEX bands, while clearly there is no emission above the background
level. The opposite is seen in MIPS measurements, when clearly visible objects
are missing from the catalog (see Fig.~\ref{70}).  In these cases, we do not
consider the GALEX data, while for MIPS we provide a new estimate to the 24
$\mu$m flux.

We also find an apparent error in the reported UKIRT J-band photometric
points. We obtain our own measurements on a sample of stars present in the
COSMOS field \citep{wright10} using the PSF-matched and PSF-original J-band
images\footnote{The PSF-matched images are obtained by convolving each
  PSF-homogenized image with a Gaussian kernel that produced the same flux
  ratio between a 3\arcsec and 10\arcsec aperture for a point source to avoid
  PSF-matching problems in the multiband catalog. The PSF-original images are
  the pure images without any PSF matching. Both the images are provided by the
  COSMOS catalog.}  and we compare them to the magnitudes given by the
catalog. This test reveals that the zero-point mag of the PSF-matched J-band
images is higher than the corresponding value of the PSF-original J-band
images, taking into account of the effect of the PSF convolution. This
systematic difference of the zero-point magnitudes is $\Delta$J =
0.90$\pm$0.06, intriguingly similar to the offset between Vega and AB mag
system for the UKIRT J-band (J$_{AB}$ = J$_{\rm Vega}$ + 0.94 for UKIRT,
\citealt{hewett06}). We obviate this problem by performing 3\arcsec-aperture
photometry on the J-band counterparts on the UKIRT images smoothed with a
Gaussian kernel to a 1.2-1.5\arcsec\ FWHM to reproduce the effect of the PSF
matching.

Once we obtain the correct aperture photometry of the various multi-band
counterparts of the radio galaxies, we apply the appropriate aperture
corrections. While this is not needed for the Subaru, CFHT, and UKIRT, for
GALEX data we apply an aperture correction by multiplying the total flux by a
factor of 0.759 \citep{capak07}. Similarly, since we select from the IRAC
catalog the 2\farcs9 aperture photometry (because it is closest to the
3\arcsec\ aperture of the COSMOS broadband catalog among the different
optional apertures), the IRAC aperture flux was converted to total source flux
by using the aperture correction factors (from the IRAC manual) of 1.19, 1.27,
1.48, and 1.27 for the four channels. The MIPS data instead already include
the aperture correction. Since the CFHT telescope is more sensitive and with
higher resolution than the images from the NOAO telescopes, this usually
results in far smaller errors for the K-band magnitudes. In such cases, we
prefer to use only the CFHT K-band data. The results of the corrected
3\arcsec-aperture photometric measurements are tabulated in Tables~\ref{tab1}
and \ref{tab2}.

\section{SED fitting}
\label{fittingsed}

The SEDs are derived by collecting multiband data from the FUV to the MIR.
Since not all of the objects are detected in the entire set of available
bands, the number of detections used to constrain the SED fitting ranges from
15 to 19.

\begin{figure*}
\includegraphics[angle=90,scale=0.35]{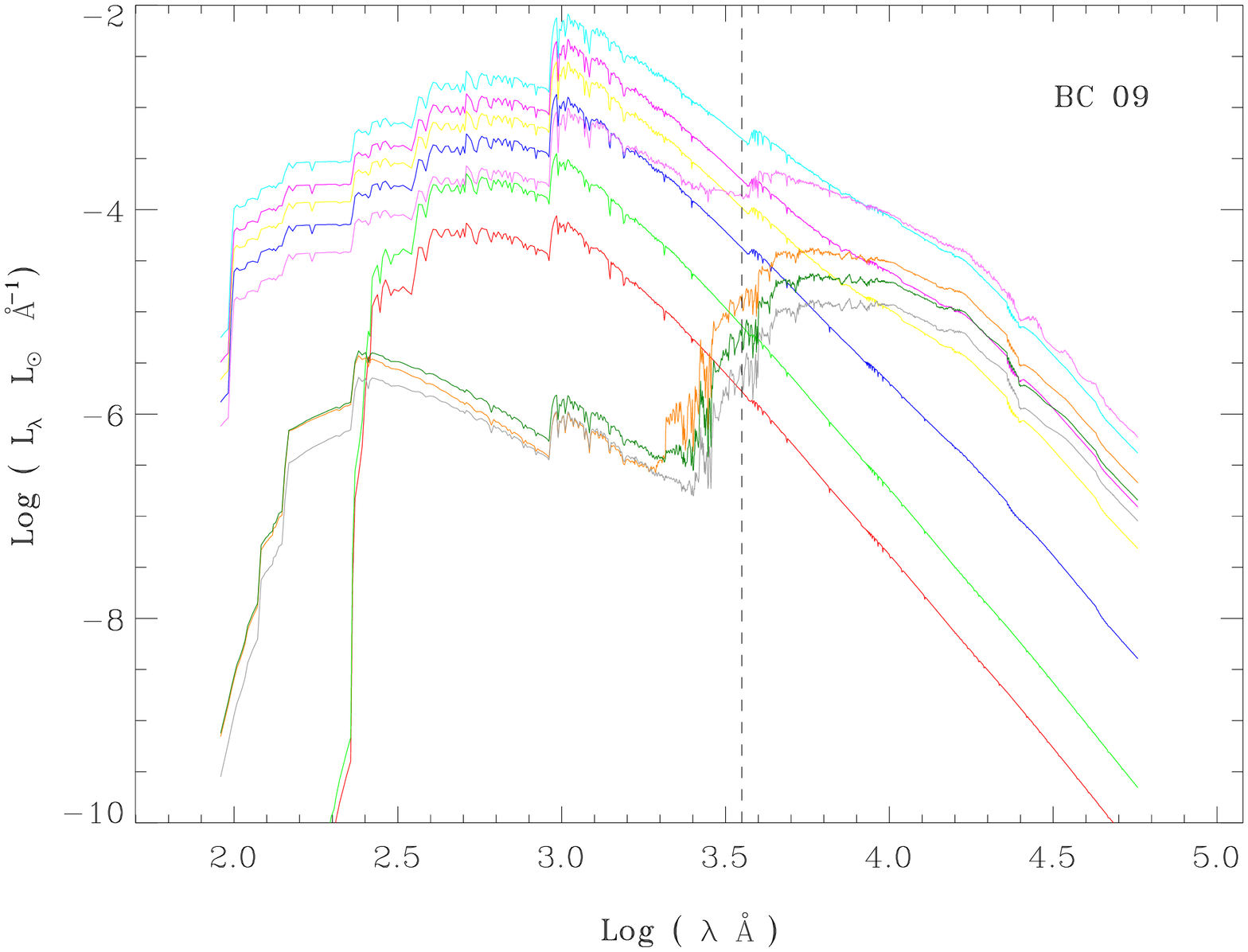}
\includegraphics[angle=90,scale=0.35]{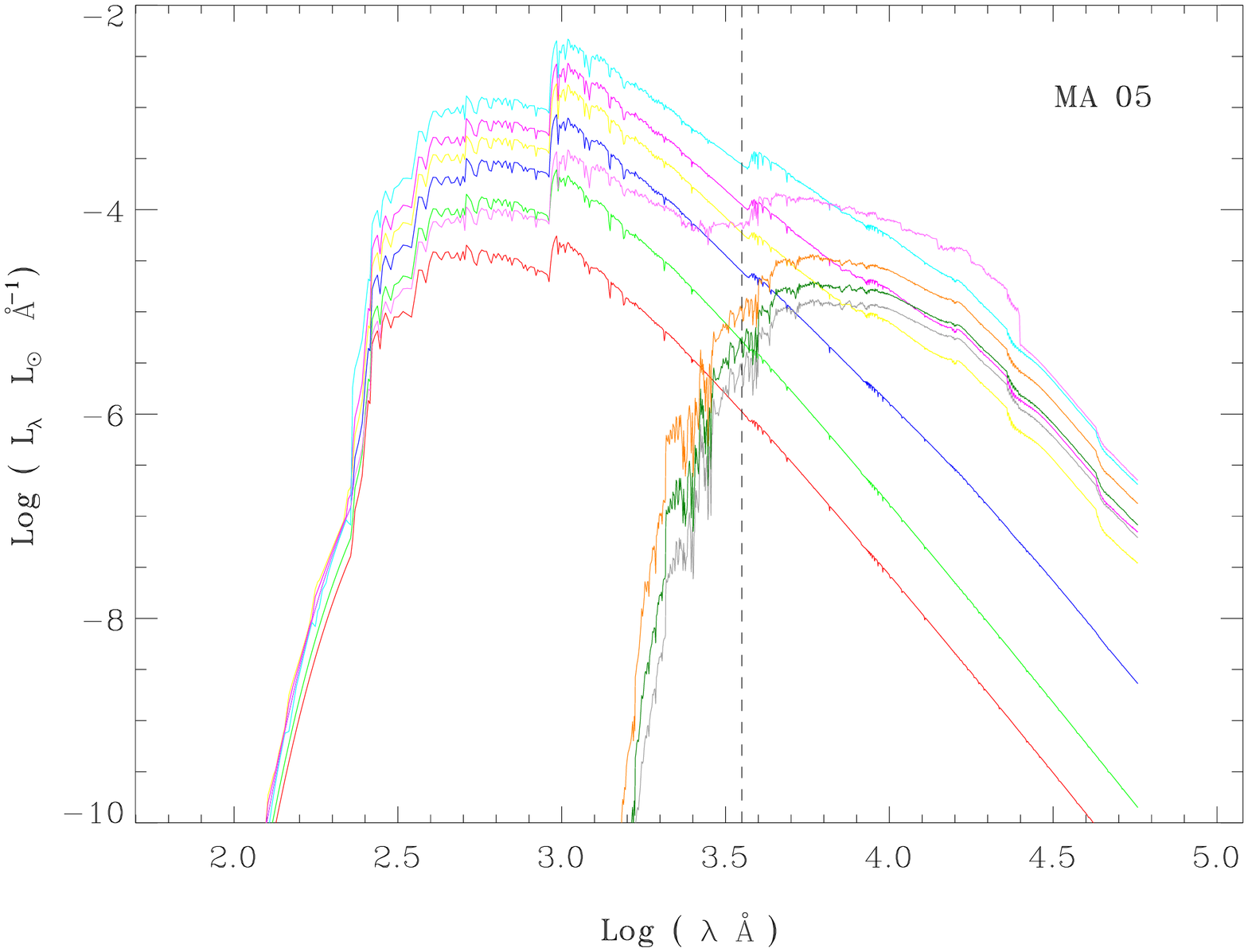}
\caption{Examples of SED templates from BC09 \citep{bc09} (left panel) and
  MA05 \citep{ma05} (right panel) in units of L$_{\sun}$ $\AA^{-1}$. The
  templates correspond to different ages and different star formation
  histories. Arranging the BC09 and MA05 templates with increasing luminosity
  at 3550 \AA\, (following the dashed line), the different plotted models
  correspond to $\mu$=15 Gyr decaying CSP of 1.1 Myr (red line); single
  stellar population of 20 Gyr (gray line); $\mu$=0.1 Gyr decaying CSP of 10
  Gyr (dark green line); $\mu$=10 Gyr decaying CSP of 2.8 Myr (light green
  line); $\mu$=0.3 Gyr decaying CSP 5.25 Gyr (orange line); $\mu$=5 Gyr
  decaying CSP of 6.9 Myr (blue line); $\mu$=3 Gyr decaying CSP of 17 Myr
  (yellow line); $\mu$=0.6 Gyr decaying CSP of 1.7 Gyr (pink line); $\mu$=2
  Gyr decaying CSP of 38 Myr (fuchsia line); and $\mu$=1 Gyr decaying CSP of
  0.2 Gyr (light blue line). We invite the reader to download the color
  version of the plot to distinguish the different lines.}
\label{templates}
\end{figure*}

The synthetic stellar templates used to model the observed SEDs are the
\citet{bc09} (priv. comm.) and \citet{ma05} templates. These templates are
defined with different Initial Mass Function (IMF)
\citep{salpeter55,kroupa01,chabrier03} with solar metallicity (Z$_{\odot}$ =
0.02). These libraries contain composite stellar population (CSP) computed
with different star formation histories: a constant star-forming system (with
constant star formation rate of 1 M$_{\odot}$/yr); a single burst of star
formation; and ten $\mu$-models\footnote{These templates correspond to
  synthetic spectra computed with exponentially decaying star formation rate,
  $\psi(t)$ $\propto$ e$^{(-t/\tau)}$ where $\tau$ is the star formation
  timescale.} with time-decays of 0.1, 0.2, 0.3, 0.6, 1.0, 2.0, 3.0, 5.0,
10.0, and 15.0 Gyr. These templates include 221 tracks of ages from 0.1 Myr to
20 Gyr and cover the wavelength range from 91 \AA\, to 160
$\mu$m. Fig.~\ref{templates} shows the stellar templates used for SED
modeling.

The \citet{bc09} models differs from those of \citet{ma05} mainly because of a
different recipe for the inclusion of thermally pulsing asymptotic giant
branch (TP-AGB) phase. These stars significantly contribute to the IR emission
for ages higher than $\sim$1 Gyr. The resulting effect is that the Bruzual \&
Charlot templates are bluer and give slightly older ages than \citet{ma05}
models.

\subsection{SED fitting method: {\it Hyperz}}
\label{hyperzsec}

We firstly estimate the photometric redshift for our sample by using the
template fitting technique, {\it Hyperz} (Version 1.3 , obtained from
M. Bolzonella in priv. comm.), described in detail by
\citet{bolzonella00}. This SED fitting procedure is based on reproducing the
overall shape of the SEDs and recognizing strong spectral properties, such as
the 4000 \AA\, Ca break or Lyman break at 912 \AA. This code consists of a
convolution of templates, which represent the rest-frame SEDs for galaxies
with different star formation histories, with the filter response functions
given by the COSMOS survey.

The basic assumption of this SED modeling is that the host galaxies of the
radio sources in our sample are dominated by stellar emission. Thus the MIPS
24 $\mu$m photometric point is excluded because this emission is obviously not
of stellar origin. Furthermore, object 236, a spectroscopically-confirmed QSO
(see Sect. \ref{5.3}), is excluded from this analysis. The code also takes
into account extinction (by using the reddening law from \citealt{calzetti00})
which is applied to the templates. The reddened templates are then shifted in
wavelength searching for the correct redshift. The best fit is obtained
through $\chi^{2}$ minimization. The grid spacings in redshift and in $A_{V}$
are 0.01 and 0.1, respectively. The {\it Hyperz} code gives the probability
P($\chi_{rid}^{2}$) associated with $\chi_{rid}^{2}$ (i.e., the reduced
$\chi^{2}$, which is defined as $\chi^{2}_{min}$/$\nu$, where $\nu$ is the
number of degrees of freedom). Because we are essentially interested in the
redshift (the other parameters are affected by the degeneracy), the degrees of
freedom are N$_{filters}$ - 1.

Similarly to \citet{ilbert09}, we increase the flux errors by 10\% when the
probability associated with the $\chi_{rid}^{2}$ is less than 0.01. This
factor does not shift the best-fit photo-z value but broadens the $\chi^{2}$
peak and derived redshift uncertainty. The program provides the redshift of
the object measured at the different confidence intervals defined by the
values of $\chi_{rid}^{2}$. We conservatively choose the solution at a 99\%
confidence level. \citet{bolzonella00} discussed the effects of degeneracy in
the parameter space defined by star formation history, age, and
reddening. However, they show that this does not strongly affect the value of
the photometric redshift. {\it Hyperzmass}, a code which works in a similar
way to {\it Hyperz}, returns the mass of the stellar population corresponding
to the best fit and we adopt again the 99\%-confidence level to quote the
errors on the mass.

Fig.~\ref{sed1a} (left panels) shows the SED fit obtained with {\it Hyperz}. The resulting
properties from the modeling are listed in Table~\ref{hyperz}. Inspection of
this figure indicates that {\it Hyperz} does not always return satisfactory
results; indeed, in several cases, it cannot reproduce the bluest part of the
spectrum (possibly due to a young stellar component) or the substantial SED
``bump'' at long wavelengths, clearly suggestive of dust emission and often
extending even to the IRAC 3.6 $\mu$m measurements.

\subsection{`2SPD' fitting technique}
\label{2SPDcode}

In order to improve the quality of the SED fitting we developed a code that
includes 2 Stellar Populations and Dust component ({\it 2SPD}). More
precisely, we take into account two different stellar populations, typically,
one younger and one older (YSP and OSP, respectively). We model the dust
component with a single (or, in a few cases, two) temperature black-body
emission. Thus, at this stage, we can include the 24 $\mu$m MIPS flux in the
fitting process.

The synthetic models used are the same as described in the previous section,
but limiting to those with single stellar population with ages ranging from 1
Myr to 12.5 Gyr. We adopt a dust-screen model for the extinction normalized
with the free parameter $A_V$, and the \citet{calzetti00} law.

The code searches for the best match between the sum of the different
components (young and old stellar templates and dust emission) and the
photometric points minimizing the appropriate $\chi^{2}$ function. {\it 2SPD}
returns the following free parameters: $z$, $A_V$, the age of the two stellar
populations, the temperature of the dust component(s), and the various
normalization factors. From these quantities we measure the stellar mass
content of the two stellar populations at 4800 \AA\ rest frame. Similarly to
the case of {\it Hyperz}, caution should be exerted before associating these
values to physical quantities because of degeneracy in the parameter
space, apart from the photometric redshifts.

To estimate the errors on the photo-z and mass derivations, we measure the
99\%- confident solutions for these quantities. This is computed by varying
the value the parameter of interest until the $\chi^{2}$ value increases by
$\Delta \chi^2$ = 6.63, corresponding to a confidence level of 99\% for that
parameter.

Fig.~\ref{sed1a} (righ panels) shows the outputs of {\it 2SPD}
code, while Table~\ref{2spd} presents the resulting parameters of the fit.

The dust emission is usually poorly constrained. In many cases no excess above
the stellar component is required at $\lambda < 24 \mu$m. In this case dust is
needed only to account for the 24 $\mu$m data-point and often this measurement
is an upper limit.  The values reported in Table~\ref{2spd} correspond to the
best fitting model for each galaxy and must be interpreted as approximate
rather than real measurements of the dust emission properties. We will return
to the dust properties in more detail in Sect. \ref{dust}.

\subsection{COSMOS photometric redshifts}
\label{5.3}

As a comparison with our results we also collect the photometric redshifts from
the COSMOS redshift catalog. \citet{mobasher07} provide photometric redshifts
for 860,000 galaxies in the COSMOS field with $i^{+}$ $<$ 25. Their technique
is based on a $\chi^{2}$ template-fitting procedure applied to the SEDs
derived with up to 16 photometric points from the $u$ to the K band. To
evaluate the reliability of the derived photometric redshifts, they compare
them with spectroscopic redshifts from a sample of 868 galaxies in zCOSMOS
with z $<$ 1.2. The rms scatter between photometric and spectroscopic
redshifts is $\sigma_{(zphot - zspec)/(1 + zspec)}$ = 0.031 with a small
fraction of outliers (2.5\%).

\citet{ilbert09} substantially improve this analysis by including data from 30
broad, intermediate, and narrow bands covering the wavelength range from UV to
MIR. Redshifts are computed for 607,617 sources (with $i^{+}<$ 26). The method
used by these authors accounts for the contribution of emission lines to the
SEDs. Comparison with 4,148 zCOSMOS spectroscopic redshifts indicates a
dispersion of $\sigma_{(zphot - zspec)/(1 + zspec)}$ = 0.007 at $i^{+}_{AB}<$
22.5. Nevertheless, the accuracy is strongly degraded at $i^{+}>$
25.5. Table~\ref{1table} shows the photometric redshifts measured by
\citet{ilbert09}.

Most of the radio galaxies of our sample are included in the COSMOS
Photometric Redshift Catalog Fall 2008, which provides the photometric
redshifts measured by \citet{ilbert09}. We adopt again a 0\farcs3 search
radius and find 29 objects. We report the results in Table~\ref{hyperz}. One
of these objects (236) turns out to be a spectroscopically-confirmed QSO at z
= 2.132 \citep{prescott06}. The attempt of \citet{ilbert09} to estimate its
photometric redshift failed because the templates used by these authors were
not suitable to fit its AGN-dominated SED. Similarly, \citet{mobasher07}
provided for this object a tentative photometric redshift, which is
z=1.23$^{+0.03}_{-0.17}$, which is clearly inconsistent with the spectroscopic
value.

\section{RESUTS}
\label{results}

The SED modeling process has been performed for all the objects (except for
the spectroscopically-confirmed QSO, object 236), by using the two
template-fitting techniques (Fig~\ref{sed1a}). As expected, {\it 2SPD} code
has turned out to more reliably model the SEDs of our sources than {\it
  Hyperz}. This is crucial in UV band.  In fact, when significant UV emission
appears in the SED, such as, in objects 31, 258 and 285, {\it Hyperz}
technique struggles to fit simultaneously such a UV component and the
remaining part of SED, which is well represented by an OSP. In this case, the
best fit result from {\it Hyperz} is obtained using an old composite stellar
population.  This is due to the fact that the $\chi^2$ is lower if only the
the old component of the SED is fitted, rather than the UV data points which
are less well represented. Conversely, as expected, {\it 2SPD} technique is
more efficient to treat this case, since it just adds a small fraction of YSP
to a dominant OSP to reproduce the UV emission, by departing from the concept
of single star formation history.

\begin{table*}
\begin{center}
\caption{{\it Hyperz} SED fitting}
\begin{tabular}{c|cc|cc|ccccc|c}
\tableline
ID   &\multicolumn{2}{c|}{redshift}                  &  \multicolumn{5}{c}{CSP Template}                 & Log M$_{*}$ \\
\tableline
    & z$_{phot,Hyperz}$            &   z$_{phot,2SPD}$       &  type & SFH        & Age    & A$_V$ &$\chi_{rid}^2$&  Log M$_{\odot} $               \\ 
\tableline                                               
1   &  0.85$^{+0.05}_{-0.08}$&0.88$^{+0.04}_{-0.05}$       &   BC  & ssp & 1.434 & 1.00   & 0.73         & 10.98$^{+0.07}_{-0.04}$ \\ 
2   &  1.31$^{+0.14}_{-0.07}$&1.33$^{+0.10}_{-0.09}$       &   BC  & $\tau$=0.3 & 0.5088  & 2.9   & 1.43         & 11.00$^{+0.09}_{-0.12}$ \\ 
3   &  2.33$^{+0.18}_{-0.18}$&2.20$^{+0.32}_{-0.44}$       &   Ma  & $\tau$=0.3 & 1.0152 & 1.0   & 11.50        &  10.43$^{+0.07}_{-0.02}$       \\
4   &  1.44$^{+0.14}_{-0.30}$&1.37$^{+0.10}_{-0.06}$       &   BC  & $\tau$=0.1 & 0.3602 & 2.4   & 0.30         & 11.10$^{+0.16}_{-0.09}$ \\  
5   &  1.93$^{+0.06}_{-0.11}$&2.01$^{+0.22}_{-0.35}$       &   Ma  & const & 2.0    & 2.3   & 9.57         & 11.68$^{+0.21}_{-0.15}$ \\
11  &  1.55$^{+0.28}_{-0.15}$&1.57$^{+0.14}_{-0.09}$       &   Ma  & $\tau$=1.0 & 1.0152 & 0.8   & 0.51         & 11.05$^{+0.11}_{-0.06}$ \\
13  &  1.12$^{+0.02}_{-0.01}$&1.19$^{+0.08}_{-0.11}$       &   BC  & ssp        & 0.1805 & 2.4   & 12.60         & 10.91$^{+0.03}_{-0.03}$    \\
16  &  1.04$^{+0.06}_{-0.15}$&0.97$^{+0.12}_{-0.07}$       &   BC  & ssp        & 0.1805 & 2.2   & 0.78      & 10.58$^{+0.10}_{-0.05}$ \\
18  &  0.92$^{+0.02}_{-0.02}$&0.92$^{+0.14}_{-0.11}$       &   BC  & ssp        & 0.1805 & 2.1   & 6.63         & 10.48$^{+0.03}_{-0.04}$ \\
20  &  0.80$^{+0.20}_{-0.05}$&0.88$^{+0.02}_{-0.02}$       &   BC  & $\tau$=0.3 & 1.7    & 1.4   & 0.76         & 10.85$^{+0.15}_{-0.08}$ \\ 
22  &  1.21$^{+0.09}_{-0.05}$&1.30$^{+0.05}_{-0.04}$       &   BC  & ssp  & 0.5088    & 3.8   & 3.83         & 10.52$^{+0.87}_{-0.08}$ \\
25  &  1.37$^{+0.09}_{-0.12}$&1.33$^{+0.11}_{-0.13}$       &   BC  & ssp        & 0.1805 & 2.2   & 2.21   & 11.08$^{+0.04}_{-0.11}$   \\
26  &  1.09$^{+0.05}_{-0.10}$&1.09$^{+0.12}_{-0.07}$       &   BC  & $\tau$=0.1 & 0.7187 & 1.8   & 1.75         & 11.29$^{+0.01}_{-0.05}$ \\
28  &  2.61$^{+0.37}_{-0.22}$&2.90$^{+0.20}_{-0.26}$       &   Ma  & ssp        & 1.0152 & 0.64  & 3.31        & 11.73$^{+0.03}_{-0.04}$ \\
29  &  1.58$^{+0.37}_{-0.17}$&1.32$^{+0.23}_{-0.24}$       &   Ma  & $\tau$=0.3 & 0.7187 & 1.0   & 0.89         & 10.36$^{+0.14}_{-0.04}$ \\ 
30  &  0.99$^{+0.37}_{-0.11}$&1.06$^{+0.11}_{-0.07}$       &   BC  & $\tau$=0.3 & 1.7    & 1.8   & 1.14         & 11.10$^{+0.14}_{-0.18}$ \\ 
31  &  0.81$^{+0.21}_{-0.18}$&0.88$^{+0.03}_{-0.05}$       &   BC  & $\tau$=0.3 & 0.7187 & 2.0   & 1.40         & 10.66$^{+0.09}_{-0.04}$ \\ 
32  &  3.11$^{+0.21}_{-0.14}$&2.71$^{+0.38}_{-0.34}$       &   BC  &  ssp       & 0.01  & 3.0   & 2.98        & 10.83$^{+0.03}_{-0.08}$\\
34  &  1.50$^{+0.56}_{-0.29}$&1.55$^{+0.41}_{-0.19}$       &   Ma  &   $\tau$=15      & 2.0  & 2.0   & 0.60      &   10.85$^{+0.64}_{-0.37}$   \\ 
36  &  0.91$^{+0.13}_{-0.08}$&1.07$^{+0.10}_{-0.04}$       &   BC  &  $\tau$=0.6      & 0.0151 & 4.6   & 1.82         & 10.05$^{+0.11}_{-0.21}$ \\ 
37  &  2.04$^{+0.15}_{-0.24}$&1.38$^{+0.43}_{-0.42}$       &   Ma  & ssp        & 0.0132 & 1.6   & 11.73        &  11.86$^{+0.13}_{-0.03}$ \\
38  &  1.34$^{+0.14}_{-0.42}$&1.30$^{+0.17}_{-0.28}$       &   Ma  & ssp        & 0.01995& 2.8   & 1.23         & 10.82$^{+0.13}_{-0.17}$ \\ 
39  &  0.80$^{+0.23}_{-0.05}$&1.10$^{+0.05}_{-0.05}$       &   BC  & const & 1.7  & 3.7   & 8.72        & 10.80$^{+0.15}_{-0.12}$ \\ 
52  &  0.75$^{+0.12}_{-0.11}$&0.74$^{+0.18}_{-0.19}$       &   BC  & $\tau$=0.3 & 0.5088 & 2.0   & 1.60     & 10.54$^{+0.19}_{-0.09}$ \\ 
70  &  2.39$^{+0.52}_{-0.14}$&2.32$^{+0.53}_{-0.20}$       &   Ma  & $\tau$=15.0 & 1.434    & 1.4   & 1.16         & 11.17$^{+0.19}_{-0.12}$ \\ 
202 &  0.95$^{+0.44}_{-0.24}$&1.31$^{+0.09}_{-0.12}$       &   BC  & $\tau$=0.3 & 1.0152 & 2.6   & 0.83         & 10.60$^{+0.28}_{-0.14}$ \\ 
219 &  1.04$^{+0.07}_{-0.15}$&1.03$^{+0.02}_{-0.04}$       &   BC  & ssp        & 0.1278 & 2.6   & 1.98         & 10.96$^{+0.12}_{-0.01}$ \\ 
224 &  1.07$^{+0.11}_{-0.14}$&1.10$^{+0.10}_{-0.04}$       &   BC  & $\tau$=0.3 & 1.0152 & 2.0   & 0.85         & 10.92$^{+0.12}_{-0.21}$ \\  
226 &  1.98$^{+0.16}_{-0.27}$&2.35$^{+0.63}_{-0.31}$       &   Ma  & ssp        & 0.01   & 2.0   & 9.57         &  10.08$^{+0.02}_{-0.05}$\\
228 &  1.30$^{+0.09}_{-0.06}$&1.31$^{+0.05}_{-0.07}$       &   BC  &  $\tau$=0.1 & 0.0263 & 4.2   & 1.25         & 10.35$^{+0.20}_{-0.40}$ \\  
234 &  1.02$^{+0.08}_{-0.04}$&1.10$^{+0.14}_{-0.08}$       &   BC  & ssp & 0.1278 & 3.0   & 1.66         & 10.74$^{+0.08}_{-0.01}$ \\
258 &  0.83$^{+0.11}_{-0.06}$&0.96$^{+0.19}_{-0.13}$       &   BC  & $\tau$=3.0 & 6.5    & 1.0   & 5.27   & 10.90$^{+0.09}_{-0.07}$\\
285 &  1.22$^{+0.13}_{-0.06}$&1.10$^{+0.13}_{-0.08}$       &   BC  &  const      & 0.0151 & 2.8   & 4.56    & 10.03$^{+0.18}_{-0.01}$ \\
\tableline
\end{tabular}
\label{hyperz}
\tablecomments{Results from the analysis of the SEDs of the sample with {\it
    Hyperz}. Column description: (1) ID number of the object; (2) photometric
  redshift as result of {\it Hyperz}; all errors are quoted at at 99\% of
  confidence level; (3) photometric redshifts obtained with {\it 2SPD}; (4)
  source library of the best template which fits the observed SED: BC from
  \citet{bc09} and MA from \citet{ma05}; (5) star formation history of the
  best template: $ssp$ corresponds to a single stellar population and $\tau$ =
  'N' corresponds to the $\mu$-models with exponentially decaying star
  formation rate with star formation timescale $\tau$ = N Gyr; (6) age of the
  best template in Gyr; (7) $A_V$ associated with the best template; (8)
  $\chi_{rid}^2$ of the best template; (9) stellar mass and its error of the
  galaxy in $Log \, M_{\odot}$ as result of {\it Hyperzmass}.}
\end{center}
\end{table*}

\begin{table*}
\begin{center}
\caption{{\it 2SPD} SED fitting}
\begin{tabular}{c|c|cccc|cc|c|cc|cc|c}
\tableline
ID  & redshift        &\multicolumn{4}{c|}{YSP} & \multicolumn{2}{c|}{OSP}
& log M$_{*}$ &   \multicolumn{2}{c}{Dust} &   \multicolumn{2}{|c|}{IR excess}   &  UV \\
\tableline
    & z$_{phot}$          &  Age &  A$_V$  & $f_{YSP}$ & Log M$_{*}$ &  Age  &  A$_V$    &            &  T$_{dust}$ & L$_{dust}$                   &   L$_{IR exc.}$   &  $\alpha_{8-24}$   &  L$_{UV}$\\        
\tableline
1  &  0.88$^{+0.04}_{-0.05}$        &  0.02 & 0.80   &  1.5\%   &  0.49\%        & 4.0  &  0.15        &  10.08$^{+0.04}_{-0.04}$      &  140         &  2.4   &     $<$43.31   &     &\\ 
2  &  1.33$^{+0.10}_{-0.09}$        &  0.05 & 0.90   &  22.1\%   &  1.3\%        & 2.0  &  0.98        &  11.00$^{+0.04}_{-0.04}$      &  156       &  7.4  &    $<$43.65     &       &   42.53  \\ 
3  &  2.20$^{+0.32}_{-0.44}$        &  0.03 & 0.12   &    6.5\%   &  0.69\%        & 1.0  &  0.00       &  10.59$^{+0.08}_{-0.10}$    &  201-347  & 250-59  & 45.27   &  -0.01  &  43.21  \\
4  &  1.37$^{+0.10}_{-0.06}$        &  0.009 & 1.00   &  6.9\%   &  0.24\%        & 3.0  &  0.57        &  11.16$^{+0.04}_{-0.03}$      &  100        &  5.2   & $<$43.55  &       &  42.71  \\ 
5  &  2.01$^{+0.22}_{-0.35}$        & 0.02 & 1.80   &  53.1\%   &  5.9\%        & 3.0  &  0.93        &  11.49$^{+0.04}_{-0.03}$     & 121       &  144.0 &  44.76  &  $>$0.49 &      \\ 
11 &  1.57$^{+0.14}_{-0.09}$        & 0.007 & 1.87  &  17.2\%   &  0.62\%       & 3.0  &  0.30      &  10.98$^{+0.10}_{-0.05}$     & 98           &  15.4   &    $<$43.80  &     &   \\ 
13  &  1.19$^{+0.08}_{-0.11}$       & 0.03 & 1.17   &  16.2\% &  3.2\%        & 1.0  &  0.58        & 10.72$^{+0.04}_{-0.03}$     &  155-406   &  7.0-6.2  &  44.23  & -0.41   &   42.65$^m$  \\  
16  &  0.97$^{+0.12}_{-0.07}$       & 0.006 & 1.16   &  14.8\%   &  0.23\%        & 2.0  &  0.52       &  10.74$^{+0.06}_{-0.06}$      &  158      &  4.8  &  43.60  &   $>$-0.12  &  42.42$^m$  \\  
18  &  0.92$^{+0.14}_{-0.11}$       & 0.004 & 2.55   &  50.3\%   &  13.5\%        & 0.2  &  1.35      &  10.02$^{+0.08}_{-0.08}$      &  116       &  13.9  &  $<$43.96  &    &   \\  
20  &  0.88$^{+0.02}_{-0.02}$       & 0.002 & 1.26   &  3.7\%   &  0.07\%        & 3.0  &  0.42        &  11.03$^{+0.02}_{-0.03}$      &  173         &  1.4   &  $<$43.02  &     &  42.25$^m$ \\  
22  &  1.30$^{+0.05}_{-0.04}$       & 0.04  & 1.18   &  10.1\%   &  0.35\%        & 2.0  & 1.50        &  11.16$^{+0.02}_{-0.03}$      &  150         &  11.8  & 43.87  &  $>$0.66 &      \\  
25  &  1.33$^{+0.11}_{-0.13}$       & 0.002 & 1.17   &  4.1\%   &  0.28\%        & 0.4  &  1.05         &  10.75$^{+0.04}_{-0.05}$      &  138        &  8.8   &  43.75  & $>$0.26   &  42.87$^m$  \\  
26  &  1.09$^{+0.12}_{-0.07}$       & 0.006 & 1.36   &  7.4\% & 0.17\%        & 1.0  &  0.98          & 11.12$^{+0.04}_{-0.04}$      &  128         &   3.9   &  $<$43.45  &      &  42.50$^m$  \\  
28  &  2.90$^{+0.20}_{-0.26}$       & 0.001 & 1.97   &  24.7\% & 3.9\%        & 0.9  &  0.00        & 11.38$^{+0.04}_{-0.04}$      &  146         &   47.4   & 44.28 & $>$0.22   &   \\  
29  &  1.32$^{+0.23}_{-0.24}$       & 0.004 & 0.38   &  11.9\% & 0.18\%        & 0.4  &  0.94      & 10.03$^{+0.05}_{-0.05}$      &  83          &   16.4    &  $<$43.68  &     &   42.83  \\  
30  &  1.06$^{+0.11}_{-0.07}$       & 0.001 & 1.77   &  3.0\% &  0.13\%        & 2.0  &  0.73       & 11.03$^{+0.05}_{-0.05}$      &  130        &   3.5   & $<$43.47  &     &  \\  
31  &  0.88$^{+0.03}_{-0.05}$       & 0.007 & 0.23   &  7.5\%    &  0.06\%       & 2.0  &  0.23          &  10.75$^{+0.03}_{-0.03}$   &   164         & 2.6  & $<$43.35 &      &   42.86  \\  
32  &  2.71$^{+0.38}_{-0.34}$       & 0.002 & 0.98   &  10.6\%    &  0.26\%        & 2.0  &  0.39        &  10.98$^{+0.04}_{-0.04}$      & 205-452  &  51-58 &  44.96  & -0.98  &  43.05   \\ 
34  &  1.55$^{+0.41}_{-0.19}$       &  0.005 & 0.55   &  11.0\%   &  0.05\%        & 3.0 &  0.51        &  10.99$^{+0.07}_{-0.07}$      &  140        &  19.7  & $<$44.04 &     &   42.82  \\ 
36  &  1.07$^{+0.10}_{-0.04}$       &  0.008 & 2.30   &  42.4\%   &  2.0\%        & 3.0 &  1.07        &  10.83$^{+0.02}_{-0.03}$      &  186         &  4.4   &  43.46 &  $>$0.20  &  \\ 
37  &  1.38$^{+0.43}_{-0.42}$       & 0.001 & 0.00   &   1.0\%   &  0.09\%        & 0.3  &  0.09      &  10.61$^{+0.11}_{-0.11}$   &  213-614 & 84-22  &  45.03  & -0.25 &  44.10  \\ 
38  &  1.30$^{+0.17}_{-0.28}$       & 0.005 & 0.95   &  27.0\%   &  0.58\%        & 0.9  &  0.86        &  10.65$^{+0.07}_{-0.07}$      &  260         &  6.2  &  43.77  & 0.45  &   43.03 \\ 
39  &  1.10$^{+0.05}_{-0.05}$       & 0.007 & 1.86   &  15.0\%   &  0.83\%       & 1.0  &  0.97        &  10.88$^{+0.03}_{-0.03}$      &  80         &  5.3   &  $<$43.20  &      &    \\ 
52  &  0.74$^{+0.18}_{-0.19}$       & 0.004 & 0.66   &  13.6\%   &  0.11\%        & 2.0  &  0.34        &  10.78$^{+0.10}_{-0.10}$      &  189        &  2.6  &  43.33  &  0.61   &   42.96   \\ 
70  &  2.32$^{+0.53}_{-0.20}$       & 0.009 & 0.48   &  12.8\%   &  0.87\%        & 0.4  &  0.74        &  10.65$^{+0.07}_{-0.03}$       &  180      &  27.5 &   44.07  &  $>$-0.50  &   43.54 \\ 
202 &  1.31$^{+0.09}_{-0.12}$       & 0.005 & 0.14   &  1.2\%   &  0.004\%        & 3.0  &  0.14        &  10.86$^{+0.03}_{-0.06}$       &  124         &  7.4  &  $<$43.66   &    &   42.21  \\ 
219 &  1.03$^{+0.02}_{-0.04}$       & 0.02 & 0.83   &  6.2\%   &  0.69\%        & 0.3  &  1.47            &  10.67$^{+0.03}_{-0.03}$       &  112         &  2.8   &  $<$43.32  &      &   42.69$^m$ \\ 
224 &  1.10$^{+0.10}_{-0.04}$       & 0.004 & 0.63   &  3.8\%   &  0.04\%        & 1.0  &  0.76       &  10.71$^{+0.03}_{-0.04}$       &  155         &  4.5  &  $<$43.53   &      &    42.53$^m$  \\ 
226 &  2.35$^{+0.63}_{-0.31}$       & 0.001 & 0.23   &  2.8\%    &  0.12\%      & 0.9  &   0.00         &  10.59$^{+0.07}_{-0.09}$    &  212-467  &  88-36  &  45.01 &  -0.46   &    43.34   \\  
228 &  1.31$^{+0.05}_{-0.07}$       & 0.002 & 2.29   &  37.1\%    &  0.72\%        & 2.0  &  2.21       &  11.06$^{+0.03}_{-0.03}$      & 133          &  6.9  &  $<$43.64  &     &  \\  
234 &  1.10$^{+0.14}_{-0.08}$       & 0.006 & 2.30   &  46.3\%    &  2.2\%        & 2.0  &  1.05       &  10.83$^{+0.04}_{-0.04}$      & 104           &  5.8  &  $<$43.52   &      &   \\  
258 &  0.96$^{+0.19}_{-0.13}$       & 0.001 & 0.08   &  1.6\%    &  0.05\%        & 0.08  &  1.72       &  10.64$^{+0.08}_{-0.09}$      & 134         & 3.0   &  $<$43.41    &    &    43.27 \\ 
285 &  1.10$^{+0.13}_{-0.08}$        & 0.001 & 0.13   &  2.1\%    & 0.05\%        & 0.04  &  1.56        &  10.43$^{+0.04}_{-0.04}$      &  135-448  &  1.9-1.8  &  43.73  &  -0.22  &    43.08  \\           
\tableline
\end{tabular}
\label{2spd}
\tablecomments{Results from the analysis of the SEDs with {\it 2SPD}. Column
  description: (1) ID number of the object; (2) photometric redshift measured
  with {\it 2SPD}; (3)-(4)-(5)-(6) age in Gyr, A$_V$, flux fraction and mass
  fraction of the young stellar population (YSP) at 4800 \AA rest frame;
  (7)-(8) age in Gyr and A$_V$ of the old stellar population (OSP); (9) the
  total stellar mass of the galaxy in M$_{\odot}$; (10)-(11) the effective
  temperature (in K) of the one or two dust components and their luminosities,
  L$_{dust}$ (in units of 10$^{9}$ L$_{\odot}$); (12)-(13) the infrared excess
  luminosity (in erg s$^{-1}$) defined in the text (Section~\ref{dust}) and
  the spectral index measured on the infrared excess at 8 and 24$\mu$m; (14)
  UV luminosity at 2000 \AA\, in the rest frame in erg s$^{-1}$. The marginal
  UV excesses are marked with a $^m$.}
\end{center}
\end{table*}

\begin{table}
\begin{center}
  \caption{'Final' redshifts and radio properties of the sample}
\begin{tabular}{c|c|cc|c}
\tableline\tableline
ID  & z & L$_{NVSS}$ & L$_{FIRST}$ & class      \\
\tableline
  1   & 0.88$^s$ &   $<$31.78     &   31.93        & LP  \\    
  2   & 1.33     &       32.02   &      32.4          & LP  \\     
  3   & 2.20     &        33.1   &     33.19          & HP  \\     
  4   & 1.37     &       32.77   &     32.86          & HP  \\     
  5   & 2.01     &       32.47   &     32.89          & HP  \\     
 11   & 1.57     &  $<$32.18   &     32.52        & LP  \\   
 13   & 1.19     &       32.02   &     32.22          & LP  \\     
 16   & 0.97$^s$  &    32.38     &   32.27           & LP  \\    
 18   & 0.92      &      32.22   &     32.28           & LP  \\    
 20   & 0.88    &        $<$31.66   &     31.93       & LP  \\    
 22   & 1.30     &       $<$32.38   &     32.34        & LP  \\   
 25   & 1.33     &       32.29   &     32.39         & LP  \\     
 26   & 1.09     &       32.02   &     32.25          & LP  \\     
 28   & 2.90     &       32.99   &     33.13          & HP  \\     
 29   & 1.32     &       32.28   &     32.31          & LP  \\     
 30   & 1.06     &       31.83   &     32.11          & LP  \\     
 31   & 0.91$^s$  &   32.14     &   32.18          & LP  \\     
 32   & 2.71     &        32.8   &     33.17         & HP  \\     
 34   & 1.55     &       32.84   &     32.77          & HP  \\     
 36   & 1.07     &       32.23   &     32.25          & LP  \\     
 37   & 1.38     &       32.26   &     32.33         & LP  \\     
 38   & 1.30     &       32.94   &      33.00         & HP  \\     
 39   & 1.10     &        $<$31.90  &      32.16        & LP  \\   
 52   & 0.74$^s$  &    $<$31.54     &   31.73         & LP  \\  
 70   & 2.32     &       33.12   &     33.18         & HP  \\     
202   & 1.31     &       31.98   &     32.46        & LP  \\     
219   & 1.03     &       $<$31.96   &     32.09        & LP  \\   
224   & 1.10     &       32.28   &     32.27         & LP  \\     
226   & 2.35     &       $<$32.61   &     32.94        & HP  \\   
228   & 1.31     &       32.25   &     32.51         & LP  \\     
234   & 1.10     &       32.41   &     32.48          & LP  \\     
236   & 2.13     &       33.29   &     33.29         & HP   \\    
258   & 0.90$^s$ &     31.90     &   32.12        & LP \\     
285   & 1.10     &       32.23   &     32.31        & LP  \\     
\tableline
\end{tabular}
\label{controparti}
\tablecomments{Column description: (1) ID number of the object; (2) redshift
  of the object used throughout the work, with the spectroscopic redshifts
  marked with a $^s$; (3)-(4) K-corrected radio luminoisty at 1.4 GHz from
  NVSS (from http://www.cv.nrao.edu/nvss/NVSSlist.shtml) and FIRST (taken from
  \citealt{chiaberge09}) in erg s$^{-1}$ Hz$^{-1}$; (5) classification based on the
  radio power: low or high power (LP or HP) radio sources.}
\end{center}
\end{table}

\subsection{Photometric redshifts}
\label{photored}

The first test on the accuracy of our photo-z derivation is a comparison with
the spectroscopic redshifts, available for 6 objects (namely 1, 16, 25, 31,
52, and 258).  They are all compatible with each other within the errors with
the only exception of object 25. For this object the photometric redshifts
measured with {\it Hyperz} and {\it 2SPD} (z = 1.37$^{+0.09}_{-0.12}$ and z =
1.33$^{+0.11}_{-0.13}$, respectively) are significantly different from the
redshift inferred from its Magellan spectrum (z = 0.7917). However, the
Magellan spectrum does not match with the COSMOS measurements, showing a large
offset, of $\sim$0.8 dex. Apparently, the object observed by Magellan is not
the radio galaxy 25 and we do not consider its spectroscopic-z as
reliable. Reassuringly, we checked that the spectra and photometric
data-points agree for the other 5 objects with available spectra.

\begin{figure}
\centerline{
\includegraphics[angle=90,scale=0.40]{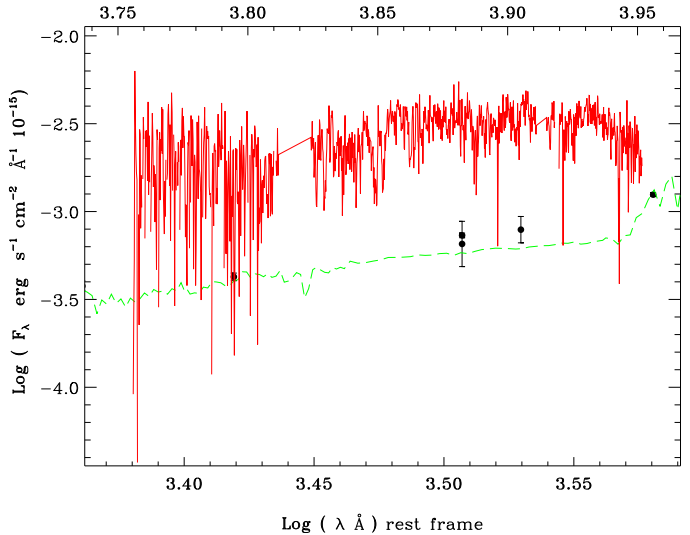}}
\caption{Spectrum of object 25 (red solid line) from the Magellan survey
  \citep{trump07} which provides a spectroscopic redshift of 0.7917. We
  overlap the synthetic SED (green dashed line) derived from {\it Hyperz} at
  redshift z = 1.37, fitting the photometric points. The wavelengths on the
  top corresponds to the observed wavelengths, and at the bottom to rest
  frame.}
\label{25spec}
\end{figure}

\begin{figure*}
\includegraphics[scale=0.45]{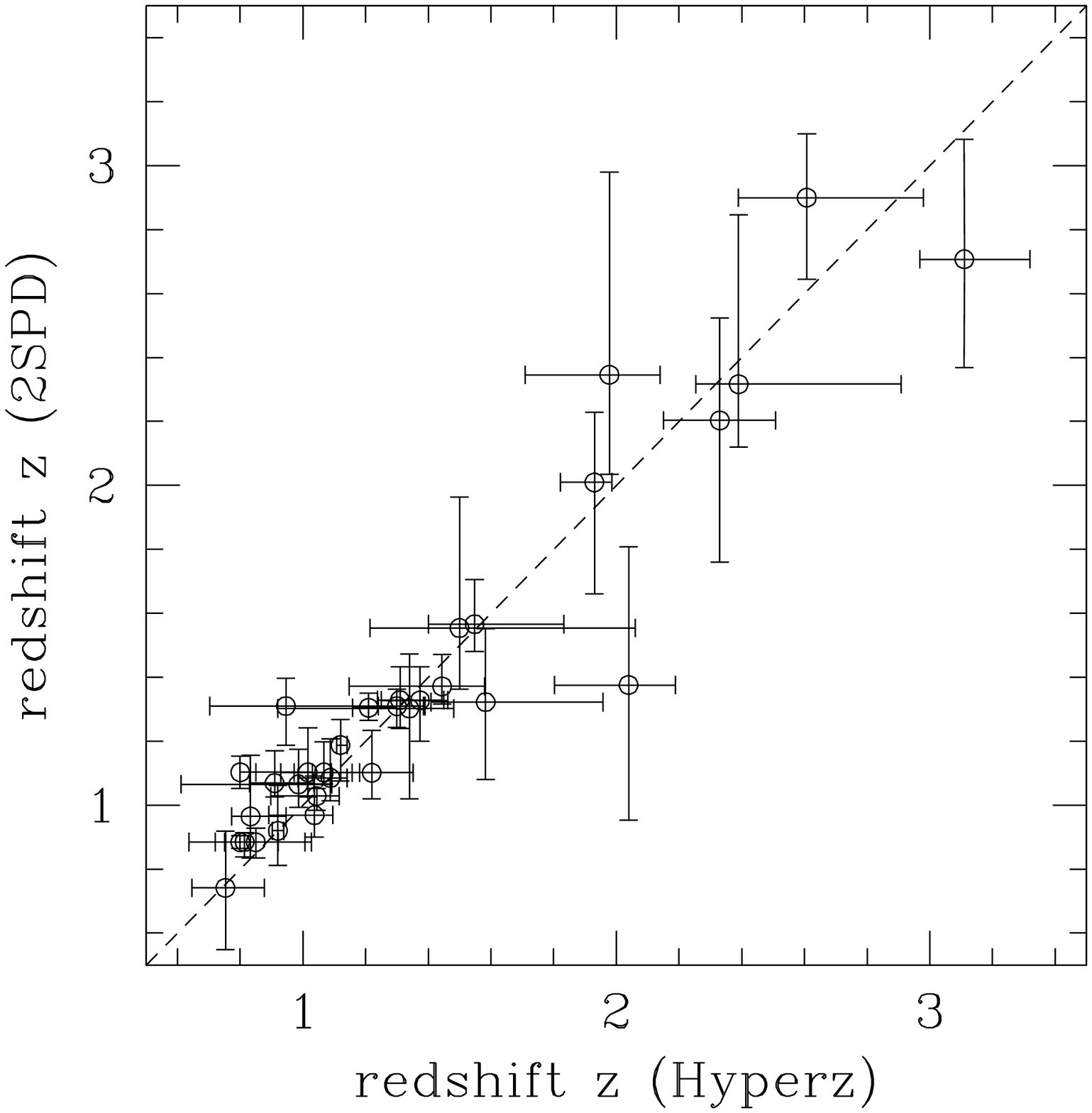}
\includegraphics[scale=0.45]{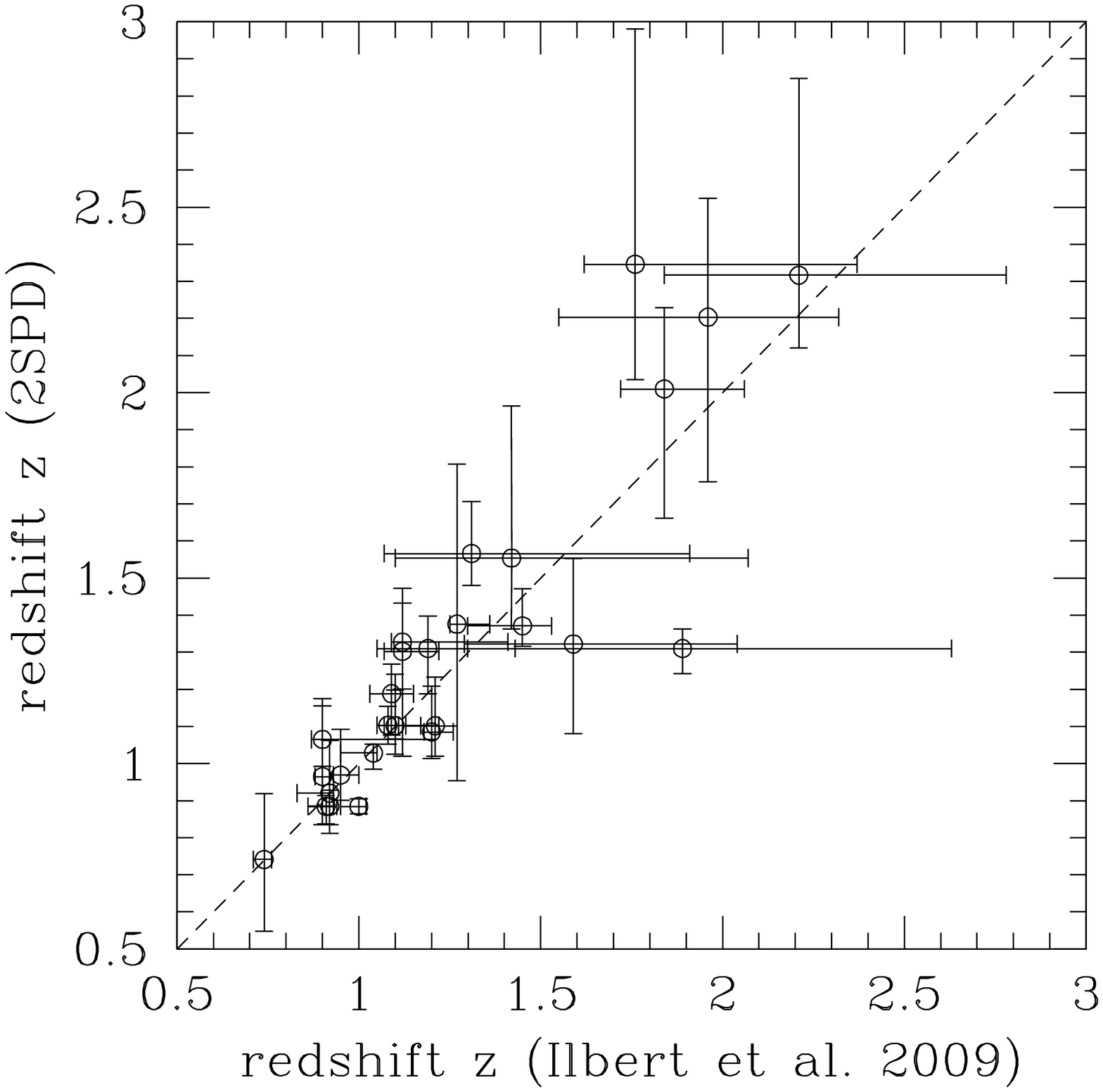}
\caption{Comparison of the photometric-z measured with {\it 2SPD} with those
  obtained with {\it Hyperz} (left) and by \citet{ilbert09} (right).  }
\label{myz}
\end{figure*}

We then compare the photometric redshifts derived with the two SED fitting
techniques. The overall range of $z$ is in both cases from $\sim$0.7 to $\sim$
3. The median photo-z are 1.21 and 1.30 for {\it Hyperz} and {\it 2SPD}
respectively. Generally, the photometric redshifts derived with the two
methods are consistent with each other within the errors (Fig.~\ref{myz},
left panel). The normalized redshift differences ($\Delta z/(1+z)$) are smaller
than 0.08 for all but 5 objects that reach $\Delta z/ z \sim 0.12-0.15.$, and a
single strong outlier (object 37, $\Delta z/ z \sim 0.28$). For this galaxy,
the {\it Hyperz} fit to the SED is particularly weak as it is not able to
fit its photometric points at both FUV and MIR wavelengths.

Furthermore, we compare the photometric redshifts measured with our method and
with the template-fitting technique of \citet{ilbert09}. Objects 2, 22, 28,
32, and 36 are not involved in this comparison since they are not included in
the \citet{ilbert09} sample. Fig.~\ref{myz} (right panel) shows the comparison
between the photo-z resulting from {\it 2SPD} and from
\citet{ilbert09}. Again, the two methods yield similar photometric redshifts
with $\Delta z/(1+z) \lesssim 0.11$ for all but 2 objects, namely 226 and 228
for which we obtain $\Delta z/(1+z) = 0.18$ and 0.25 respectively. In the case
of object 226, similarly to the object 37 discussed above, the SED obtained
with {\it 2SPD} reveals a strong excess at both UV and MIR wavelengths.
Object 228 is instead a case of misidentification in the COSMOS multiband
catalog. This result stresses the importance of our detailed work of
identification to infer the genuine properties of the galaxies.

Therefore, unless a spectroscopic redshift is available, we generally prefer
to use the photometric redshift obtained with {\it 2SPD} because our code also
includes the dust component which turns out to be important in the SED
modeling. Table~\ref{controparti} shows the redshifts used throughout all the
paper.

\subsection{Radio power distribution}
\label{radio}

\begin{figure}
\includegraphics[scale=0.45]{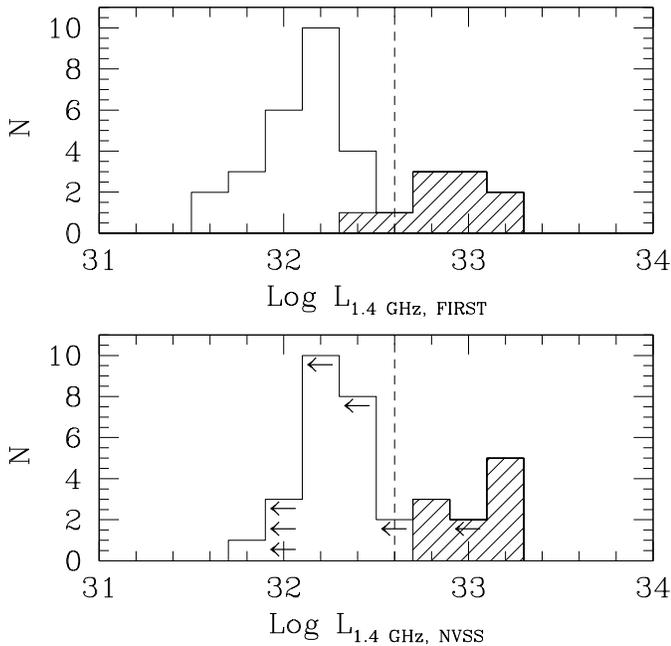}
\caption{Distribution of the K-corrected total radio luminosity (in erg
  s$^{-1}$ Hz$^{-1}$) at 1.4 GHz from the FIRST data (upper panel) and NVSS
  data (lower panel). The dashed lines corresponds to the local FR~I/FR~II
  break used to separate the sample in high power (HP, shaded histograms) and
  low power (LP) sources.}
\label{histoLr}
\end{figure}

\citet{chiaberge09} selected the sample in the COSMOS field looking for radio
sources with radio luminosities below the FR~I/FR~II break. Since we now have
improved estimates of the photometric redshifts of the objects, we can derive
a more accurate value for their radio power. We show in Fig.~\ref{histoLr} the
histograms of K-corrected (using a radio spectral index $\alpha$ = 0.8) radio
powers at 1.4 GHz measured with NVSS and FIRST. The resulting 1.4 GHz
luminosities are in the range 10$^{31.5}-10^{33.3}$ erg s$^{-1}$ Hz$^{-1}$.
The local FR~I/FR~II break luminosity \citep{fanaroff74}, converted from 178
MHz to 1.4 GHz adopting $\alpha$ = 0.8, is L$_{1.4 \,\rm GHz} \sim 10^{32.6}$
erg s$^{-1}$ Hz$^{-1}$. Therefore, the radio distribution of our sample
actually straddles the FR~I/FR~II break. Nonetheless we remind the reader that
radio sources with a clear FR~II morphologies were previously excluded by
\citet{chiaberge09}.

As already discussed by \citet{chiaberge09}, the high-frequency radio
observations available for this sample (their rest frame frequency are in the
$\sim$ 3 - 5 GHz range) are not ideal to observe and detect their total radio
structures because they miss diffuse/steep-spectrum radio emission. Lower
frequency radio observations are needed for a more accurate measurement of
their intrinsic total radio power distribution.

\begin{figure}
\includegraphics[scale=0.45]{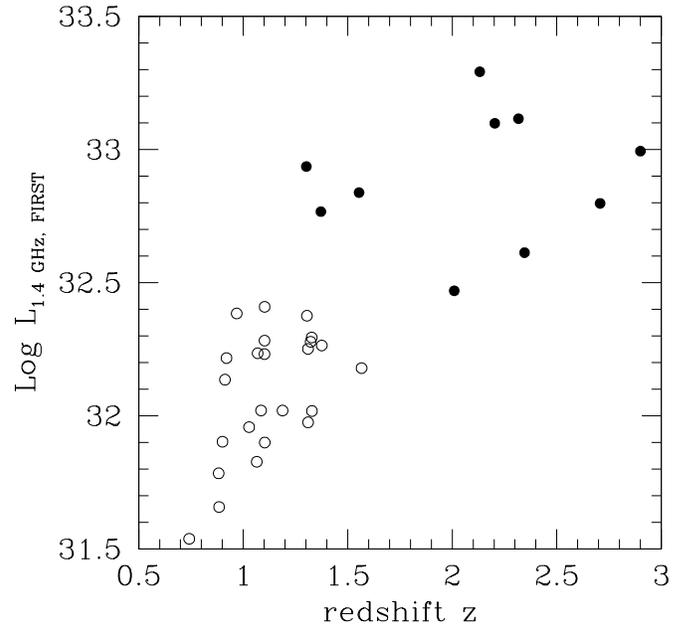}
\caption{Redshifts versus the K-corrected FIRST radio
  luminosity at 1.4 GHz (in erg s$^{-1}$ Hz$^{-1}$). The empty points are the
  LPs and the full points are the HPs.}
\label{zfirst}
\end{figure}

Since the radio power distribution is rather broad (covering two orders of
magnitudes), we separate the sample into two groups, including
sources of high and low power (HP and LP, respectively) using the local
FR~I/FR~II break as divide (see Table~\ref{controparti}). This operative
definition enables us to explore the role of radio power in defining the
overall properties of these galaxies. Two third of the sample shows radio
luminosities below the break. The LP and HP classes are, not surprisingly,
roughly separated also in redshifts (Fig.~\ref{zfirst}), since our sample is
limited in radio flux, and indeed the HP sources generally lie at z $\gtrsim$
1.5.

\subsection{Host galaxy properties inferred from SED modeling}
\label{host}

We now focus on the properties of the host galaxies inferred from SED
modeling and in particular on their stellar populations.

\begin{figure*}
  \includegraphics[scale=0.45]{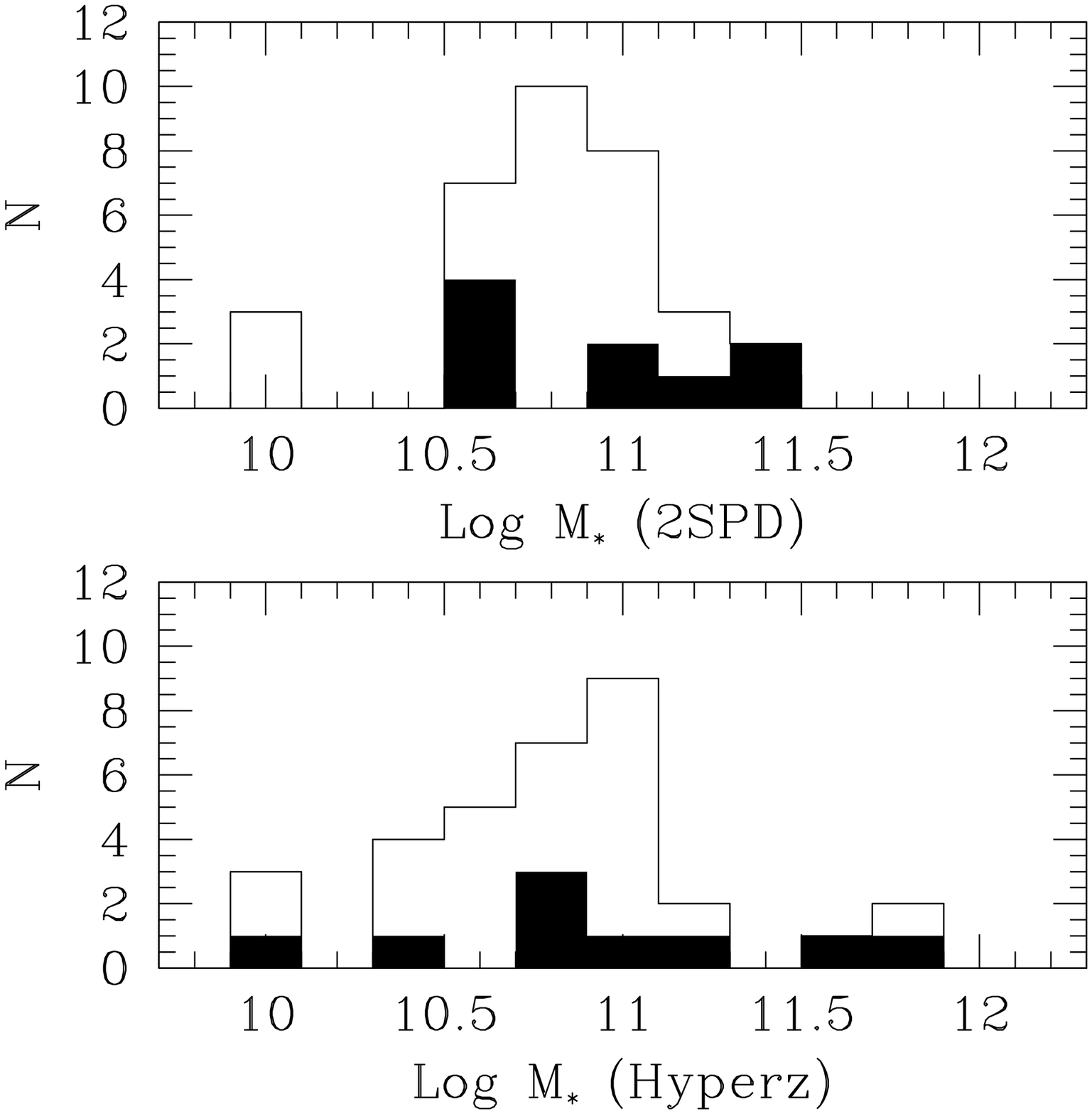}
\includegraphics[scale=0.45]{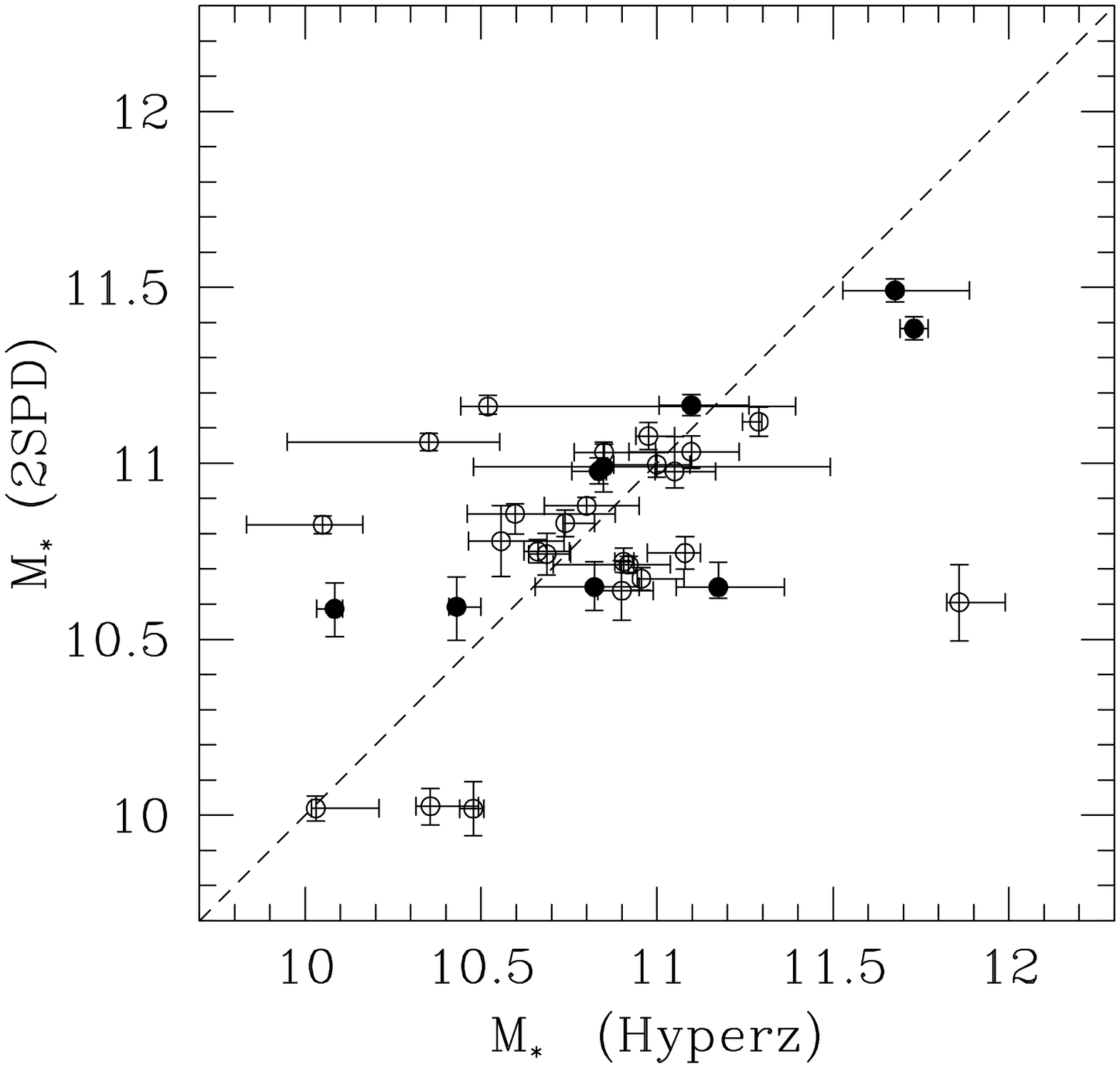}
\caption{Left panel: distribution of stellar masses (in M$_{\odot}$) of our
  sample obtained with {\it 2SPD} (upper panel) and with {\it Hyperz} (lower
  panel). Right panel: comparison between the stellar masses (in M$_{\odot}$)
  inferred from SED modeling using {\it Hyperz} and {\it 2SPD} techniques. The
  black histogram (points) represents the HPs, while the empty one the LPs.}
\label{Msisto}
\end{figure*}

One of the key and more robust result is the measurement of the stellar
content, M$_{*}$. The inferred mass range is $\sim$10$^{10} - 10^{12}$
M$_{\odot}$. The masses measured with {\it Hyperzmass} and with {\it 2SPD} are
compared in Fig.~\ref{Msisto}. The two techniques return generally consistent
values of the stellar masses for most sources (the median values are in both
cases $\sim$ 7 $\times$ 10$^{10}$ M$_{\odot}$), but with a few evident
outliers. The stellar masses derived from single- and two- stellar components
are know to differ significantly (e.g. \citealt{papovic06}). Furthermore, the
inclusion of dust components by {\it 2SPD} has two effects by changing i) the
stellar mass since thermal emission largely contributes to observed infrared
emission, and ii) the photometric redshift. The largest outlier is object 37
whose SED is poorly fitted by {\it Hyperz} with a very high mass 7.0 $\times$
10$^{11}$ M$_{\odot}$, while a strong dust component and a mass 10 times lower
are required by {\it 2SPD}. At the opposite end of the mass distribution, we
find two objects (36 and 226) in which again {\it 2SPD} finds a strong dust
component, but where the {\it Hyperz} is a factor 3 to 7 lower than that
obtained with {\it 2SPD}. Apparently, the presence of dust components has a
stronger impact on the stellar mass estimate than the inclusion of a young
stellar population.

\begin{figure*}
\includegraphics[scale=0.45]{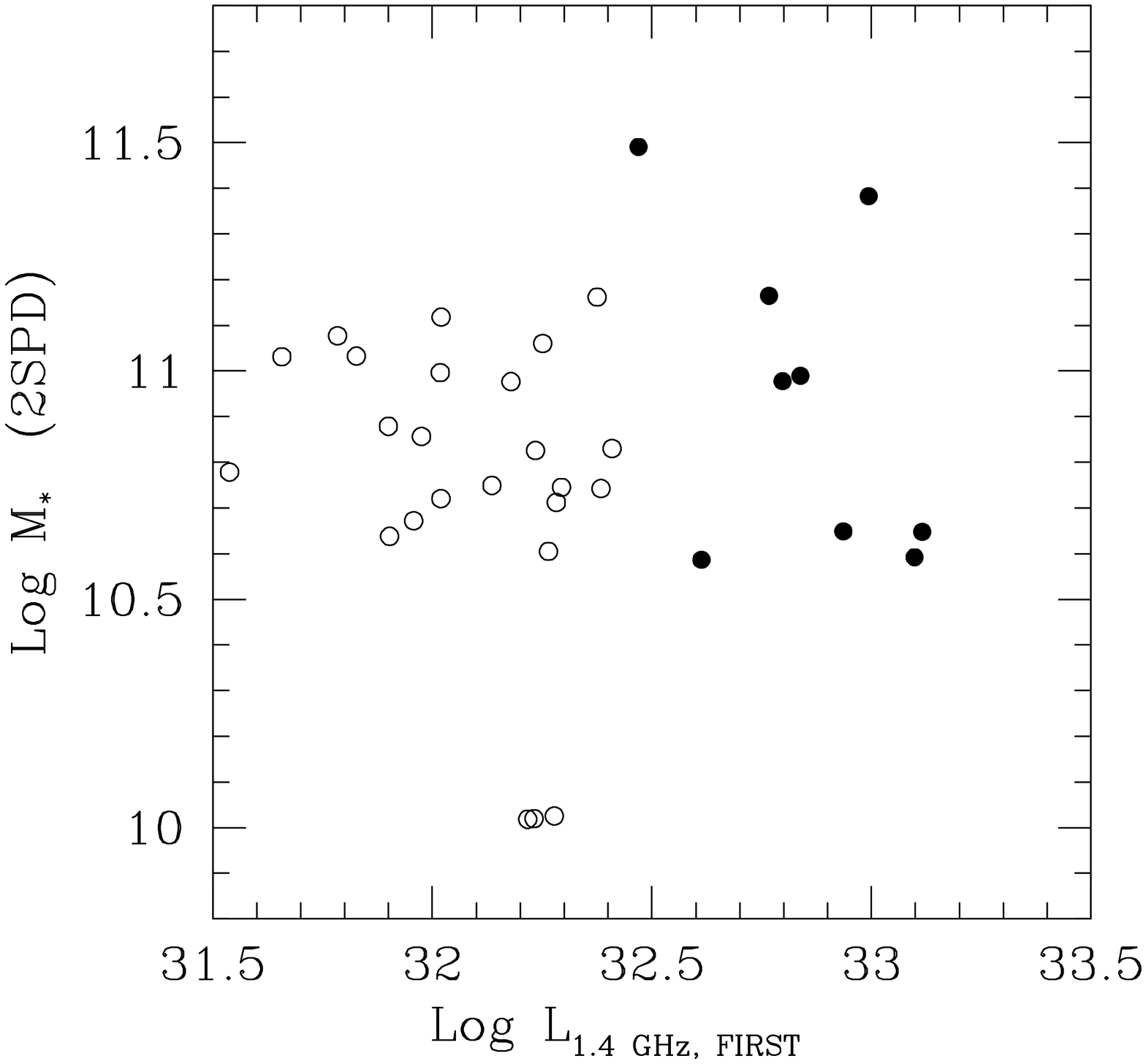}
\includegraphics[scale=0.45]{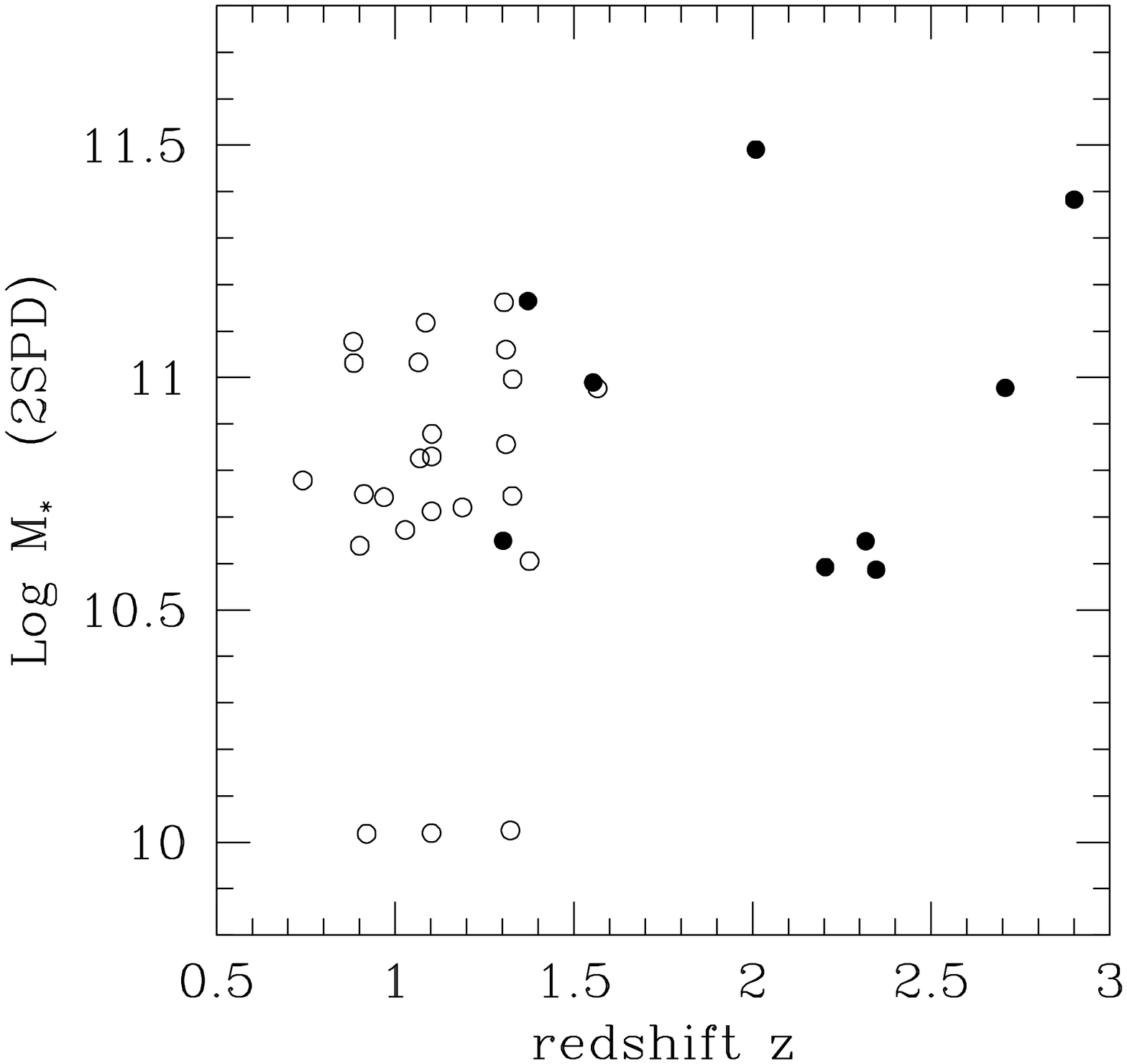}
\caption{Stellar masses (in M$_{\odot}$) measured with {\it 2SPD} in relation with the
  K-corrected FIRST radio powers (in erg s$^{-1}$ Hz$^{-1}$) (left panel) and
  redshifts (right panel) of the sample. Empty points are LPs, while filled
  points are HPs.}
\label{Ms}
\end{figure*}

Considering the radio power of the sources (see Fig.~\ref{Ms}, left panel),
HPs show slightly larger stellar mass content than LPs. The median values
(obtained from {\it 2SPD} code) for LPs is 5.8 $\times$ 10$^{10}$ M$_{\odot}$,
while for HPs is 8.7 $\times$ 10$^{10}$ M$_{\odot}$. However, the two
distributions are very broad and show an almost complete overlap. A very
similar result is found looking for differences in mass between objects at
different redshifts (see Fig.~\ref{Ms}, right panel).

Due to the degeneration inherent to the modeling of SEDs the remaining
results of the stellar populations are relatively unconstrained. Overall, if
we concentrate in the whole sample, the SED fitting shows that the hosts are
mainly dominated by a massive OSP with an age of $\sim$ 1$-$3 $\times$
10$^{9}$ years. Although less reliable, the YSPs required to fit the UV
excesses show ages of $1 - 30$ Myr, with a contribution to the total mass of
the galaxy of $\lesssim 1 \%$ for most sources. The flux contribution at 4800
\AA\ (rest frame) of the YSP is $<$30\% for most of the objects.

\subsection{Dust emission}
\label{dust}

\begin{figure}
\centerline{
\includegraphics[angle=90,scale=0.40]{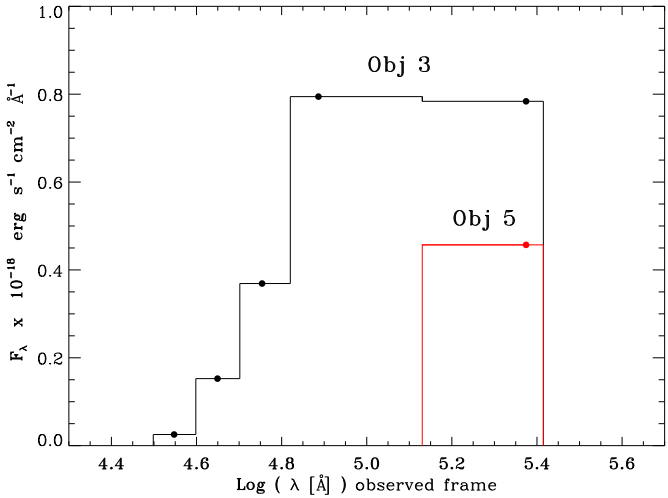}}
\caption{Residuals obtained by subtracting the stellar component from the SED
  in the 5-bands Spitzer data which we ascribe to the dust emission. We show
  the results for object 3 (black) and 5 (red) as an example. The step
  function connecting the data-points is used to obtain the IR excess flux,
  i.e., the total dust flux.}
\label{toy}
\end{figure}

As discussed in Section~\ref{2SPDcode}, the parameters related to the dust
emission must be taken as approximate rather than a real measurements of the
dust component. Nonetheless, it is still significant that dust emission is
required to adequately model the SEDs of 15 objects due to the detection of
emission at 24 $\mu$m. In addition significant excesses above the stellar
emission are observed also at shorter infrared wavelengths in 8 of these
galaxies. In order to explore the dust properties we estimated the residuals
between the best fitting model including only the stellar component and the
data-points, looking for an excess at long wavelengths due to dust. We then
integrated the residuals to obtain the total dust luminosity in the range
covered by the Spitzer data, i.e. $\sim$ 3 - 26 \mum. The integration is
performed by assuming that the spectrum is represented by a multiple step
function (see Fig. \ref{toy}).

\begin{figure*}
\includegraphics[scale=0.45]{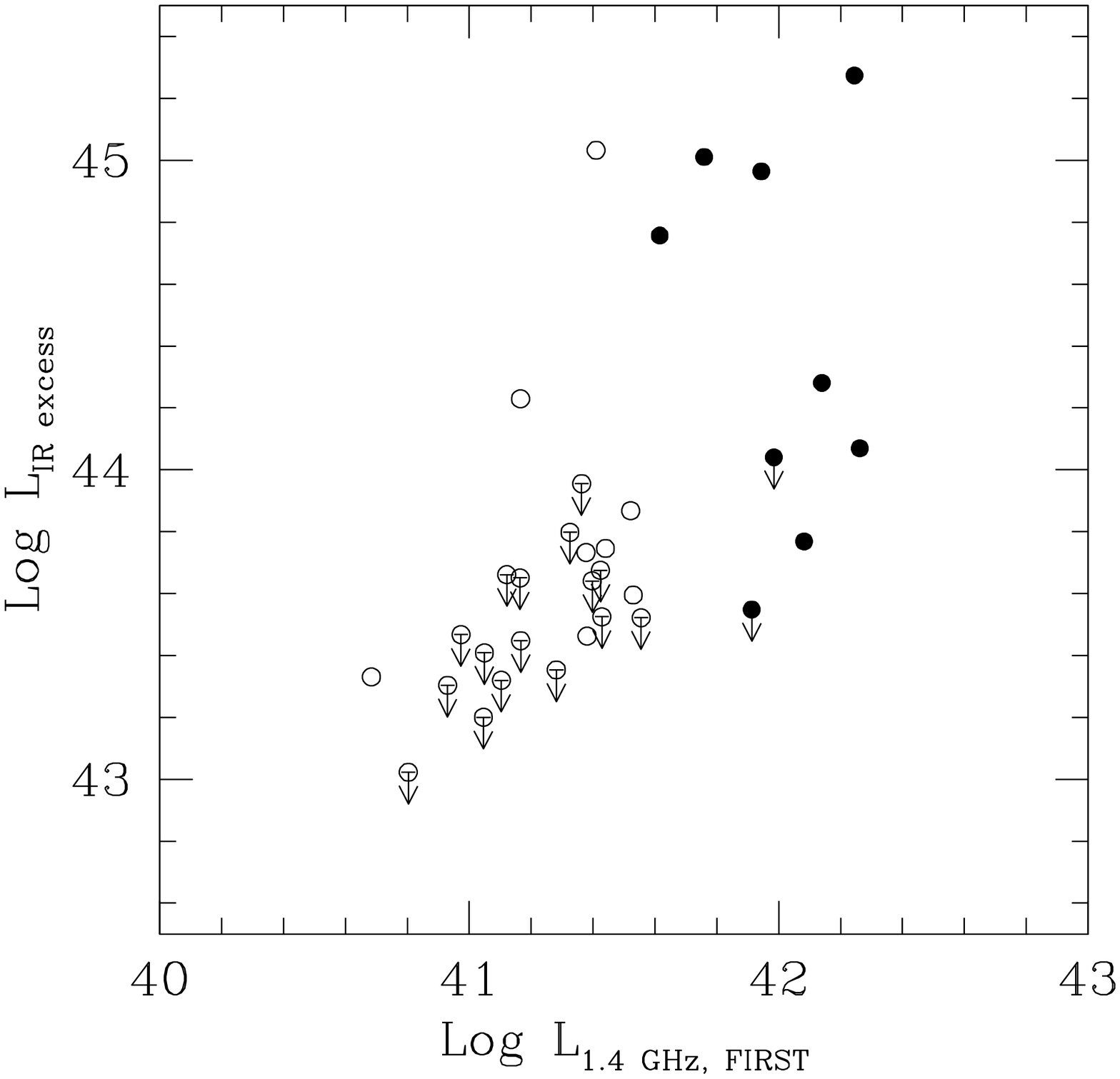}  
\includegraphics[scale=0.45]{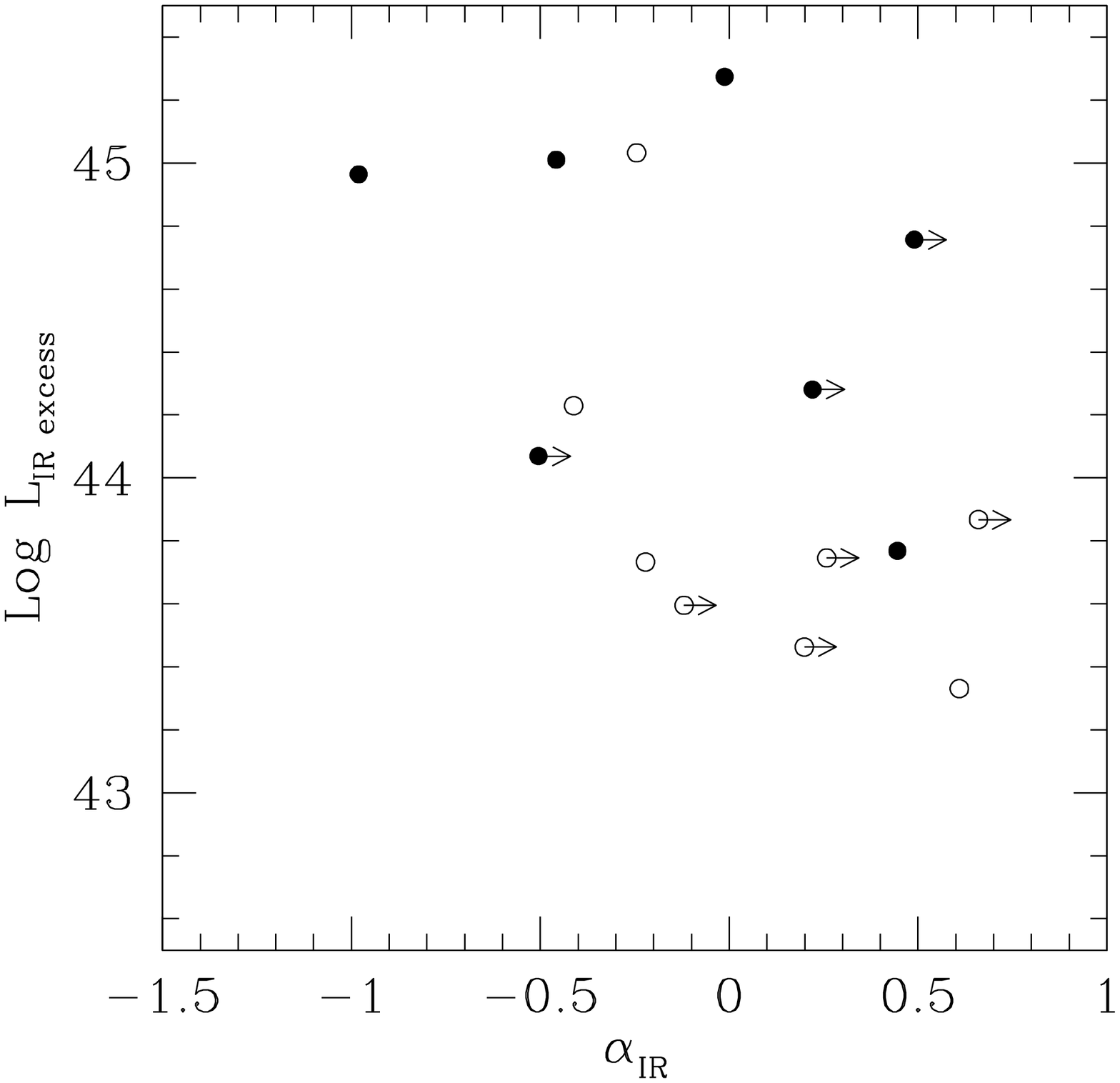}  
\caption{Infrared excess luminosity ($\ergs$) versus: (left panel) K-corrected FIRST radio
  luminosity ($\ergs$), and (right panel) spectral index from 8 to 24 \mum\
  estimated from the IR excess in the SED above the stellar emission. Empty
  points are the LPs and full points are the HPs.}
\label{dustfigs}
\end{figure*}

The resulting dust luminosities (see Table \ref{2spd}), estimated as IR excess
luminosities, are in the range $ L_{\rm dust} \sim10^{43.5} -10^{45.5}$ erg
s$^{-1}$. A trend links radio and dust emission (see Fig. \ref{dustfigs}, left
panel) with most (7/9) HP radio-sources showing a significant dust emission
with luminosities larger than $L_{\rm dust} \sim10^{44}$ erg s$^{-1}$. The LPs
are instead (with only 2 exceptions) of lower L$_{\rm dust}$ and in many cases
(16 galaxies) only upper limits can be derived. 

We check the statistical significance of such a trend, by using a censored
statistical analysis (ASURV, \citealt{lavalley92}) which takes into account of
the presence of upper limits.  Using the generalized Kendall's $\tau$ test
\citep{kendall83}, the probability that a fortuitous correlation appears is
0.0003. However, the common dependence of the two luminosities on redshift
might play an important role. Therefore, we perform a partial correlation
analysis \citep{akritas96} to examine the linear relation between the
luminosities excluding the dependence on redshift. Operatively, we use the
partial Kendall's $\tau$ test, whose null hypothesis is the absence of the
correlation excluding the redshift variable. The partial Kendall's coefficient
$\tau$ is 0.20 and the standard deviation is 0.081. The null hypothesis is
then rejected at the level of 0.05.

We also measured the spectral index of the residuals between 8 \mum\ and 24
\mum, $\alpha_{(8-24)}$.  Taking into account only significant excesses at
8 \mum\ (residuals larger than 3 $\sigma$), this value can be estimated in 8
cases, with values spanning between $\alpha_{(8-24)} \sim$ 1 and -1. For the
remaining objects with only a 24 \mum\ detection, the limit to the 8 \mum\
flux translates into a lower limit of $\alpha_{(8-24)} \gtrsim 1$. This
parameter can be crudely related to the overall dust temperature. By assuming
a single black-body dust component, the values of $\alpha_{(8-24)}$ translate
into a temperature range of 500-850 K and 300-550 K for $\alpha_{(8-24)} = -1$
and $1$, respectively. The derived temperature depends on redshift,
with the lower (upper) values of T being derived for $z=0.75$ ($z=2$).

Fig.~\ref{dustfigs} (middle panel) shows a broad relation between the IR
spectral index and the IR excess luminosities. This would mean that the
decrease of $\alpha_{(8-24)}$ is ascribable to the increase of the
high-temperature dust component. Since a large radio power also
implies a large dust luminosity, this may suggest a possible AGN nature of the
dust heating source for the objects with largest IR excess.

\subsection{UV excess}
\label{uv}

As noted in Sect.~\ref{hyperzsec}, the {\it Hyperz} code often does not
reproduce satisfactorily the bluest part of the SED and we chose to model such
part of spectrum with a young stellar population, included by our {\it
  2SPD}. The inclusion of this component does not alter significantly the
photometric redshift or the galaxies mass, but it is clearly of great
importance to understand the nature of these radio galaxies. Nonetheless,
inspection of the SEDs obtained with {\it 2SPD} indicates that the UV excesses
(above the contribution of the old stellar component) are usually very poorly
constrained. Furthermore, the very stellar origin is not granted and the UV
excess might hide an AGN contribution.

It is necessary to introduce a model-independent criterion to assess which
sources actually show an UV excess and to estimate its luminosity. We visually
inspected all SEDs, searching for sources with a substantial flattening in the
SED at short wavelengths or with a change of the slope between the OSP and the
emission in the UV band. Fifteen sources show a clear UV excess (namely object
2, 3, 4, 29, 31, 32, 34, 37, 38, 52, 70, 202, 226, 258, and 285) and
additional 7 objects show a marginal UV excess (namely object 13, 16, 20, 25,
26, 219, and 224). In the remaining galaxies the SEDs in UV bands drop sharply
and are well reproduced by the emission from OSPs. In order to quantify the UV
contribution, we measure the flux at 2000 \AA\ in the rest frame, L$_{UV}$,
from the best fitting model, for those objects showing the UV excess. The UV
luminosities range is $10^{42}-10^{44}$ erg s$^{-1}$. HPs show larger UV
luminosities than LPs by a factor 2.5, even though the strongest UV excess is
seen in the LP radio galaxy 37. If we concentrate the sources which do not
show an UV excess, the 2000 \AA\ luminosity is $\lesssim 10^{43}$ erg s$^{-1}$
(mostly, $\sim10^{42}$ erg s$^{-1}$).

\begin{figure*}
\includegraphics[scale=0.45]{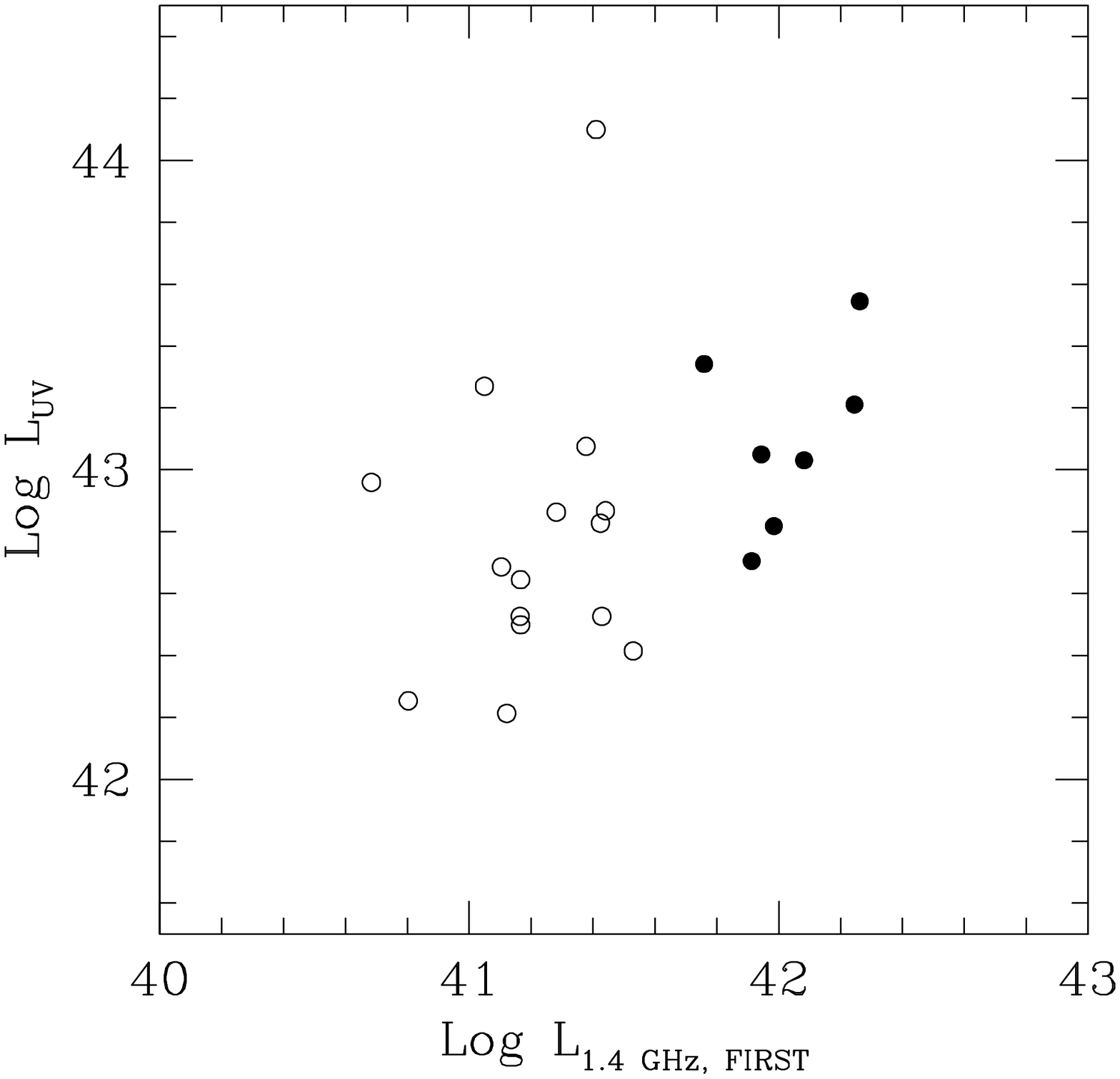}  
\includegraphics[scale=0.45]{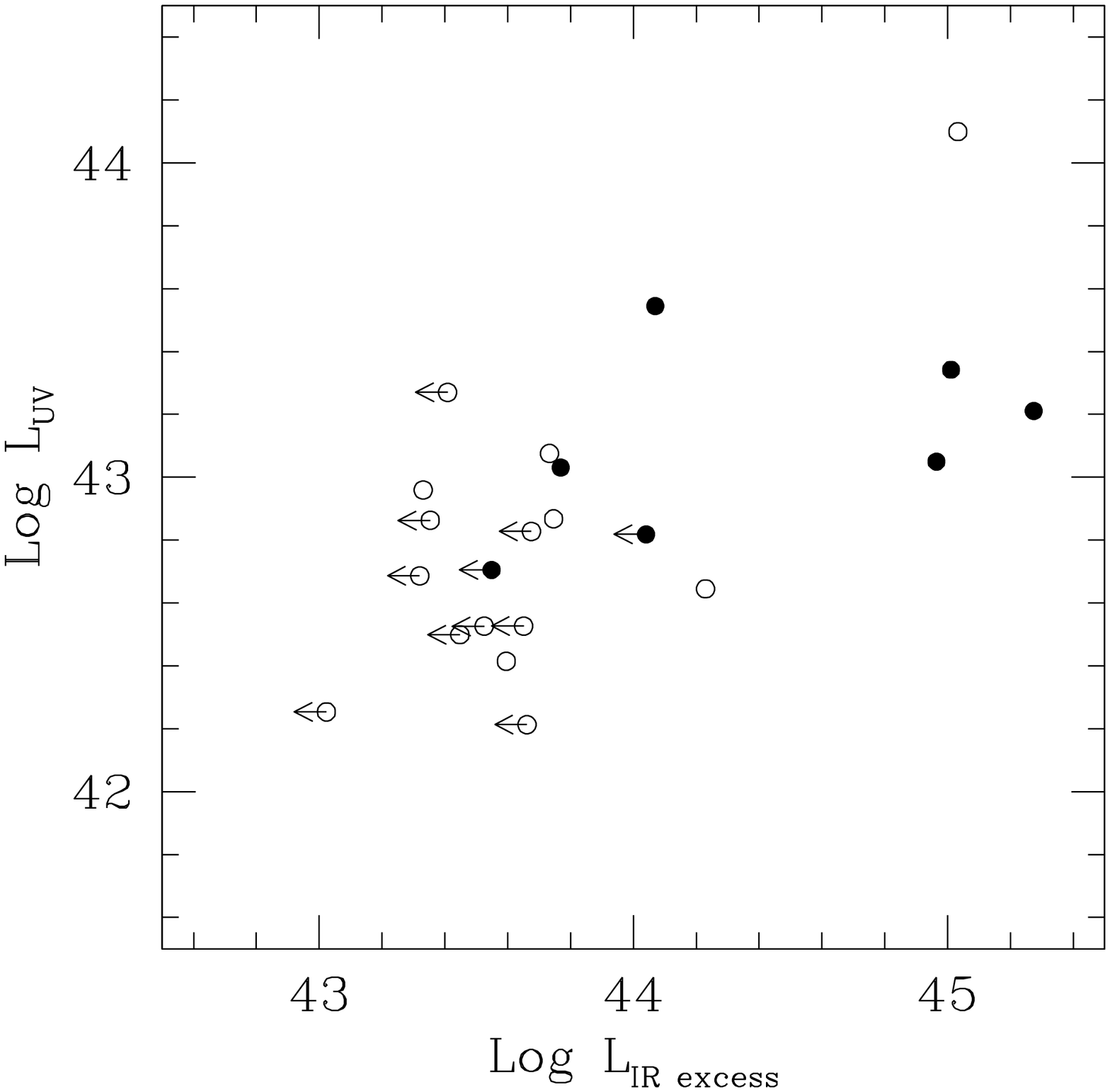}  
\caption{UV luminosity ($\ergs$) measured at 2000 \AA\, versus: (left panel)
  K-corrected FIRST radio luminosity ($\ergs$) , and (right panel) infrared excess
  luminosity ($\ergs$) . Empty points are the LPs and full points are the
  HPs.}
\label{UVplot}
\end{figure*}

In order to explore the nature of the UV emission, it is useful to compare the UV
luminosity with the multiband properties obtained in the previous sections. No
relations appear to link the UV excess to the redshift and the mass of the
galaxy. Conversely, the UV luminosity appears to be linked to the radio power
and IR excess luminosity (Fig.~\ref{UVplot}).

For the radio and UV luminosities the linear correlation coefficient is $r =
0.404$ and the probability of the presence of a fortuitous relation is P=
0.062. The partial Kendall's coefficient $\tau$ is 0.17 and the standard
deviation is 0.15: the null hypothesis, i.e. the absence of the correlation,
can not be rejected.

For the IR-UV relation, the generalized Kendall's $\tau$ test returns a
probability of no correlation of P= 0.0072. We obtain a partial Kendall's
coefficient $\tau$ = 0.41 and a standard deviation of 0.069: the null
hypothesis, i.e. the absence of the correlation, is rejected at the level of
0.05.

\section{Discussion and Conclusions}
\label{discussion}

We used the multiwavelength data provided by the COSMOS survey to analyze
the first sizeable sample of 34 low-power radio galaxies at high redshifts (z
$\gtrsim$ 1), selected by \citet{chiaberge09}. We performed a careful
visual inspection of all their multiwavelength counterparts to identify their
genuine emission and to infer their SEDs. Those would have been compromised in
case of a blind use of the COSMOS catalog.  Taking advantage of the FUV-MIR
spectral coverage, we modeled their SEDs using two different template
fitting techniques: i) the {\it Hyperz} code that uses a single stellar
population and ii) our own code {\it 2SPD} that includes also dust
component(s) and a young stellar population. We analyzed the properties
of the SEDs of the sample. We here summarize the main results and
briefly discuss them:

\begin{enumerate}[(i)]

\item The photometric redshifts of these radio sources range from $\sim$0.7 to
  3. The photo-z measured with the two techniques are generally consistent
  with each other, with those measured by \citet{ilbert09}, and with the
  available spectroscopic redshifts. In addition, we measured the
  photometric redshifts for five objects which were not included in the
  \citet{ilbert09} sample.

\item The new and accurate measurements of the photometric redshifts
  enable us to infer the radio power distribution of these objects. The
  resulting K-corrected 1.4 GHz luminosities are in the range
  10$^{31.5}-10^{33.3}$ erg s$^{-1}$ Hz$^{-1}$, straddling the FR~I/FR~II
  break luminosity (L$_{1.4 \,\rm GHz} \sim 10^{32.6}$ erg s$^{-1}$ Hz$^{-1}$)
  as defined for low redshifts galaxies.

\item The resulting stellar masses are mostly confined in the range
  $\sim$10$^{10.5-11.5}$ M$_{\odot}$, with a median value of 7 $\times$
  10$^{10}$ M$_{\odot}$. 

\item The SED of most objects of the sample is consistent with the
  presence of a dominant contribution from an old stellar population with an
  age $\sim 1 - 3  \times$ 10$^{9}$ years. However, significant excesses are
  often observed at the shortest and/or longest wavelengths.

\item A dust component is needed to account for the 24 \mum\ emission in 15
  galaxies and significant excesses above the stellar emission are observed
  also at shorter infrared wavelengths in 8 of these galaxies. Estimates of
  the dust luminosity yield values in the range $L_{\rm dust} \sim10^{43.5}
  -10^{45.5}$ erg s$^{-1}$. The overall dust temperature,  estimated
  for the 8 radio-galaxies with a substantial dust excess at $\lambda \lesssim
  8$\mum, is in the range 300-850 K.

\item Inspection of the SED obtained with {\it 2SPD} indicates that the UV
  excesses (above the contribution of the old stellar component) are often
  present (significantly in 15 sources and marginally in 7 sources), but they
  are usually weakly constrained. The UV luminosities measured at 2000
  \AA\ (rest frame) is in the range $10^{42}-10^{44}$ erg s$^{-1}$.

\item Although the censored analysis does not provide significantly high
    statistical parameters, we can tentatively confirm the presence of positive
    links between the dust emission with both the radio and UV
    luminosities. For these relations, the possibility of a common luminosity
    dependence on distances is rejected at the level of 0.05.

\end{enumerate}

The selection performed by \citet{chiaberge09} with the aim of searching for
FR~I candidates at z $\sim1-2$ turned out to be successful. In fact, our work
confirms i ) their location in the range of redshifts aimed with the selection
process, although extending slightly below 1 and up to z $\sim$ 3 and ii) the
low radio luminosity of the sample, generally consistent with those of the
local FR~I, although 1/3 of the sample exceeds the local FR~I/FR~II luminosity
break. The extension of their radio power to larger luminosities does not
necessarily imply a FR~II nature for those sources. In fact the sources with a
clear FR~II morphology were excluded from the analysis. In addition, in the
local Universe, the radio distribution of FR~Is is broad and overlaps with the
FR~II distribution \citep{zirbel95}. Furthermore, this overlap increases at
higher radio frequencies.

Overall, the hosts of these high-z low-luminosity radio-sources are similar to
those of the local FR~I which usually live in red massive early-type
galaxies (e.g., \citealt{zirbel96,best05b,baldi08,smolcic09,baldi10a}).

However, the SED modeling reveals that additional components to the old
stellar population have to be included to account for the emission at the SED
extremes, i.e. in either the UV or in the MIR band, in most of the sources of
the sample. This behavior is not seen in the low-z FR~I that are generally
faint in UV (both from the point of view of star formation and nuclear
emission, \citealt{chiaberge02b,baldi08}). Similarly, their MIR luminosities
exceed by a very large factor (between 30 and 3000 for the MIR detected
sources) the typical low-z FR~I luminosities ($\sim 10^{42} \ergs$,
\citealt{hardcastle09}). Conversely, the UV and MIR properties are somewhat
similar to those of local FR~IIs, which show bluer color (e.g.,
\citealt{baldi08,smolcic09}) and large dust amount (e.g.,
\citealt{dekoff00,dicken10}) than FR~Is.

The origin of the MIR and UV emission can not be firmly established based on
the available data. The estimate of the dust temperature (possible for only 8
objects) is in the range expected from dust heating from a quasar-like nucleus
($T\gtrsim 300$ K, e.g. \citealt{siebenmorgen04,ogle06}) and far larger than the
dust associated with star formation (e.g.,
\citealt{hwang10,sreenilayam11,boquien11,patel11}). The high end of the MIR
luminosity reaches values similar to those found in high power radio-galaxies
and QSOs. However, dust emission is seen only in less than half the sources of
the sample.

Similarly, a UV component in excess to the old stellar population is firmly
detected in the same number of sources (but there is not a one-to-one
correspondence between UV and MIR emission). The observed UV luminosities are
much larger that the faint non-thermal UV-nuclei seen in FR~I and, instead,
similar to those FR~IIs \citep{chiaberge99,chiaberge02b}, dominated by the
accretion disk emission.

Summing up, we find that the sources of the sample display a wide variety of
properties, despite the relatively narrow range in radio luminosities. The 8
objects with the strongest MIR excess, and with high dust temperature
indicative of a quasar-like nature, also show a significant UV excess. To this
group we must obviously add the object 236, the spectroscopically confirmed
QSO. We must note, however, that the SED of these 8 sources is much redder
than that of object 236 and that their HST images do not provide any evidence
for the presence of bright unresolved nuclei, as described by
\citet{chiaberge09} based on visual inspection of the data.

At the opposite end of the radio luminosity distribution, there are 7 galaxies
not showing any UV excess nor any 24 \mum\ emission (and this number raises to
11 by including also the objects with only marginal UV excesses).

The remaining galaxies, amounting to about half of the sample, show an excess
only in either the UV or MIR band. For those objects which do not show an UV
excess or are not detected at 24 $\mu$m, the limits to the UV/MIR ratio are
broadly consistent with the values obtained for the remaining galaxies of the
sample. In particular we do not find objects with a high MIR luminosity
without an UV excess that might be expected in the case of an obscured
QSO. Unfortunately, with the available information we cannot distinguish
between an origin related to star formation or to an active nucleus. In
addition, relatively large amounts of dust, suggested by the dust
luminosities, indicate that possible obscuration may prevent us from
detecting UV emission.

A further detailed analysis of the nuclear properties is needed to understand
which type of AGN are associated with these high-z radio galaxies. This will
include their X-ray emission \citep{tundo11} and the radio core flux,
available from the COSMOS/VLA data, but which we defer to a future study. This
might also provide new insights on the controversial FR~I-QSO association
(e.g., \citealt{falcke95a,baum95,cao04,blundell01}, see also
\citealt{blundell03} for review on this subject).

Summarizing, this work validates the first sizeable sample of low-luminosity
radio galaxies at high redshifts, 0.7 $\lesssim$ z $\lesssim$ 3. This opens
the possibility to perform a detailed comparison of the host and nuclear
properties of these sources with those of i) the local low-luminosity
radio-galaxies, ii) the powerful radio-galaxies in the same redshift range,
and iii) the population of non active galaxies at $z\sim 1 - 3$. These issues
will be addressed in a forthcoming paper.

\acknowledgments

R.D.B. acknowledges the financial support (grant DDRF D0001.82439) from Space
Telescope Science Institute, Baltimore.  We are grateful to M. Bolzonella,
C. Maraston, and J. Pforr.  which significantly help with the SED modeling. We
also thank the referee and A. Celotti for their contributions to improve the
paper.

\bibliography{my}

\begin{figure*}[h]
\begin{center}$
\begin{array}{ccc}
\vspace{2em}
\includegraphics[scale=0.33,angle=90]{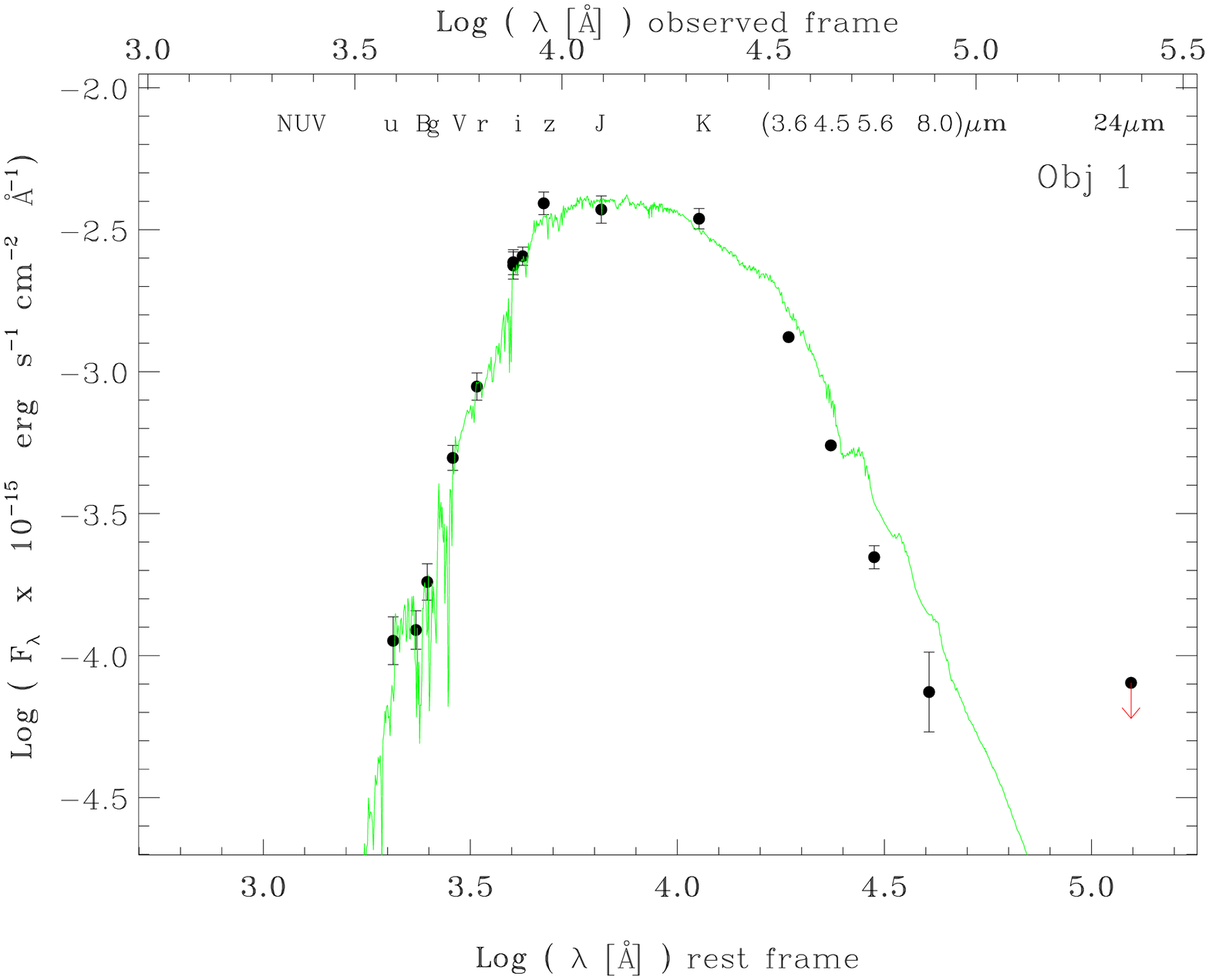} &
\includegraphics[scale=0.33,angle=90]{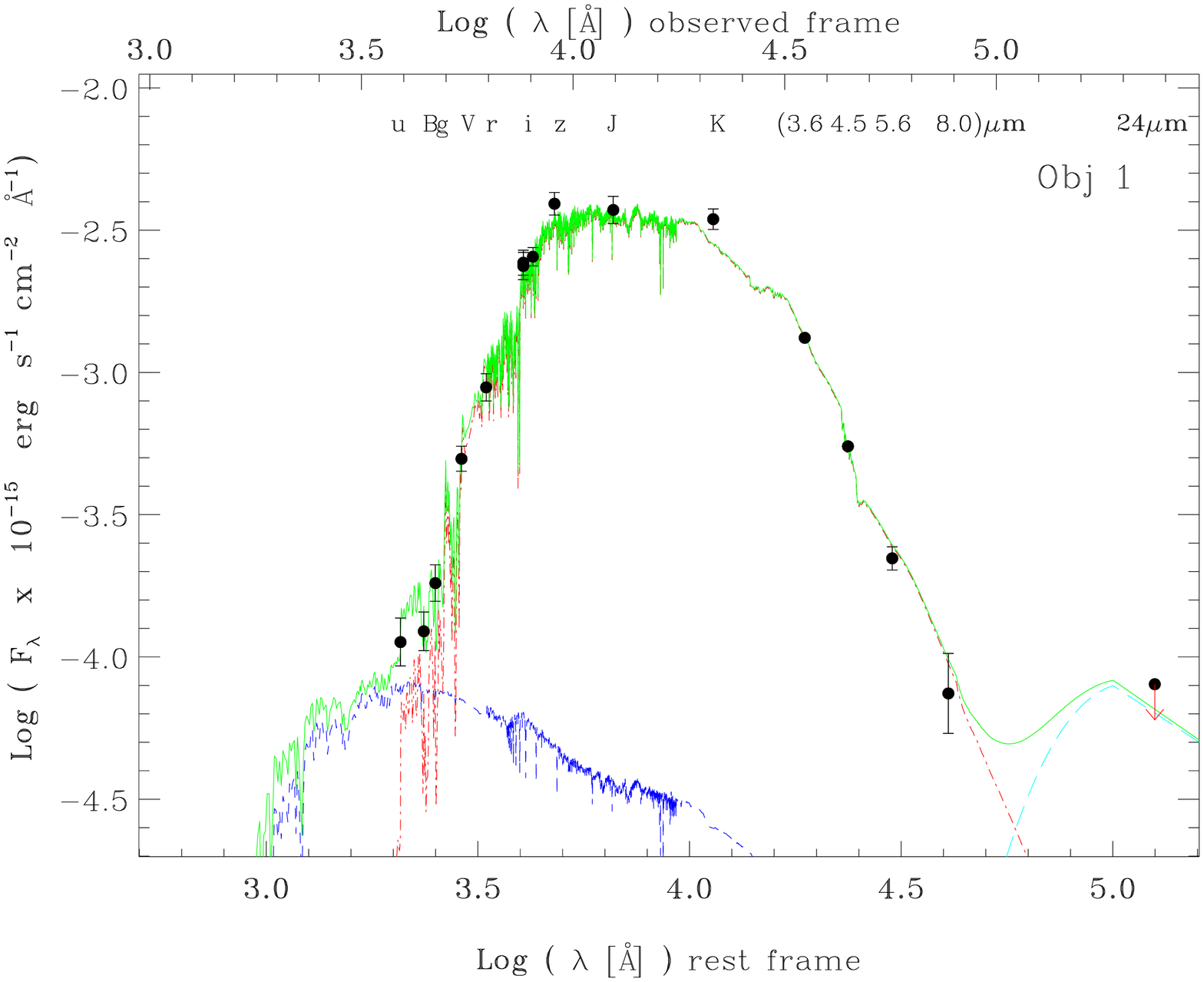} \\
\vspace{1em}
\includegraphics[scale=0.33,angle=90]{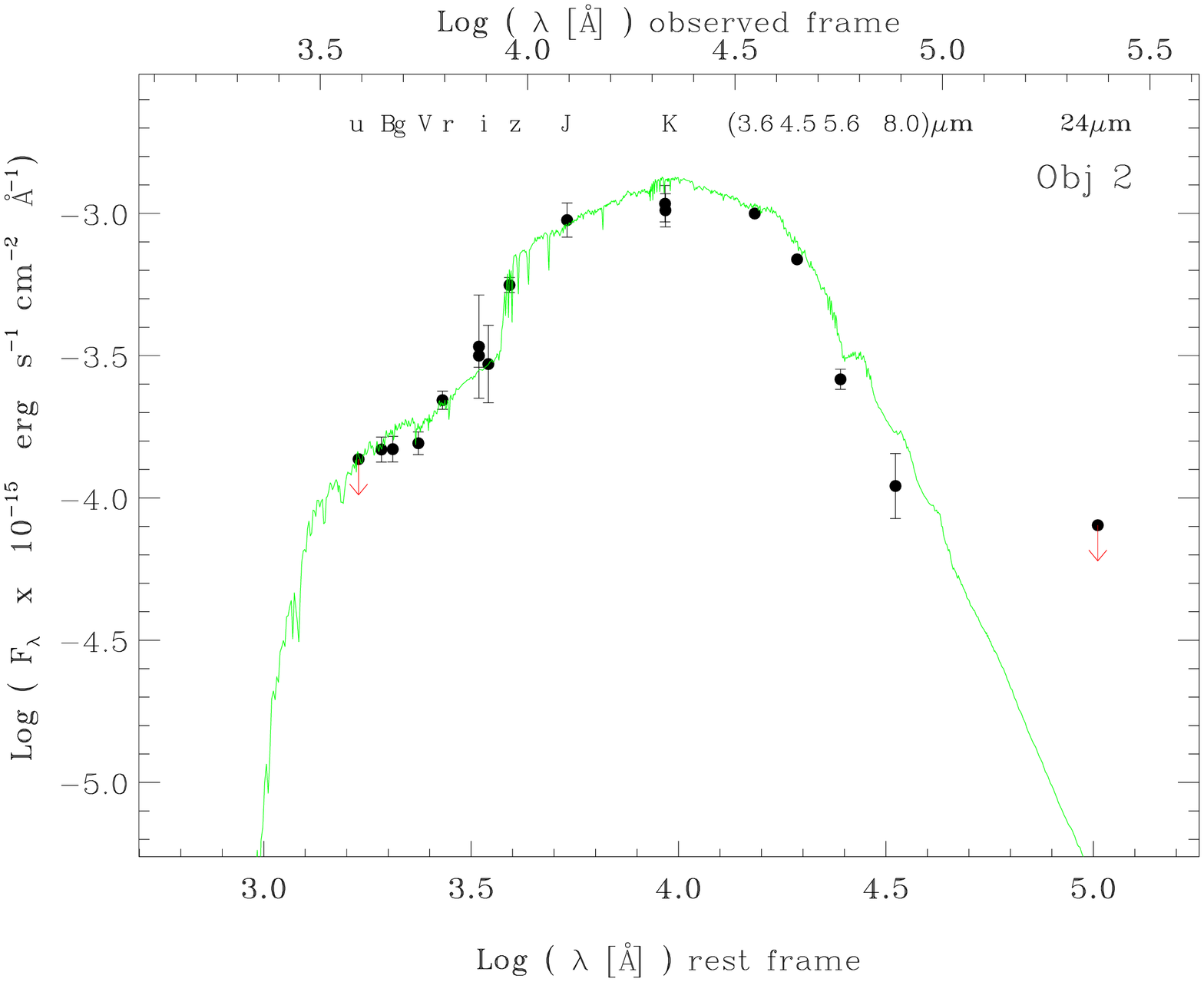} &
\includegraphics[scale=0.33,angle=90]{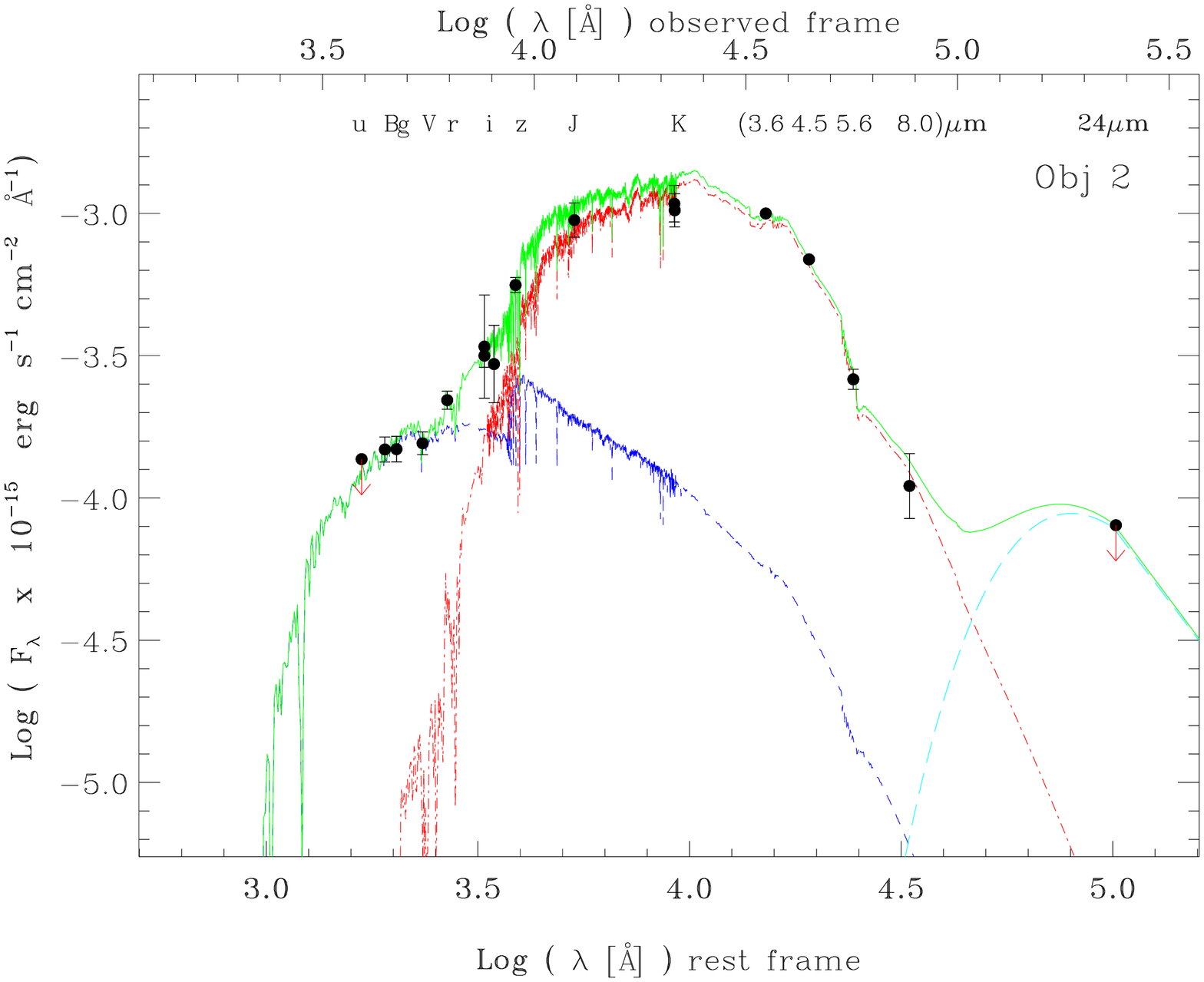} \\
\vspace{1em}
\includegraphics[scale=0.33,angle=90]{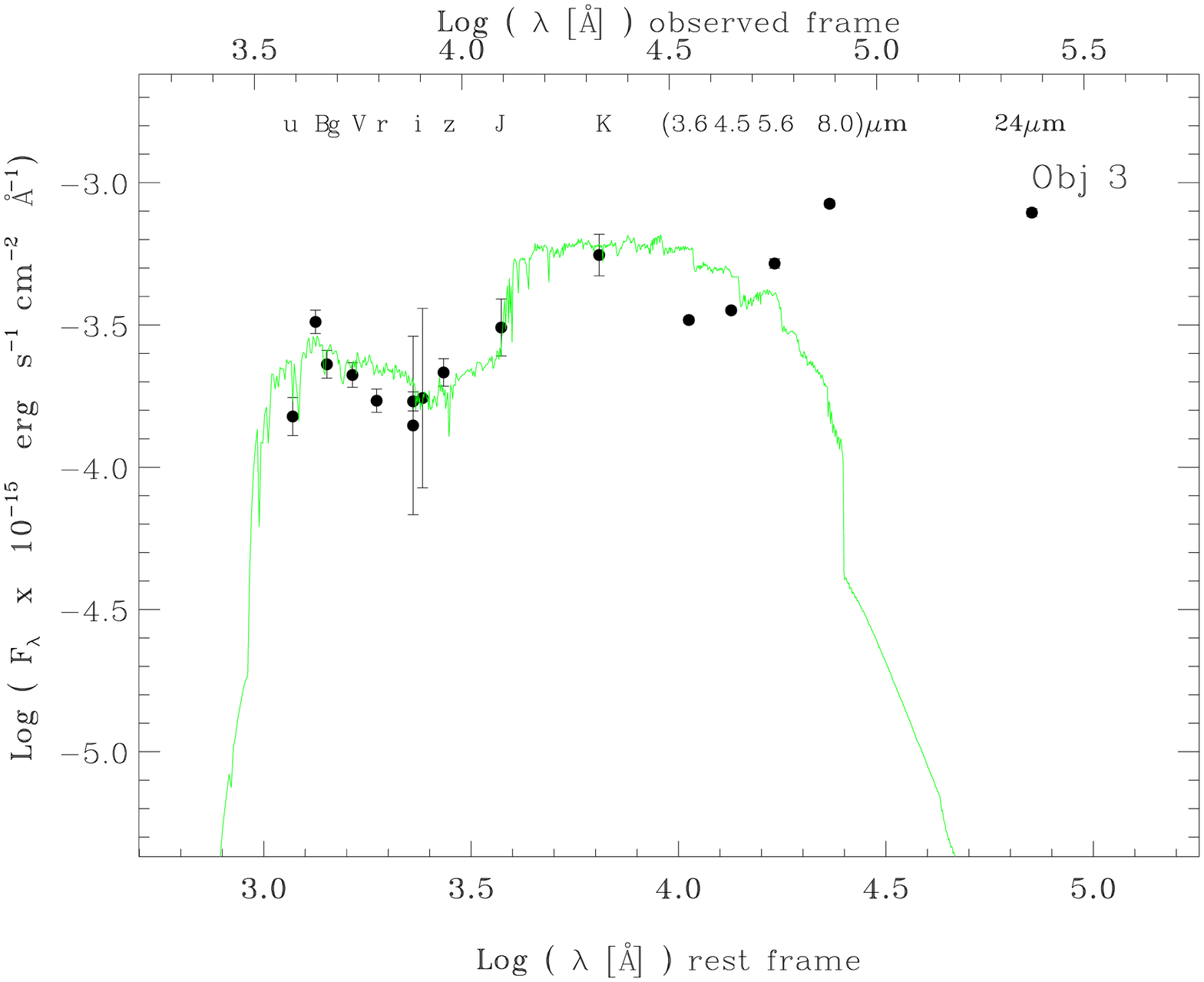} &
\includegraphics[scale=0.33,angle=90]{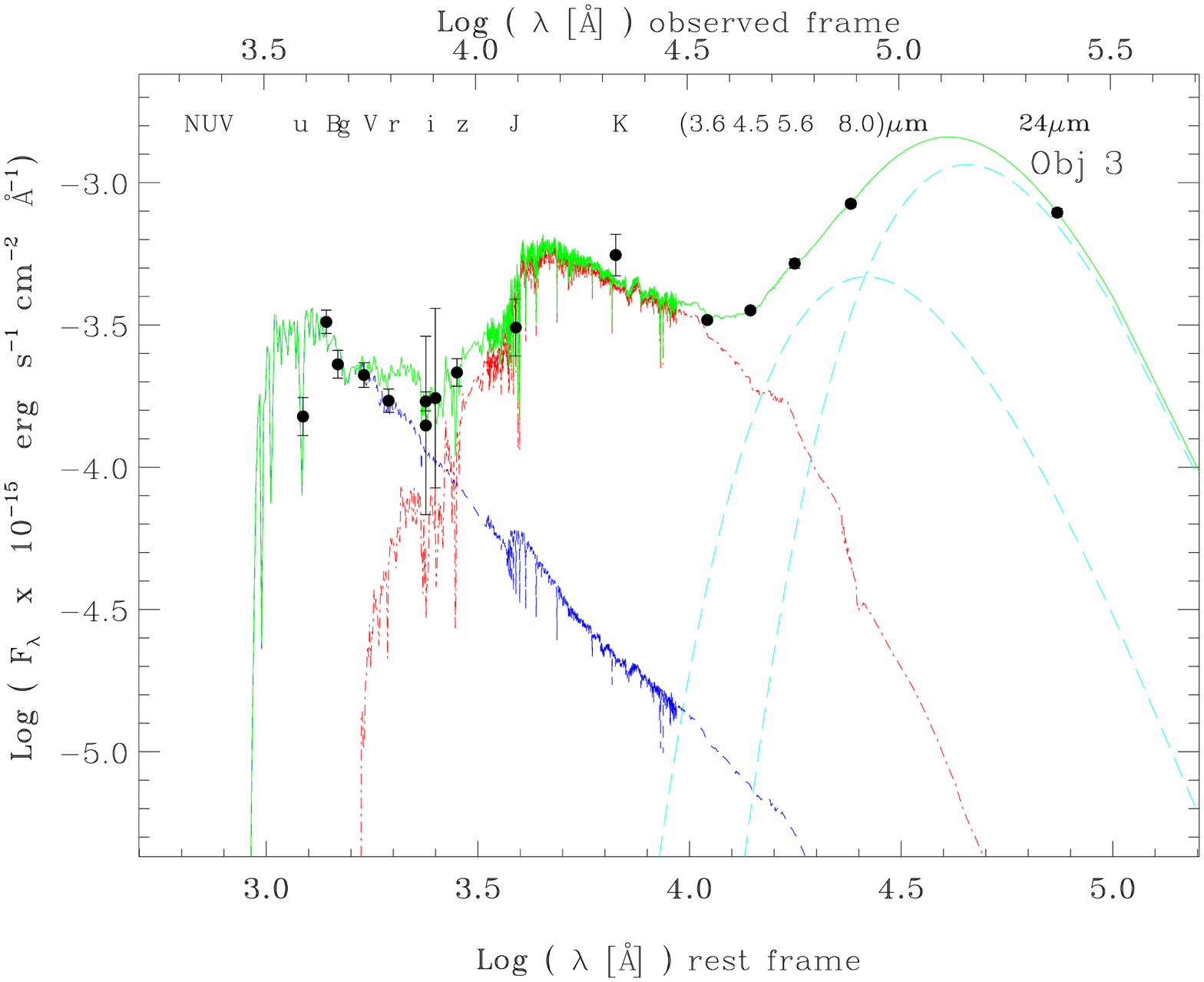} \\
\vspace{1em}
\includegraphics[scale=0.33,angle=90]{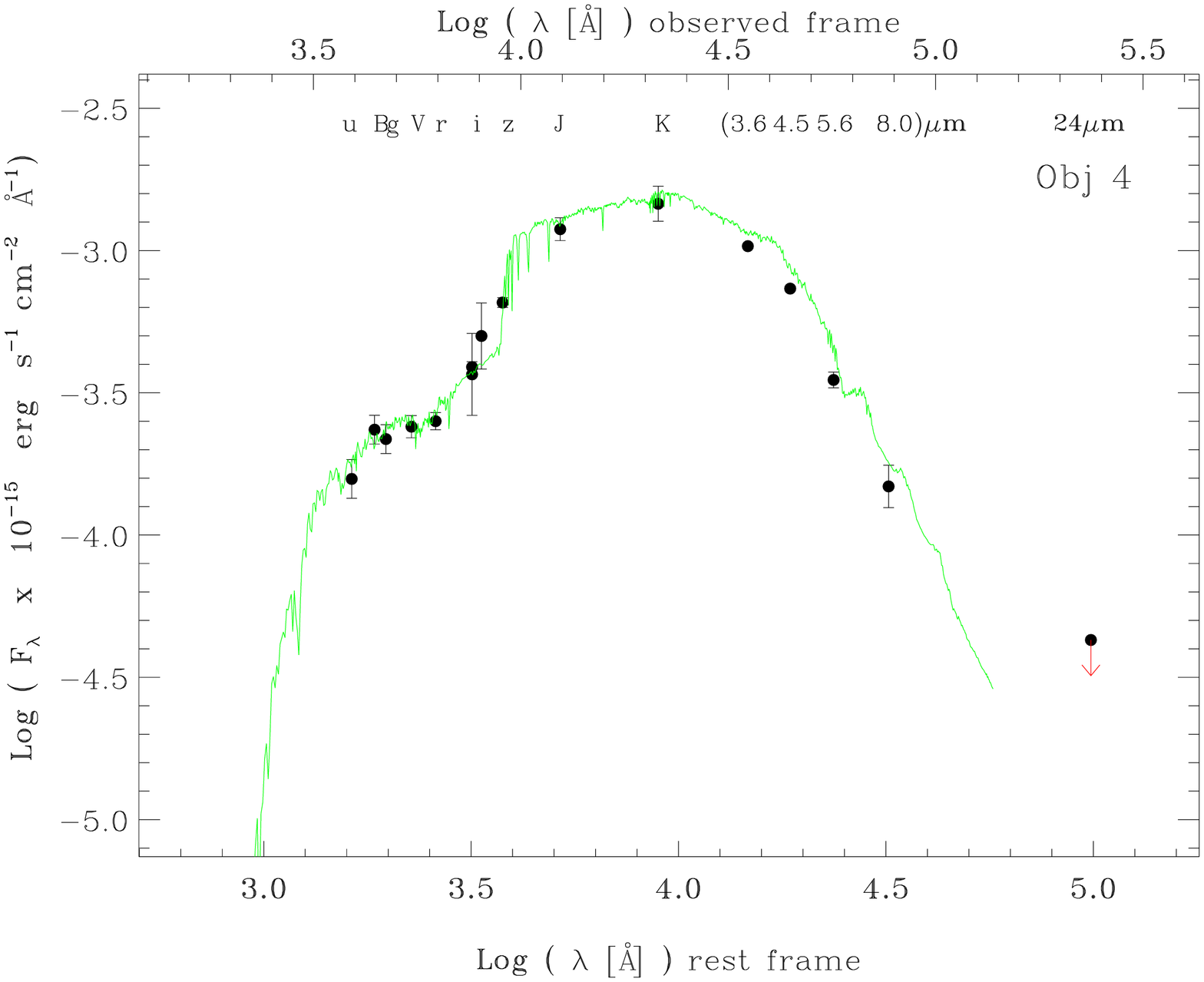} &
\includegraphics[scale=0.33,angle=90]{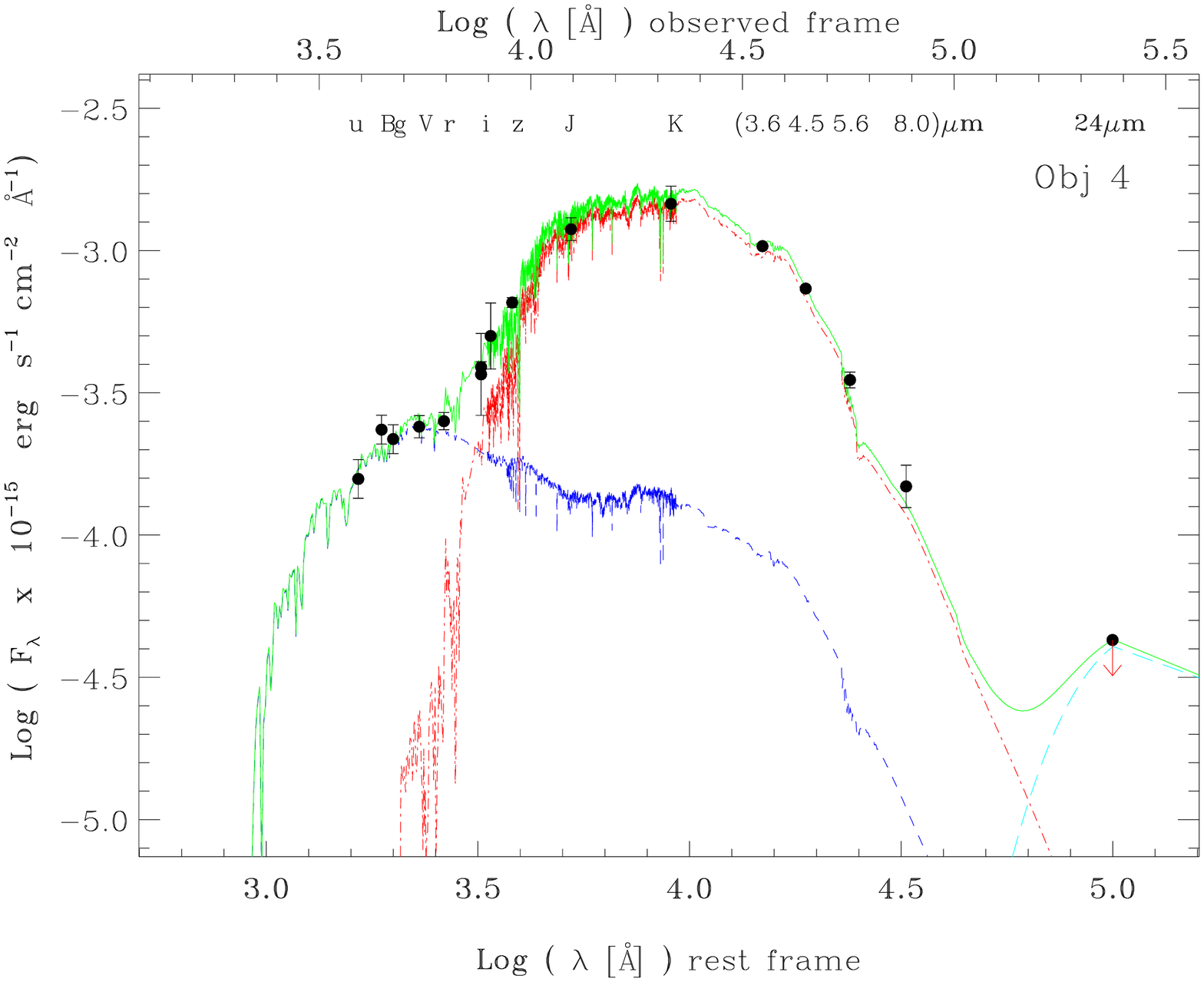} 
\end{array}$
\end{center}
\caption{Left panels: SEDs of the sample (green line) fitting the photometric
  points, as result of {\it Hyperz}. We plot the photometric point at 24
  $\mu$m (Spitzer/MIPS), if available, but is not included in the
  modeling. Right panels: SEDs of the objects fitted with {\it 2SPD}. The total
  model is the green line, the YSP is the blue line, the OSP is the red line,
  and the dust component(s) is the light blue line. For both the panels, the
  wavelengths on top of the plots correspond to observed wavelengths, while
  those on bottom are at rest frame. For object 236 (QSO) we only show the
  photometric points.}
\label{sed1a}
\end{figure*}

\addtocounter{figure}{-1}
\begin{figure*}[h]
\begin{center}$
\begin{array}{ccc}
\vspace{2em}
\includegraphics[scale=0.33,angle=90]{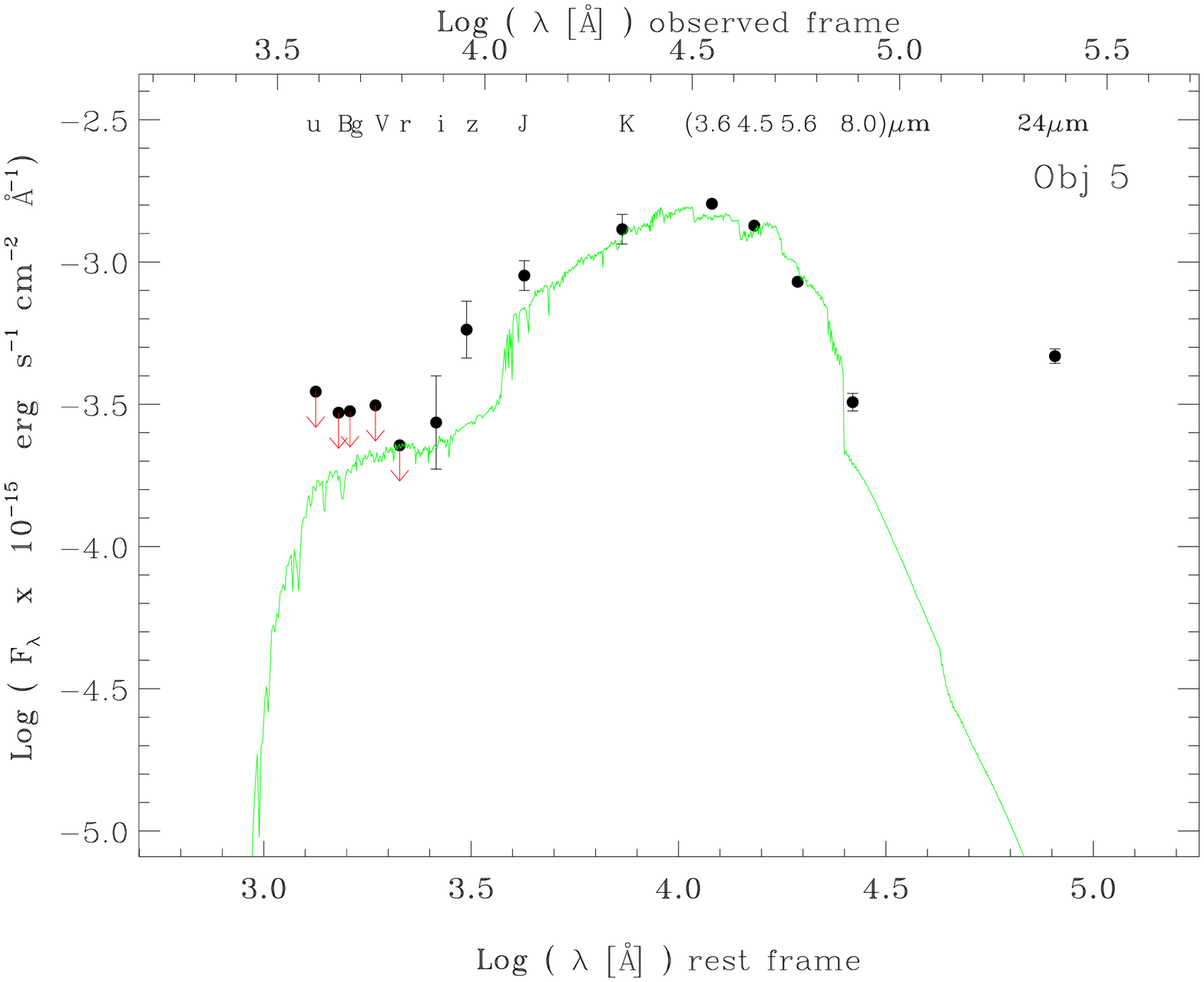} &
\includegraphics[scale=0.33,angle=90]{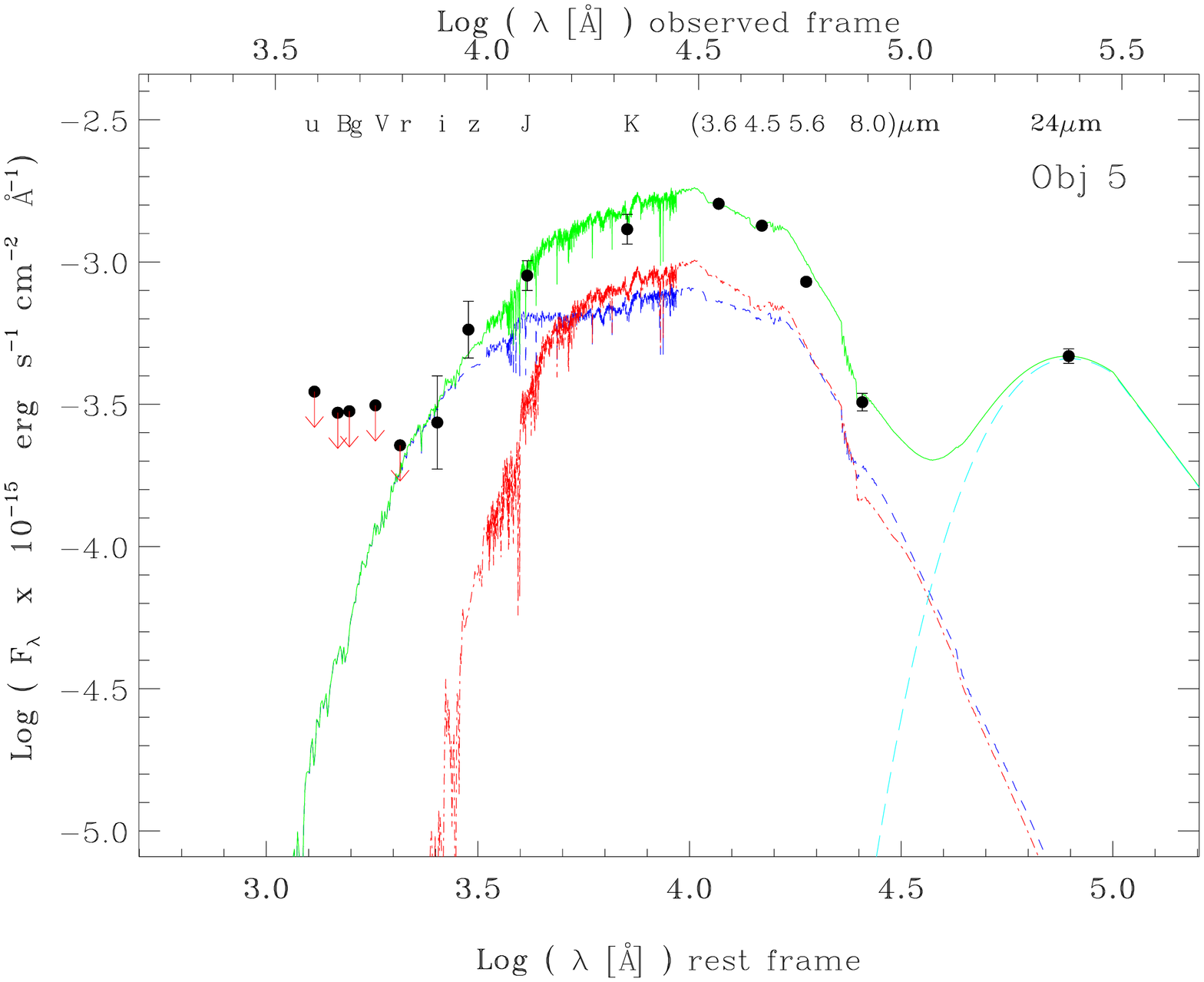} \\
\vspace{1em}
\includegraphics[scale=0.33,angle=90]{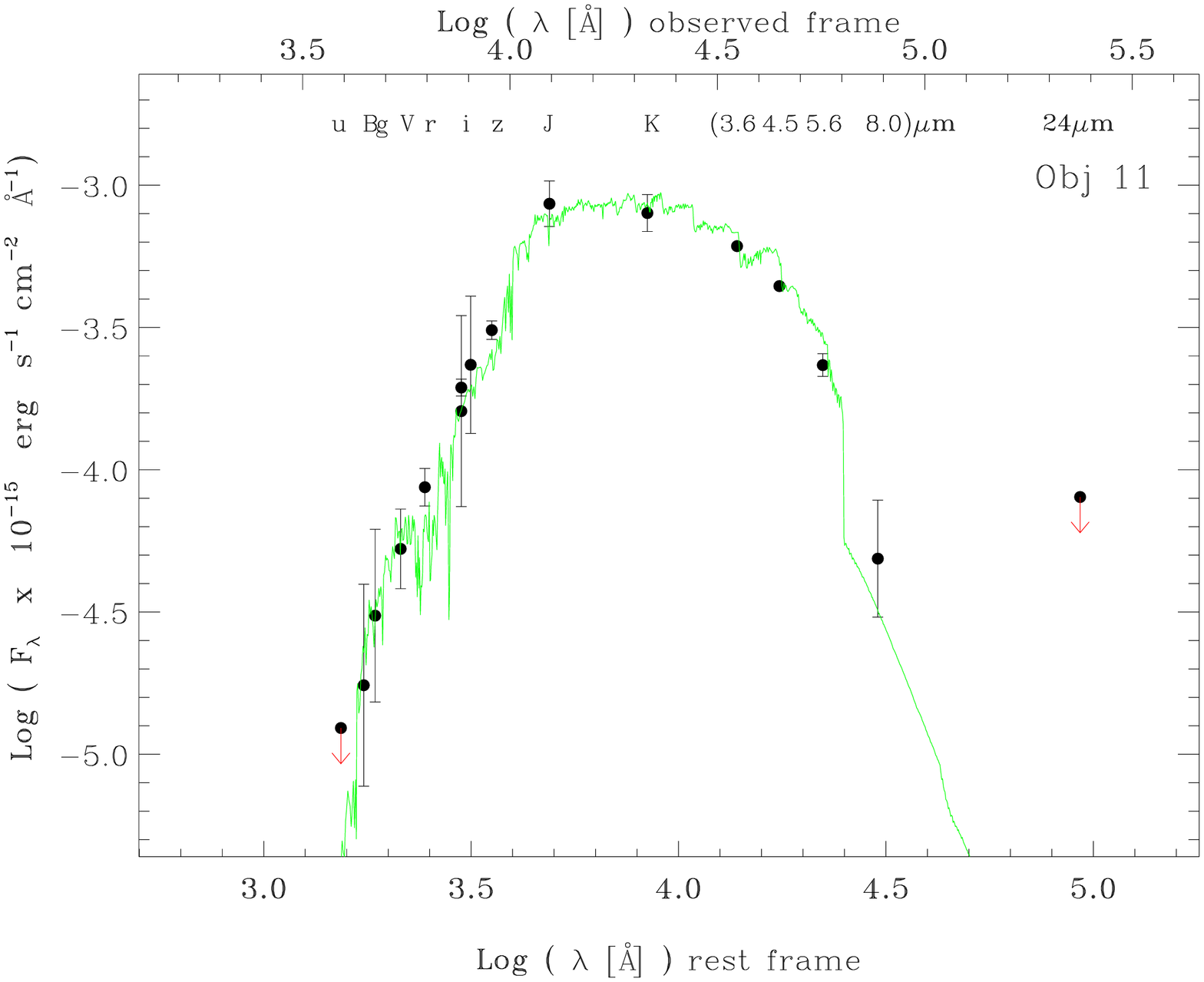} &
\includegraphics[scale=0.33,angle=90]{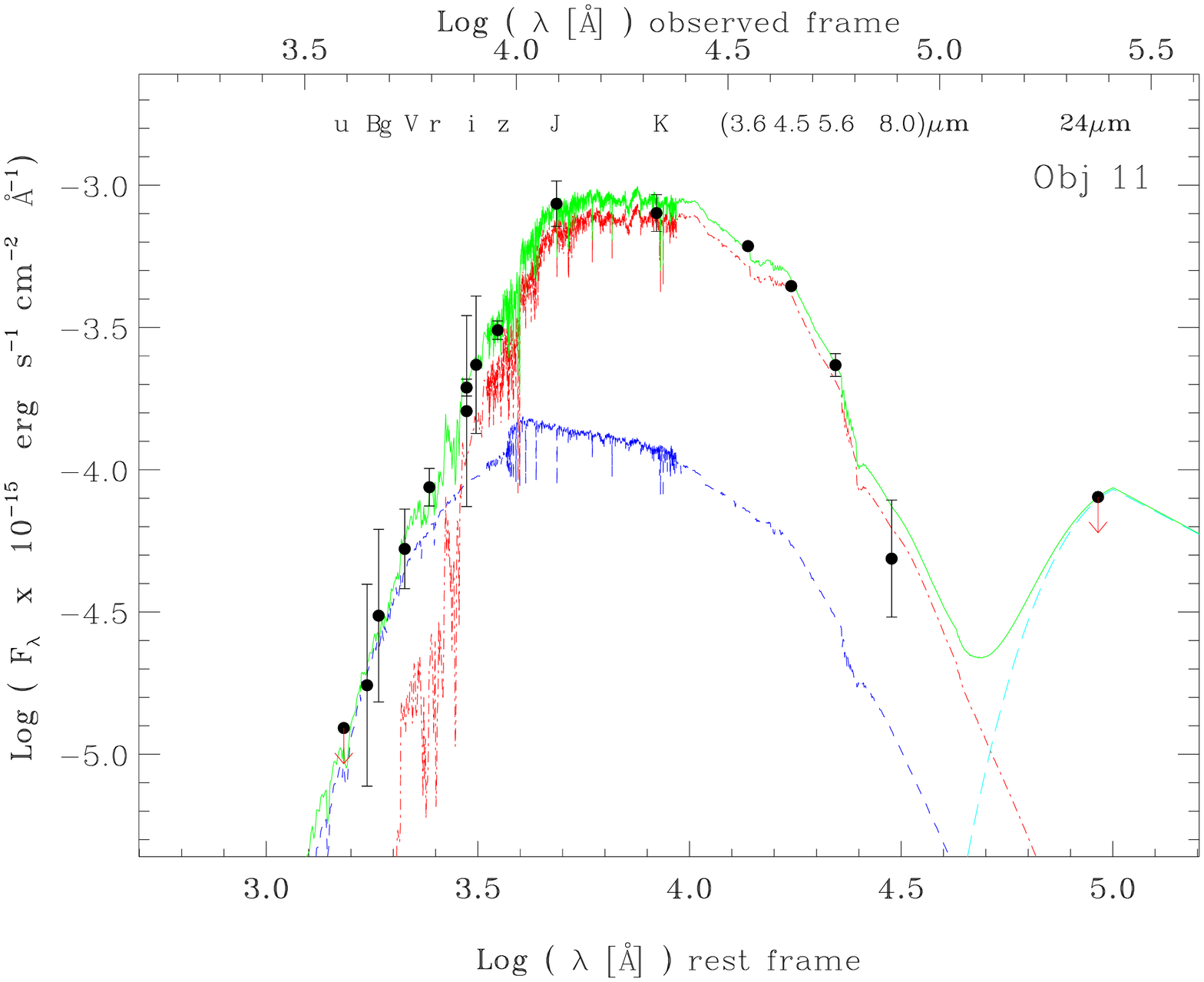} \\
\vspace{2em}
\includegraphics[scale=0.33,angle=90]{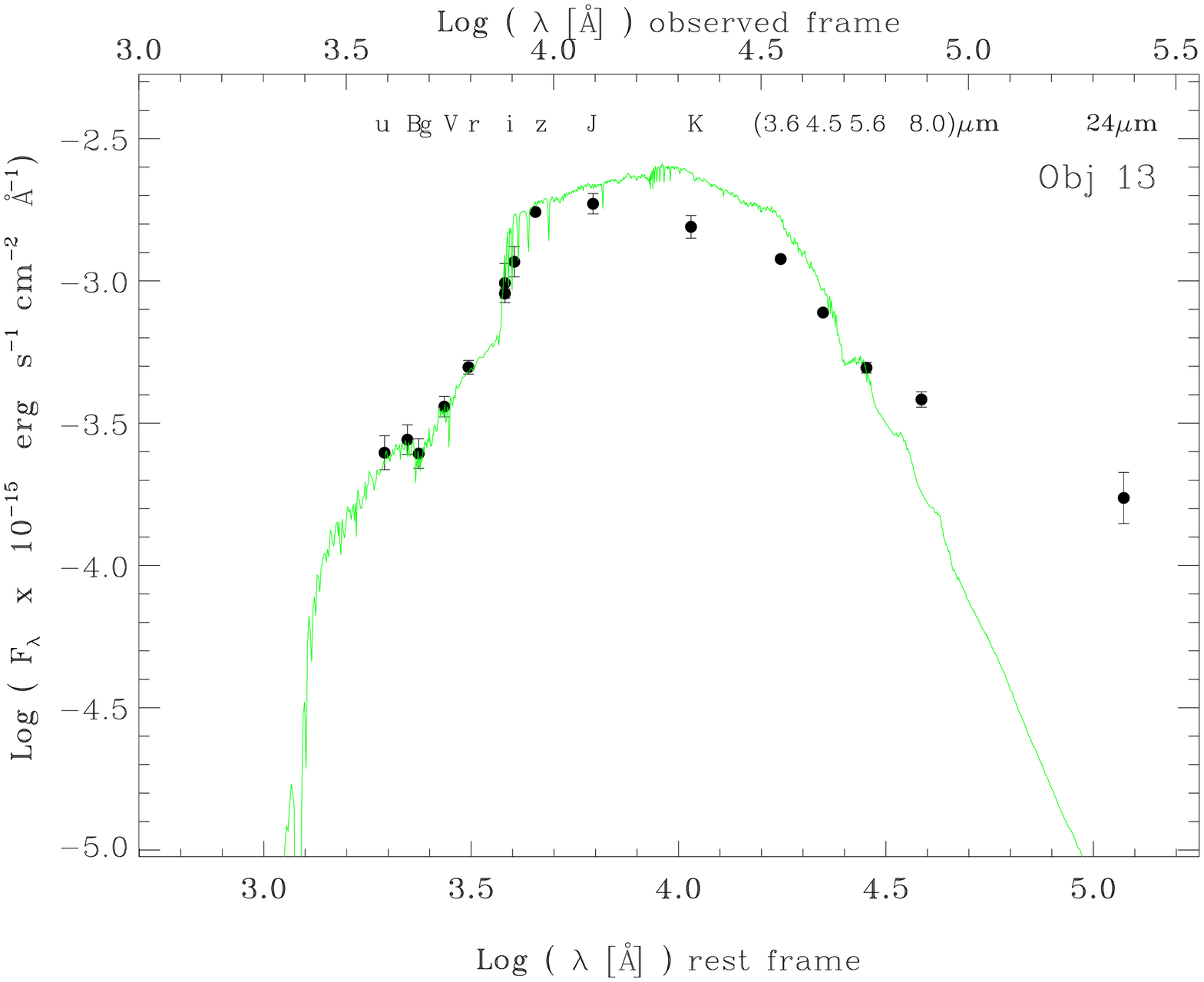} &
\includegraphics[scale=0.33,angle=90]{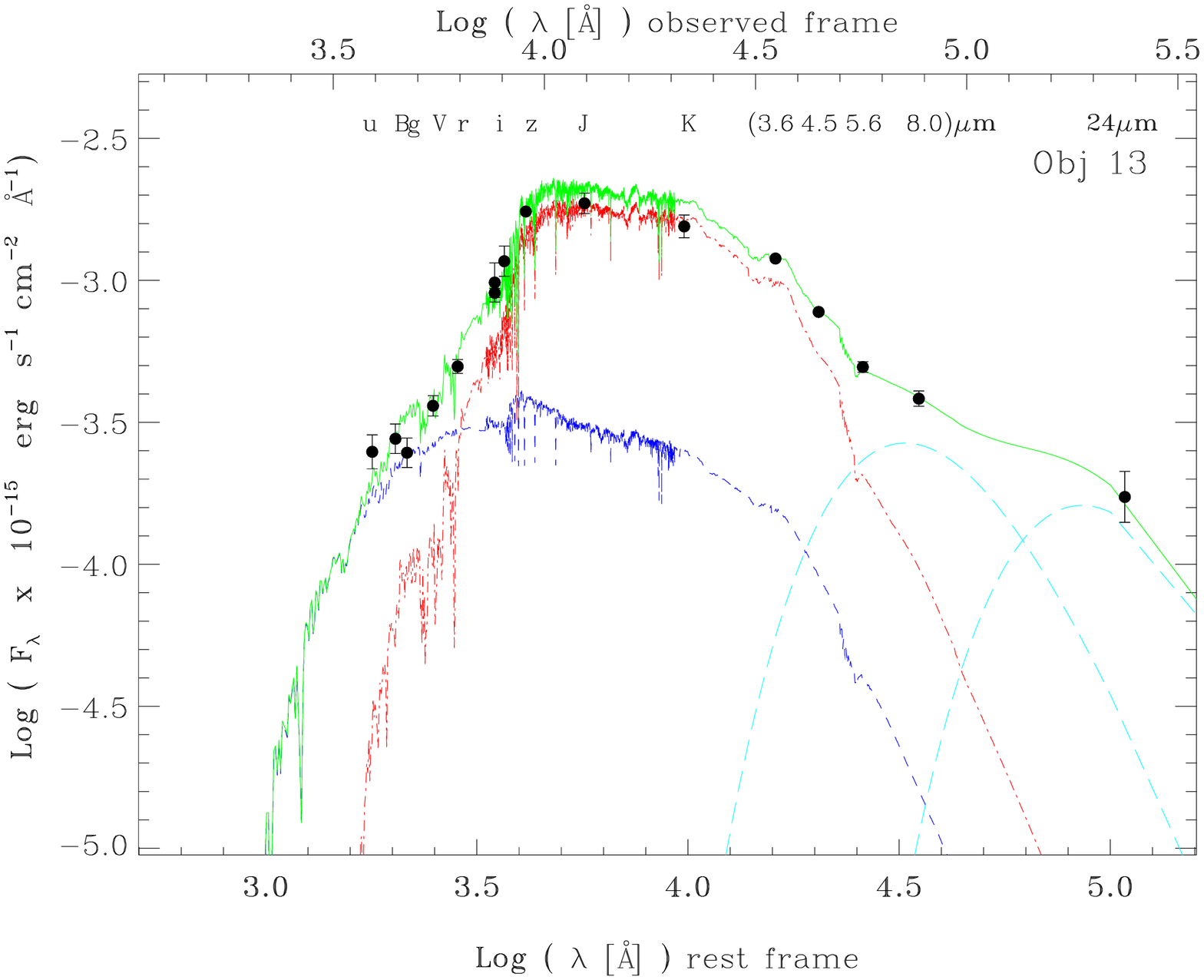} \\
\vspace{1em}
\includegraphics[scale=0.33,angle=90]{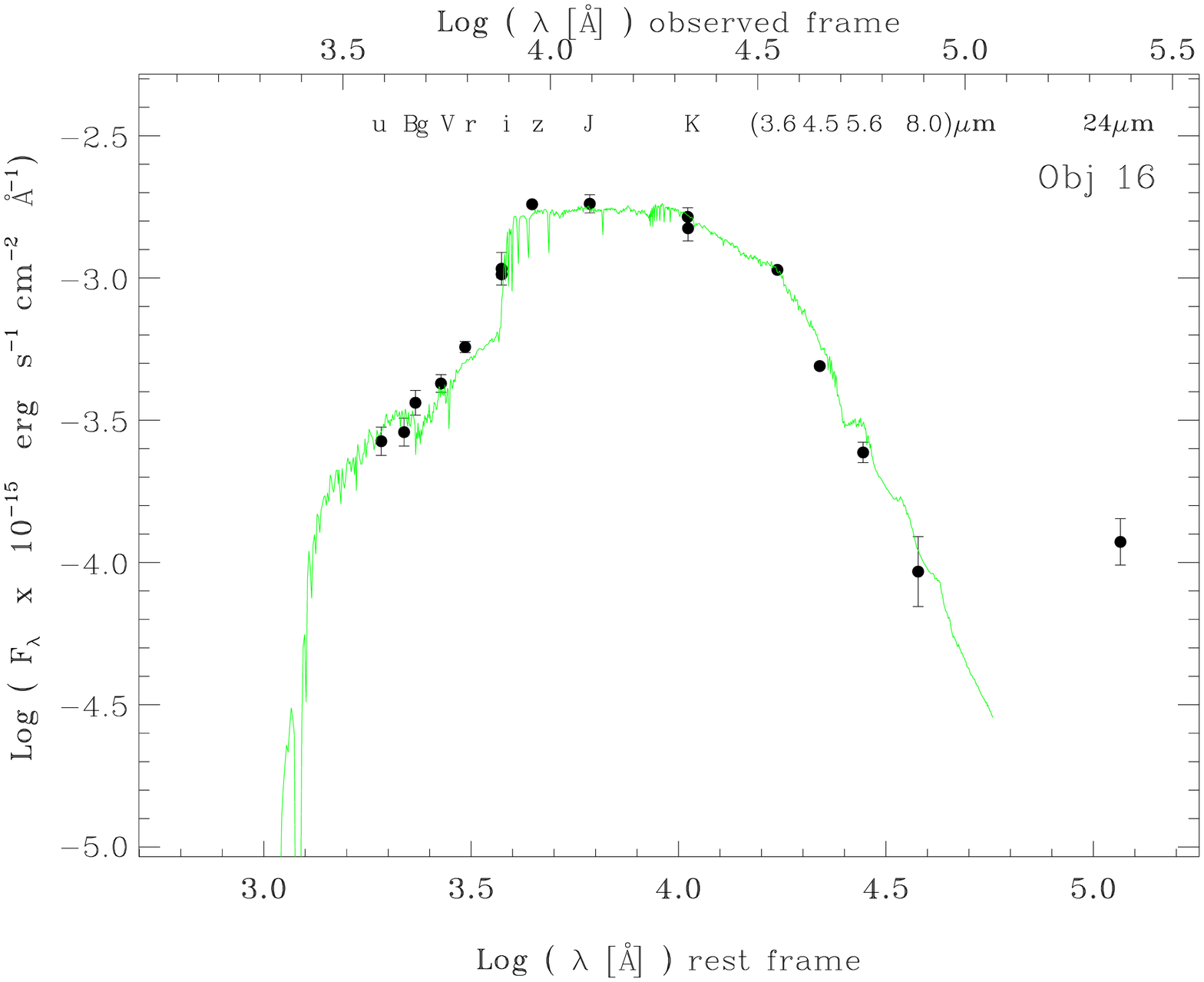} &
\includegraphics[scale=0.33,angle=90]{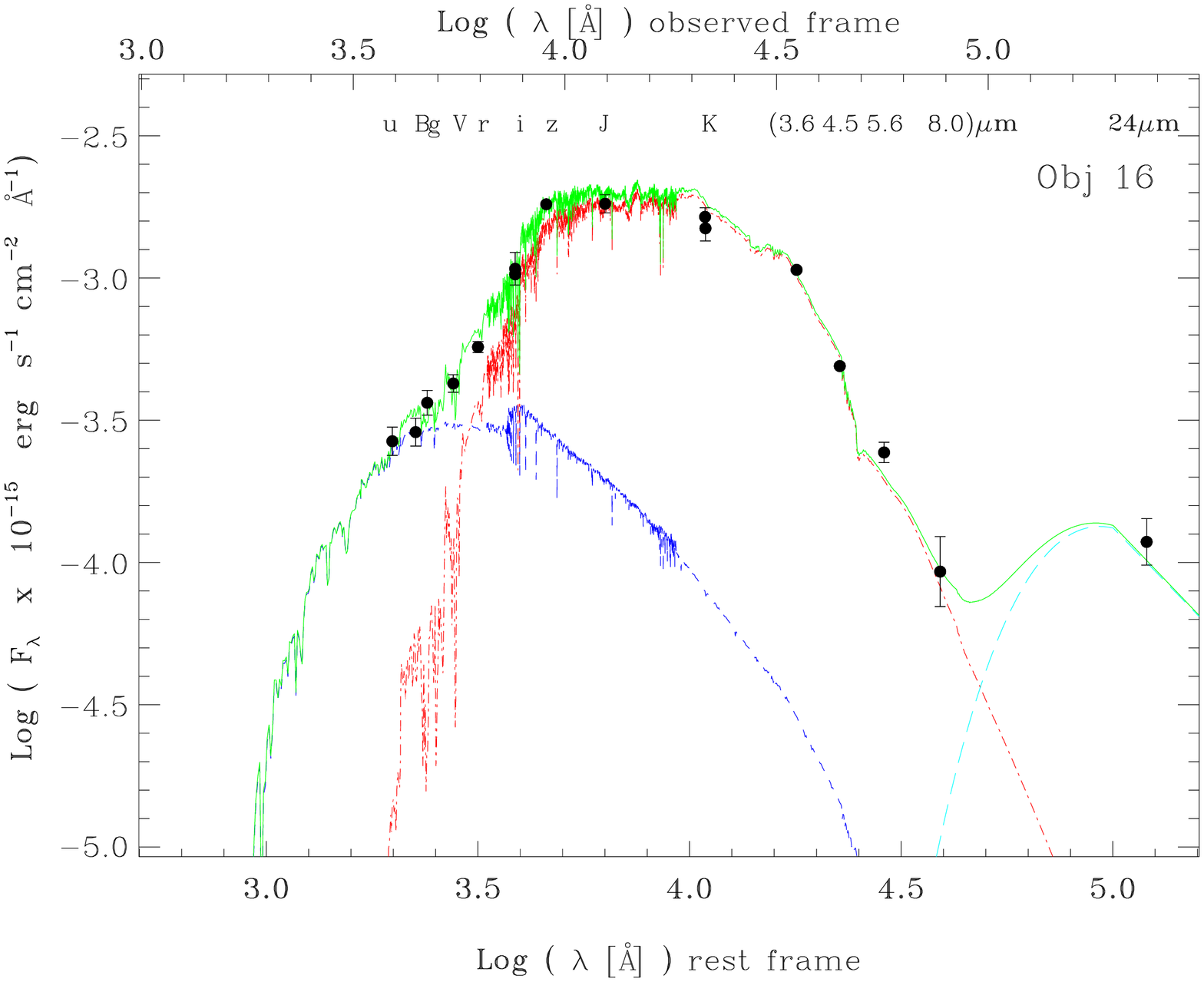} 
\end{array}$
\end{center}
\caption{Continued}
\end{figure*}

\addtocounter{figure}{-1}
\begin{figure*}[h]
\begin{center}$
\begin{array}{ccc}
\vspace{2em}
\includegraphics[scale=0.33,angle=90]{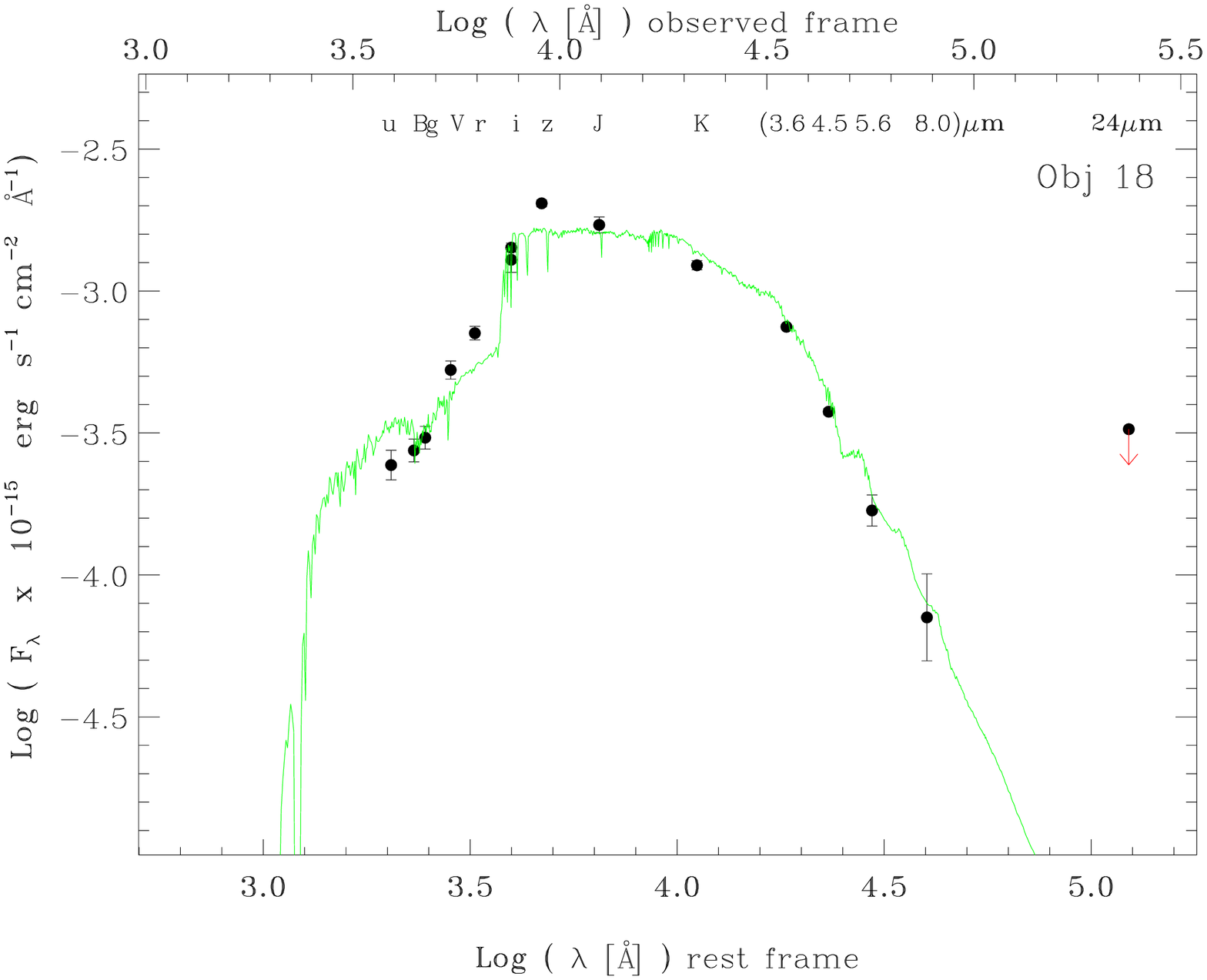} &
\includegraphics[scale=0.33,angle=90]{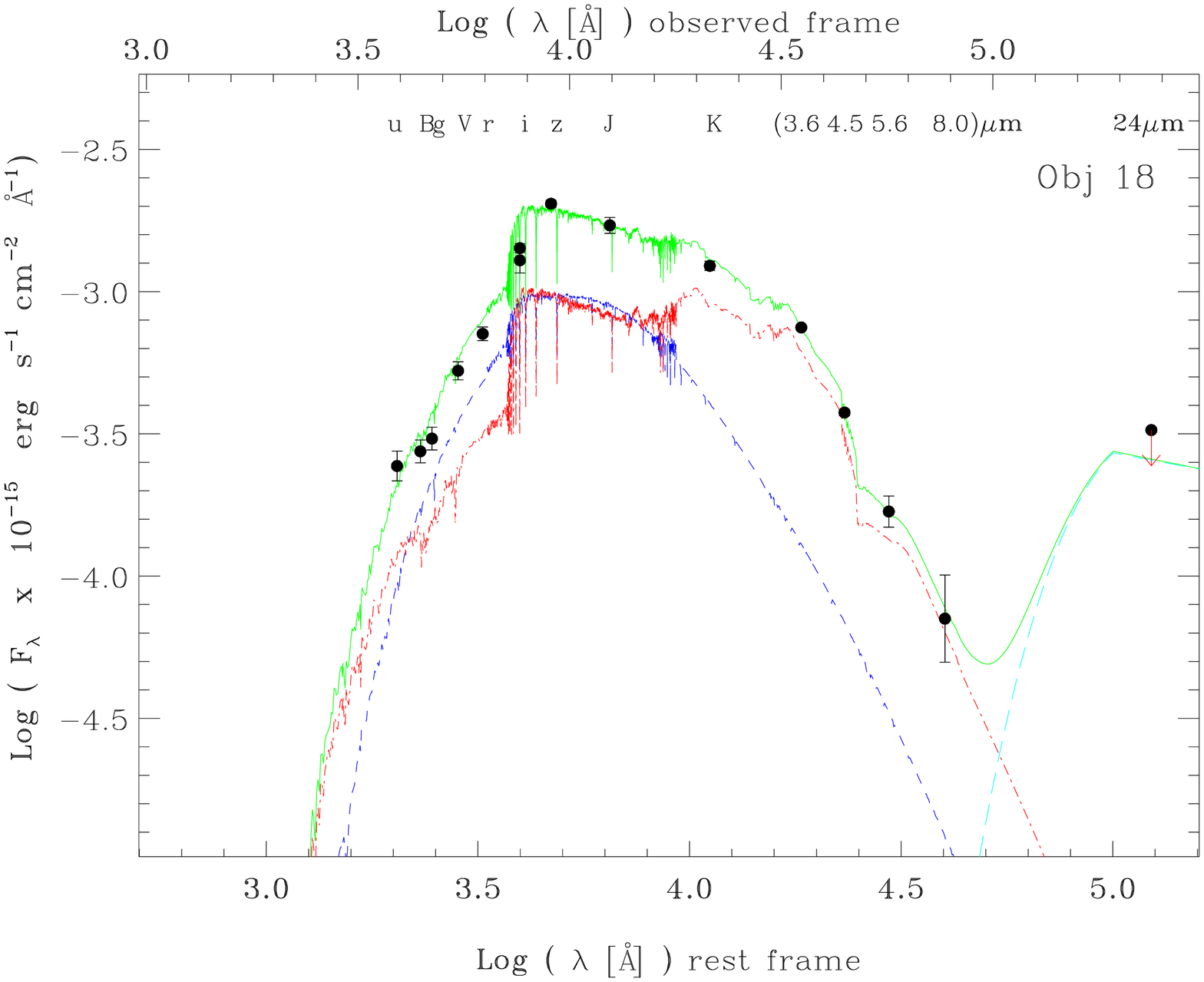} \\
\vspace{1em}
\includegraphics[scale=0.33,angle=90]{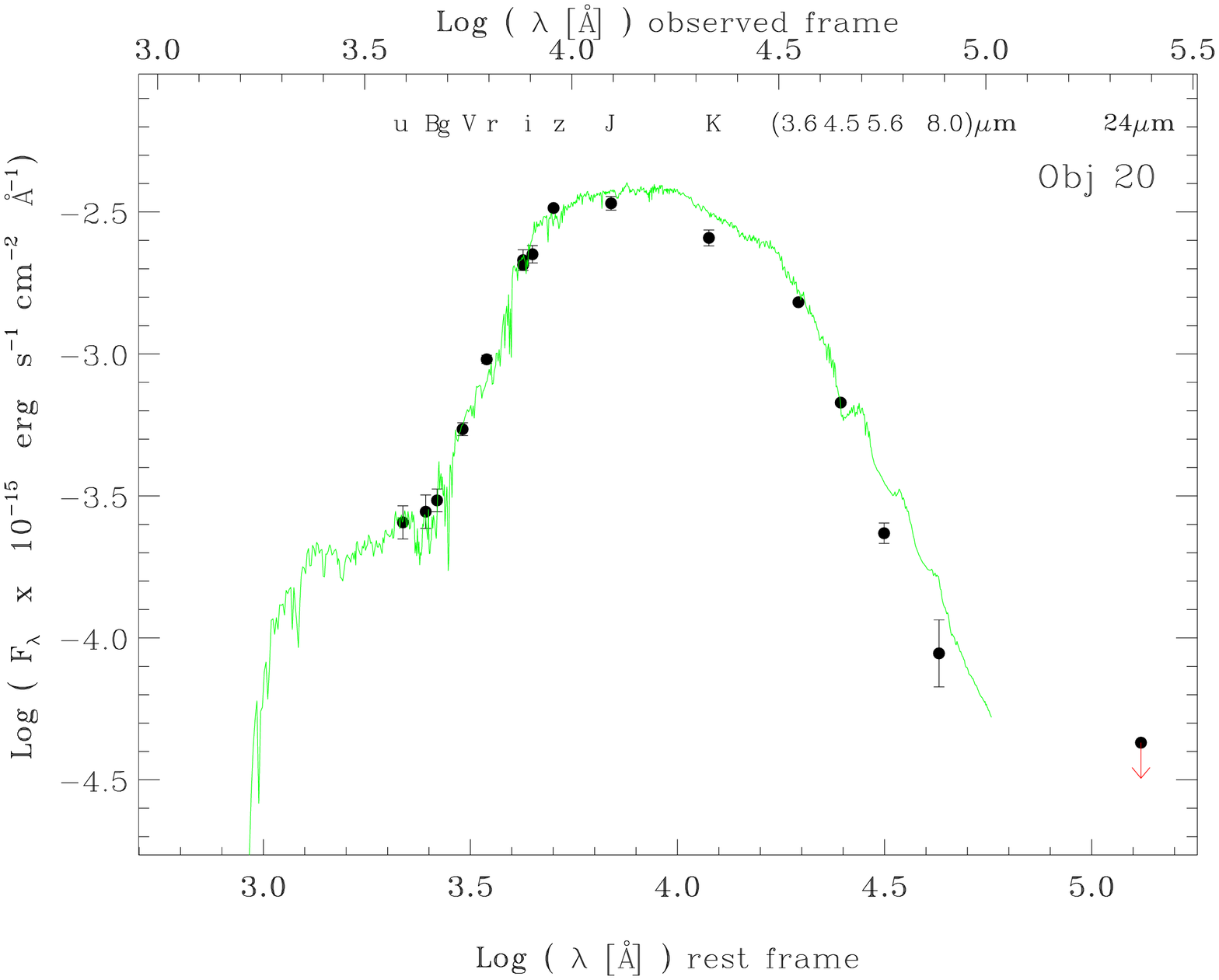} &
\includegraphics[scale=0.33,angle=90]{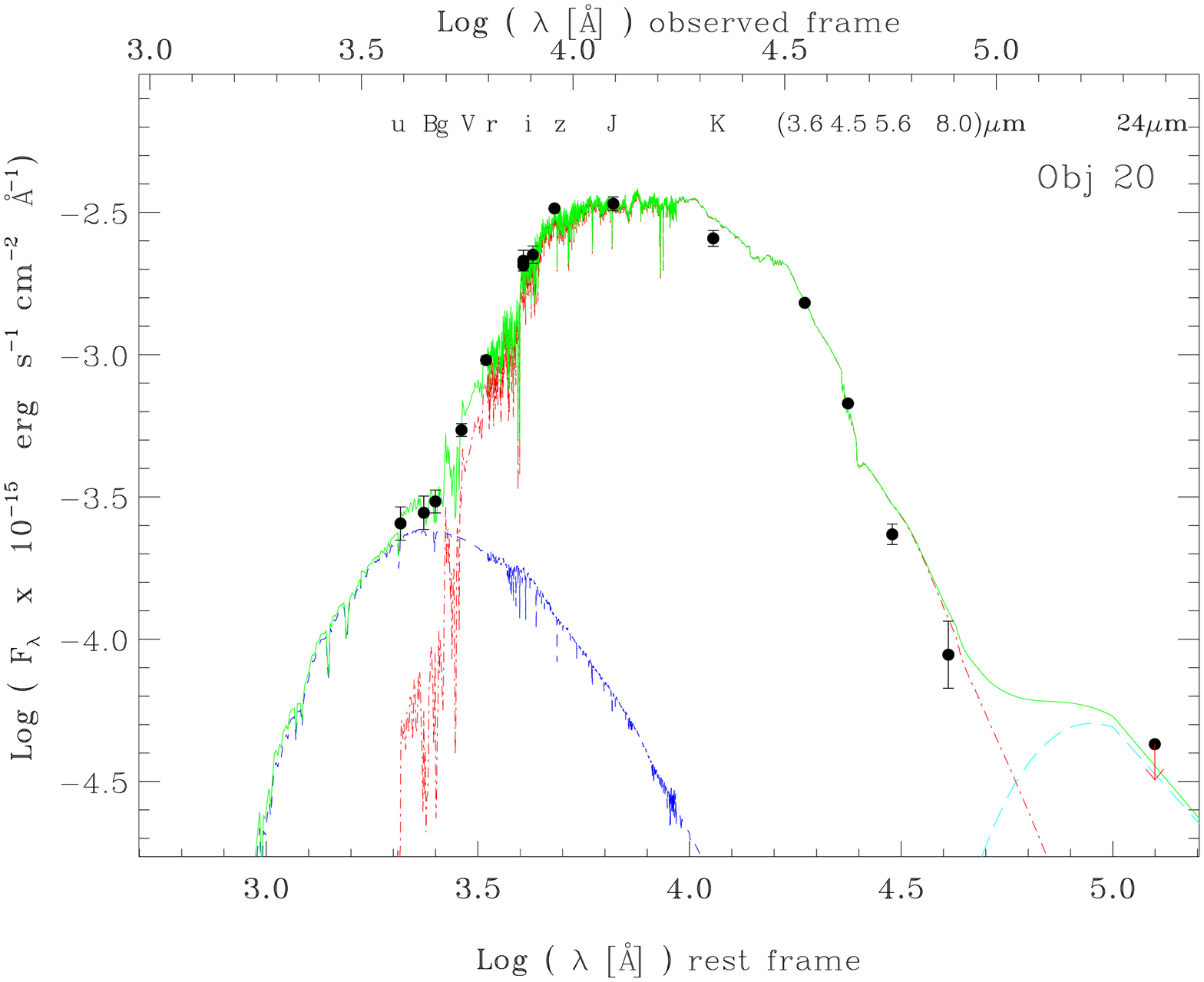} \\
\vspace{1em}
\includegraphics[scale=0.33,angle=90]{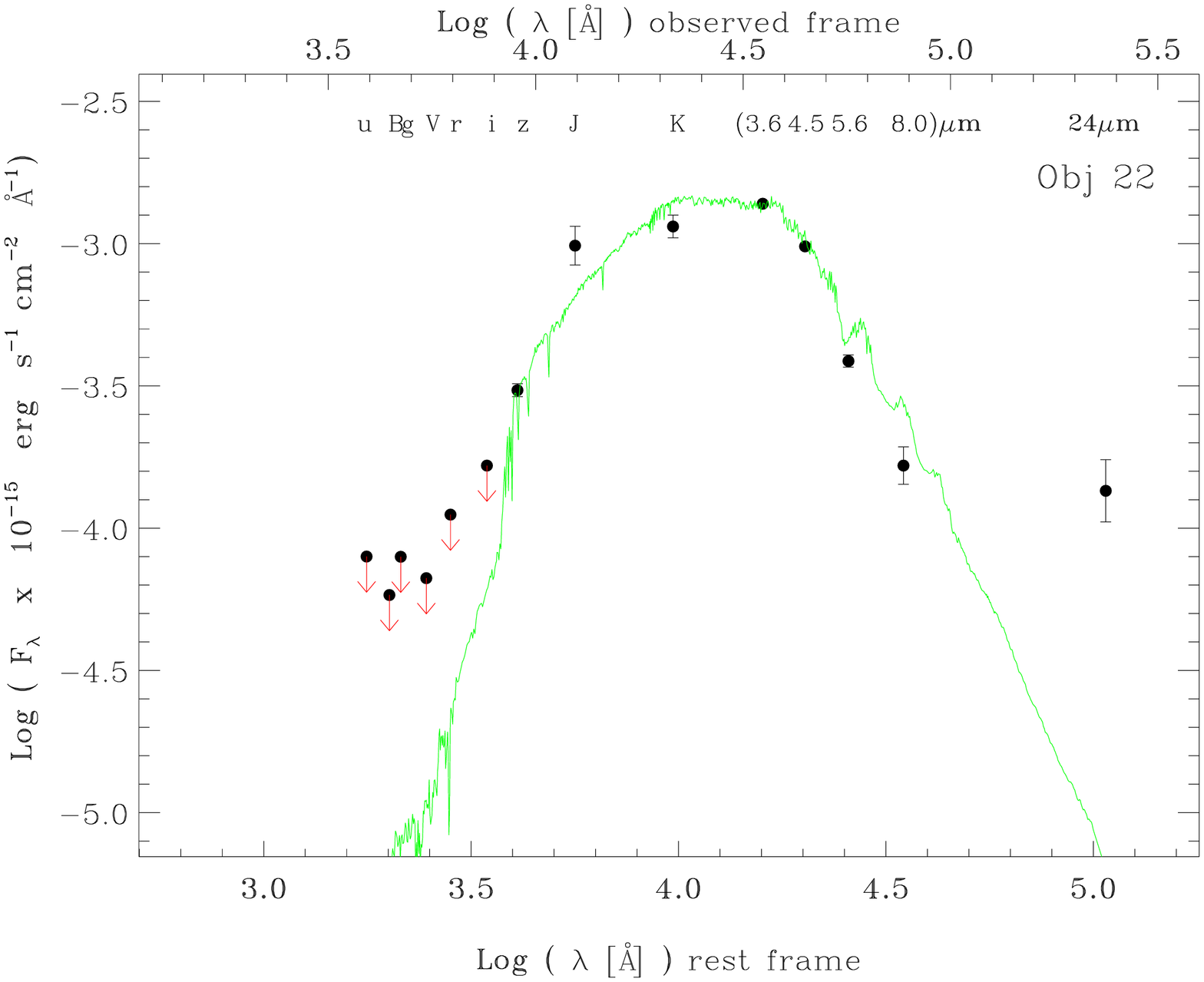} &
\includegraphics[scale=0.33,angle=90]{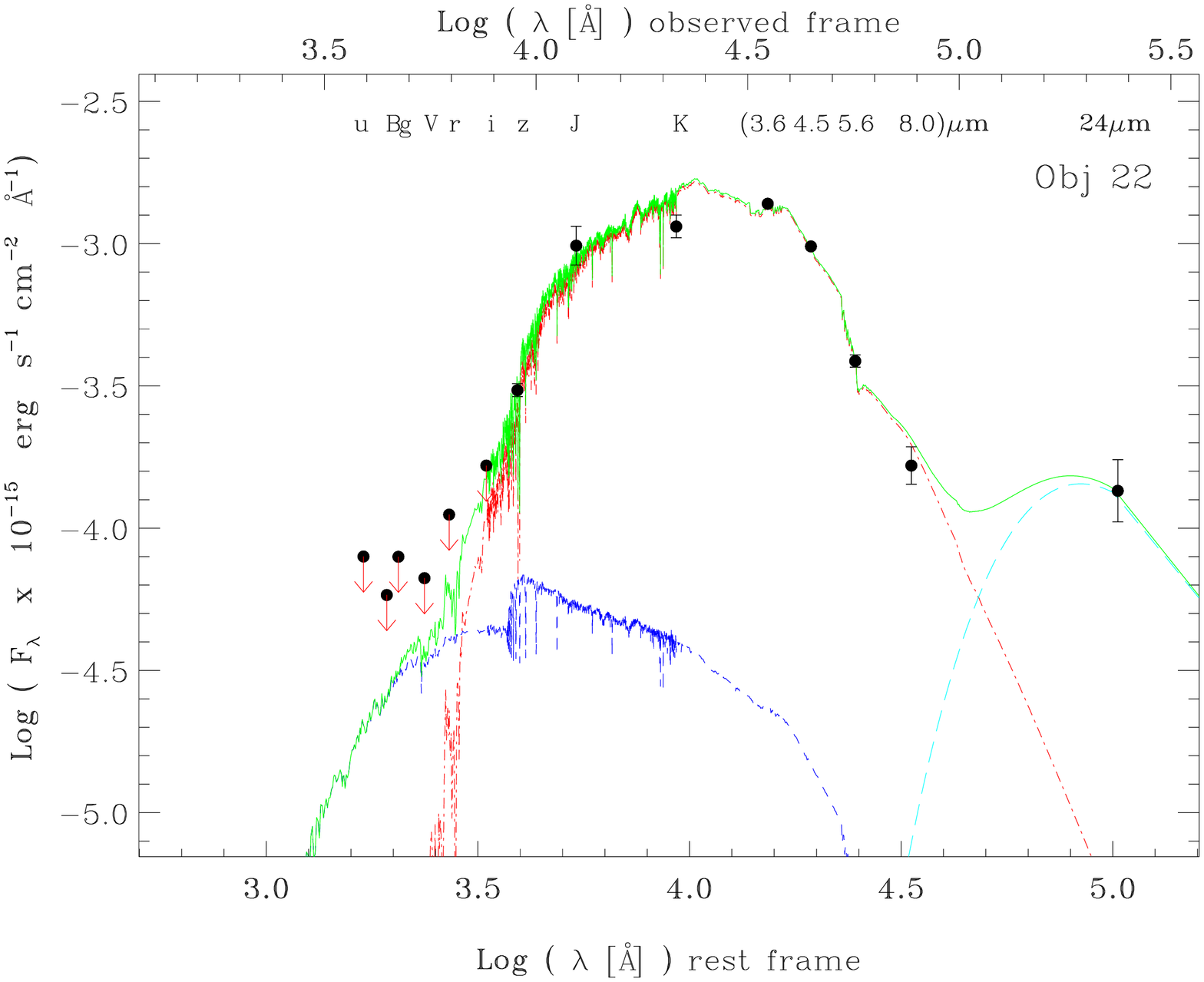} \\
\vspace{1em}
\includegraphics[scale=0.33,angle=90]{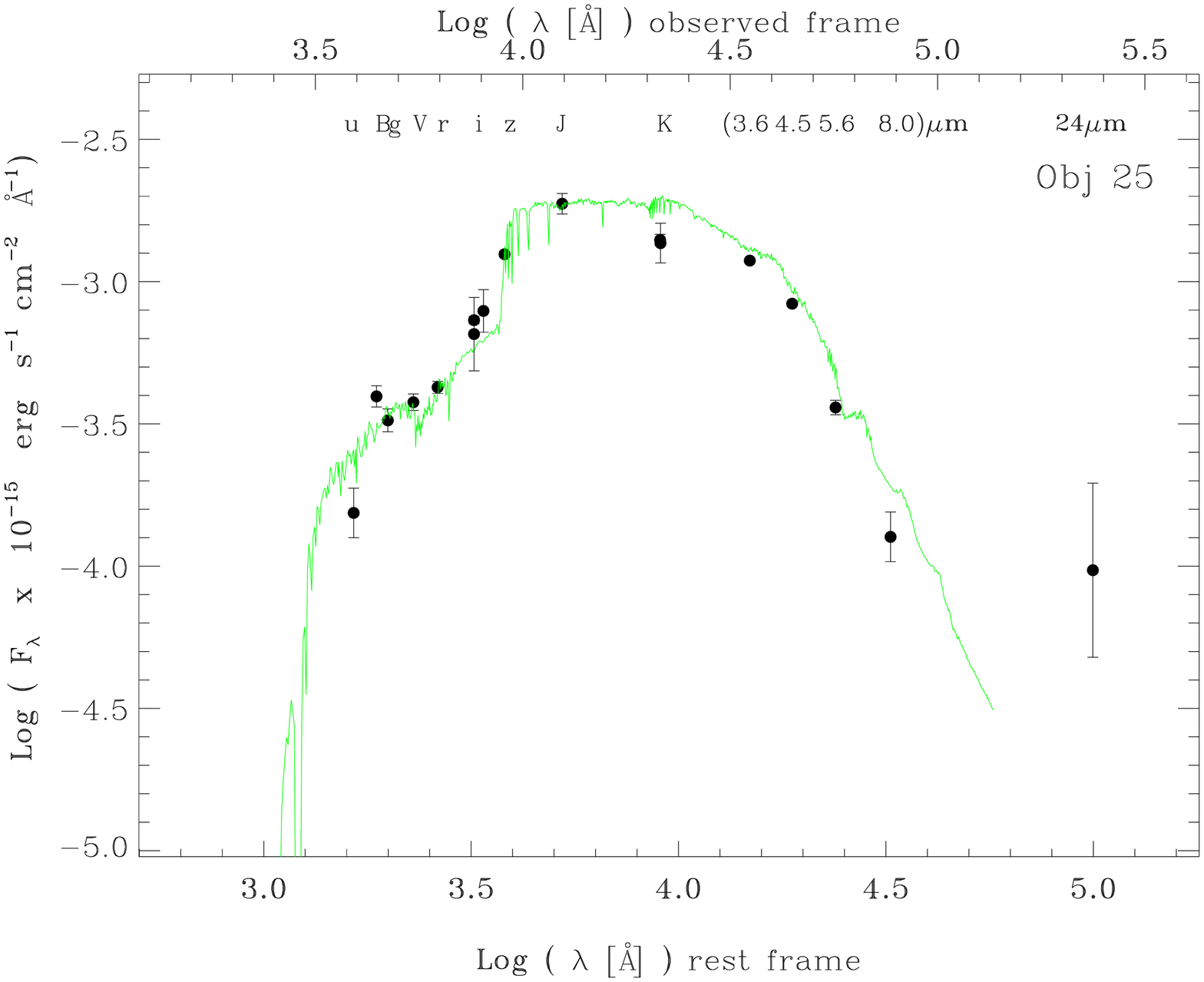} &
\includegraphics[scale=0.33,angle=90]{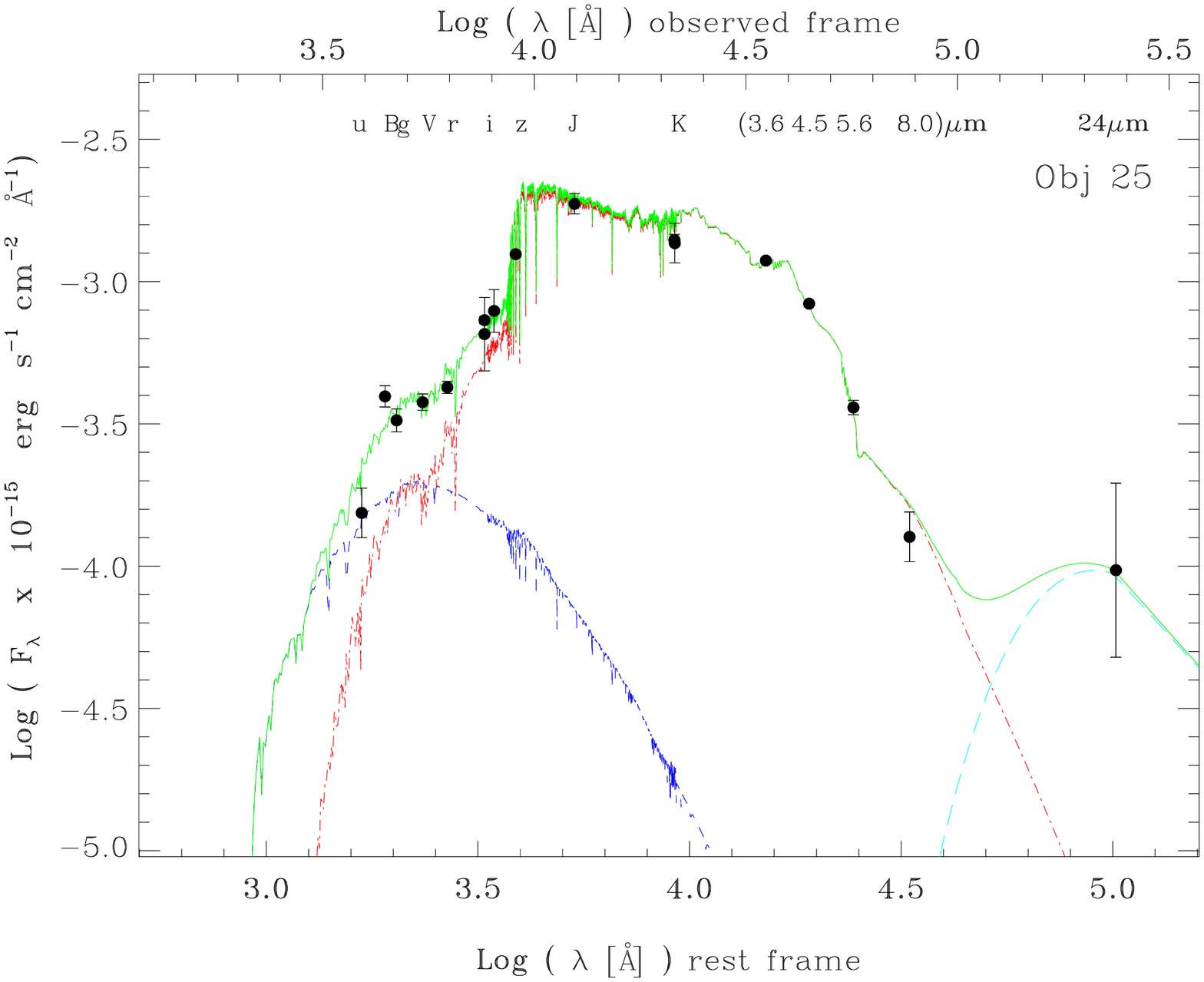} 
\end{array}$
\end{center}
\caption{Continued}
\end{figure*}

\addtocounter{figure}{-1}
\begin{figure*}[h]
\begin{center}$
\begin{array}{ccc}
\vspace{2em}
\includegraphics[scale=0.33,angle=90]{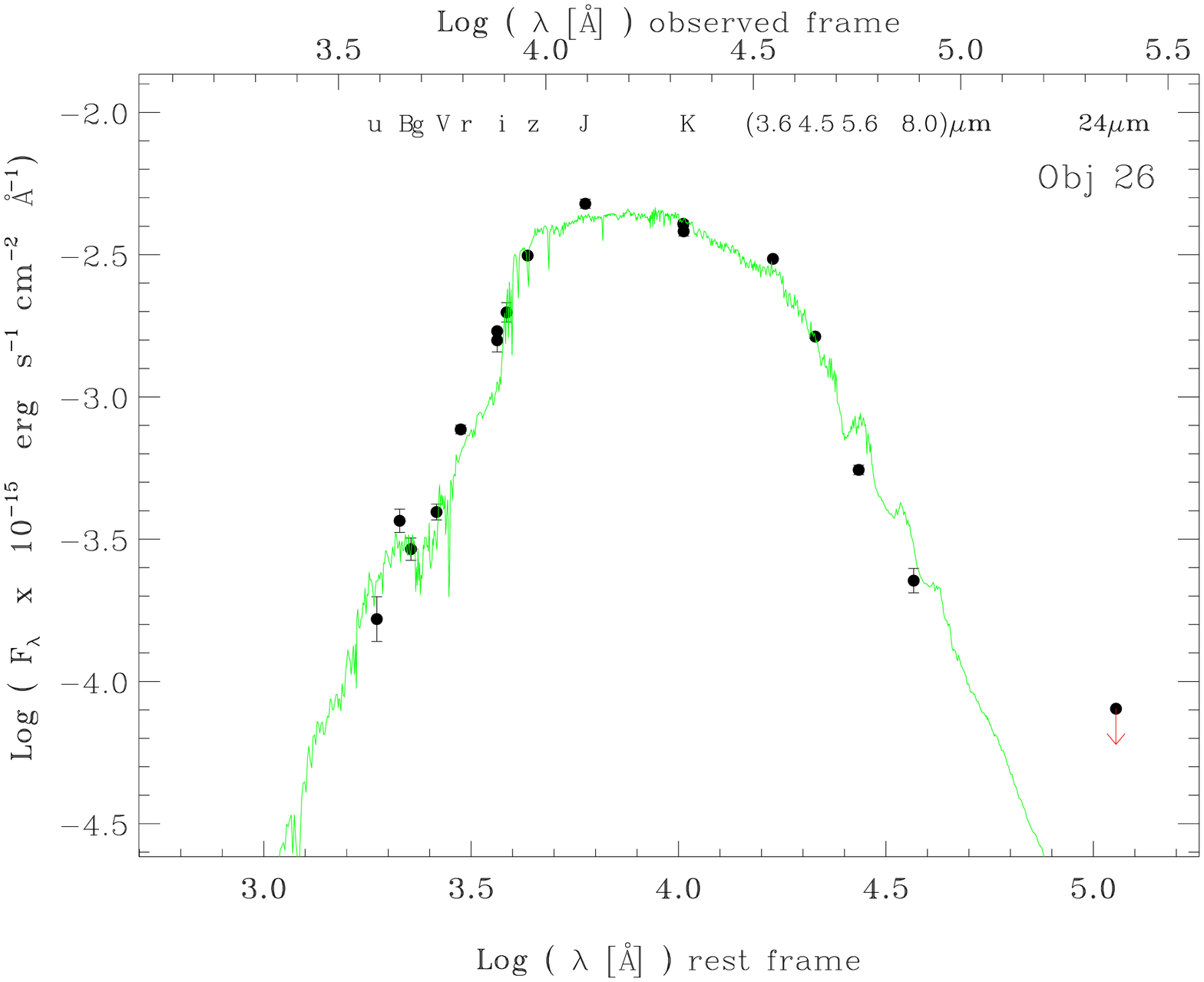} &
\includegraphics[scale=0.33,angle=90]{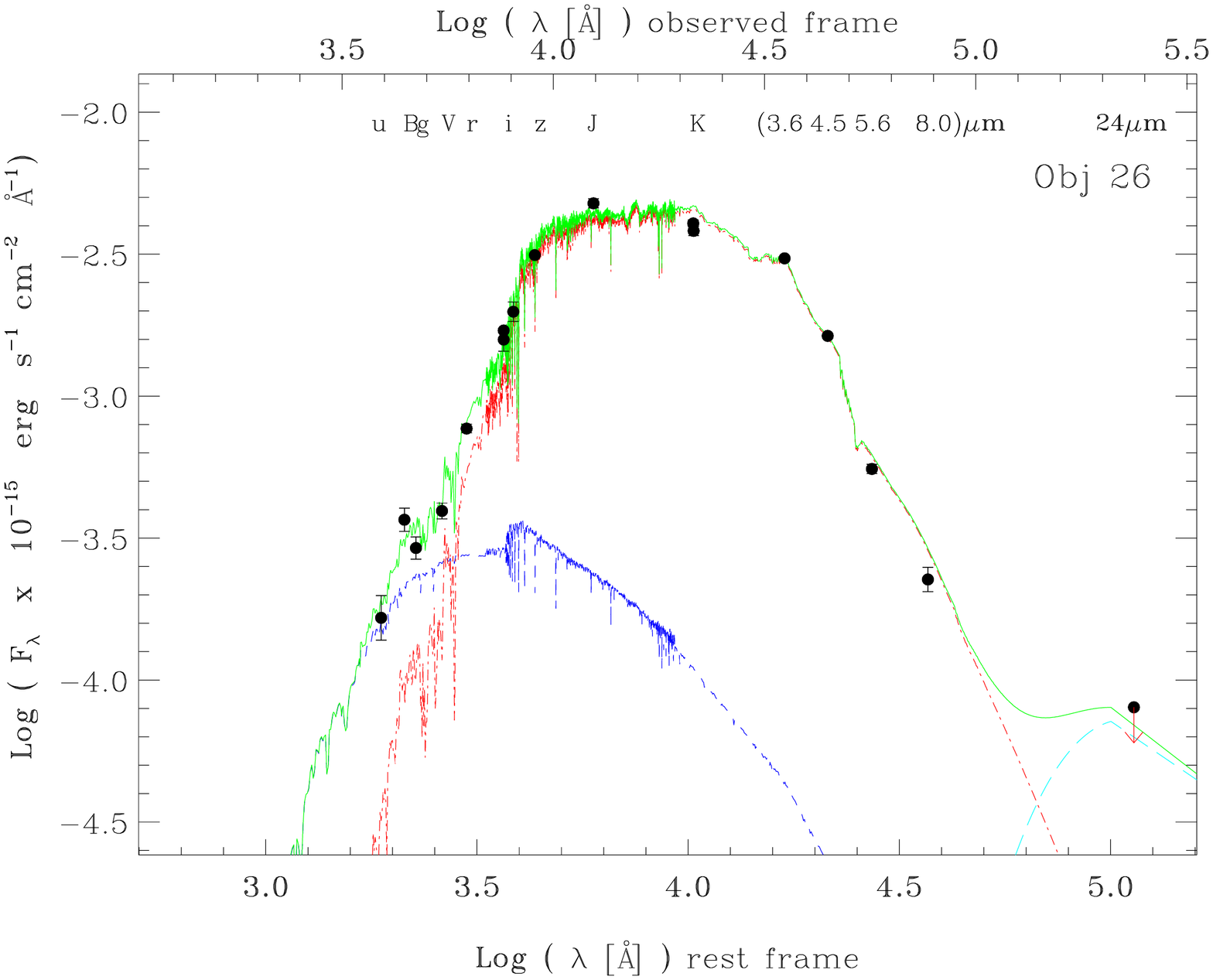} \\
\vspace{1em}
\includegraphics[scale=0.33,angle=90]{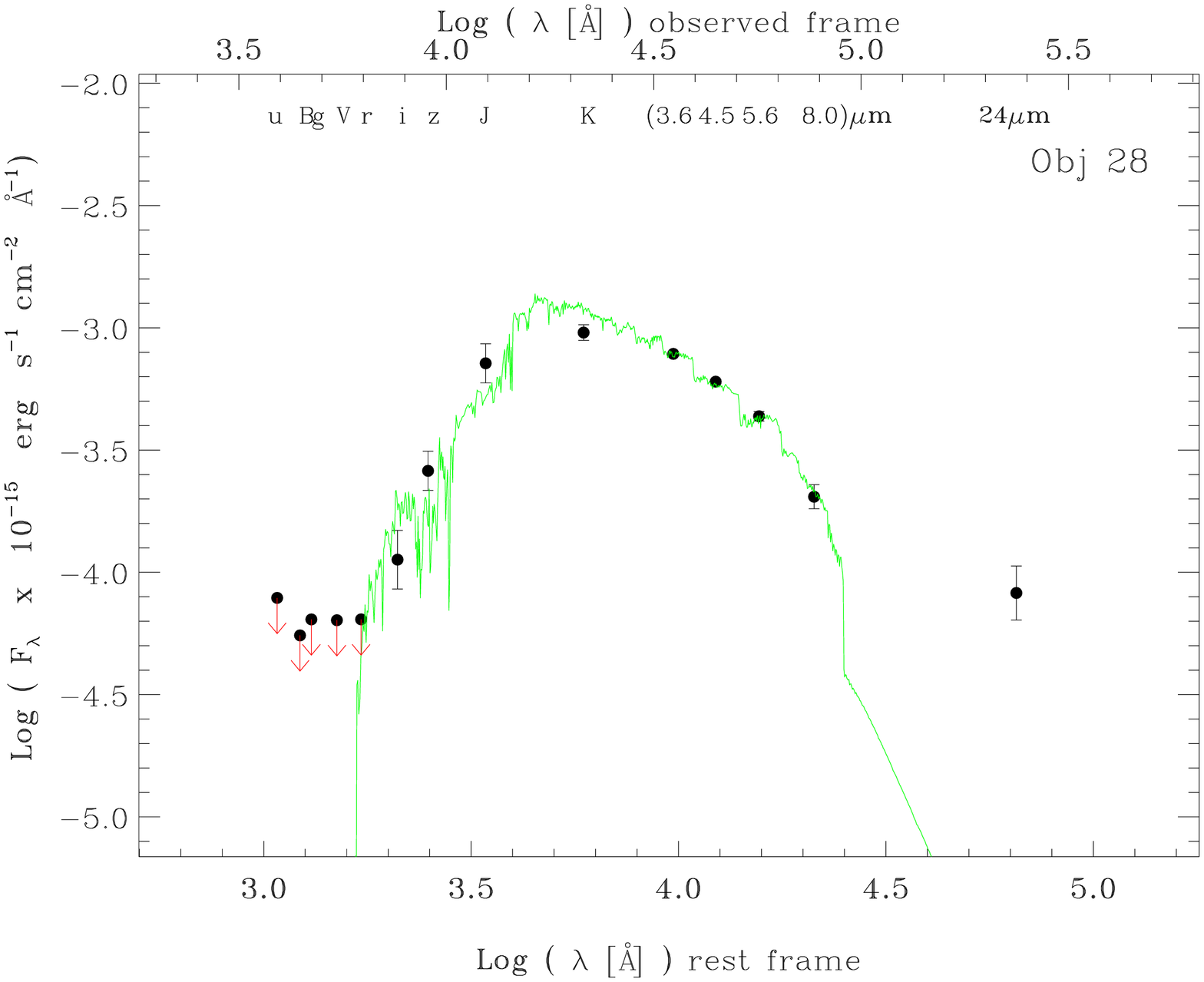} &
\includegraphics[scale=0.33,angle=90]{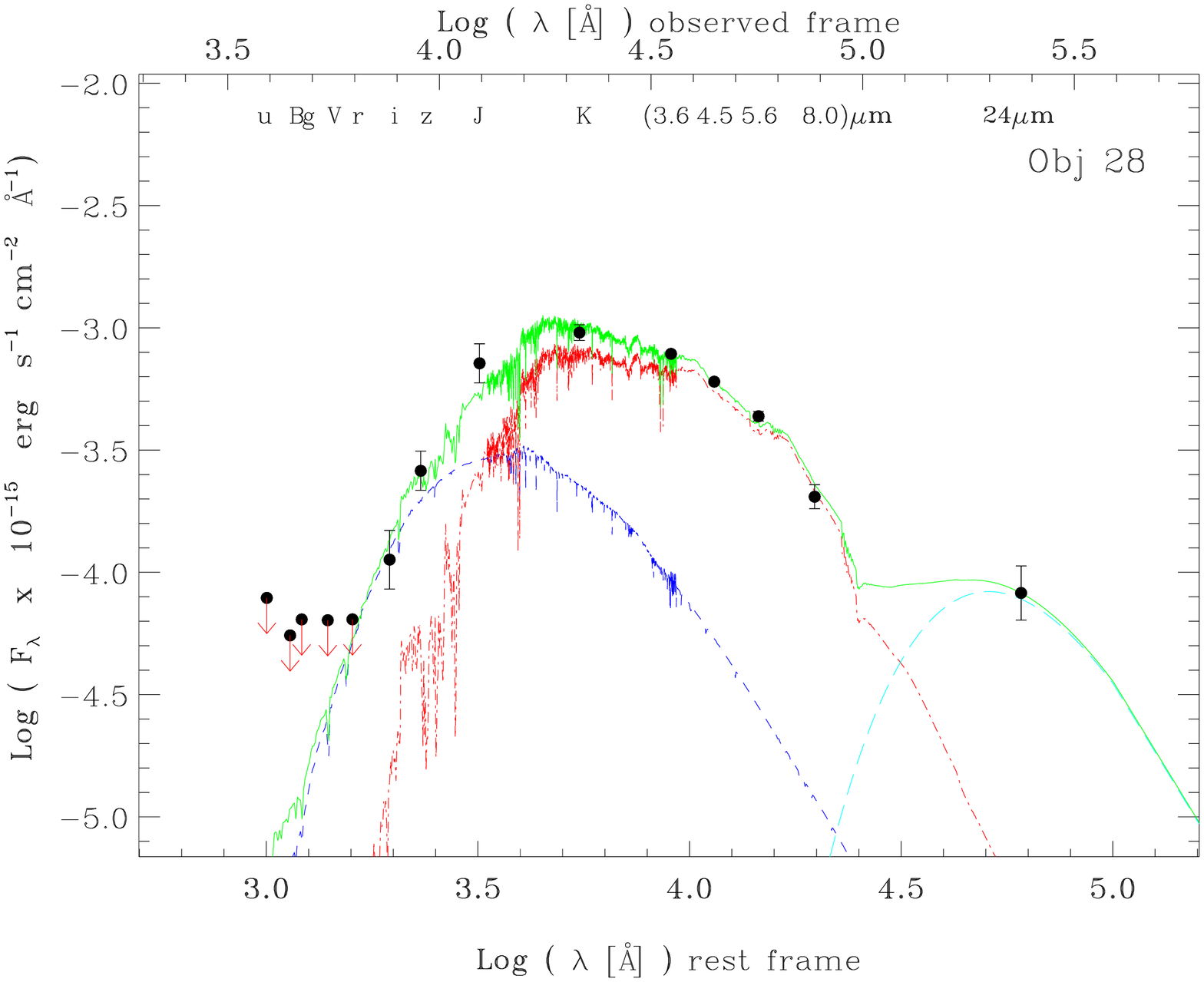} \\
\vspace{1em}
\includegraphics[scale=0.33,angle=90]{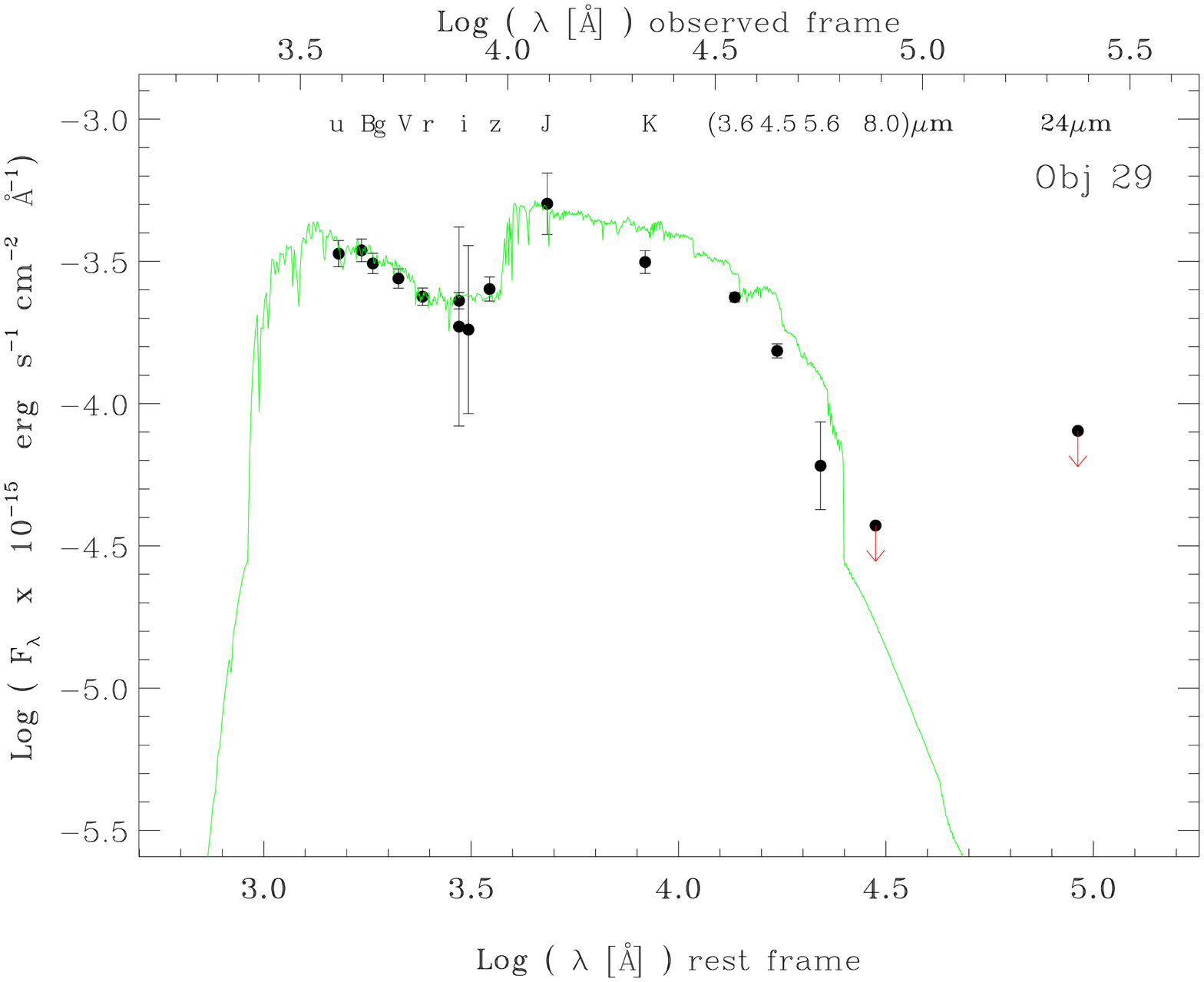} &
\includegraphics[scale=0.33,angle=90]{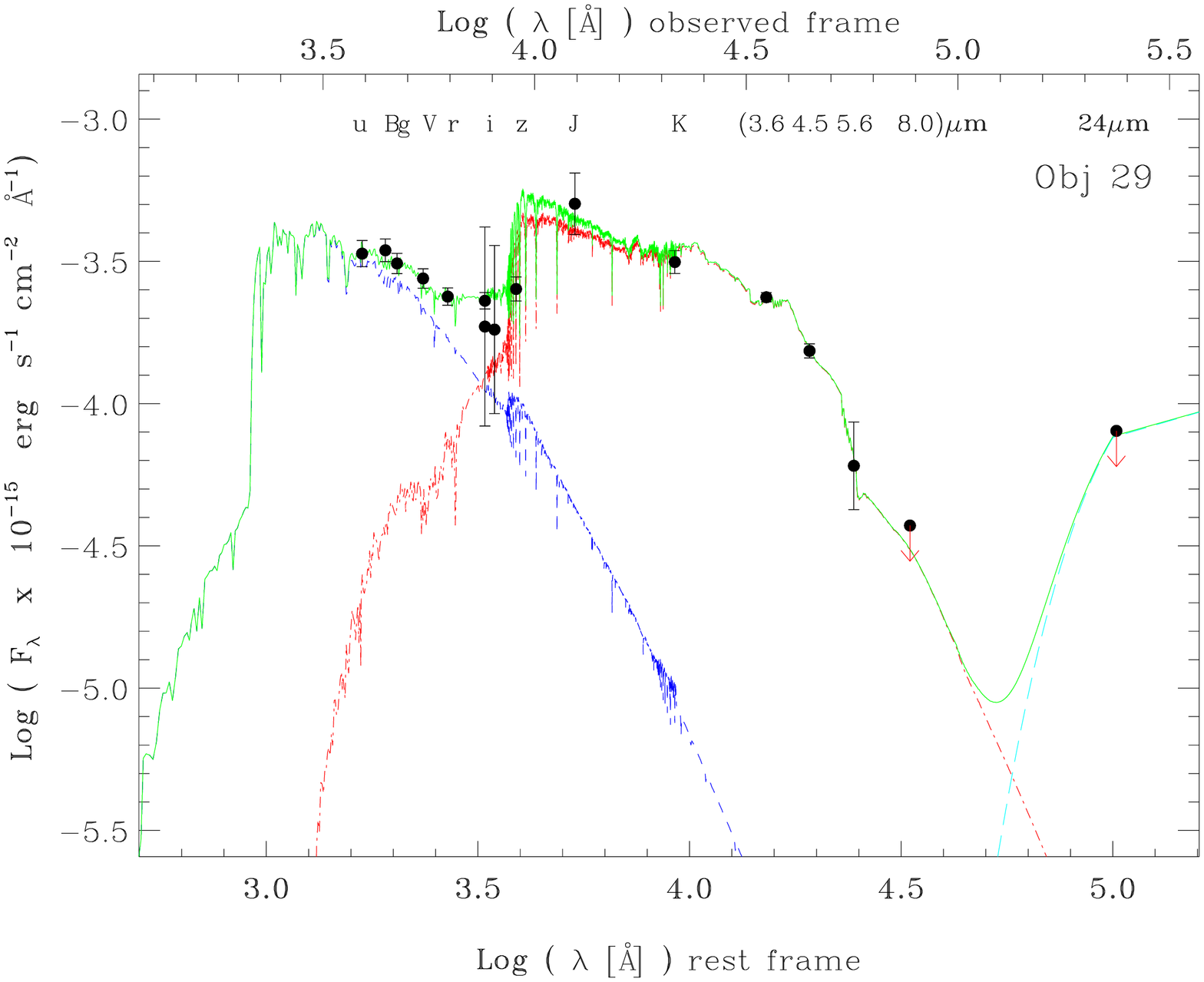} \\
\vspace{2em}
\includegraphics[scale=0.33,angle=90]{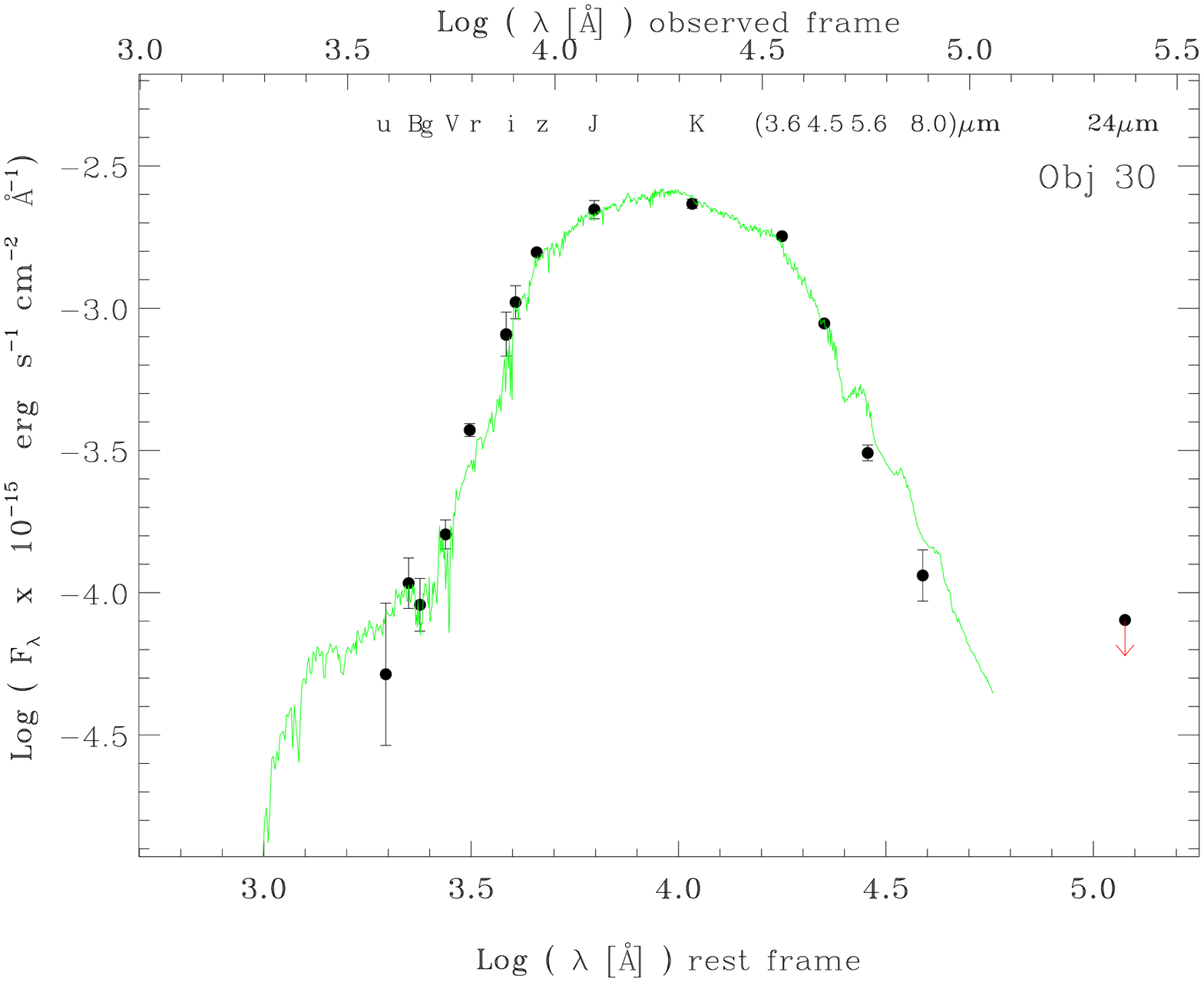} &
\includegraphics[scale=0.33,angle=90]{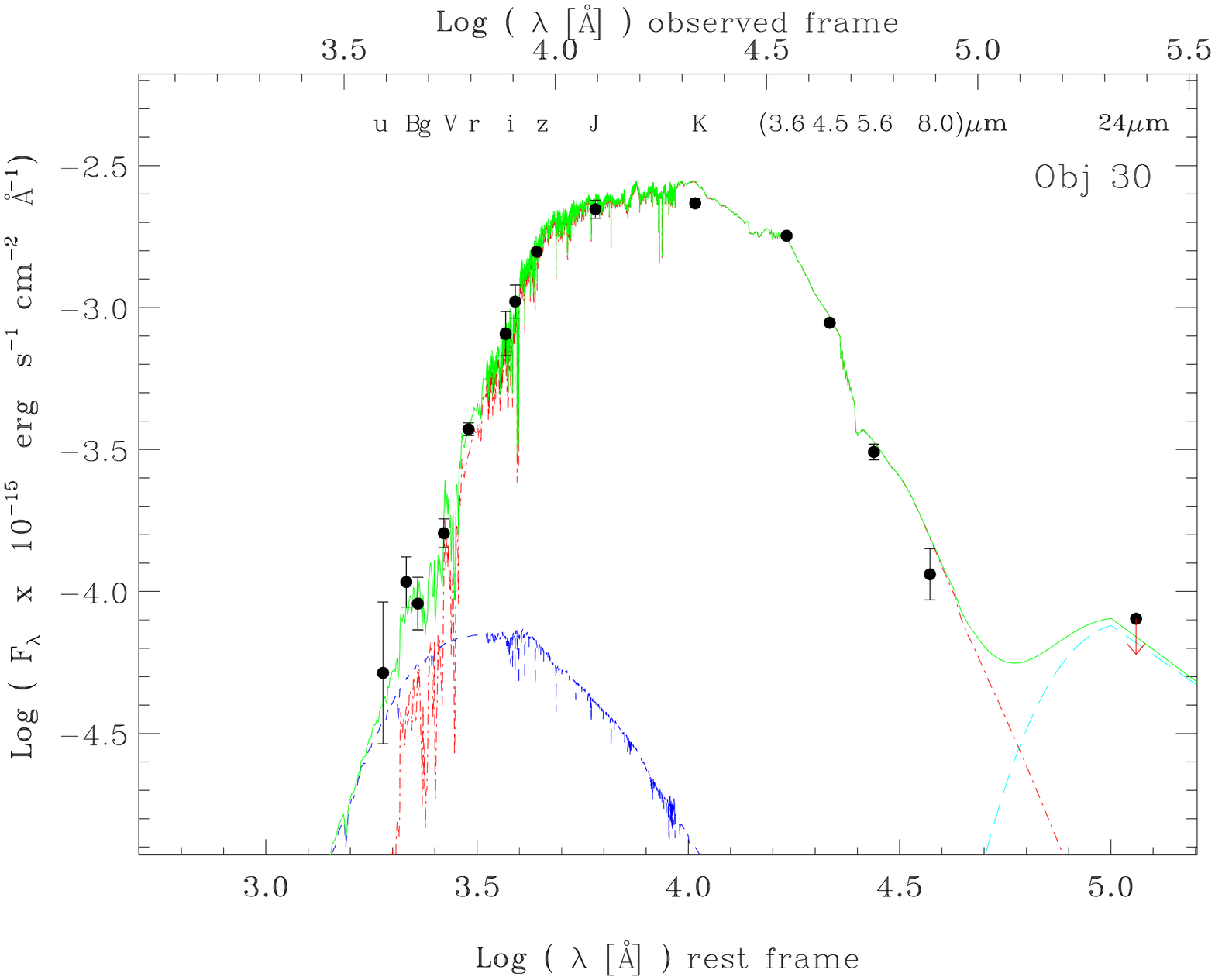} 
\end{array}$
\end{center}
\caption{Continued}
\end{figure*}

\addtocounter{figure}{-1}
\begin{figure*}[h]
\begin{center}$
\begin{array}{ccc}
\vspace{2em}
\includegraphics[scale=0.33,angle=90]{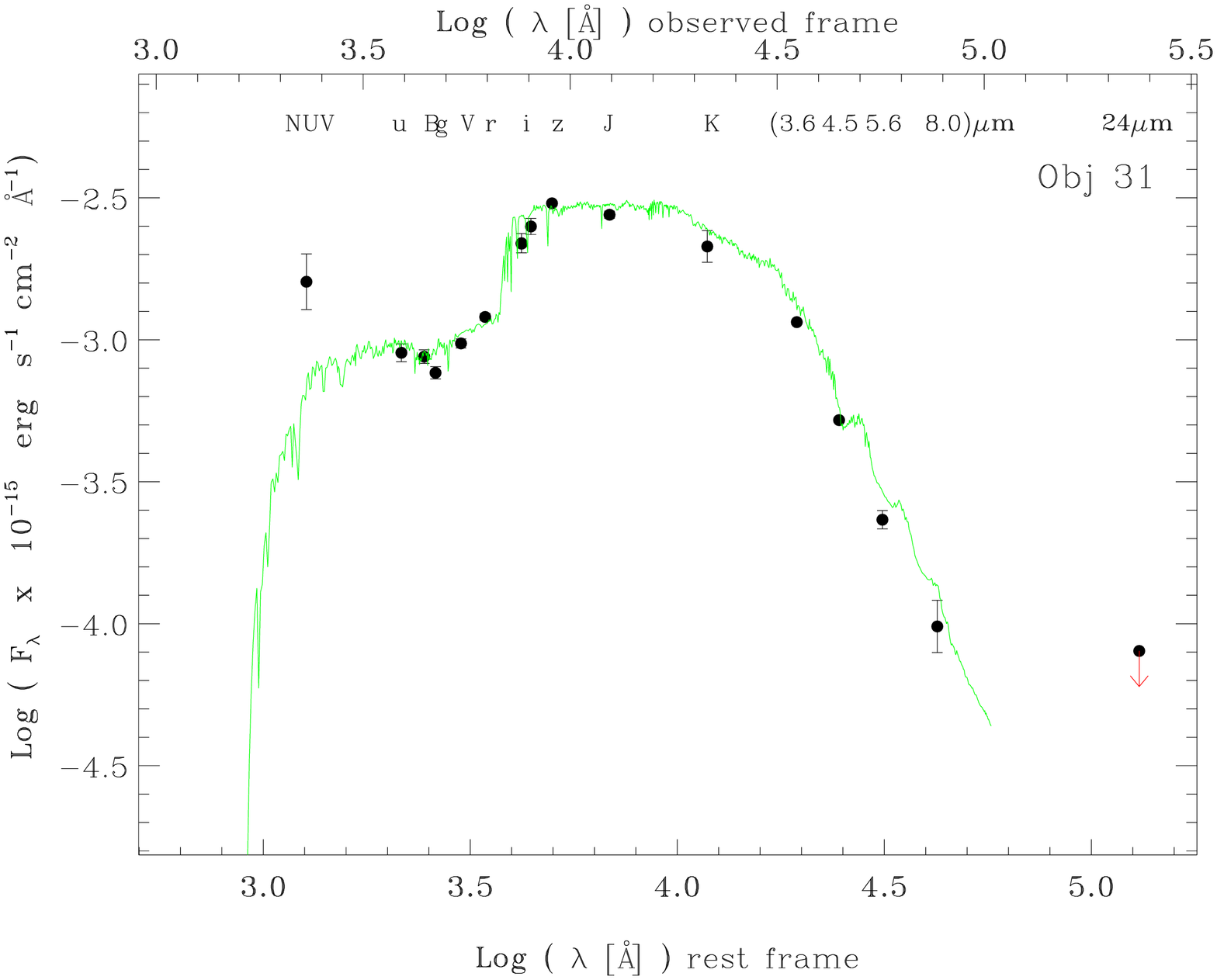} &
\includegraphics[scale=0.33,angle=90]{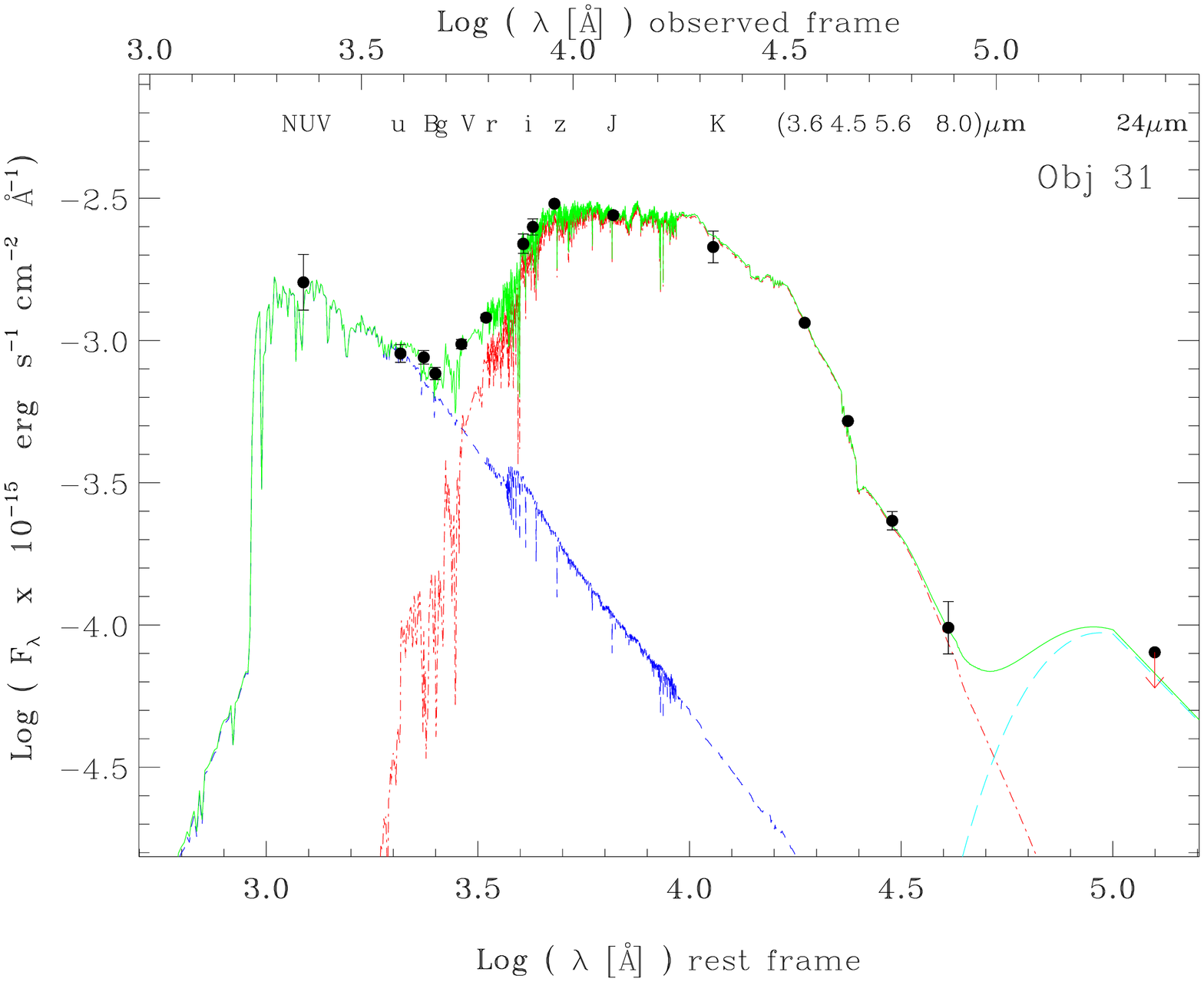} \\
\vspace{1em}
\includegraphics[scale=0.33,angle=90]{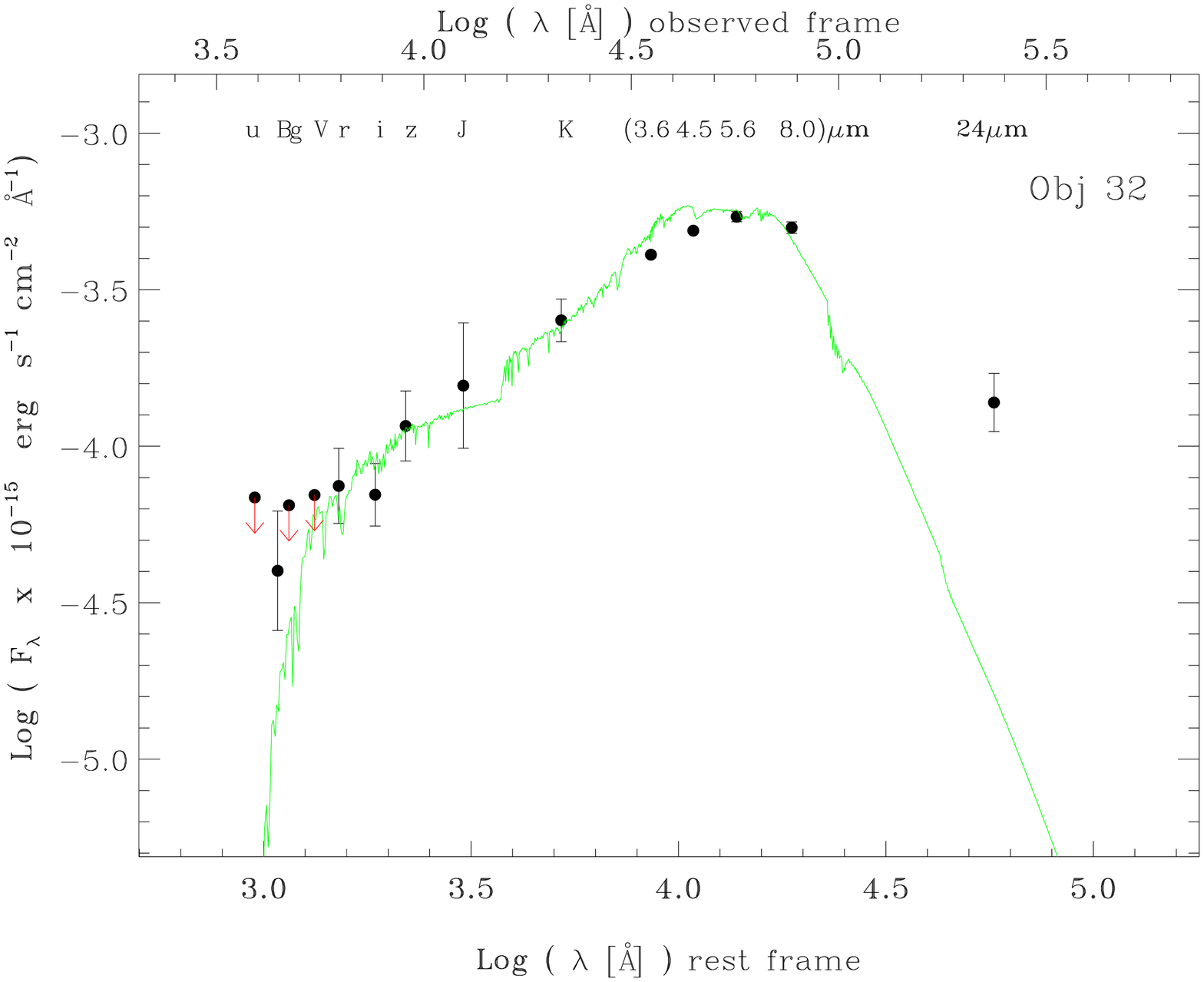} &
\includegraphics[scale=0.33,angle=90]{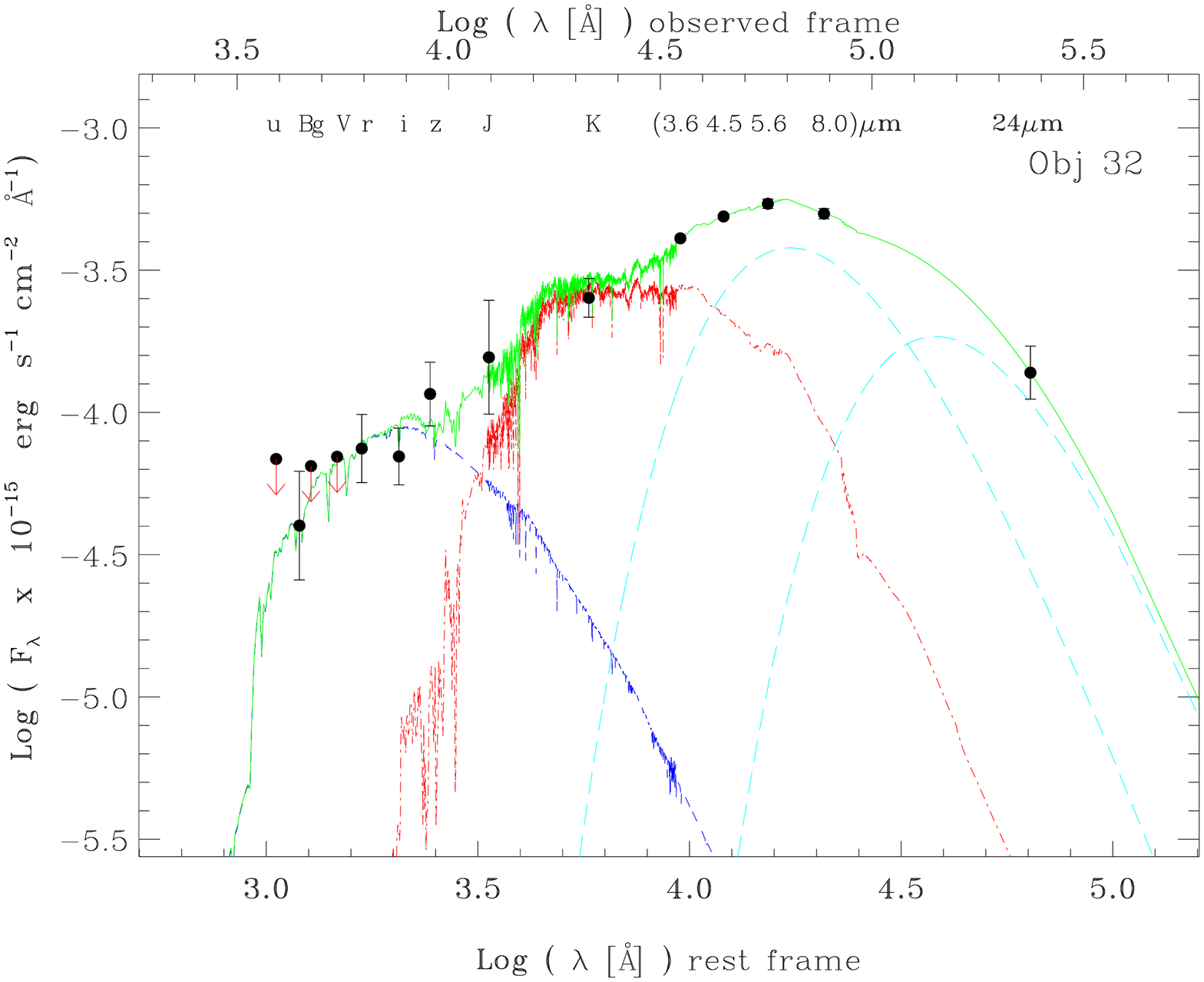} \\
\vspace{2em}
\includegraphics[scale=0.33,angle=90]{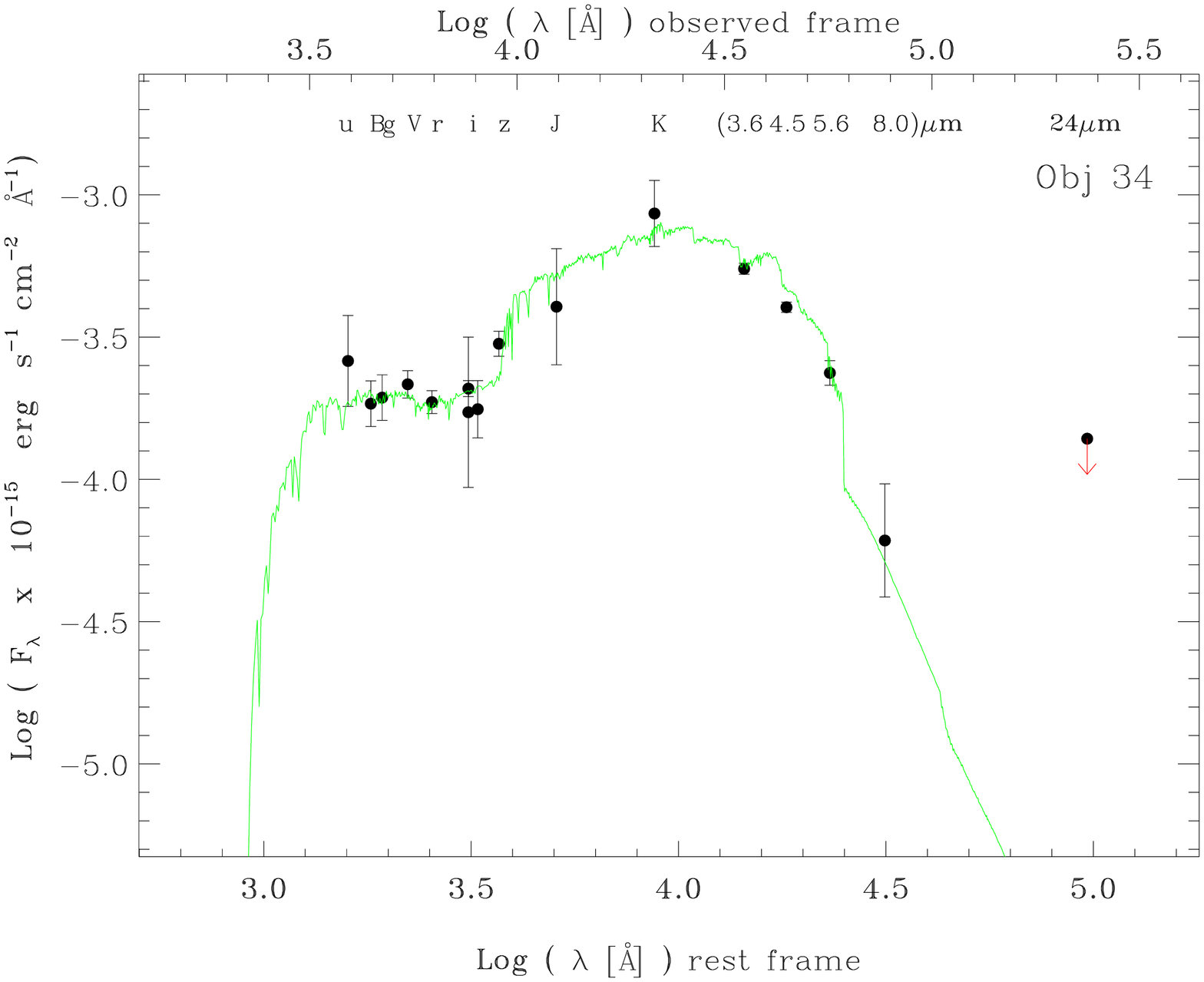} &
\includegraphics[scale=0.33,angle=90]{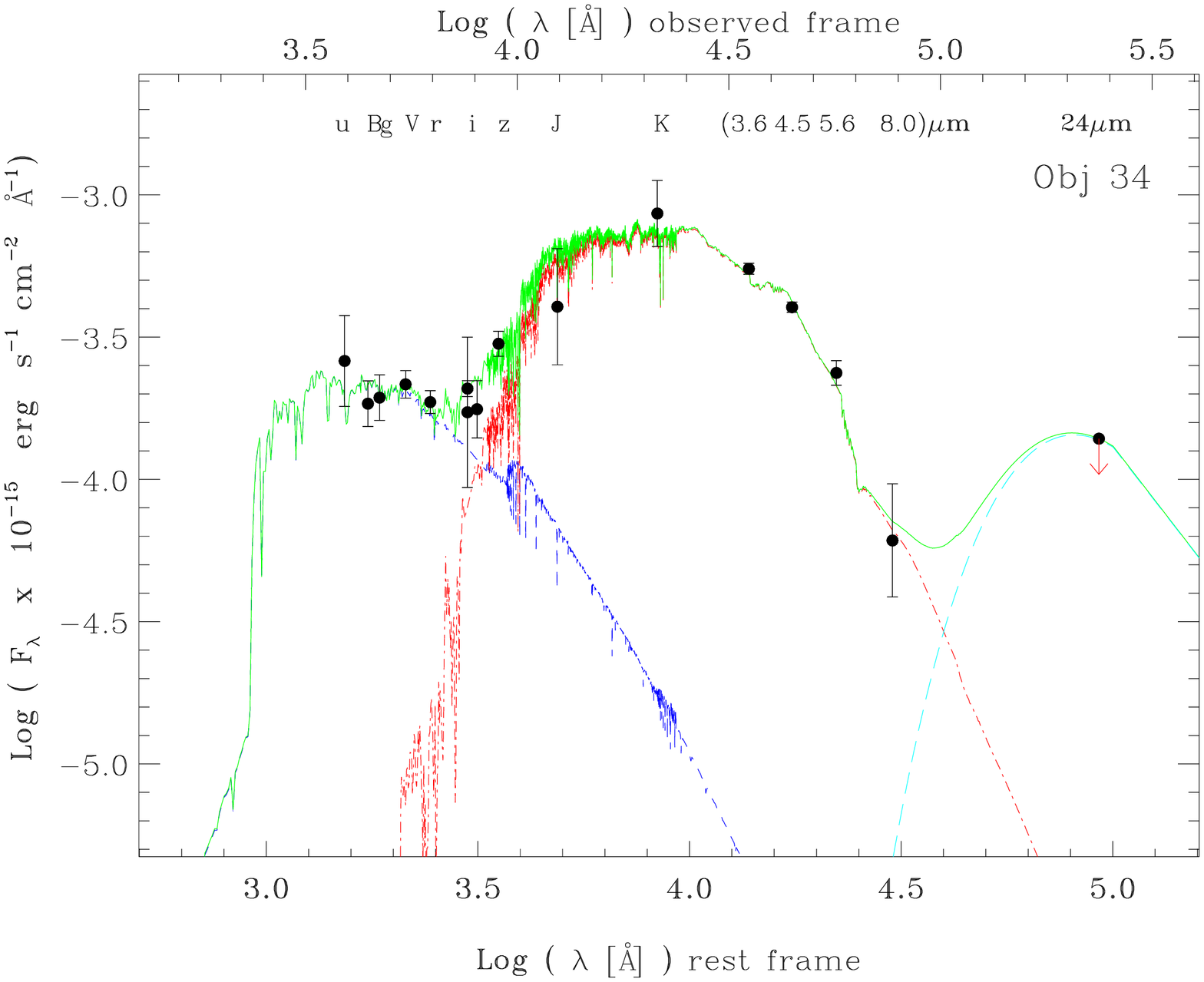} \\
\vspace{1em}
\includegraphics[scale=0.33,angle=90]{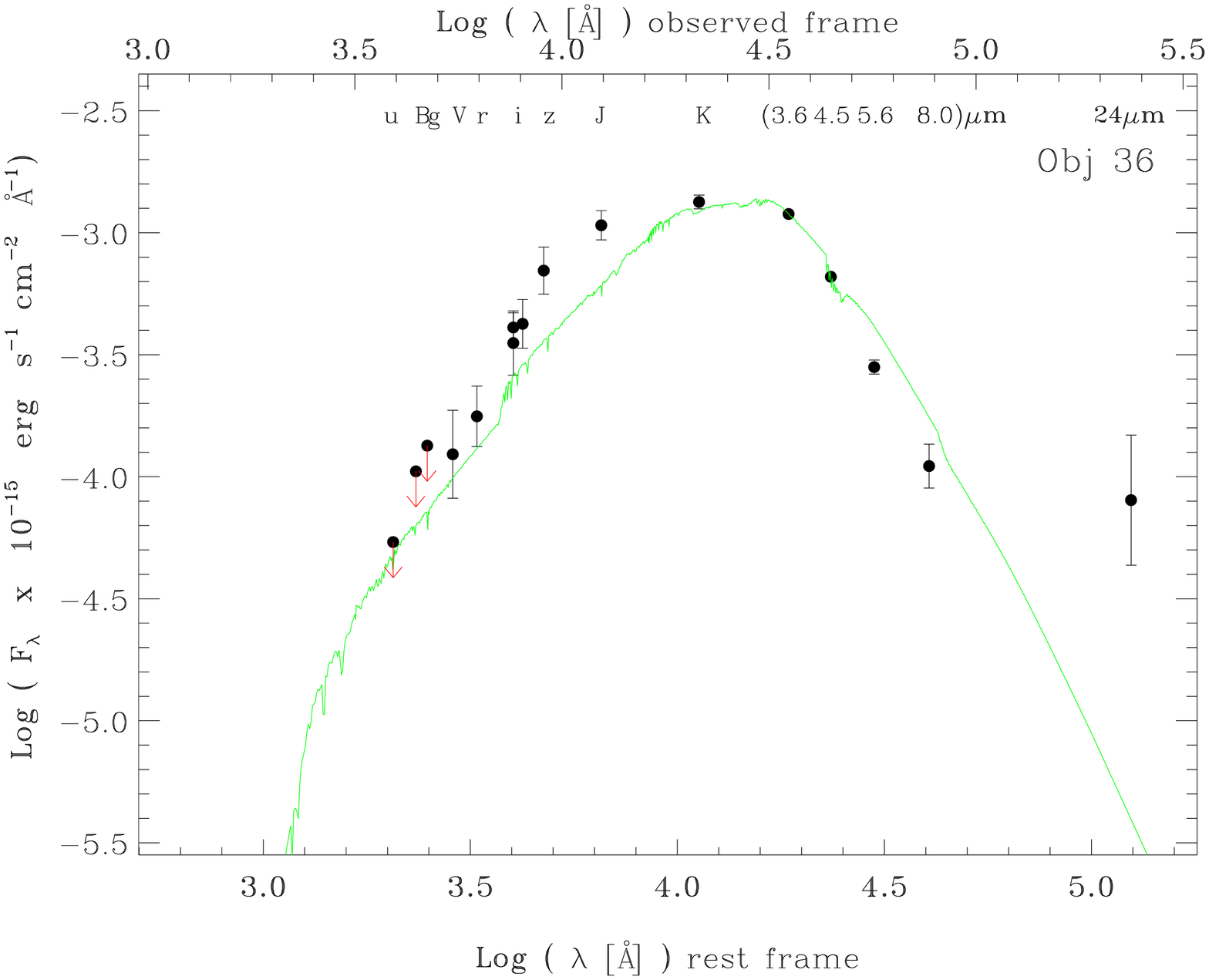} &
\includegraphics[scale=0.33,angle=90]{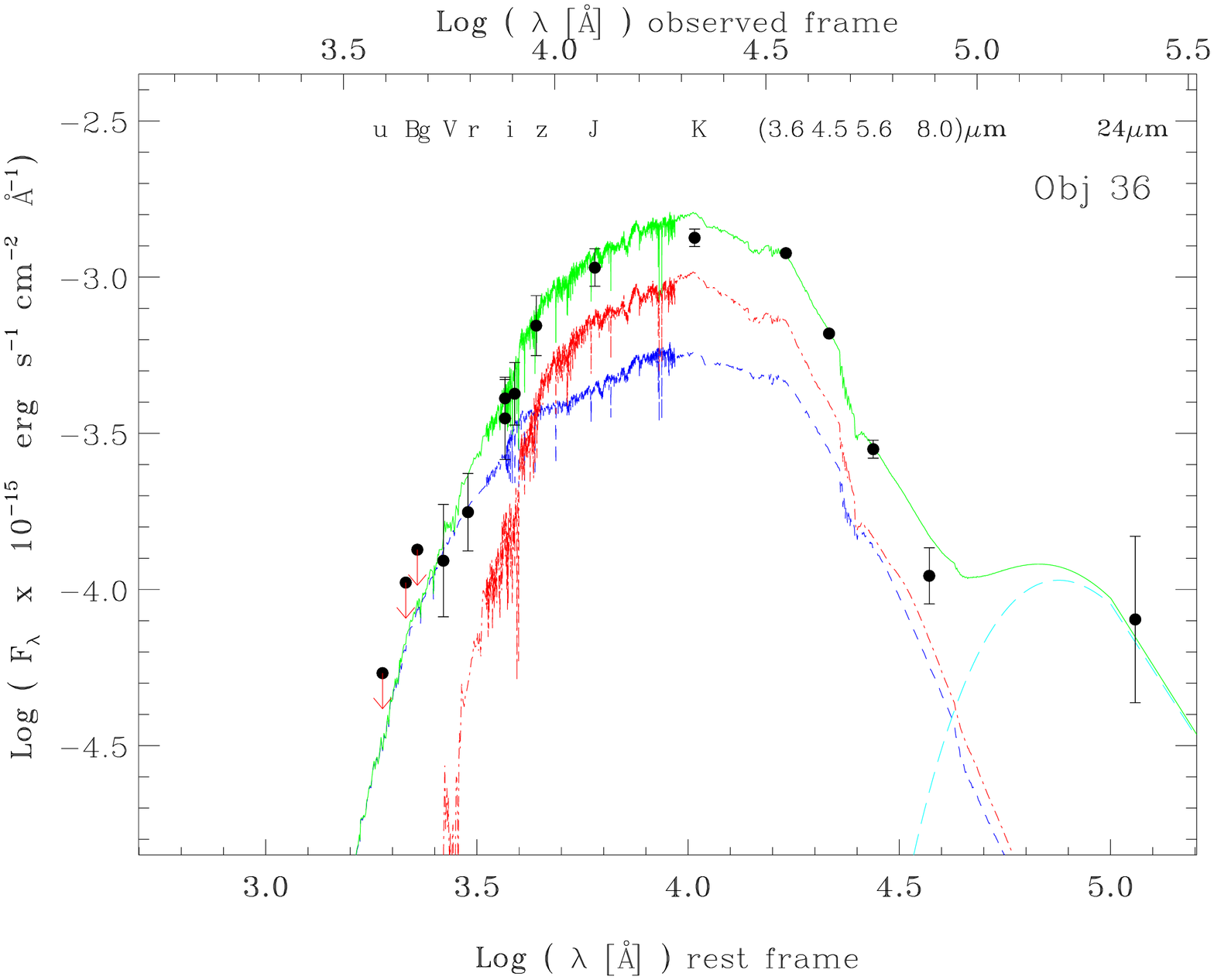} 
\end{array}$
\end{center}
\caption{Continued}
\end{figure*}

\addtocounter{figure}{-1}
\begin{figure*}[h]
\begin{center}$
\begin{array}{ccc}
\vspace{2em}
\includegraphics[scale=0.33,angle=90]{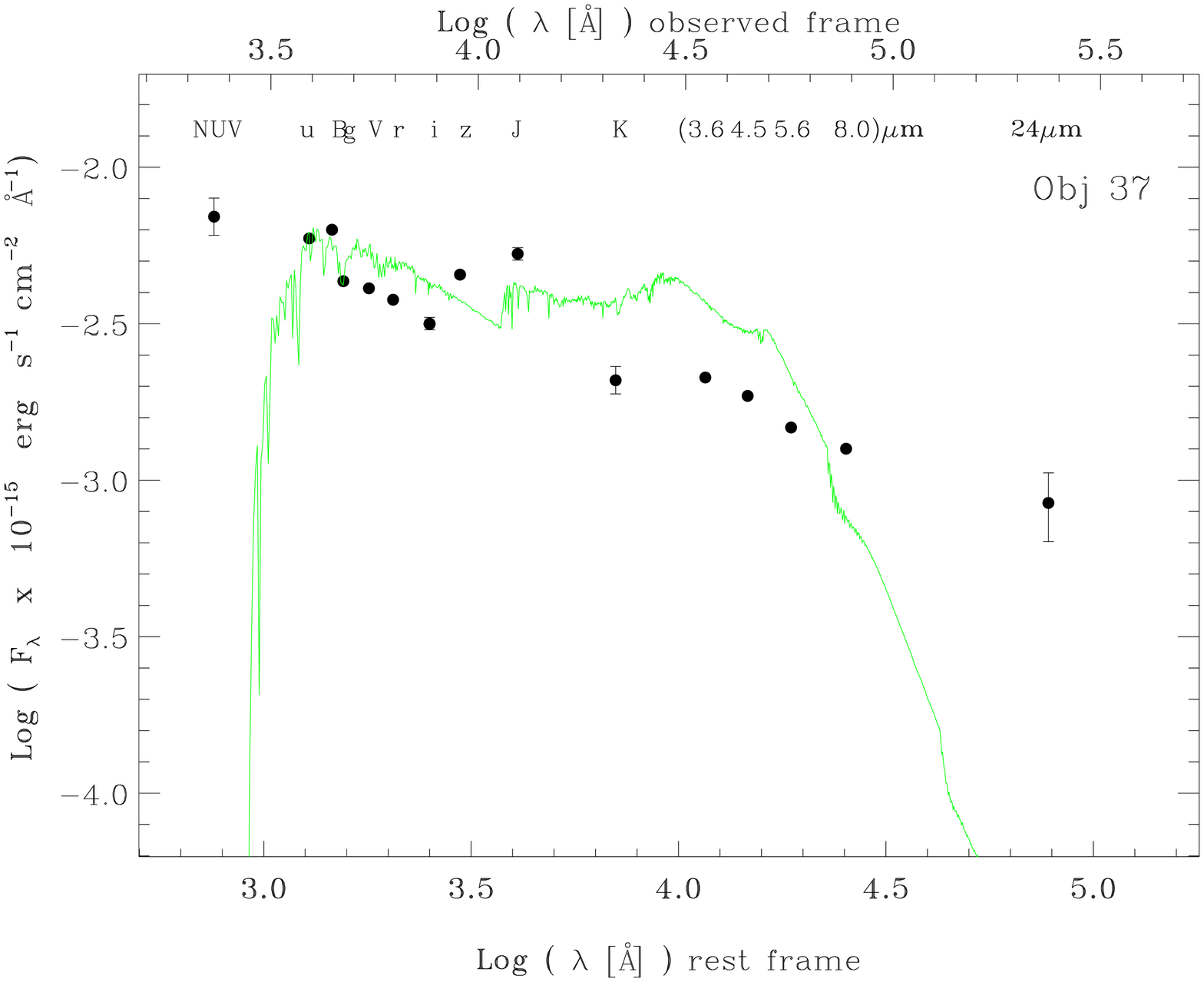} &
\includegraphics[scale=0.33,angle=90]{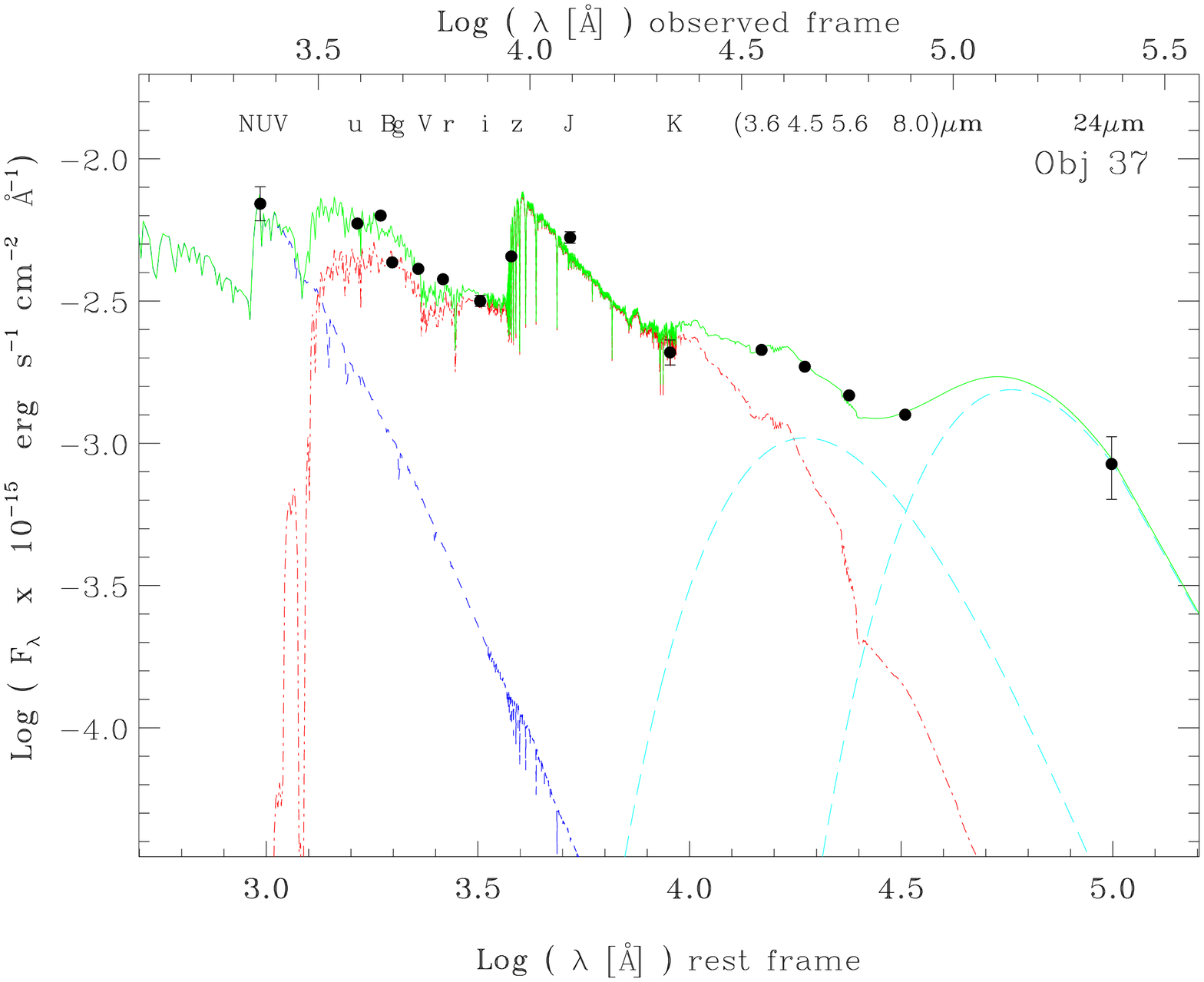} \\
\vspace{1em}
\includegraphics[scale=0.33,angle=90]{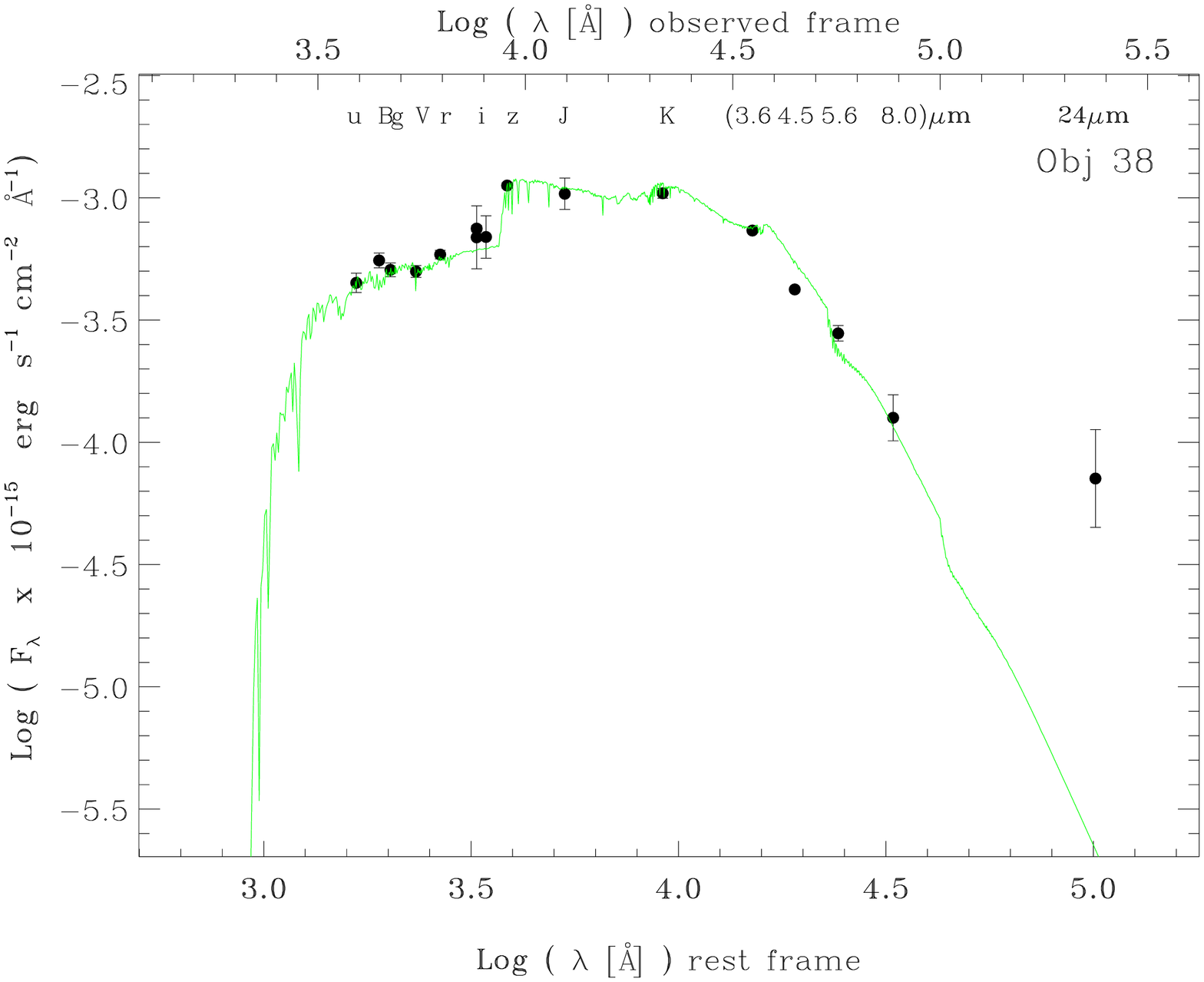} &
\includegraphics[scale=0.33,angle=90]{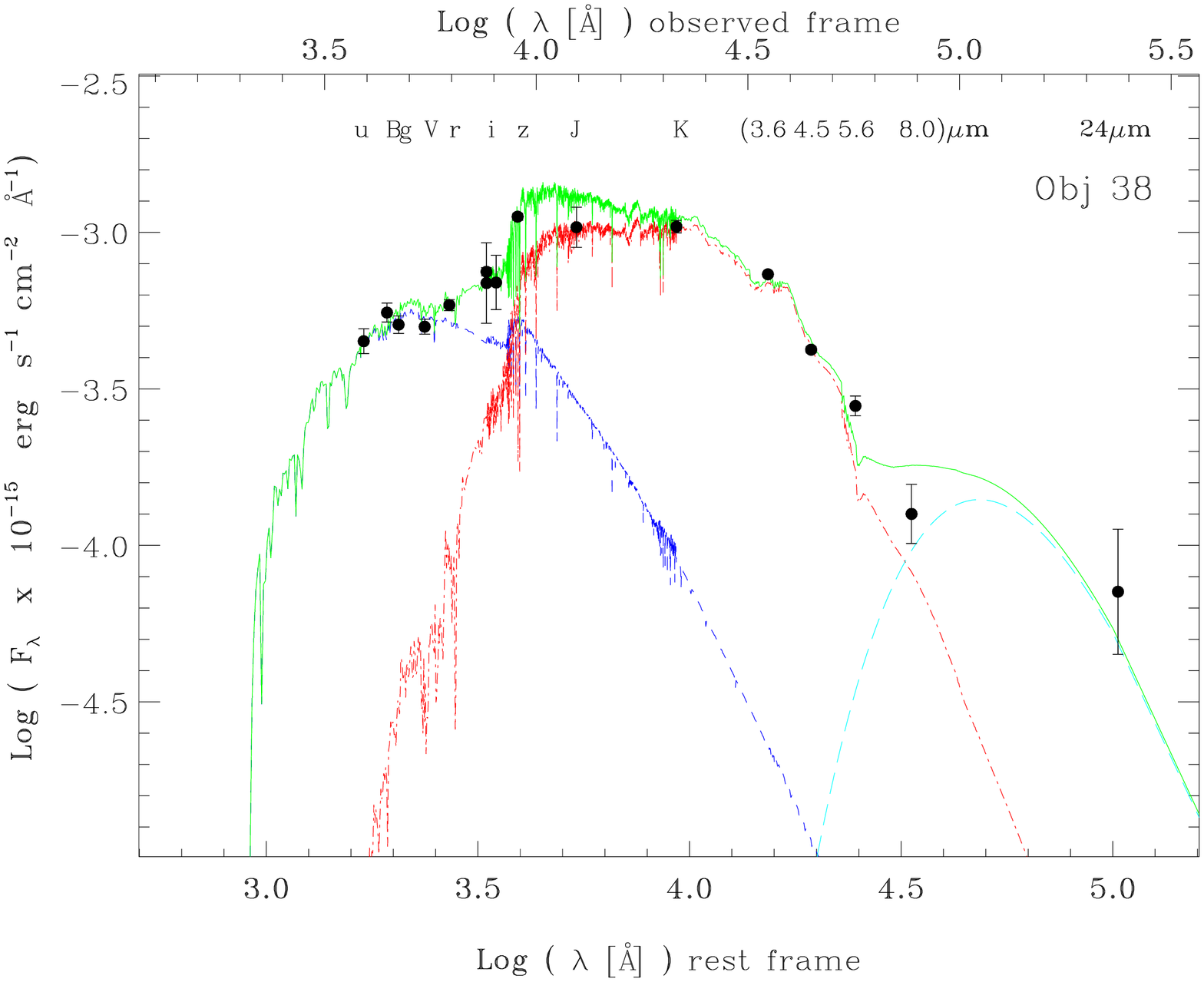} \\
\vspace{1em}
\includegraphics[scale=0.33,angle=90]{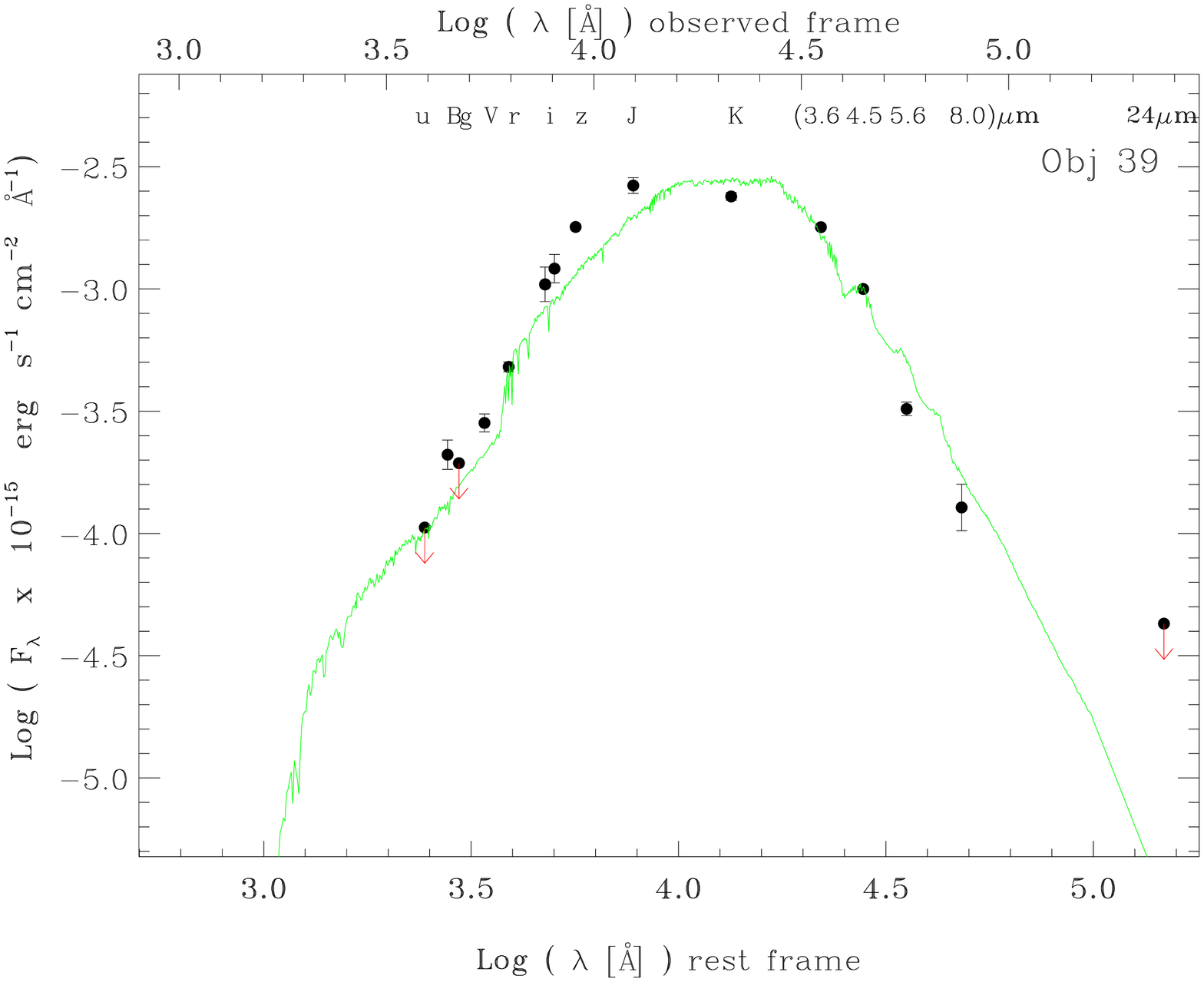} &
\includegraphics[scale=0.33,angle=90]{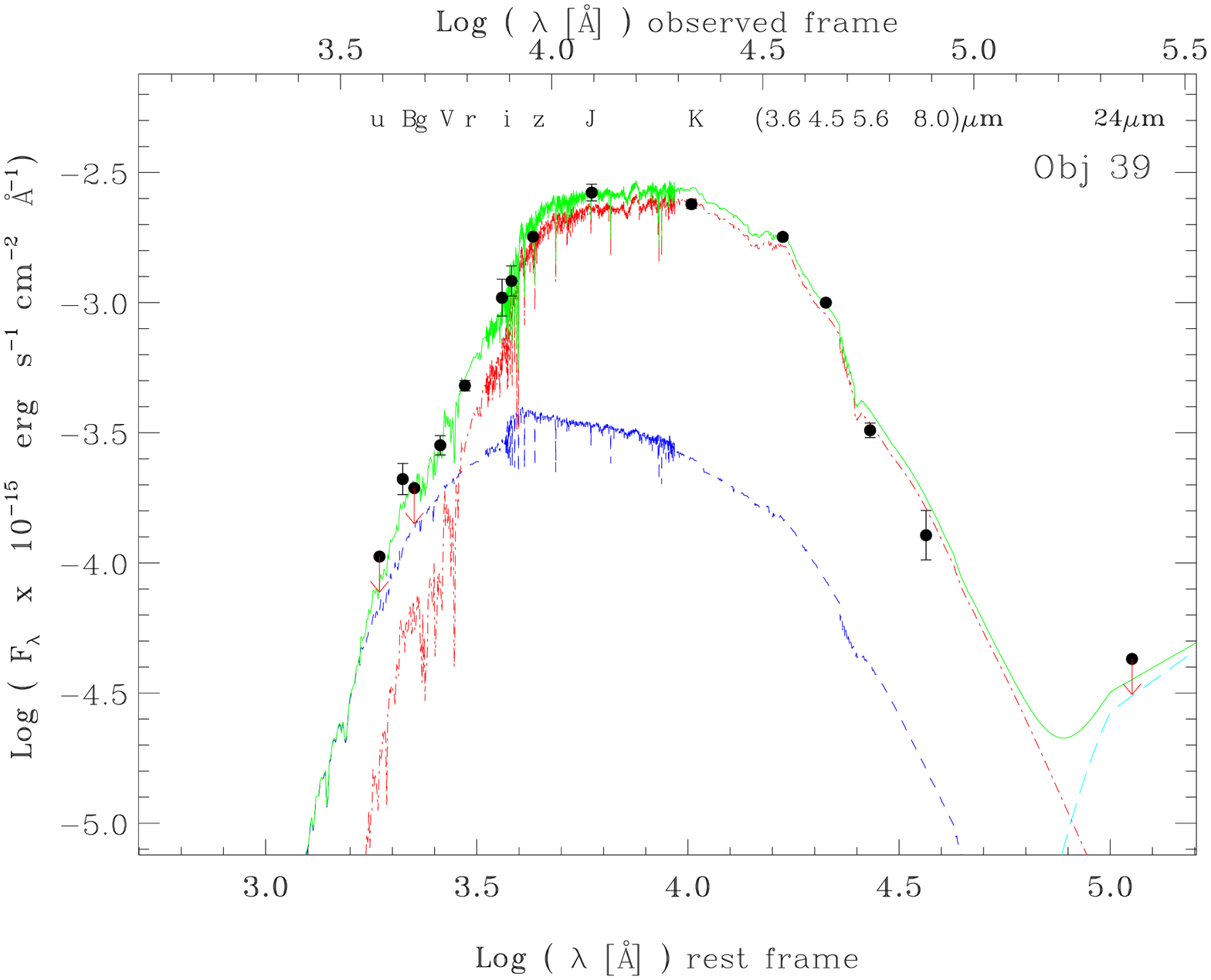} \\
\vspace{1em}
\includegraphics[scale=0.33,angle=90]{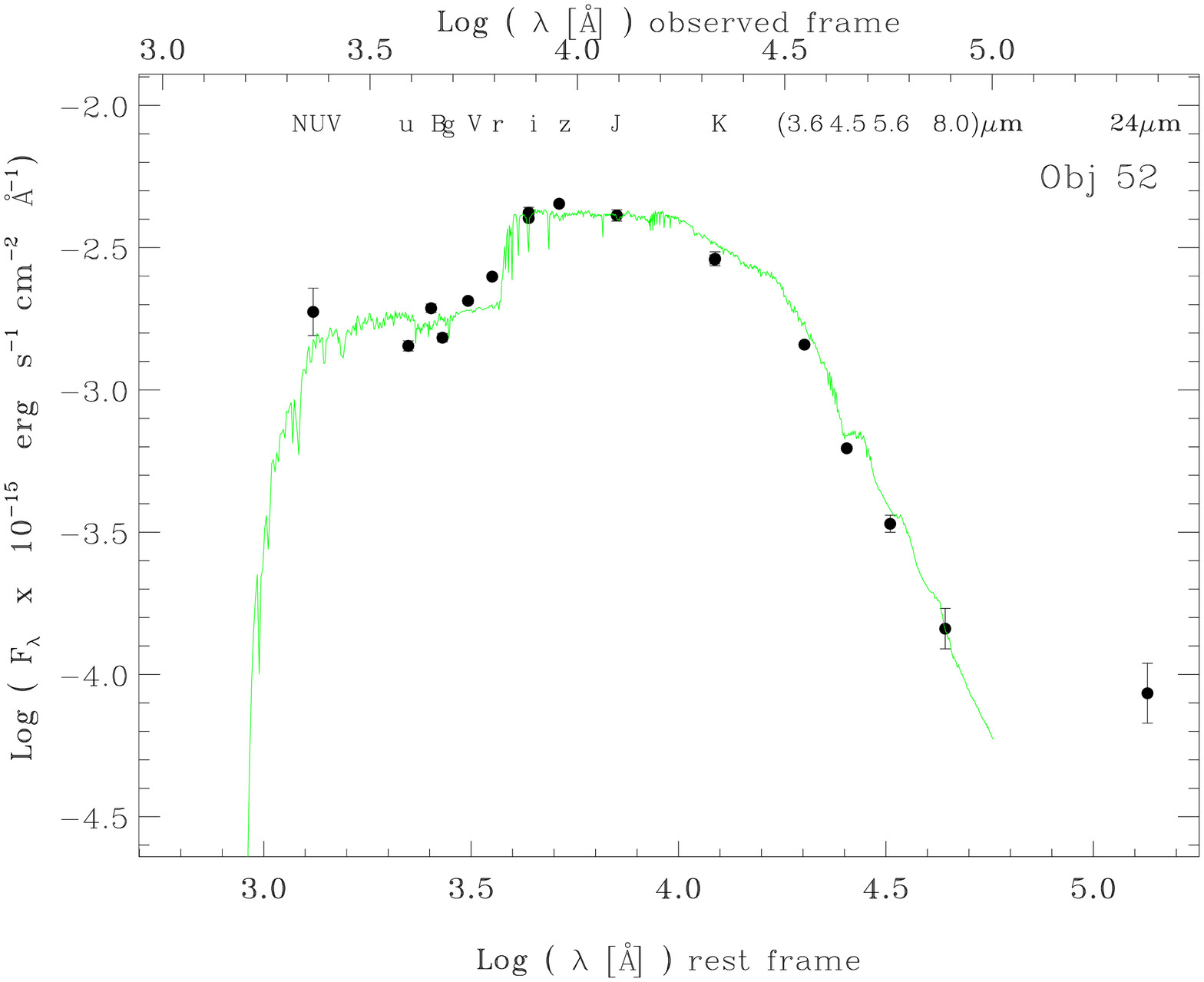} &
\includegraphics[scale=0.33,angle=90]{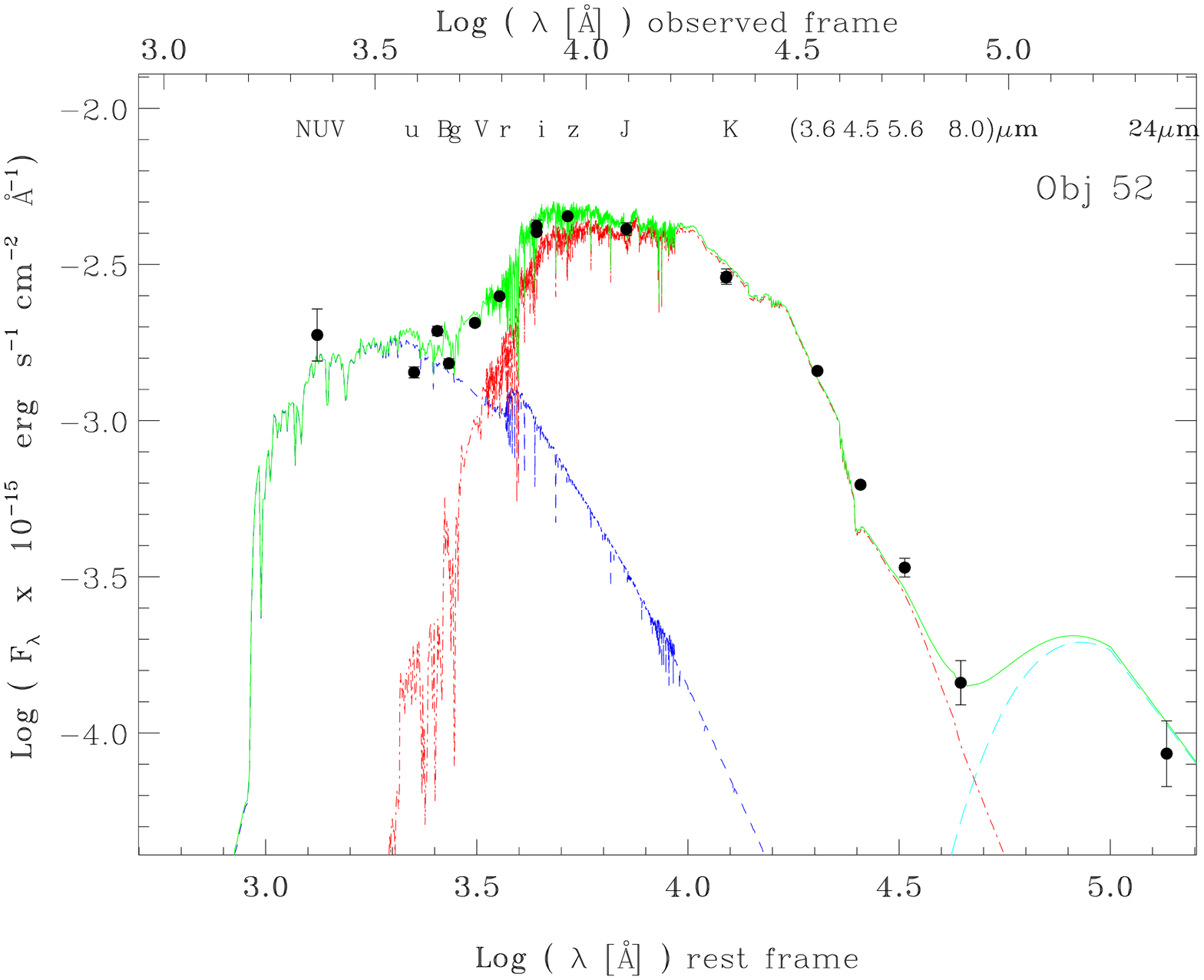} 
\end{array}$
\end{center}
\caption{Continued}
\end{figure*}

\addtocounter{figure}{-1}
\begin{figure*}[h]
\begin{center}$
\begin{array}{ccc}
\vspace{2em}
\includegraphics[scale=0.33,angle=90]{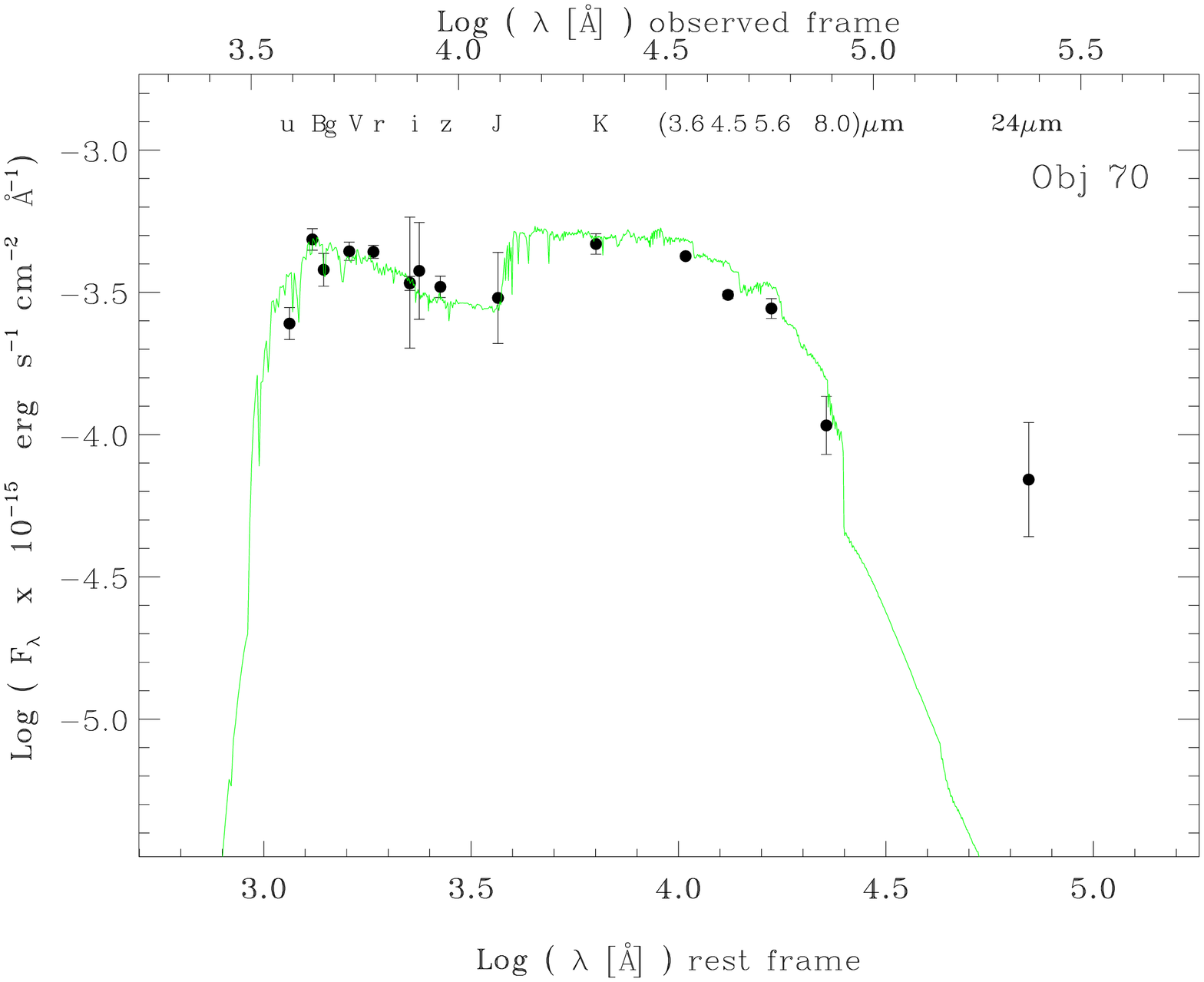} &
\includegraphics[scale=0.33,angle=90]{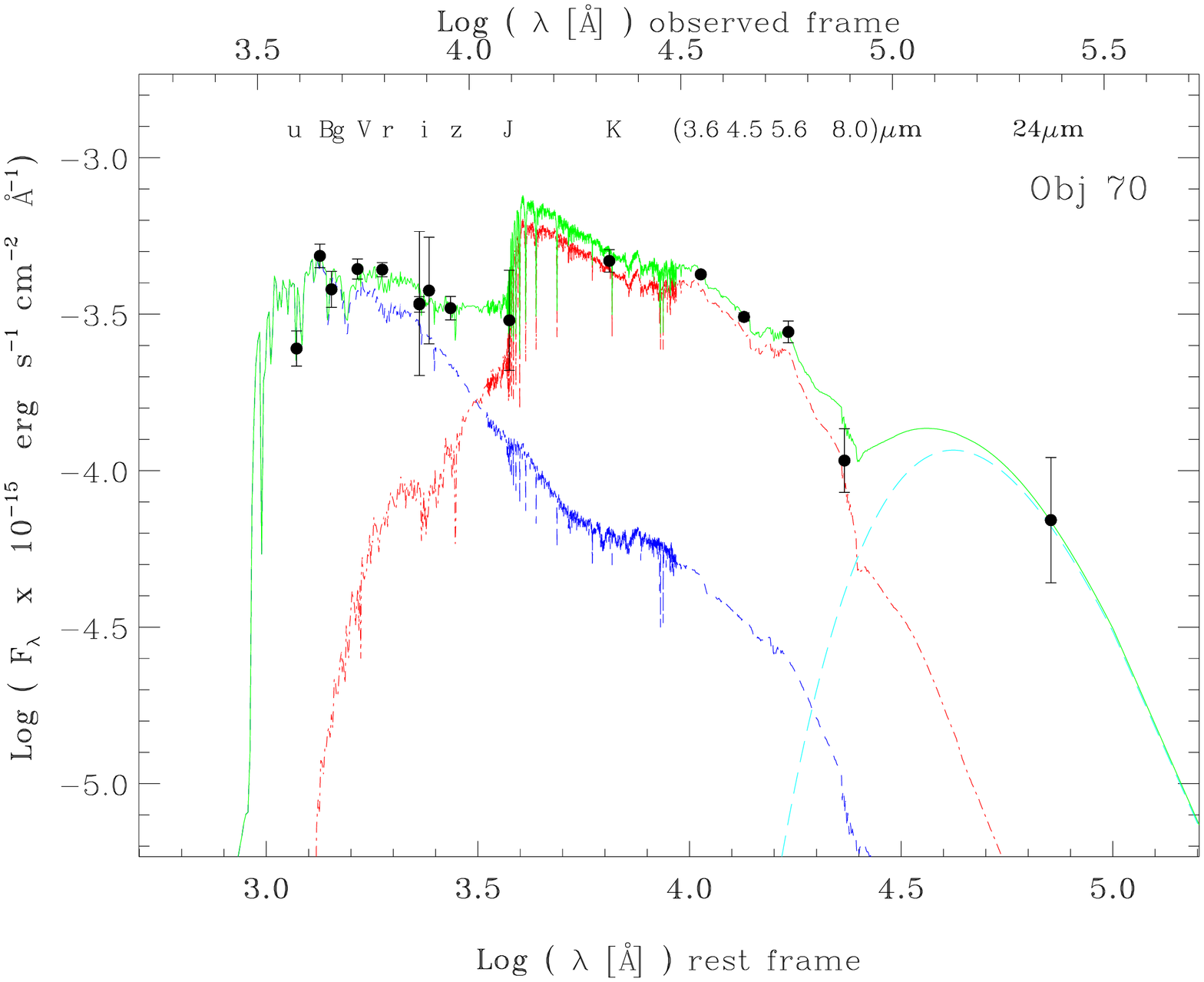} \\
\vspace{1em}
\includegraphics[scale=0.33,angle=90]{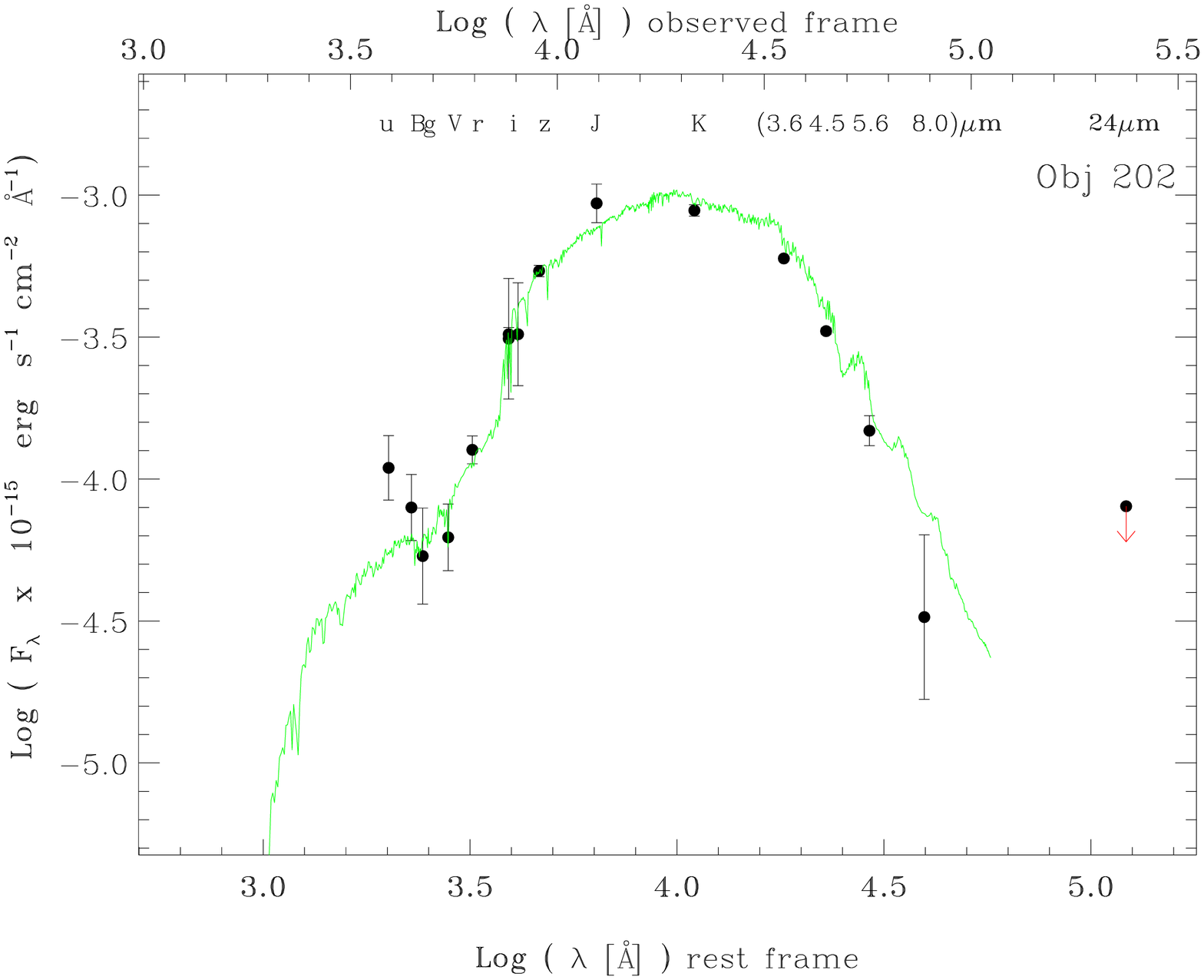} &
\includegraphics[scale=0.33,angle=90]{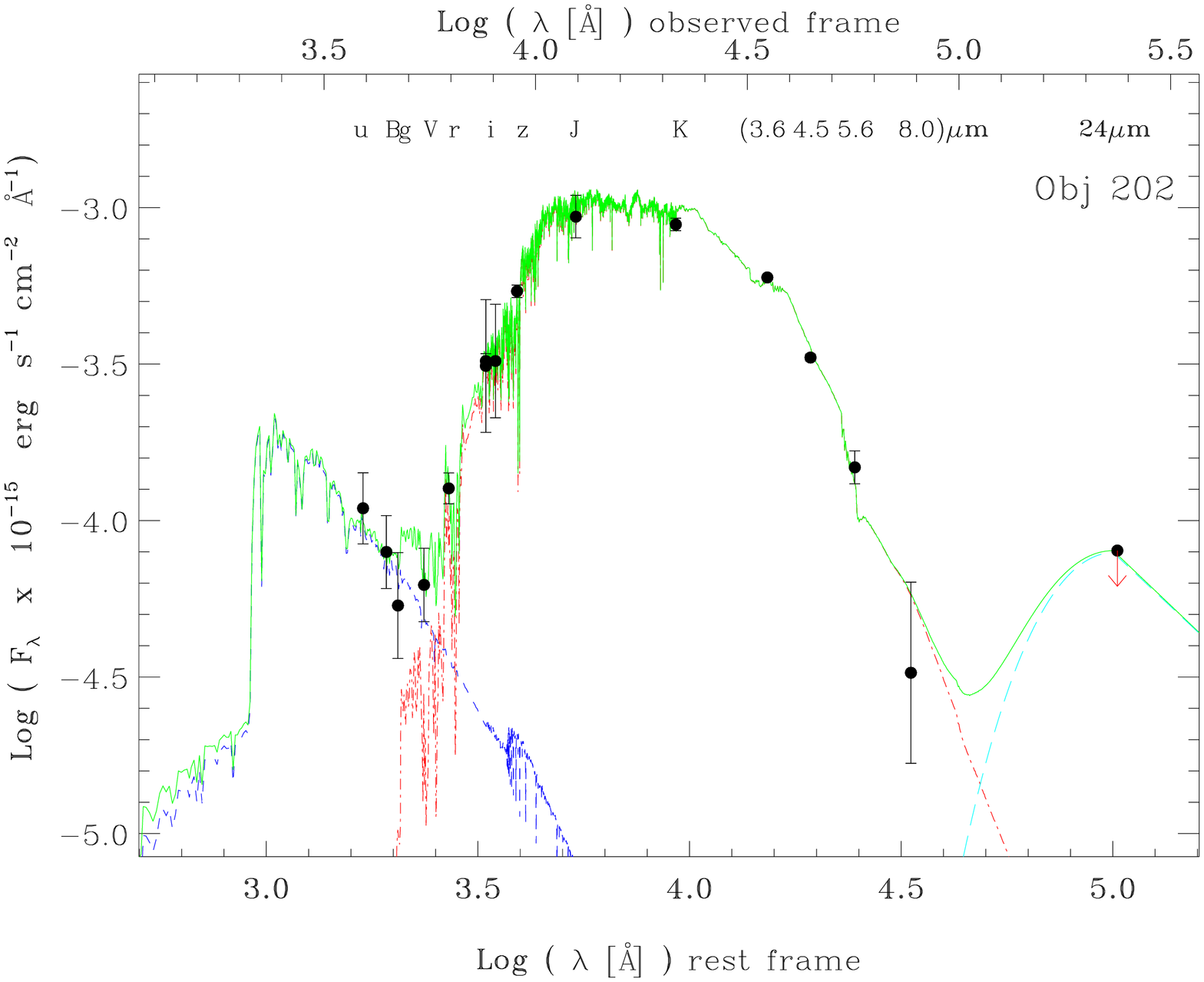} \\
\vspace{1em}
\includegraphics[scale=0.33,angle=90]{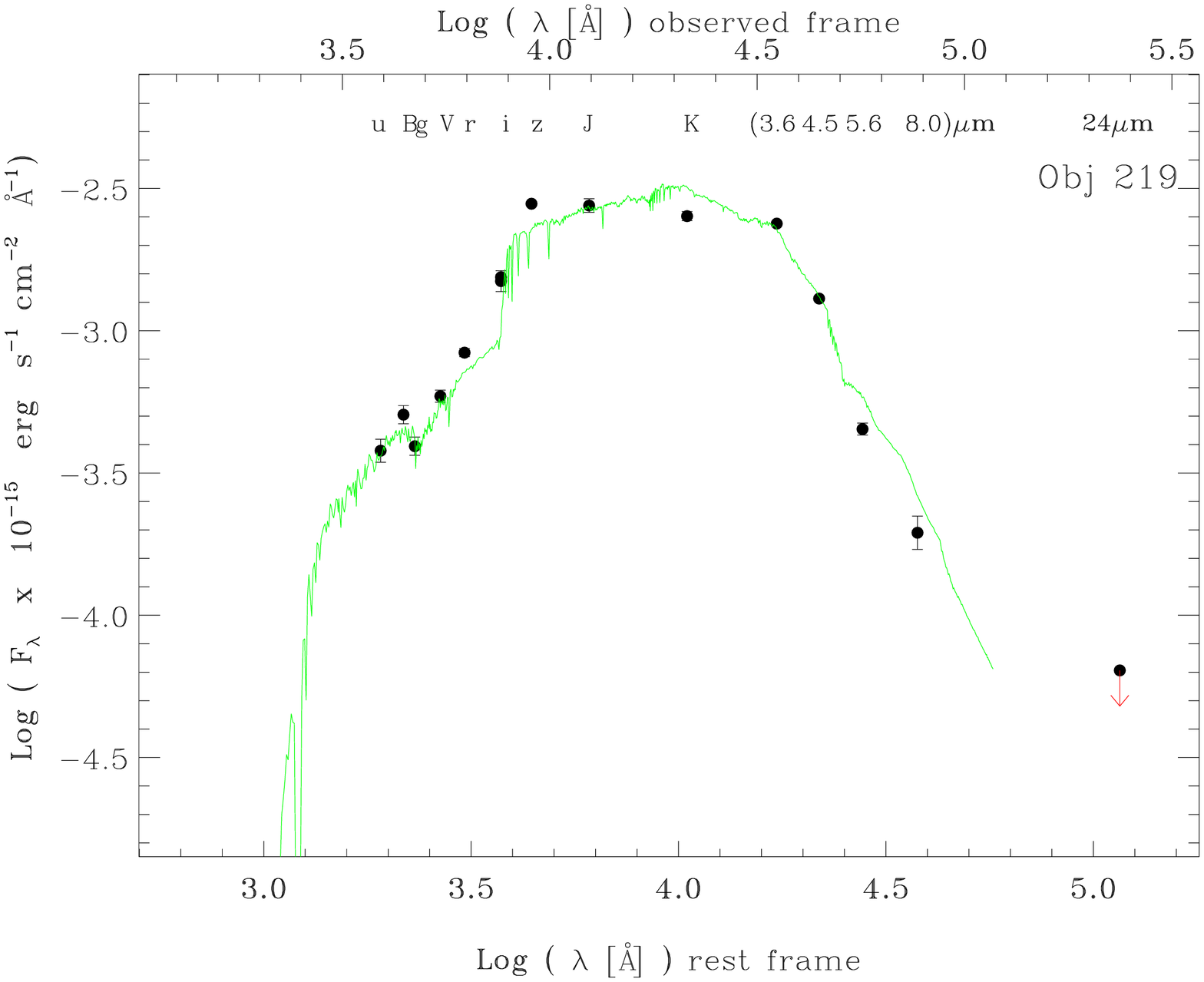} &
\includegraphics[scale=0.33,angle=90]{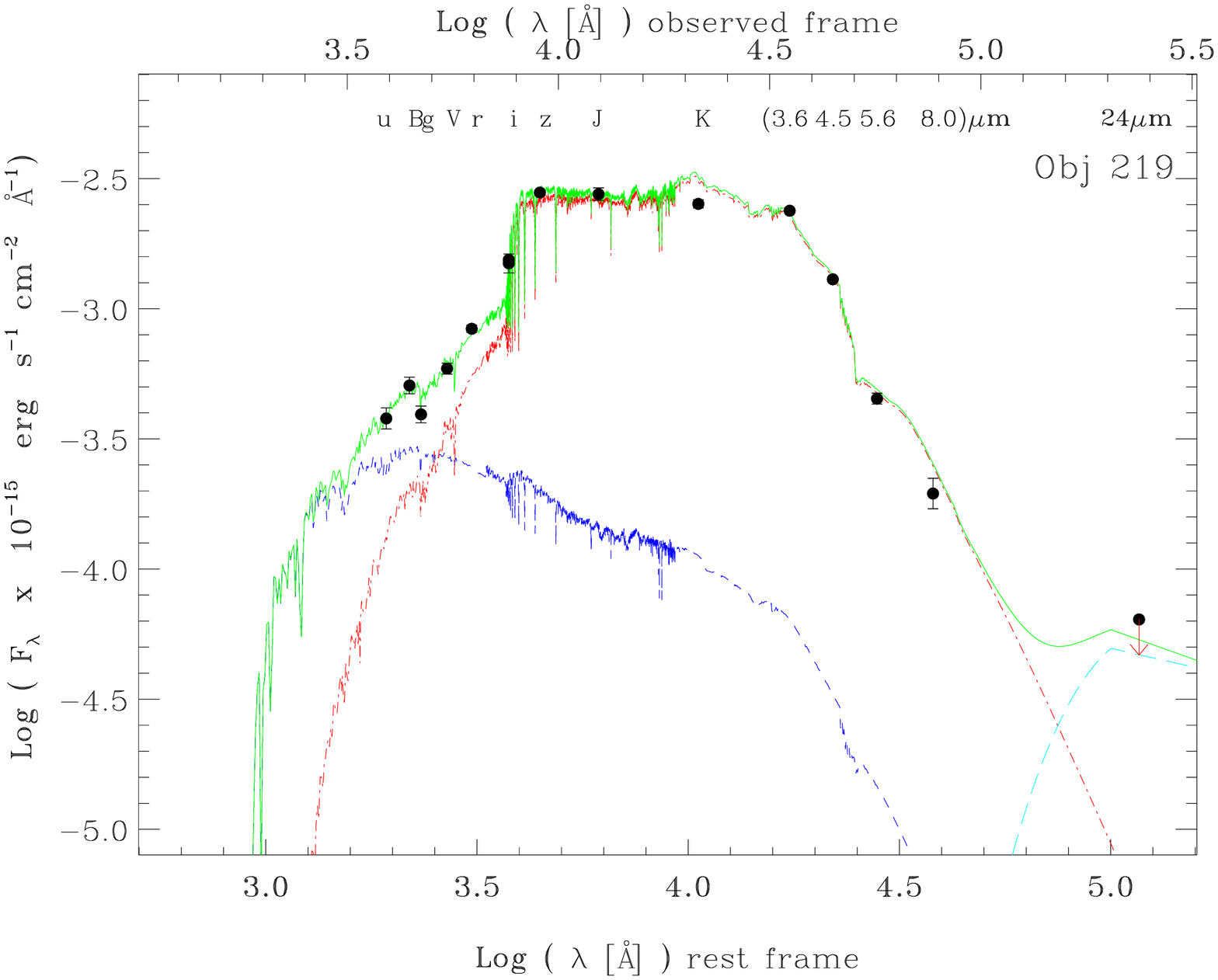} \\
\vspace{2em}
\includegraphics[scale=0.33,angle=90]{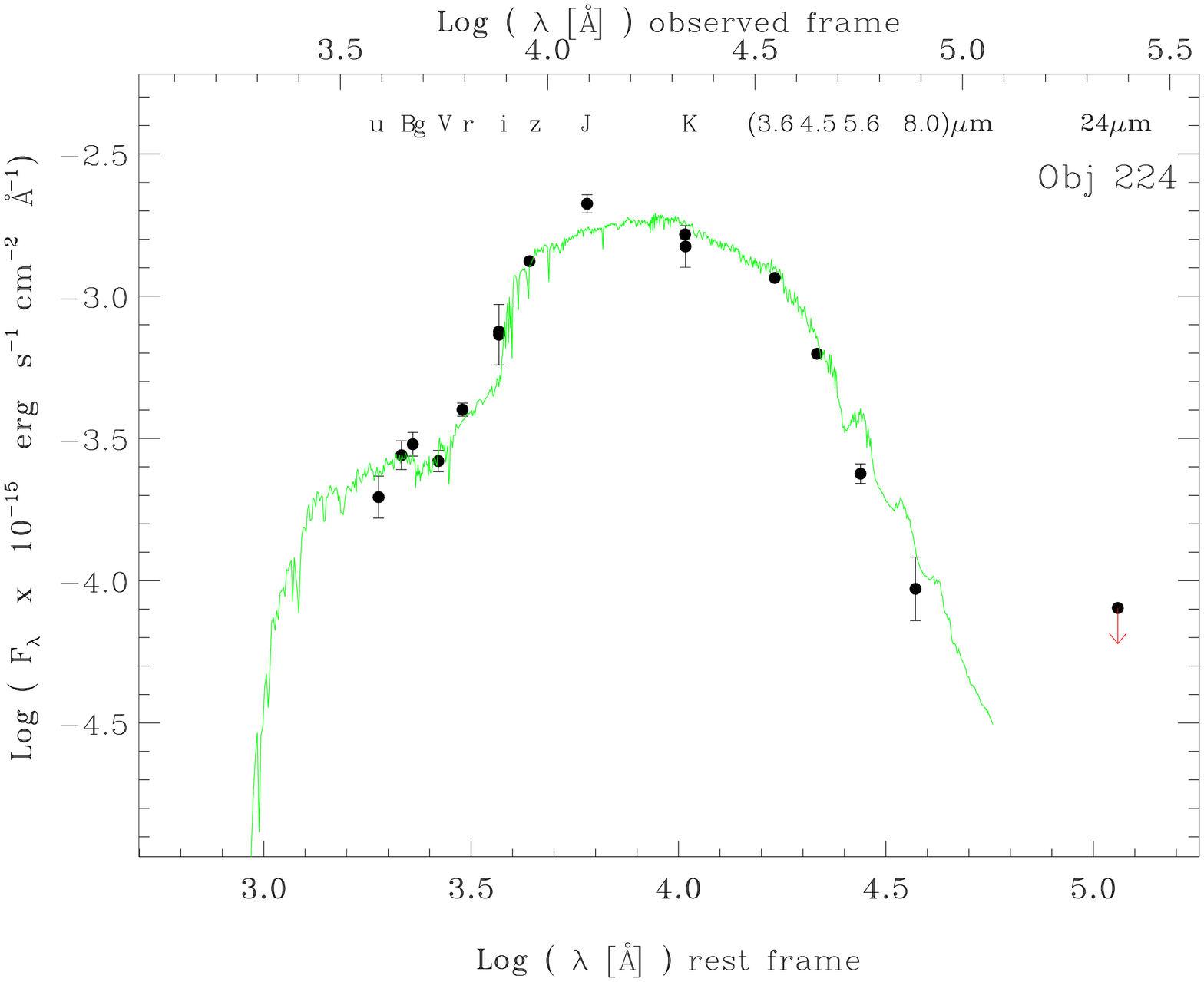} &
\includegraphics[scale=0.33,angle=90]{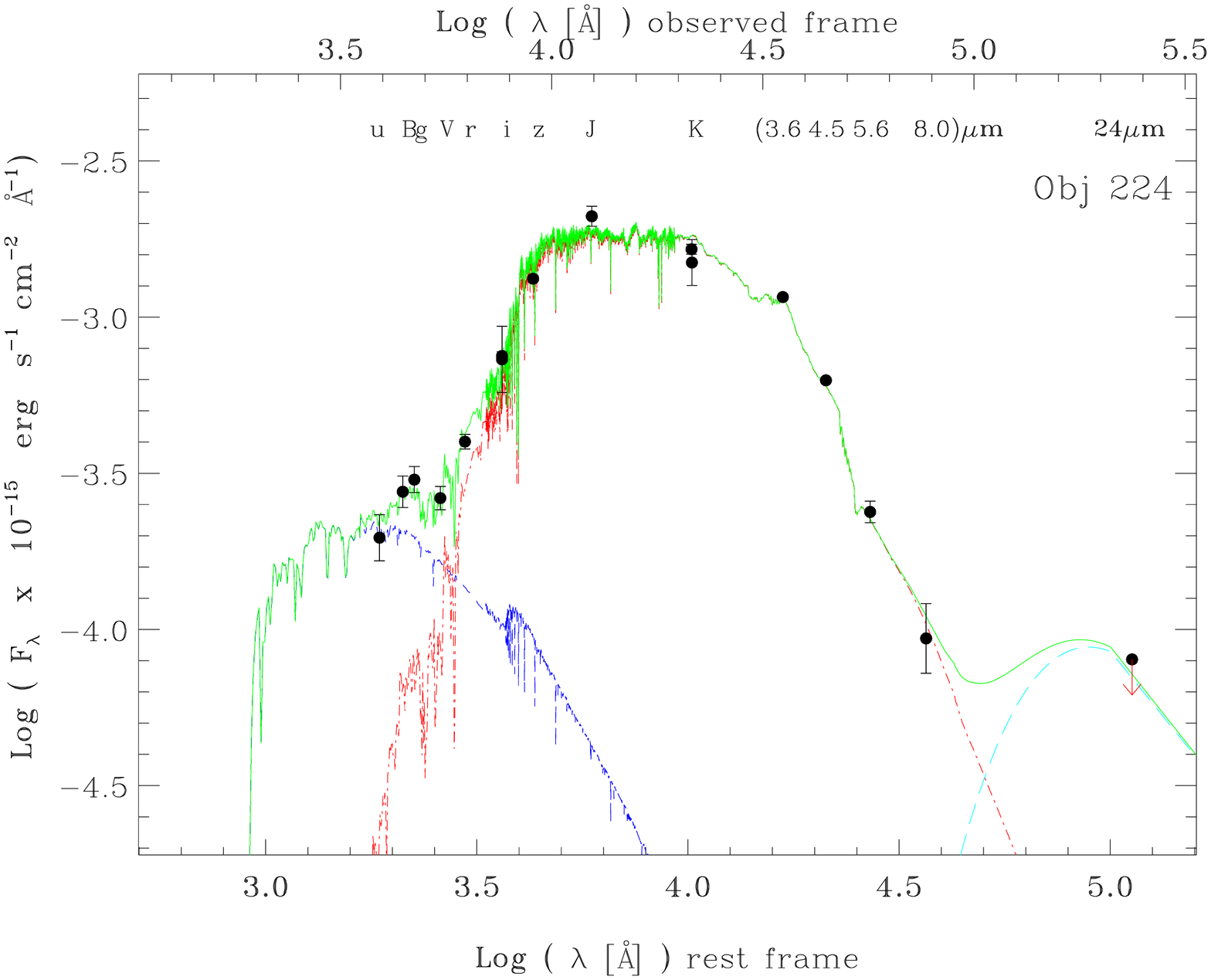} 
\end{array}$
\end{center}
\caption{Continued}
\end{figure*}

\addtocounter{figure}{-1}
\begin{figure*}[h]
\begin{center}$
\begin{array}{cccc}
\vspace{2em}
\includegraphics[scale=0.33,angle=90]{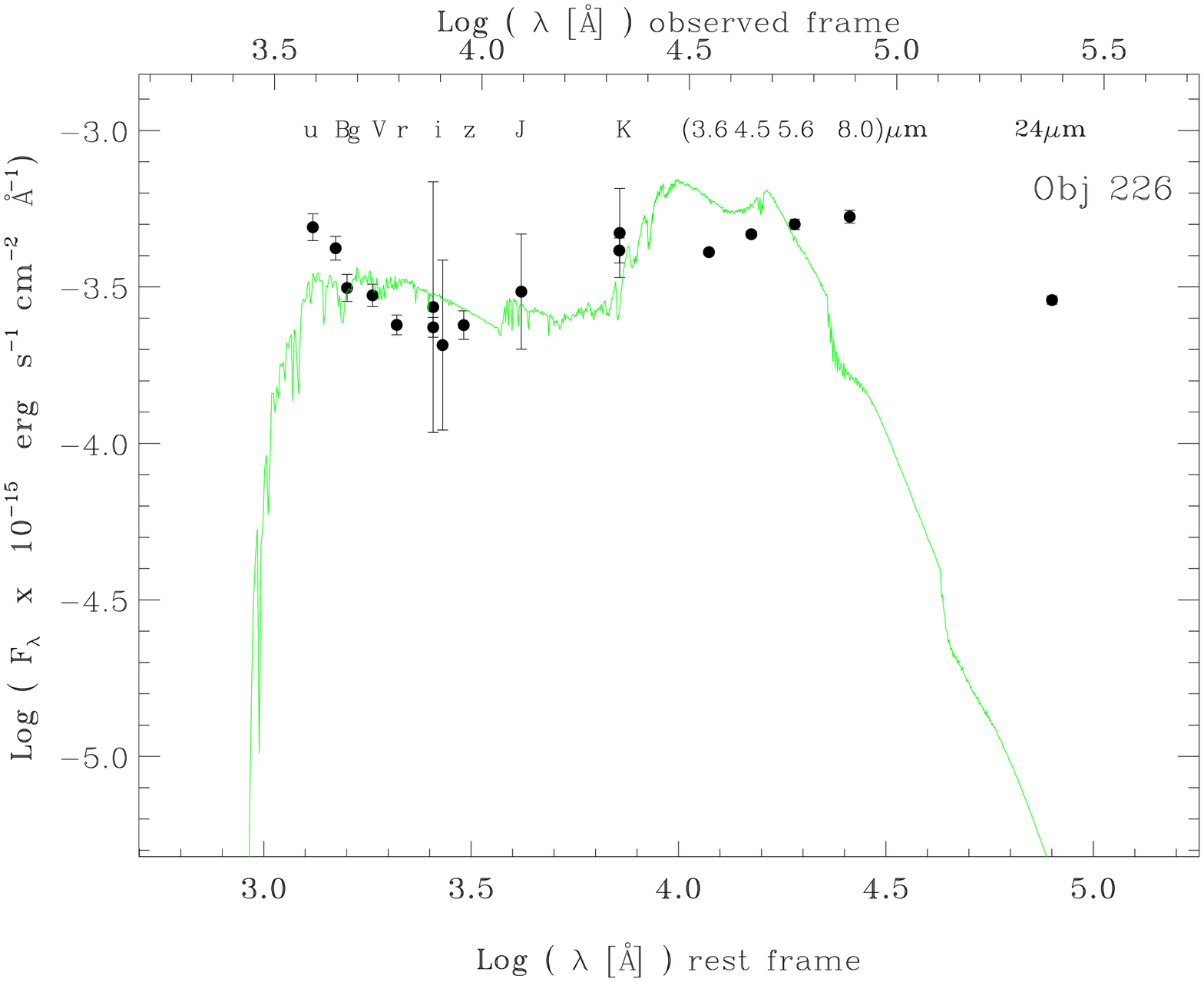} &
\includegraphics[scale=0.33,angle=90]{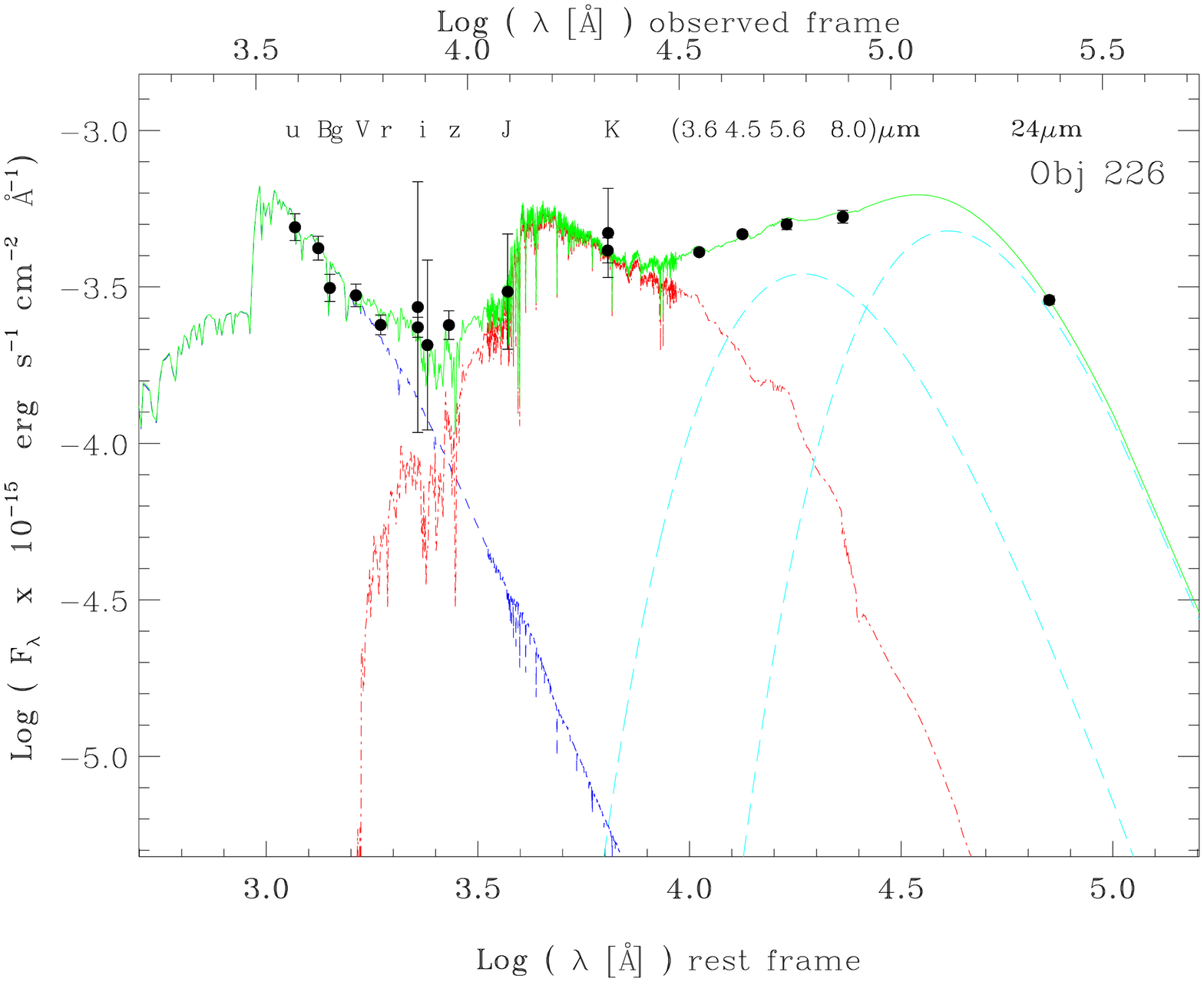} \\
\vspace{1em}
\includegraphics[scale=0.33,angle=90]{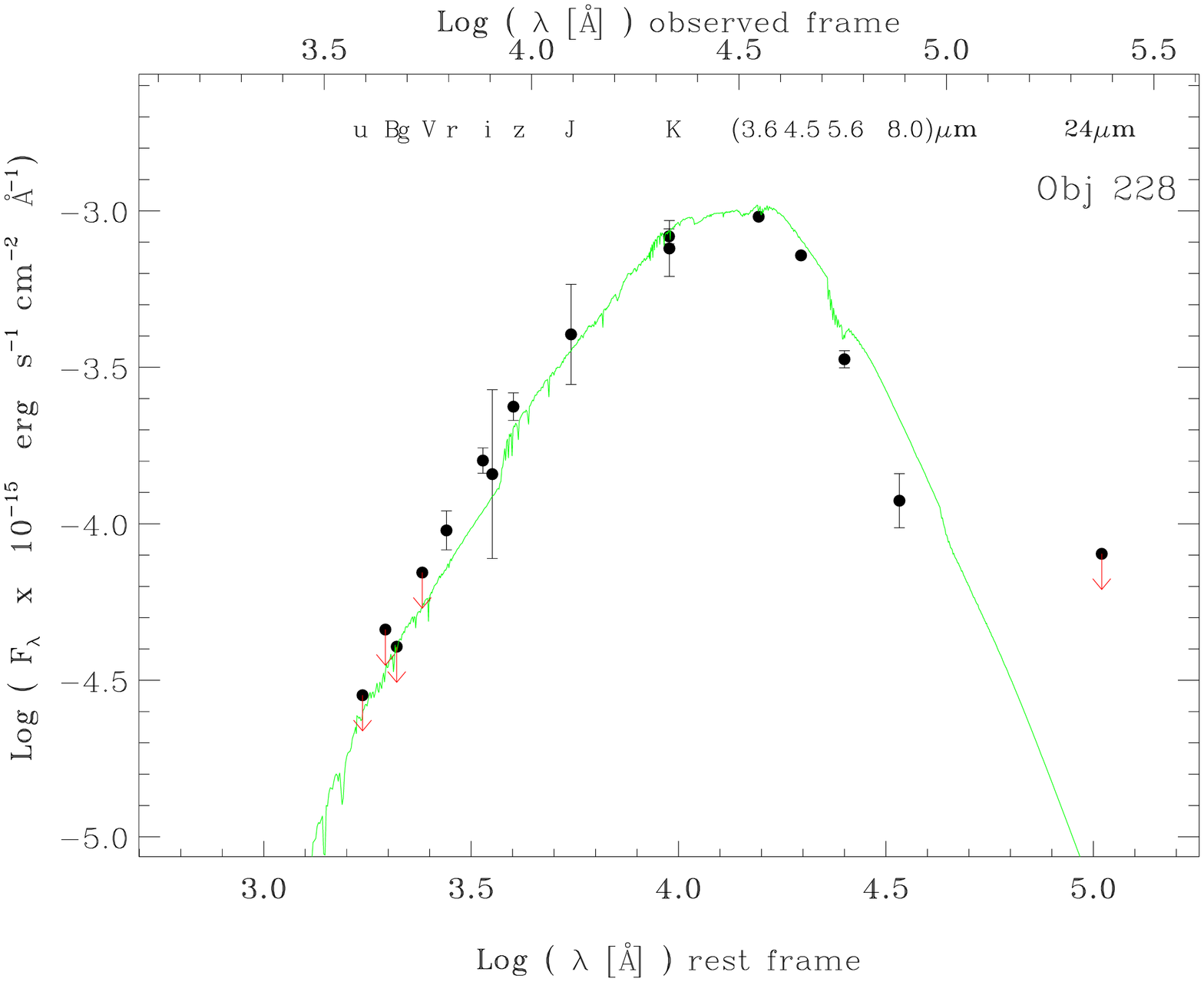} &
\includegraphics[scale=0.33,angle=90]{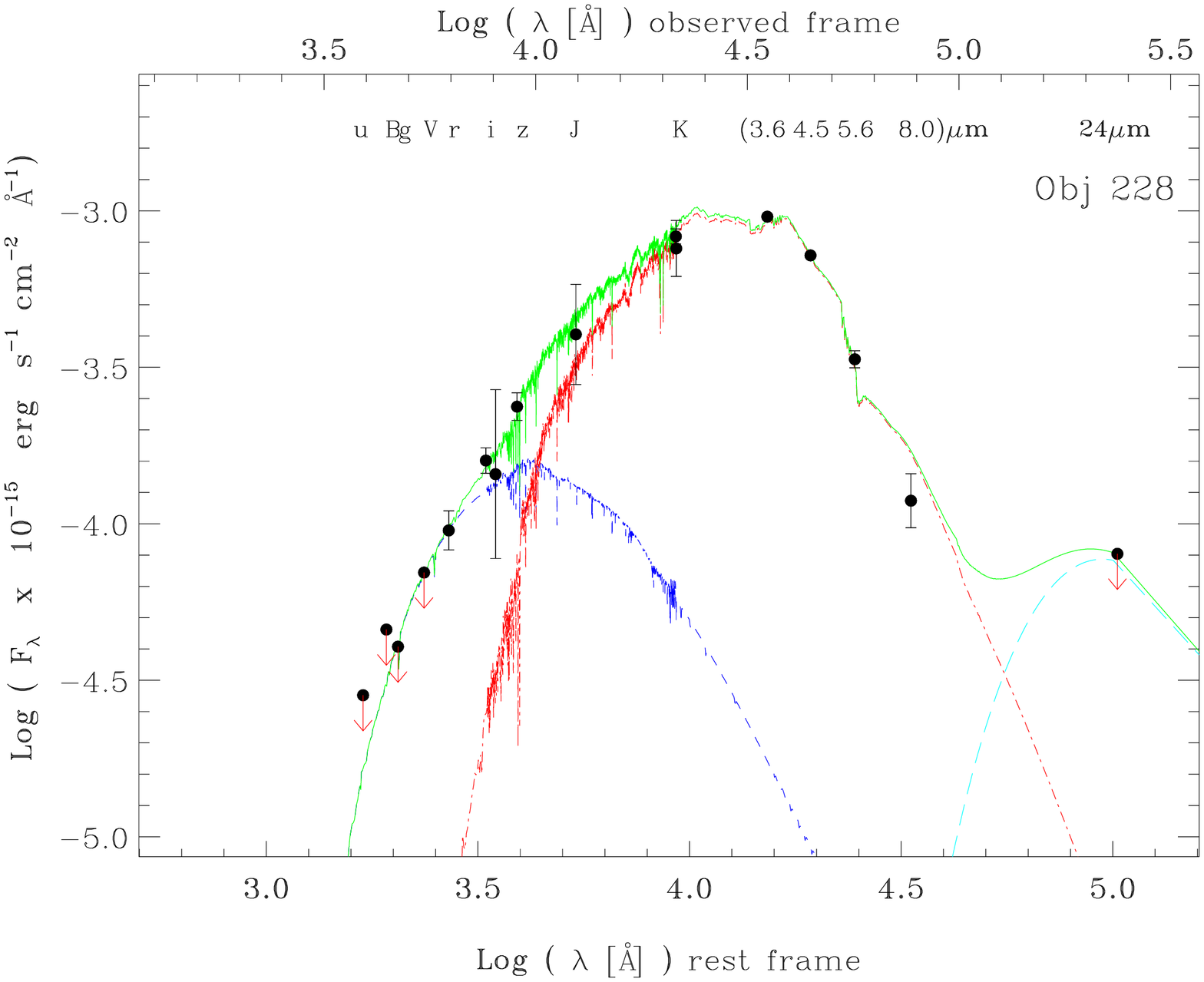} \\
\vspace{2em}
\includegraphics[scale=0.33,angle=90]{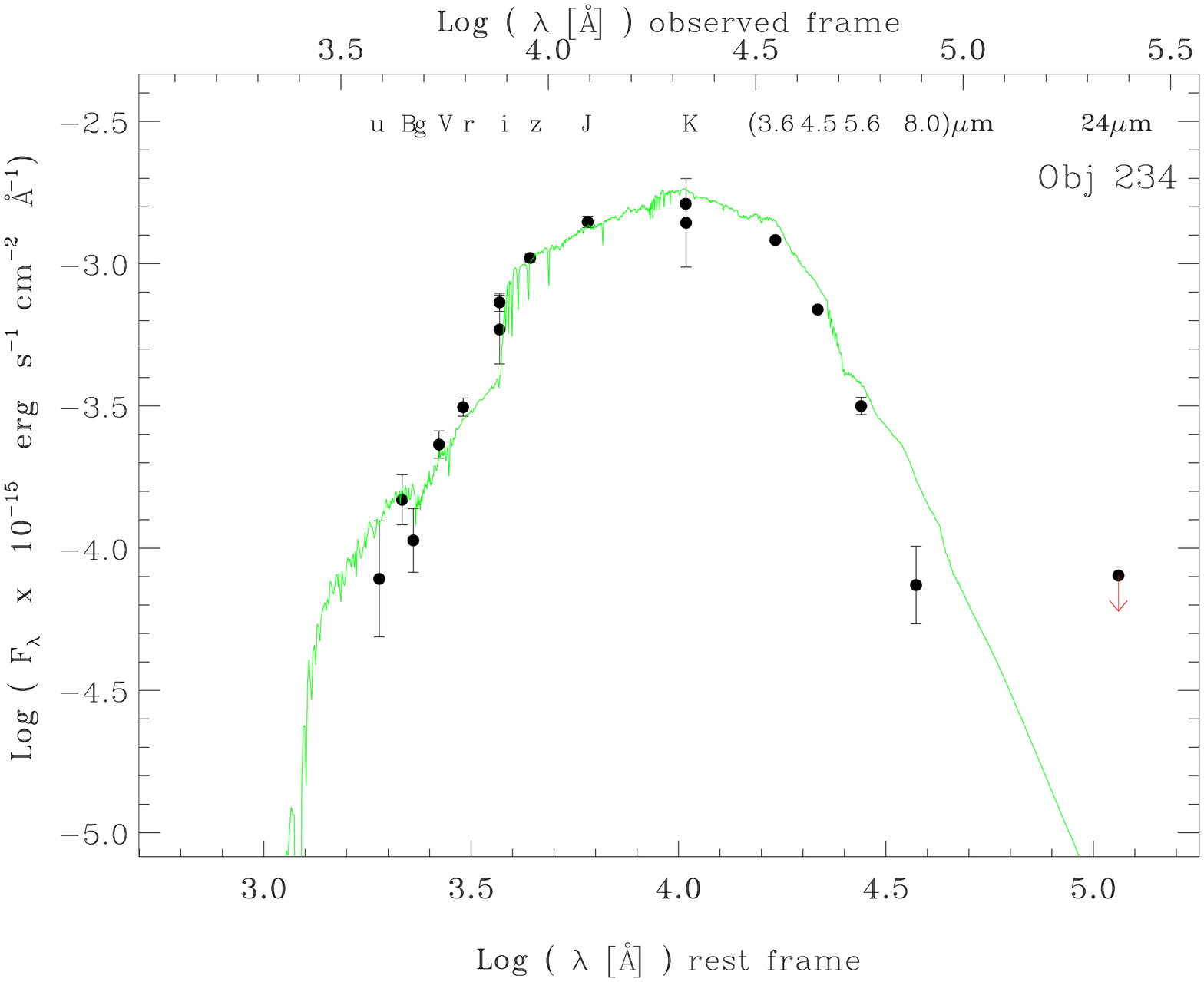} &
\includegraphics[scale=0.33,angle=90]{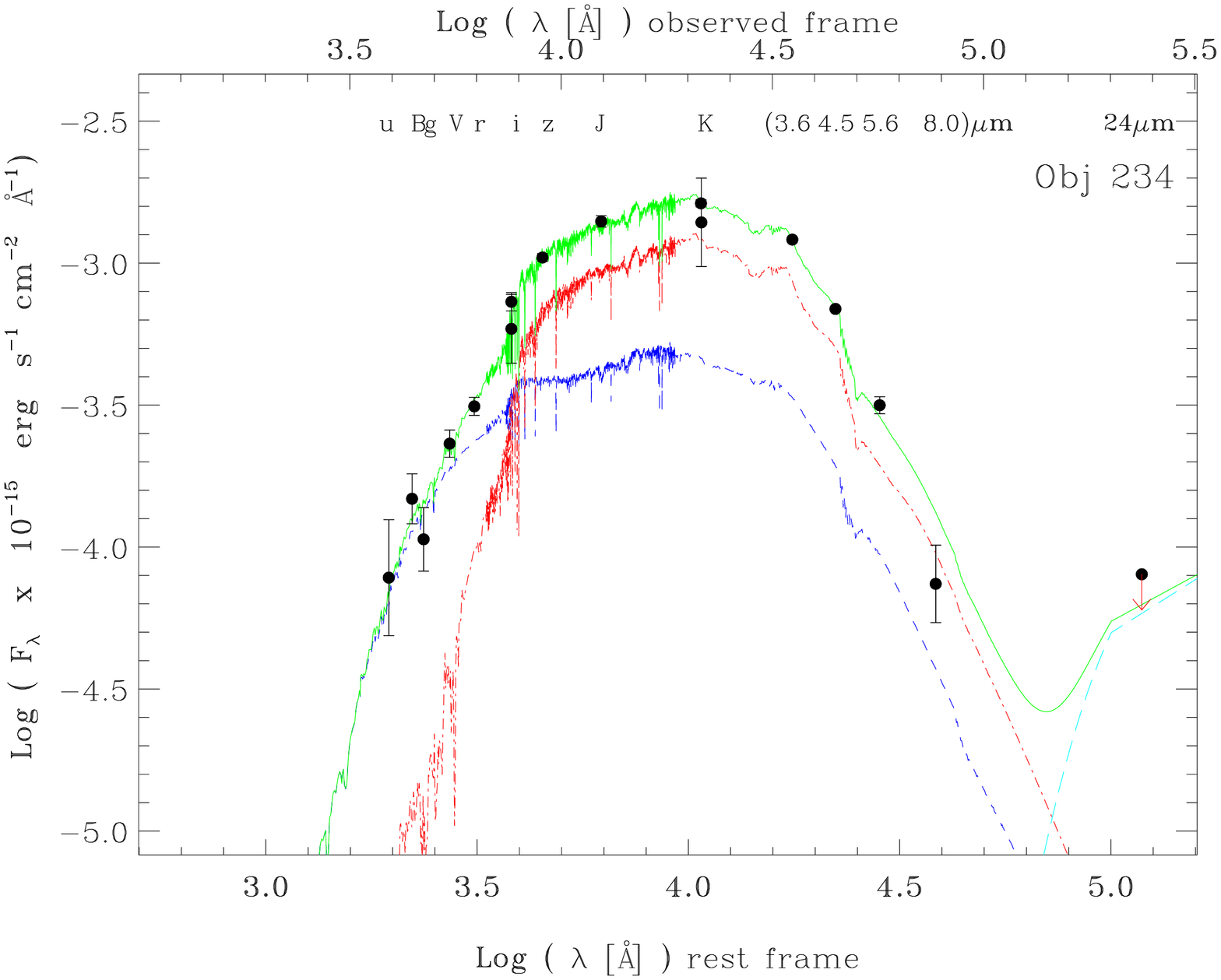} \\
\vspace{1em}
\includegraphics[scale=0.33,angle=90]{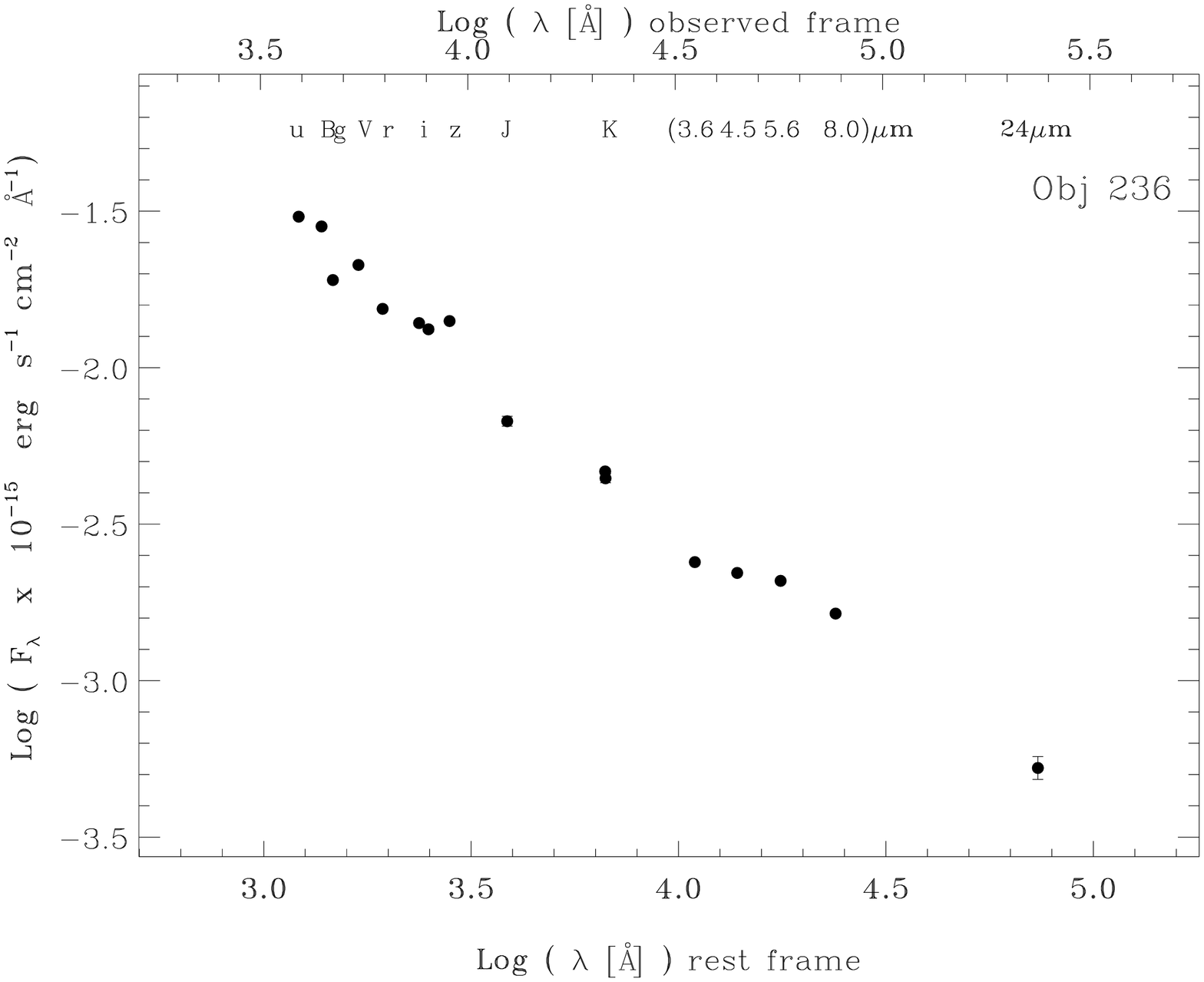} &  
\end{array}$
\end{center}
\caption{Continued}
\end{figure*}

\addtocounter{figure}{-1}
\begin{figure*}[h]
\begin{center}$
\begin{array}{cc}
\vspace{2em}
\includegraphics[scale=0.33,angle=90]{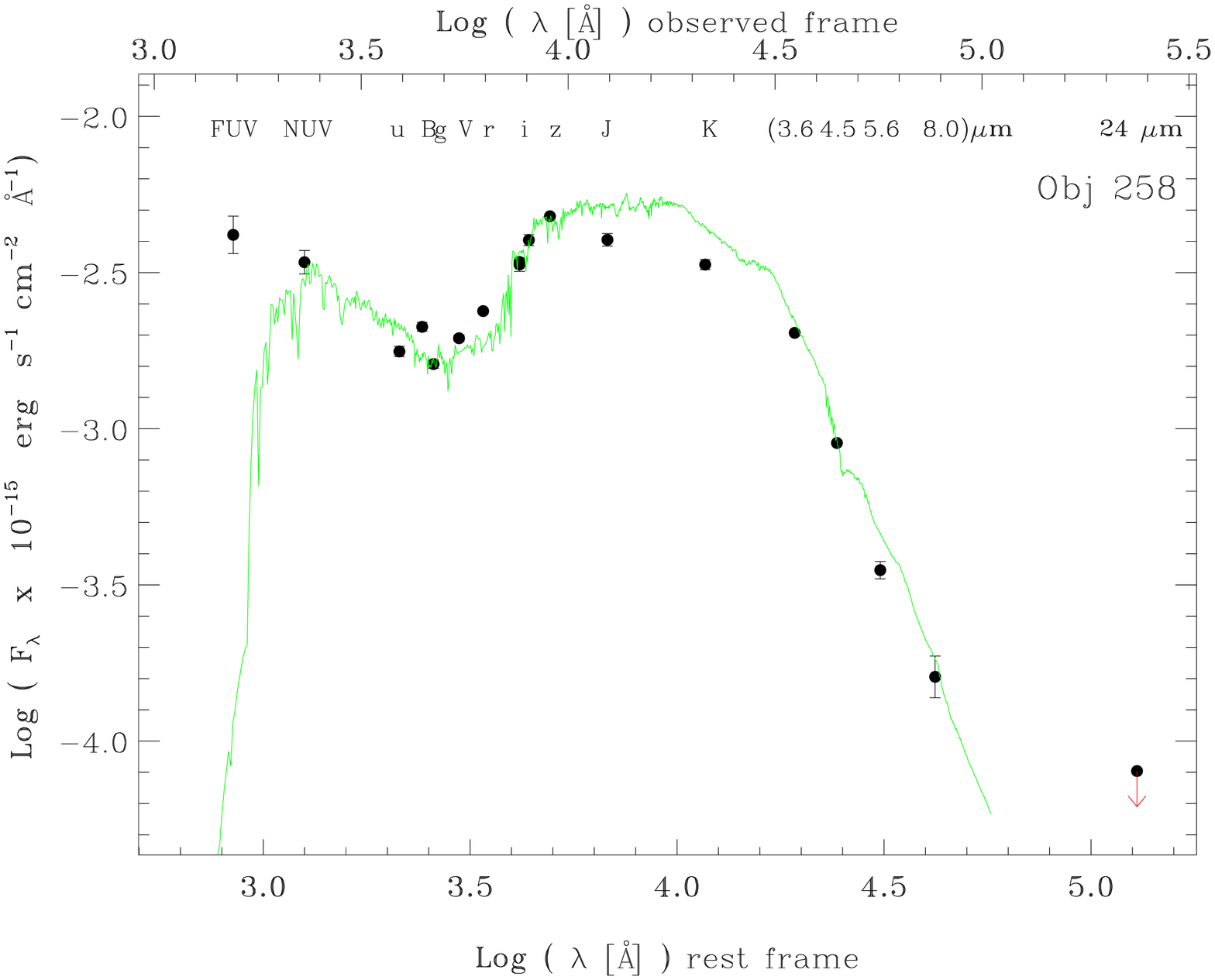} &
\includegraphics[scale=0.33,angle=90]{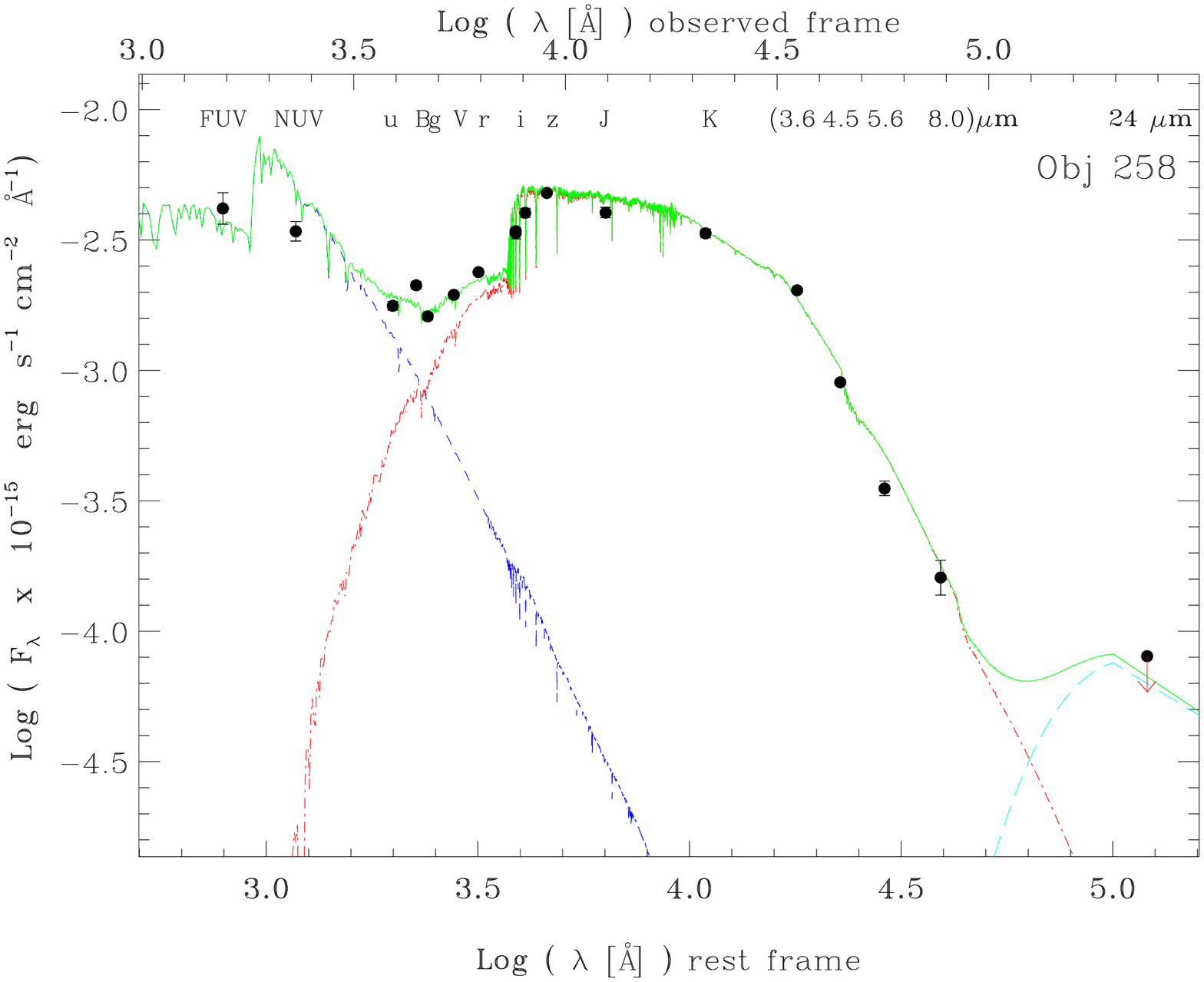} \\
\vspace{2em}
\includegraphics[scale=0.33,angle=90]{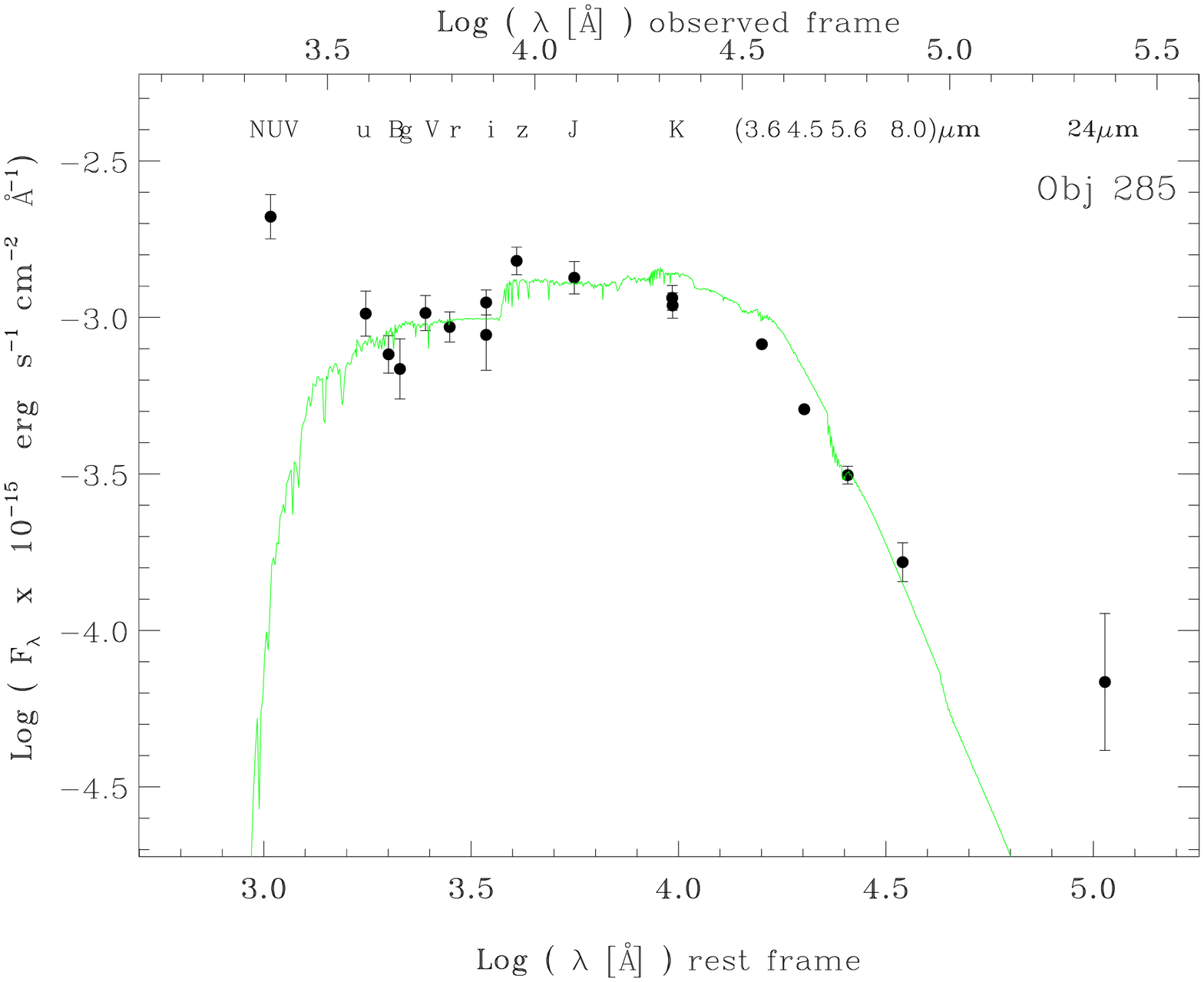} &
\includegraphics[scale=0.33,angle=90]{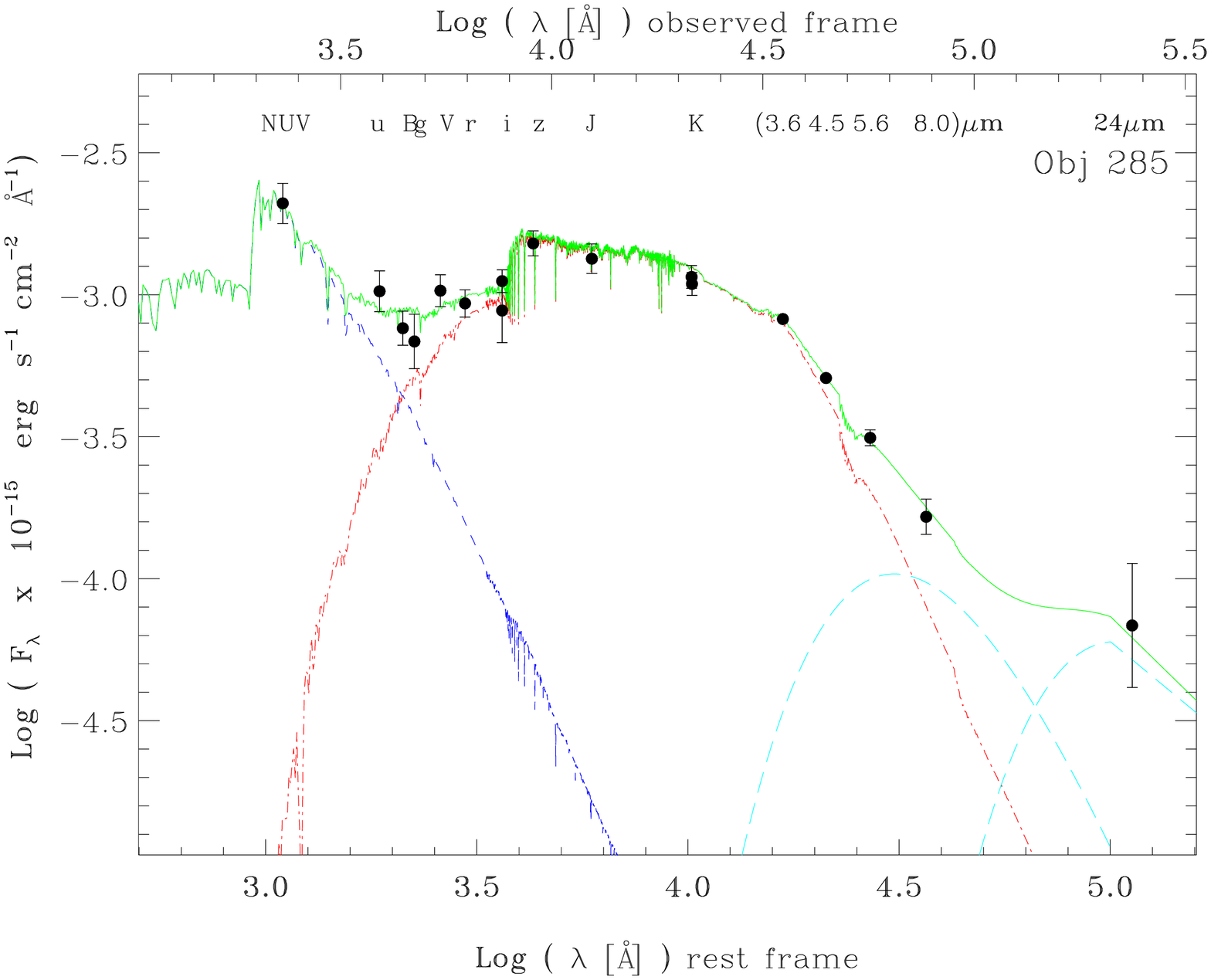} 
\end{array}$
\end{center}
\caption{Left panels: SEDs of the sample (green line) fitting the photometric
  points, as result of {\it Hyperz}. We plot the photometric point at 24
  $\mu$m (Spitzer/MIPS), if available, but is not included in the
  modeling. Right panels: SEDs of the objects fitted with {\it 2SPD}. The total
  model is the green line, the YSP is the blue line, the OSP is the red line,
  and the dust component(s) is the light blue line. For both the panels, the
  wavelengths on top of the plots correspond to observed wavelengths, while
  those on bottom are at rest frame. For object 236 (QSO) we only show the
  photometric points}
\end{figure*}

\clearpage

\LongTables
\begin{landscape}
\begin{deluxetable*}{r|cccccccccccc}
\tabletypesize{\tiny}
\tablewidth{0pt}
\tablecaption{\label{tab1}COSMOS multiwavelength counterparts of the sample}
\tablehead{
\colhead{{\tiny ID }}&
\colhead{{\tiny $u^{*}$}}&
\colhead{{\tiny $B_{J}$ }}&
\colhead{{\tiny $g^{+}$}}&
\colhead{{\tiny $V_{J}$}}&
\colhead{{\tiny $r^{+}$}}&
\colhead{{\tiny $i^{*}$}}&
\colhead{{\tiny $i^{+}$}}&
\colhead{{\tiny $F814W$}}&
\colhead{{\tiny $z^{+}$}}&
\colhead{{\tiny $J$}}&
\colhead{{\tiny $K_{S}$}}&
\colhead{{\tiny $K$}}
}
\startdata
{\tiny (1)}  & {\tiny (2)}  & {\tiny (3)}  & {\tiny (4)}  & {\tiny (5)}  & {\tiny (6)}  & {\tiny (7)}  & {\tiny (8)}  & {\tiny (9)}  & {\tiny (10)}  & {\tiny (11)}  & {\tiny (12)}  & {\tiny (13)}  \\
{\tiny 1}  &  {\tiny 26.56$\pm$0.15} & {\tiny 25.89$\pm$0.13} & {\tiny 25.61$\pm$0.10} & {\tiny 24.42$\pm$0.05} & {\tiny 23.55$\pm$0.03} & {\tiny 22.24$\pm$0.05} & {\tiny 22.29$\pm$0.11$^{*}$} & {\tiny 21.87$\pm$0.06} & {\tiny 21.32$\pm$0.01} & {\tiny 20.46$\pm$0.10$^{*}$} & {\tiny 19.59$\pm$0.09$^{*}$} & {\tiny 19.57$\pm$0.03} \\ 
{\tiny 2}  &  {\tiny 26.79$\pm$0.20$^{*}$} & {\tiny 26.43$\pm$0.11$^{*}$} & {\tiny 26.29$\pm$0.11$^{*}$} & {\tiny 25.93$\pm$0.10$^{*}$} & {\tiny 25.26$\pm$0.08$^{*}$} & {\tiny 24.35$\pm$0.45$^{*}$} & {\tiny 24.43$\pm$0.10$^{*}$} & {\tiny 24.49$\pm$0.34$^{*}$} & {\tiny 23.44$\pm$0.07$^{*}$} & {\tiny 22.18$\pm$0.15$^{*}$} & {\tiny 20.85$\pm$0.02} & {\tiny 20.90$\pm$0.15} \\
{\tiny 3}  &  {\tiny 26.69$\pm$0.17} & {\tiny 25.58$\pm$0.10} & {\tiny 25.81$\pm$0.12} & {\tiny 25.60$\pm$0.11} & {\tiny 25.53$\pm$0.10} & {\tiny 25.31$\pm$0.78} & {\tiny 25.10$\pm$0.08} & {\tiny 24.96$\pm$0.79} & {\tiny 24.48$\pm$0.12} & {\tiny 23.39$\pm$0.25$^{*}$} & {\tiny 21.57$\pm$0.03} & {\tiny 22.84$\pm$0.96}  \\
{\tiny 4}  &  {\tiny 26.64$\pm$0.17} & {\tiny 25.93$\pm$0.13} & {\tiny 25.88$\pm$0.13} & {\tiny 25.46$\pm$0.10} & {\tiny 25.12$\pm$0.08} & {\tiny 24.27$\pm$0.36} & {\tiny 24.20$\pm$0.05} & {\tiny 23.82$\pm$0.29} & {\tiny 23.27$\pm$0.04} & {\tiny 21.93$\pm$0.10$^{*}$} & {\tiny 20.53$\pm$0.01} & {\tiny 20.74$\pm$0.16}  \\
{\tiny 5}  &  {\tiny $<$25.77$^{*}$} & {\tiny $<$25.68$^{*}$} & {\tiny $<$25.53$^{*}$} & {\tiny $<$25.17$^{*}$} & {\tiny $<$25.23$^{*}$} & {\tiny 24.12$\pm$0.27} & {\tiny 24.59$\pm$0.41$^{*}$} & {\tiny 23.98$\pm$0.33} & {\tiny 23.41$\pm$0.25} & {\tiny 22.24$\pm$0.13$^{*}$} & {\tiny 20.65$\pm$0.12} & {\tiny 20.82$\pm$0.15} \\
{\tiny 11} &  {\tiny $<$29.40$^{*}$} & {\tiny 28.75$\pm$1.14} & {\tiny 28.00$\pm$0.76} & {\tiny 27.11$\pm$0.35} & {\tiny 26.27$\pm$0.17} & {\tiny 25.16$\pm$0.84} & {\tiny 24.96$\pm$0.07} & {\tiny 24.64$\pm$0.60} & {\tiny 24.09$\pm$0.08} & {\tiny 22.28$\pm$0.07$^{*}$} & {\tiny 21.18$\pm$0.16} & {\tiny 21.75$\pm$0.49}  \\
{\tiny 13} &  {\tiny 26.14$\pm$0.15$^{*}$} & {\tiny 25.75$\pm$0.13$^{*}$} & {\tiny 25.74$\pm$0.13} & {\tiny 25.02$\pm$0.09$^{*}$} & {\tiny 24.38$\pm$0.06$^{*}$} & {\tiny 23.20$\pm$0.17} & {\tiny 23.29$\pm$0.04$^{*}$} & {\tiny 22.90$\pm$0.13} & {\tiny 22.21$\pm$0.02} & {\tiny 21.44$\pm$0.09$^{*}$} & {\tiny 20.46$\pm$0.10} & {\tiny 20.54$\pm$0.16}\\
{\tiny 16} &  {\tiny 26.07$\pm$0.12} & {\tiny 25.71$\pm$0.12} & {\tiny 25.32$\pm$0.11} & {\tiny 24.84$\pm$0.08} & {\tiny 24.23$\pm$0.05} & {\tiny 23.10$\pm$0.14} & {\tiny 23.15$\pm$0.03} & {\tiny 22.85$\pm$0.10} & {\tiny 22.16$\pm$0.02} & {\tiny 21.47$\pm$0.08$^{*}$} & {\tiny 20.40$\pm$0.08} & {\tiny 20.50$\pm$0.11} \\
{\tiny 18} &  {\tiny 26.16$\pm$0.13$^{*}$} & {\tiny 25.76$\pm$0.10$^{*}$} & {\tiny 25.51$\pm$0.10} & {\tiny 24.61$\pm$0.08$^{*}$} & {\tiny 23.99$\pm$0.06$^{*}$} & {\tiny 22.91$\pm$0.11$^{*}$} & {\tiny 22.80$\pm$0.04$^{*}$} & {\tiny 22.51$\pm$0.10$^{*}$} & {\tiny 22.04$\pm$0.03$^{*}$} & {\tiny 21.54$\pm$0.07$^{*}$} & {\tiny 20.71$\pm$0.04} & {\tiny 20.92$\pm$0.18} \\
{\tiny 20} &  {\tiny 26.11$\pm$0.15} & {\tiny 25.74$\pm$0.15} & {\tiny 25.51$\pm$0.10} & {\tiny 24.57$\pm$0.06} & {\tiny 23.67$\pm$0.04} & {\tiny 22.35$\pm$0.09} & {\tiny 22.40$\pm$0.02} & {\tiny 22.19$\pm$0.08} & {\tiny 21.53$\pm$0.02} & {\tiny 20.79$\pm$0.06$^{*}$} & {\tiny 19.91$\pm$0.07} & {\tiny 20.10$\pm$0.07} \\
{\tiny 22} &  {\tiny $<$27.38$^{*}$} & {\tiny $<$27.44$^{*}$} & {\tiny $<$26.97$^{*}$} & {\tiny $<$26.85$^{*}$} & {\tiny $<$26.00$^{*}$} & {} & {\tiny $<$25.13$^{*}$} & {} & {\tiny 23.93$\pm$0.06} & {\tiny 22.14$\pm$0.17$^{*}$} & {\tiny 20.79$\pm$0.10} & {\tiny 21.24$\pm$0.15} \\
{\tiny 25} &  {\tiny 26.66$\pm$0.22} & {\tiny 25.36$\pm$0.09} & {\tiny 25.44$\pm$0.10} & {\tiny 24.97$\pm$0.07} & {\tiny 24.55$\pm$0.05} & {\tiny 23.64$\pm$0.32} & {\tiny 23.52$\pm$0.03} & {\tiny 23.32$\pm$0.19} & {\tiny 22.57$\pm$0.03} & {\tiny 21.43$\pm$0.09$^{*}$} & {\tiny 20.57$\pm$0.05} & {\tiny 20.59$\pm$0.17} \\
{\tiny 26} &  {\tiny 26.58$\pm$0.20} & {\tiny 25.44$\pm$0.10} & {\tiny 25.56$\pm$0.10} & {\tiny 24.92$\pm$0.07} & {\tiny 23.90$\pm$0.04} & {\tiny 22.68$\pm$0.10} & {\tiny 22.60$\pm$0.02} & {\tiny 22.32$\pm$0.09} & {\tiny 21.57$\pm$0.01} & {\tiny 20.42$\pm$0.04$^{*}$} & {\tiny 19.42$\pm$0.03} & {\tiny 19.48$\pm$0.04} \\
{\tiny 28} &  {\tiny $<$27.39$^{*}$}  &  {\tiny $<$27.50$^{*}$} & {\tiny $<$27.20$^{*}$}  & {\tiny $<$26.90$^{*}$} & {\tiny $<$26.80$^{*}$} &                        & {\tiny 25.55$\pm$0.30$^{*}$} &                        & {\tiny 24.27$\pm$0.20$^{*}$}  & {\tiny 22.48$\pm$0.20$^{*}$}  &  {\tiny 20.99$\pm$0.08$^{*}$}   &   \\
{\tiny 29} &  {\tiny 25.81$\pm$0.12} & {\tiny 25.51$\pm$0.10} & {\tiny 25.49$\pm$0.09} & {\tiny 25.31$\pm$0.09} & {\tiny 25.18$\pm$0.08} & {\tiny 25.00$\pm$0.87} & {\tiny 24.78$\pm$0.07} & {\tiny 24.92$\pm$0.74} & {\tiny 24.30$\pm$0.11} & {\tiny 22.86$\pm$0.27$^{*}$} & {\tiny 22.19$\pm$0.10} & {\tiny 22.89$\pm$0.80} \\
{\tiny 30} &  {\tiny 27.85$\pm$0.63} & {\tiny 26.77$\pm$0.22} & {\tiny 26.82$\pm$0.23} & {\tiny 25.90$\pm$0.13} & {\tiny 24.69$\pm$0.06} & {\tiny 23.41$\pm$0.19} & {\tiny 23.41$\pm$0.03} & {\tiny 23.01$\pm$0.15} & {\tiny 22.32$\pm$0.02} & {\tiny 21.25$\pm$0.08$^{*}$} & {\tiny 20.02$\pm$0.04} & {\tiny 20.19$\pm$0.09} \\
{\tiny 31} &  {\tiny 24.75$\pm$0.08} & {\tiny 24.50$\pm$0.06} & {\tiny 24.51$\pm$0.05} & {\tiny 23.94$\pm$0.04} & {\tiny 23.42$\pm$0.03} & {\tiny 22.33$\pm$0.08} & {\tiny 22.33$\pm$0.02} & {\tiny 22.07$\pm$0.07} & {\tiny 21.61$\pm$0.02} & {\tiny 20.97$\pm$0.08$^{*}$} & {\tiny 20.12$\pm$0.05} & {\tiny 20.32$\pm$0.09} \\
{\tiny 32} &  {\tiny $<$27.54$^{*}$} & {\tiny 27.85$\pm$0.48$^{*}$} & {\tiny $<$27.19$^{*}$} & {\tiny $<$27.00$^{*}$} & {\tiny 26.44$\pm$0.30$^{*}$} &                        & {\tiny 26.07$\pm$0.25$^{*}$} &                        & {\tiny 25.15$\pm$0.28$^{*}$} & {\tiny 24.13$\pm$0.50$^{*}$} & {\tiny 22.43$\pm$0.17$^{*}$} & {\tiny 23.24$\pm$1.44} \\
{\tiny 34} &  {\tiny 26.09$\pm$0.40$^{*}$} & {\tiny 26.19$\pm$0.20$^{*}$} & {\tiny 26.00$\pm$0.19$^{*}$} & {\tiny 25.58$\pm$0.12$^{*}$} & {\tiny 25.44$\pm$0.10$^{*}$} & {\tiny 25.09$\pm$0.66$^{*}$} & {\tiny 24.88$\pm$0.07$^{*}$} & {\tiny 24.95$\pm$0.25$^{*}$} & {\tiny 24.12$\pm$0.11$^{*}$} & {\tiny 23.10$\pm$0.51$^{*}$} & {\tiny 21.10$\pm$0.29$^{*}$} & {\tiny 20.96$\pm$0.14} \\
{\tiny 36} &  {\tiny $<$27.8$^{*}$} & {\tiny $<$26.80$^{*}$} & {\tiny $<$26.40$^{*}$} & {\tiny 26.18$\pm$0.45$^{*}$} & {\tiny 25.50$\pm$0.31$^{*}$} & {\tiny 24.31$\pm$0.33$^{*}$} & {\tiny 24.15$\pm$0.15$^{*}$} & {\tiny 24.00$\pm$0.25$^{*}$} & {\tiny 23.20$\pm$0.24$^{*}$} & {\tiny 22.04$\pm$0.15$^{*}$} & {\tiny 20.62$\pm$0.07} & {\tiny 20.44$\pm$0.10} \\
{\tiny 37} &  {\tiny 22.70$\pm$0.02} & {\tiny 22.35$\pm$0.02} & {\tiny 22.63$\pm$0.02} & {\tiny 22.38$\pm$0.02} & {\tiny 22.18$\pm$0.02} & {\tiny 21.93$\pm$0.05} & {\tiny 21.93$\pm$0.01} &                        & {\tiny 21.17$\pm$0.01} & {\tiny 20.31$\pm$0.05$^{*}$} & {\tiny 20.14$\pm$0.11$^{*}$} & {\tiny 20.06$\pm$0.08} \\
{\tiny 38} &  {\tiny 25.50$\pm$0.10} & {\tiny 25.00$\pm$0.08} & {\tiny 24.96$\pm$0.07} & {\tiny 24.66$\pm$0.06} & {\tiny 24.20$\pm$0.04} & {\tiny 23.58$\pm$0.32} & {\tiny 23.49$\pm$0.03} & {\tiny 23.47$\pm$0.22} & {\tiny 22.69$\pm$0.03} & {\tiny 22.08$\pm$0.16$^{*}$} & {\tiny 20.89$\pm$0.05} & {\tiny 21.34$\pm$0.22} \\
{\tiny 39} &  {\tiny $<$27.07$^{*}$} & {\tiny 26.05$\pm$0.15$^{*}$} & {\tiny $<$26.00$^{*}$} & {\tiny 25.28$\pm$0.09} & {\tiny 24.42$\pm$0.05} & {\tiny 23.13$\pm$0.18} & {\tiny 23.13$\pm$0.03} & {\tiny 22.86$\pm$0.15} & {\tiny 22.18$\pm$0.02} & {\tiny 21.06$\pm$0.08$^{*}$} & {\tiny 19.99$\pm$0.04} & {\tiny 20.27$\pm$0.09} \\
{\tiny 52} &  {\tiny 24.24$\pm$0.04} & {\tiny 23.64$\pm$0.04} & {\tiny 23.76$\pm$0.04} & {\tiny 23.13$\pm$0.03} & {\tiny 22.62$\pm$0.02} & {\tiny 21.62$\pm$0.05} & {\tiny 21.67$\pm$0.01} & {\tiny 21.52$\pm$0.05} & {\tiny 21.18$\pm$0.01} & {\tiny 20.58$\pm$0.05$^{*}$} & {\tiny 19.79$\pm$0.04} & {\tiny 19.78$\pm$0.06} \\
{\tiny 70} &  {\tiny 26.16$\pm$0.14$^{*}$} & {\tiny 25.14$\pm$0.09$^{*}$} & {\tiny 25.27$\pm$0.14$^{*}$} & {\tiny 24.79$\pm$0.08$^{*}$} & {\tiny 24.51$\pm$0.06} & {\tiny 24.34$\pm$0.58} & {\tiny 24.35$\pm$0.06} & {\tiny 24.13$\pm$0.43} & {\tiny 24.01$\pm$0.09} & {\tiny 23.42$\pm$0.50$^{*}$} & {\tiny 21.76$\pm$0.09} & {\tiny 21.40$\pm$0.27} \\
{\tiny 202}&  {\tiny 27.03$\pm$0.28} & {\tiny 27.11$\pm$0.29} & {\tiny 27.40$\pm$0.42} & {\tiny 26.92$\pm$0.29} & {\tiny 25.86$\pm$0.12} & {\tiny 24.44$\pm$0.53} & {\tiny 24.41$\pm$0.06} & {\tiny 24.29$\pm$0.45} & {\tiny 23.48$\pm$0.05} & {\tiny 22.19$\pm$0.17$^{*}$} & {\tiny 21.07$\pm$0.05} & {\tiny 21.62$\pm$0.30} \\
{\tiny 219}&  {\tiny 25.68$\pm$0.10} & {\tiny 25.09$\pm$0.08} & {\tiny 25.23$\pm$0.08} & {\tiny 24.49$\pm$0.05} & {\tiny 23.81$\pm$0.04} & {\tiny 22.74$\pm$0.09} & {\tiny 22.71$\pm$0.02} & {\tiny 22.37$\pm$0.09} & {\tiny 21.70$\pm$0.02} & {\tiny 21.02$\pm$0.06$^{*}$} & {\tiny 19.93$\pm$0.04} & {\tiny 20.15$\pm$0.37} \\
{\tiny 224}&  {\tiny 26.40$\pm$0.18} & {\tiny 25.75$\pm$0.13} & {\tiny 25.52$\pm$0.10} & {\tiny 25.36$\pm$0.09} & {\tiny 24.62$\pm$0.06} & {\tiny 23.52$\pm$0.27} & {\tiny 23.49$\pm$0.03} &                        & {\tiny 22.50$\pm$0.03} & {\tiny 21.31$\pm$0.08$^{*}$} & {\tiny 20.39$\pm$0.04} & {\tiny 20.49$\pm$0.18} \\
{\tiny 226}&  {\tiny 25.40$\pm$0.11} & {\tiny 25.30$\pm$0.09} & {\tiny 25.48$\pm$0.11} & {\tiny 25.23$\pm$0.09} & {\tiny 25.17$\pm$0.08} & {\tiny 24.59$\pm$1.00} & {\tiny 24.75$\pm$0.08} & {\tiny 24.78$\pm$0.68} & {\tiny 24.37$\pm$0.11} & {\tiny 23.41$\pm$0.46$^{*}$} & {\tiny 21.90$\pm$0.10} & {\tiny 21.75$\pm$0.36} \\
{\tiny 228}&  {\tiny $<$28.50$^{*}$} & {\tiny $<$27.70$^{*}$} & {\tiny $<$27.70$^{*}$} & {\tiny $<$26.80$^{*}$} & {\tiny 26.17$\pm$0.16} &                        & {\tiny 25.17$\pm$0.10} & {\tiny 25.17$\pm$0.67} & {\tiny 24.38$\pm$0.11} & {\tiny 23.10$\pm$0.40$^{*}$} & {\tiny 21.14$\pm$0.06} & {\tiny 21.23$\pm$0.22} \\
{\tiny 234}&  {\tiny 27.40$\pm$0.51$^{*}$} & {\tiny 26.43$\pm$0.22$^{*}$} & {\tiny 26.65$\pm$0.28$^{*}$} & {\tiny 25.50$\pm$0.12$^{*}$} & {\tiny 24.88$\pm$0.08$^{*}$} & {\tiny 23.76$\pm$0.30} & {\tiny 23.52$\pm$0.08$^{*}$} &                        & {\tiny 22.76$\pm$0.03} & {\tiny 21.38$\pm$0.20$^{*}$} & {\tiny 20.31$\pm$0.11$^{*}$} & {\tiny 20.57$\pm$0.39} \\
{\tiny 236}&  {\tiny 20.92$\pm$0.01} & {\tiny 20.73$\pm$0.01} & {\tiny 21.02$\pm$0.01} & {\tiny 20.59$\pm$0.01} & {\tiny 20.65$\pm$0.01} & {\tiny 20.32$\pm$0.02} &                        & {\tiny 20.26$\pm$0.02} & {\tiny 19.94$\pm$0.01} & {\tiny 20.05$\pm$0.04$^{*}$} & {\tiny 19.27$\pm$0.02} & {\tiny 19.32$\pm$0.04} \\
{\tiny 258}&  {\tiny 24.01$\pm$0.04} & {\tiny 23.54$\pm$0.04} & {\tiny 23.70$\pm$0.03} & {\tiny 23.19$\pm$0.03} & {\tiny 22.68$\pm$0.02} & {\tiny 21.87$\pm$0.06} & {\tiny 21.85$\pm$0.01} & {\tiny 21.56$\pm$0.04} & {\tiny 21.11$\pm$0.01} & {\tiny 20.61$\pm$0.05$^{*}$} & {\tiny 19.62$\pm$0.04} & {\tiny 19.71$\pm$0.04} \\
{\tiny 285}&  {\tiny 24.60$\pm$0.18$^{*}$} & {\tiny 24.65$\pm$0.15$^{*}$} & {\tiny 24.63$\pm$0.24$^{*}$} & {\tiny 23.88$\pm$0.14$^{*}$} & {\tiny 23.70$\pm$0.12$^{*}$} & {\tiny 23.32$\pm$0.28} & {\tiny 23.06$\pm$0.10$^{*}$} &                        & {\tiny 22.36$\pm$0.11$^{*}$} & {\tiny 21.80$\pm$0.13$^{*}$} & {\tiny 20.78$\pm$0.10$^{*}$} & {\tiny 20.84$\pm$0.10} \\
\enddata
\tablecomments{Column description: (1) ID number of the object; (2) CFHT
  $u^{*}$ magnitude with its error; (3)-(4)-(5)-(6) Subaru $B_{J}$, $g^{+}$,
  $V_{J}$, $r^{+}$ magnitudes with their errors; (7) CFHT $i^{*}$ magnitude
  with its error; (8) Subaru $i^{+}$ magnitude with its error; (9) HST/ACS
  $F814W$ magnitude with its error; (10) Subaru $z^{+}$ magnitude with its
  error; (11) UKIRT $J$ magnitude with its error; (12) CFHT $K$ magnitude with
  its error; (13) NOAO$K_{S}$ with its error. The values marked by $*$ are
  measured by our 3\arcsec-aperture photometry on the images.}
\end{deluxetable*}
 \clearpage

\end{landscape}

\begin{deluxetable*}{r|cc|cccc|c}
\tablewidth{0pt}
\tablecaption{\label{tab2}COSMOS multiwavelength counterparts of the sample}
\tablehead{
\colhead{{\tiny ID}}&
\colhead{{\tiny $FUV$}}&
\colhead{{\tiny $NUV$ }}&
\colhead{{\tiny $IRAC1$}}&
\colhead{{\tiny $IRAC2$ }}&
\colhead{{\tiny $IRAC3$}}&
\colhead{{\tiny $IRAC4$}}&
\colhead{{\tiny $MIPS$}}
}
\startdata
1  &                 &                &  54.89$\pm$0.23$^{*}$ & 36.50$\pm$0.29$^{*}$ & 23.87$\pm$0.97$^{*}$ & 14.73$\pm$2.07$^{*}$ & $<$0.15                \\	 
2  &                 &                &  41.44$\pm$0.19 & 45.75$\pm$0.30 & 28.07$\pm$0.99 & 21.8$\pm$2.49 &  $<$0.15                  \\   	 
3  &                 &                &  13.65$\pm$0.15 & 23.62$\pm$0.25 & 55.88$\pm$0.94 & 166.89$\pm$1.98 & 1.47$\pm$0.02 \\	 
4  &                 &                &  42.97$\pm$0.19 & 48.78$\pm$0.29 & 37.71$\pm$1.04 & 29.34$\pm$2.19 &   $<$0.08             \\	 
5  &                 &                &  66.44$\pm$0.19 & 89.03$\pm$0.29 & 91.57$\pm$0.99 & 63.65$\pm$1.97 & 0.87$\pm$0.02  \\	 
11 &                 &                &  25.34$\pm$0.17 & 29.34$\pm$0.26 & 25.07$\pm$1.00 & 9.64$\pm$1.98 &  $<$0.15                      \\	 
13 &                 &                &  49.51$\pm$0.18 & 51.39$\pm$0.26 & 53.24$\pm$0.98 & 75.84$\pm$2.04 & 0.32$\pm$0.03    \\	 
16 &  	          &                & 44.29$\pm$0.18 & 32.54$\pm$0.28 & 26.19$\pm$0.94 & 18.39$\pm$2.26 & 0.22$\pm$0.02  \\ 	     
18 &                 &                &  31.03$\pm$0.17 & 24.94$\pm$0.28 & 18.13$\pm$0.99 & 14.03$\pm$2.15 &      $<$0.61$^{*}$           \\	 
20 &                 &                &  63.11$\pm$0.18 & 44.73$\pm$0.27 & 25.12$\pm$0.90 & 17.46$\pm$2.06 &   $<$0.08                \\	 
22 &                 &                &  57.27$\pm$ 0.17 & 64.90$\pm$0.29 & 41.55$\pm$0.89 & 32.84$\pm$2.16&  0.25$\pm$0.03$^{*}$                \\	 
25 &                 &                &  49.17$\pm$0.19 & 55.52$\pm$0.28 & 38.82$\pm$0.99 & 25.08$\pm$2.19 &   0.18$\pm$0.05$^{*}$               \\	 
26 &                 &                &  126.74$\pm$0.25 & 108.17$\pm$0.29 & 59.54$\pm$0.98 & 44.75$\pm$1.92 &   $<$0.15             \\	 
28 &                 &                &  32.46$\pm$0.16 & 39.98$\pm$0.26 & 46.73$\pm$0.91 & 40.34$\pm$1.98 & 0.15$\pm$0.02   \\	 
29 &                 &                &  9.81$\pm$0.16 & 10.17$\pm$0.25 & 6.50$\pm$1.00$^{*}$ & $<$7.38$^{*}$        &   $<$0.15              \\	 
30 &                 &                &  74.26$\pm$0.19 & 58.67$\pm$0.28 & 33.32$\pm$0.91 & 22.76$\pm$2.05 &    $<$0.15                 \\	 
31 &                 & 25.15$\pm$0.24 &  47.86$\pm$0.16 & 34.60$\pm$0.25 & 24.98$\pm$0.81 & 19.36$\pm$1.78 &   $<$0.15                  \\	 
32 &                 &                &  16.97$\pm$0.15 & 32.41$\pm$0.24 & 58.17$\pm$0.95 & 98.81$\pm$1.81 & 0.26$\pm$0.02   \\	 
34 &                 &                &  22.80$\pm$0.44$^{*}$ & 26.73$\pm$0.48$^{*}$ & 25.44$\pm$1.10$^{*}$ & 12.07$\pm$2.40$^{*}$ &  $<$0.26$^{*}$                \\	 
36 &                 &                &  49.50$\pm$0.16 & 43.80$\pm$0.27 & 30.24$\pm$0.87 & 21.89$\pm$1.97 &  0.15$\pm$0.04$^{*}$                \\	 
37 &                 & 23.55$\pm$0.15 &  88.32$\pm$0.22 & 123.39$\pm$0.32 & 158.41$\pm$1.10 & 249.52$\pm$2.18&1.58$\pm$0.39   \\	 
38 &                 &                &  30.47$\pm$0.17 & 28.00$\pm$0.28 & 29.99$\pm$0.95 & 24.94$\pm$2.35 & 0.13$\pm$0.06$^{*}$   \\	 
39 &                 &                &  74.19$\pm$0.21 & 66.35$\pm$0.32 & 34.74$\pm$0.96 & 25.29$\pm$2.40 &    $<$0.08             \\	 
52 &                 & 24.97$\pm$0.21 &  59.84$\pm$0.21 & 41.39$\pm$0.28 & 36.37$\pm$1.09 & 28.67$\pm$2.04 & 0.16$\pm$0.02   \\	 
70 &                 &                &  17.58$\pm$0.16 & 20.57$\pm$0.26 & 29.83$\pm$1.04 & 21.32$\pm$2.17 & 0.13$\pm$0.03$^{*}$   \\	 
202&                 &                &  24.81$\pm$0.14 & 22.02$\pm$0.24 & 15.90$\pm$0.84 & 6.46$\pm$1.87 &  $<$0.15                 \\	 
219&                 &                &  98.71$\pm$0.23 & 86.12$\pm$0.32 & 48.50$\pm$1.02 & 38.60$\pm$2.26 &  0.12$\pm$0.01                \\	 
224&                 &                &  48.09$\pm$0.17 & 41.65$\pm$0.26 & 25.56$\pm$0.88 & 18.53$\pm$2.07 &   $<$0.15              \\	 
226&                 &                &  16.94$\pm$0.15 & 30.93$\pm$0.28 & 53.89$\pm$0.89 & 104.92$\pm$2.16 & 0.54$\pm$0.02 \\	 
228&                 &                &  39.72$\pm$0.18 & 47.82$\pm$0.27 & 36.06$\pm$0.99 & 23.46$\pm$2.03 &  $<$0.15               \\	 
234&                 &                &  50.19$\pm$0.19 & 45.78$\pm$0.26 & 33.95$\pm$1.03 & 14.69$\pm$2.00 &   $<$0.15               \\	 
236&                 &                &  99.23$\pm$0.20 & 146.61$\pm$0.36 & 224.02$\pm$1.04 & 324.18$\pm$2.50&0.98$\pm$0.04   \\	 
258&   24.97$\pm$0.15& 24.32$\pm$0.09 &  84.11$\pm$0.23 & 59.75$\pm$0.30 & 37.90$\pm$1.06 & 31.78$\pm$2.12 &     $<$0.15            \\	 
285&                 & 24.85$\pm$0.18 & 34.06$\pm$0.17 & 33.78$\pm$0.26 & 33.68$\pm$0.95 & 32.70$\pm$2.03 & 0.13$\pm$0.03$^{*}$    \\   
\enddata
\tablecomments{Column description: (1) ID number of the object; (2)-(3)
GALEX FUV and NUV magnitudes with their errors; (4)-(5)-(6)-(7) Spitzer/IRAC
4-channel (3.6, 4.5, 5.8, and 8.0 $\mu$m) fluxes with their
errors; (8) Spitzer/MIPS flux at 24$\mu$m with its error. The values marked by
$*$ are measured by our 3\arcsec-aperture photometry on the images.}
\end{deluxetable*}

\end{document}